\documentclass[11pt,a4paper]{article}
\pdfoutput=1
\usepackage[dvipsnames]{xcolor}
\PassOptionsToPackage{backref=page}{hyperref}
\PassOptionsToPackage{colorlinks=true}{hyperref}
\usepackage{jheppub}
\usepackage{mathrsfs}
\usepackage{braket}
\usepackage{amsmath}
\usepackage{mathrsfs}
\usepackage{verbatim}
\usepackage{amssymb}
\usepackage{bbold}
\usepackage[normalem]{ulem}
\usepackage{inputenc, array}
\usepackage{textcomp}
\usepackage{appendix}
\usepackage{epsfig}
\usepackage{pifont}
\usepackage{multirow}
\usepackage{makecell}
\usepackage{upgreek}
\usepackage{multirow}
\usepackage{jheppub}
\usepackage{enumerate}
\usepackage{relsize}
\usepackage{mathrsfs}
\usepackage{inputenc}
\usepackage{braket}
\usepackage{graphicx}
\usepackage{chemarrow}
\usepackage{mathtools}
\usepackage[font=small]{caption}
\usepackage{blindtext}
\usepackage{bbold,amsmath,lipsum,amssymb,leftindex}
\newcommand{\n}{\nu}
\newcommand{\m}{\mu}

\newcommand{\xcentcolon}

\usepackage{color}
\newcommand{\e}{\epsilon}
\newcommand{\be}[1]{\begin{equation}\label{#1} }
\newcommand{\ee}{\end{equation}}
\newcommand{\bea}[1]{\begin{eqnarray}\label{#1} }
\newcommand{\eea}{\end{eqnarray}}
\newcommand{\p}{\partial}
\newcommand{\refb}[1]{(\ref{#1})}

\renewcommand{\O}{{\mathcal{O}}}
\renewcommand{\L}{{\mathcal{L}}}

\newcommand{\bL}{\bar{{\mathcal{L}}}}

\renewcommand{\>}{\rangle}
\newcommand{\<}{\langle}
\newcommand{\w}{\omega}

\newcommand{\eps}{\varepsilon}

\newcommand{\de}{\delta}

\newcommand{\zc}{|0\rangle_c}

\newcommand{\g}{\gamma}

\renewcommand{\a}{\alpha}
\newcommand{\ta}{\tilde{\alpha}}

\newcommand{\C}{\tilde{C}}
\renewcommand{\b}{\beta}
\renewcommand{\t}{\tau}
\newcommand{\s}{\sigma}
\renewcommand*{\backref}[1]{}  

\renewcommand*{\backrefalt}[4]{%
  \ifcase #1 %
  \or
    [cited on page~#2]%
  \else
    [cited on pages~#2]%
  \fi
}

\newcommand\rnote[1]{\textcolor{brown}{[RC:\,#1]}}

\newcommand\pnote[1]{\textcolor{purple}{[PP:\,#1]}}

\title{The Tensionless Lives of Null Strings}

\author[a]{Arjun Bagchi,} \author[b]{Aritra Banerjee,} \author[c]{Ritankar Chatterjee,} \author[d]{and Priyadarshini Pandit}\author{\\}

\affiliation[a]{Indian Institute of Technology Kanpur, Kanpur 208016, India.\\} 

\affiliation[b]{Birla Institute of Technology and Science, Pilani Campus, Pilani, Rajasthan 333031, India.\\}

\affiliation[c]{Beijing Institute of Mathematical Sciences and Applications, Beijing 101408, China\\}

\affiliation[d]{Tata Institute of Fundamental Research, Mumbai 400005, India.\\}

\emailAdd{abagchi@iitk.ac.in, aritra.banerjee@pilani.bits-pilani.ac.in, ritankar@bimsa.cn, priyadarshini.pandit@tifr.res.in}

\preprint{}

\abstract{The tensionless limit probes the very high energy regime of string theory in contrast to the well studied point-particle limit which reduces to Einstein gravity. Tensionless  strings sweep out null worldsheets in the target space and hence are also called null strings. This article aims to provide a comprehensive review of tensionless null string theory beginning with the initial work of Schild, and continuing to the foundational work of Isberg et al (ILST) and then focussing on developments in the past decade. 

\medskip

Recent work centres on the emergence of the Carrollian Conformal Algebra as residual worldsheet symmetries of the ILST action and the identification of tensionless limit as a worldsheet Carrollian limit on the string worldsheet. Carrollian structures are used to address the classical and quantum aspects of the null string. In the classical theory, the aforementioned limit agrees with the analysis from the ILST action. Symmetries, constraints, mode expansions computed from both perspectives match nicely providing a robust cross-check of the analyses. We discuss closed and open null strings as well as their supersymmetric cousins. 

\medskip

The quantum null string comes with several surprises, the foremost of which is the emergence of three consistent quantum theories from the ILST action. We detail the canonical quantisation and the spectrum of the triumvirate of theories. We discuss the novelties of the quantum null theories and the effect compactifaction has on them. We also discuss Carroll strings, applications of these ideas to strings approaching black holes and give a quick overview of other related developments.

\bigskip

\bigskip

{\begin{center}
    {\em{Invited Review for Physics Reports.}}
\end{center}}
}

\begin{document}
\maketitle

\newpage

\section{Introduction}\label{Introduction}

Unifying classical gravity and quantum mechanics to form a unified framework of a quantum theory of gravity remains {\em the} outstanding problem in theoretical physics. Of the vying contenders of such a theory, string theory is one of the strongest contenders. 

\medskip

Of the four fundamental forces of nature, all except for gravity are amenable to methods of quantum field theory (QFT). Gravity is a fundamentally non-renormalizable theory and hence resists any attempts to shoehorn it into standard quantisation techniques of relativistic QFTs.  One of the principle causes for the uncontrollable ultraviolet divergences in quantum gravity framed in the language of relativistic QFT is the point-particle like nature of interaction in QFTs.  

\medskip

String theory offers a very unconventional solution to the problem by positing that fundamental objects are not point-like but extended one-dimensional strings. String theory was historically introduced as a theory to understand strong interactions before the advent of quantum chromodynamics, but it was later understood that this theory with the postulates of special relativity and quantum mechanics as inputs, naturally and rather remarkably generates a quantum theory of gravity. In other words, the spectrum of string theory contains a massless spin-two excitation that can be identified as the graviton, the quanta of the gravitational field. 

\medskip

One of the most attractive features of string theory is the paucity of free parameters in the theory. The spectacularly successful Standard Model of Particle Physics, which explains nature around us to a very high accuracy, in its most minimal form, has 19 free parameters. von Neumann famously said, ``With four parameters I can fit an elephant, with five I can make it wiggle its trunk.'' Theorists always prefer to have as few free parameters in their theories as possible. String theory is particularly nice in this regard, - it contains two free parameters, the length of the fundamental string $\ell_s$ and the string coupling $g_s$. 
Free string theory is thus dictated by just $\ell_s$ or equivalently $\alpha' = \ell_s^2 = \frac{1}{2\pi T}$, where $T$ is the tension of the string. 

\medskip
Bosonic free strings are described by the Polyakov action \cite{Polyakov:1981rd}. For strings propagating in a flat target spacetime, this is given by: 
\begin{align}\label{pol-action}
    S_P = -\frac{T}{2} \int d^2\xi \sqrt{-\gamma} \gamma^{\alpha\beta} \partial_\alpha X^\mu \partial_\beta X^\nu \eta_{\mu\nu}.
\end{align}
Here $\xi^\alpha = \sigma, \tau$ label the coordinates of the two dimensional worldsheet swept out by the string. $X^\mu$ are spacetime coordinates which also act as pullback scalar fields on the worldsheet. $\gamma^{\alpha\beta}$ is the worldsheet metric and $\eta_{\mu\nu}$ is the flat metric on the background spacetime. This theory has worldsheet diffeomorphism invariance which act as gauge symmetries that necessitates choosing a gauge on the worldsheet. One can choose the worldsheet metric to be flat and after accommodating for Weyl symmetries on the worldsheet, there is still a residual gauge symmetry which turns out to be two copies of the Virasoro algebra:
\begin{align}
    [\L_n^\pm, \L_m^\pm] = (n-m) \L^\pm_{n+m} + \frac{c^\pm}{12} \delta_{n+m,0} (n^3-n), \quad [\L_n^\pm, \L_m^\mp] = 0.
\end{align}
This thus signals the appearance of a two dimensional (2d) relativistic conformal field theory (CFT) on the worldsheet of the relativistic tensile string theory. The power of 2d CFTs is central to the understanding of quantum mechanics of the string. In the framework of Old Covariant quantisation (OCQ), one treats the string as a bunch of free worldsheet scalars and imposes the constraints in terms of the Virasoro algebra on the Hilbert space. The spectrum of the theory, so calculated, gives rise to a tachyonic vacuum state and a massless first excited level of states which contain a spin-two excitation that is identified as the graviton, and an infinite number of states at higher levels. 

\medskip

Given that free string theory has only one free parameter, there are a priori three different regions that we could tune this parameter to get different theories, viz.
\begin{align}\label{T}
    T = \text{finite} \, (\equiv T=1), \quad T \to \infty, \quad T \to 0.
\end{align}
The first choice is just the usual relativistic tensile string that we discussed above and is given by the Polyakov action \eqref{pol-action}. The second one, 
where the string tension goes to infinity, is the well studied point-particle limit of string theory. Here the fundamental string reduces to a point particle giving rise to (super)gravity from string theory. The final one is the tensionless limit \cite{Schild:1976vq, Isberg:1993av} and would be the focus of this review.

\medskip

The regimes described above in \eqref{T} is similar in spirit to what one could envision doing with the speed of light ($c$) in a relativistic QFT, i.e. 
\begin{subequations}\label{c}
\begin{align}
\text{Relativistic QFT:}& \quad    c = \text{finite} (\, \equiv c=1) \\
\text{Galilean QFT:}& \quad c \to \infty, \\ 
\text{Carrollian QFT:}& \quad c \to 0.
\end{align}
\end{subequations}
The first is of course the statement that the absolute value of the speed of light does not matter and for convenience this can be set to one in natural units in a relativistic theory. The $c\to \infty$ limit is the familiar non-relativistic limit where Lorentz transformations reduce to their Galilean counterparts, and time becomes absolute. The final one is the unconventional Carroll limit, where instead of time, space becomes absolute. Here lightcones close up and usual motion comes to a standstill. Carrollian symmetries, born of curiosity in the 1960's \cite{Leblond65, SenGupta:1966qer}, is now finding wide ranging applications starting from condensed matter physics to quantum gravity \cite{Bagchi:2025vri}. 

\medskip

For the string, the similarity between \eqref{T} and \eqref{c} is far from a coincidence. The free string, as described above, has only one free parameter and tension can thus be viewed in terms of a worldsheet speed of light. 

\medskip

In this review, we will be interested in the singular limit of string theory where the tension of the string goes to zero. As we will see, mirroring what happens in the QFT case, Carrollian symmetries would arise, now on the worldsheet. The analogy with point particles here is very apt. Massless point particles travel along null geodesics. The tensionless limit is the string analogue of the massless point particle. Here the string worldsheet becomes null \cite{Schild:1976vq}. Carrollian symmetries naturally arise on null surfaces \cite{Duval:2014uva, Duval:2014lpa} and hence also in this case. We now give a brief outline of this primary feature of the tensionless string. The action of these strings propagating, on a flat background geometry, following ILST \cite{Isberg:1993av} is given by 
\begin{align}\label{ilst}
    S_{\text{ILST}} = \int d^2\xi \, V^\alpha V^\beta \partial_\alpha X^\mu \partial_\beta X^\nu \eta_{\mu\nu},
\end{align}
where $V^\alpha V^\beta$ replaces the degenerate $\sqrt{-\gamma} \gamma^{\alpha\beta}$ in the limit $T\to0$. Worldsheet diffeomorphisms still exist in the limit, but an analogous covariant gauge fixing leads to a different residual symmetry algebra:
\begin{subequations} \label{introbms3}
    \begin{align}
        &[L_n, L_m] = (n-m) L_{n+m} + \frac{c_L}{12} \delta_{n+m,0} (n^3-n),\\
        &[L_n, M_m] = (n-m) M_{n+m} + \frac{c_M}{12} \delta_{n+m,0} (n^3-n),\\
        &[M_n, M_m] = 0.
    \end{align}
\end{subequations}
This was recognised as the Conformal Carroll algebra in two dimensions (2d) or equivalently the 3d Bondi-van der Burgh-Metzner-Sachs algebra \footnote{or interestingly the 2d Galilean Conformal Algebra, which is isomorphic to the Carrollian Conformal Algebra in these dimensions \cite{Bagchi:2010zz}.} in \cite{Duval:2014uva}. 

\medskip

The same algebra has made its appearance as the asymptotic symmetry algebra in 3d asymptotically flat spacetimes \cite{Barnich:2006av} and has been proposed as the symmetry dictating the 2d holographically dual field theory \cite{Bagchi:2010zz, Bagchi:2012cy}. The programme is called the Carrollian approach to flatspace holography and has gained much recent traction with generalisations to the 4d -- 3d case following \cite{Bagchi:2022emh, Donnay:2022aba}. We point the reader to the recent review \cite{Bagchi:2025vri} for more details on this. 

\medskip

For most applications, including the above mentioned best studied case of flatspace holography, Carrollian (conformal) symmetries arise as spacetime symmetries. In contrast, on the worldsheet of the tensionless string, these symmetries are residual gauge symmetries. We will see that the rather intriguing quantum aspects of the null tensionless string arise from various curious properties of the Carroll (conformal) symmetries and the algebra \eqref{introbms3} and its supersymmetric generalizations help us organise the quantum null (super)string. 

\medskip

This review aims to provide the reader an in-depth analysis of the tensionless null string {\footnote{We should mention that there has been a lot of recent remarkable progress in the context of tensionless strings in AdS spacetime, specifically superstring in AdS$_3 \times$S$^3\times$T$^4$ \cite{Eberhardt:2018ouy, Eberhardt:2019ywk}. This has been used to provide a derivation of AdS$_3$/CFT$_2$ \cite{Eberhardt:2019ywk}. These tensionless strings are {\em not} null strings. This is because in AdS, unlike flat spacetime, there is another scale, the AdS radius, with respect to which the tensionless limit can be taken without making the string null. We will not be considering these tensionless strings in our review.}}. Our review is divided into four parts. Part \ref{I} of the review provides some historical background, starting with the initial construction of the null string by Schild \cite{Schild:1976vq}. We also collect some basics of string theory we would need for constructing null strings and review aspects of Carrollian symmetries, especially Carrollian conformal field theory in $D=2$, which will be essential in what follows. 

\medskip

Part \ref{II} and \ref{III} form the main material of our review. Part \ref{II} focusses on the formulation of the classical null tensionless string. Much of the recent resurgence in the field follows the seminal work by Isberg et al \cite{Isberg:1993av}. The action of the null string that we would use in the main body of the paper is the ILST action \eqref{ilst}. The initiation of the recent body of work was the identification of \eqref{introbms3} as the BMS$_3$ algebra in \cite{Bagchi:2013bga}. This also led to understanding of the tensionless limit as an ultra-relativistic or a Carroll contraction on the worldsheet. The understanding was made more concrete in \cite{Bagchi:2015nca} where the first hints of the very rich quantum structures were also uncovered. We discuss this and the related curiosities in the quantum null string in Part \ref{III}. 

\medskip

In all our discussions in the first three parts focussed on Carrollian symmetries arising on the string worldsheet. Finally in Part \ref{IV}, Carrollian symmetries appear on the spacetime and we cover related developments of strings moving in Carrollian and generalities of Carrollian spacetimes. 

\medskip

To facilitate reading of the review, at the start of each part, we briefly summarise the different sections contained in the part and link them so that the reader can easily navigate through the pages.


\newpage

\part{Beginnings: History and Essentials}\label{I} 

\bigskip

This part of our review start with a recollection of early work on the theory of null strings beginning with the pioneering study of Schild. We will then devote a couple of sections to some basic but essential material on tensile string theory and Carrollian symmetries. 

\medskip

\subsection*{Outline of Part I}

Part I of this review contains the following:
\begin{itemize}
    \item {\em \hyperref[Null Strings]{Null Strings: A historical interlude}}: Here we take the historic route and describe Schild's seminal work \cite{Schild:1976vq} in some detail. We also mention other important advances in the field. 
    \item {\em \hyperref[String Theory]{String Theory: Collecting ingredients to go null}}: We will require some specific constructions of string theory, especially in the Hamiltonian framework, which may be somewhat unfamiliar. We discuss these aspects in a section. 
    \item {\em \hyperref[A Summary of Carrollian Symmetry]{A Summary of Carrollian Symmetry}}: Conformal Carrollian symmetries arise on the worldsheet of the tensionless string and helps organise the theory of null strings. In this section we collect the central results of two dimensional Carrollian CFTs which we will use heavily in the later parts of the review. 
\end{itemize}

This part of the review is meant to provide necessary background for the uninitiated reader. For the experts, only the historical interlude may be the interesting read in this section. 

\newpage

\section{Null Strings: A Historical Interlude}\label{Null Strings}

Our treatment of null strings in the main body of the review would be based on a modern symmetry based approach and would follow from the important work of Isberg et al \cite{Isberg:1993av} from the early 1990s and the identification of the worldsheet symmetries with the 2d Conformal Carroll (or BMS$_3$) algebra \cite{Bagchi:2013bga} in 2013 and subsequent developments in the last decade. But before this, it is important to make a historical detour and this section is an acknowledgement to the important developments preceding \cite{Isberg:1993av}. 

\medskip

We begin at the very beginning. The first formulation of strings with null worldsheets can be traced back to Schild's seminal work \cite{Schild:1976vq} in the 1970's. We now build a bit of background and delve into Schild's work in some detail. 


 \subsection{Massless point particles}
 Let us consider a free relativistic point particle of mass $m$ propagating in a $D$ dimensional Minkowski spacetime (light speed $c$ is taken to be unity). The action is the proper time between the initial and final locations of the particle in the spacetime along the worldline of the particle:
\begin{align}\label{relpar}
    S=-m\int_A^B\sqrt{-ds^2}=-m\int_{\tau_A}^{\tau_B}d\tau\sqrt{-\dot{X}^2},~~~ds^2=\eta_{\mu\nu}dX^{\mu}dX^{\nu},~~~\dot{X}=\frac{dX}{d\t}.
\end{align}
Here $\tau$ is affine parameter defined on the worldline, $X^\mu(\t)$ is the position of the particle in spacetime. This action is invariant under the reparametrization $\tau=\tau'=f(\t)$.
It is clear however that the action for massless point particles cannot be arrived at by naively taking $m\to 0$ on \eqref{relpar}.
 
\medskip
There is an alternate way to understand the problem without taking massless limit which was highlighted in \cite{Schild:1976vq}. The equation of motion (EOM) of the action \eqref{relpar} gives
\begin{align}\label{p.e.o.m.}
    &m\frac{d}{d\t}\Bigg(\frac{\dot{X}}{\sqrt{-\dot{X}^2}}\Bigg)=0.
  \end{align}
Under appropriate reparametrization, EOM simplifies to
\begin{align}\label{p.e.o.m.2}
    \sqrt{-\dot{X}^2}=c \quad \implies m\frac{d}{d\t'}\Bigg(\frac{\dot{X}}{\sqrt{-\dot{X}^2}}\Bigg)=\frac{m}{c}\ddot{X}=0,~~~\text{where}~\dot{X}=\frac{dX}{d\t'}.
\end{align}
Here $c$ is a constant. The left hand side (LHS) of \eqref{p.e.o.m.} or \eqref{p.e.o.m.2} makes sense only $\dot{X}^2\neq 0$. Clearly, this EOM isn't applicable for massless particles which move along the null direction (where $\dot{X}^2=0$). 

\medskip

We now detail the way around this problem. In \cite{Stueckelberg:1941th}, St\"uckelberg proposed a new action for the point particle which can describe particles traversing in the null direction too. The action is  
\begin{align}\label{schildpt}
    S=-\frac{1}{2}\int_A^B  d\t~\dot{X}^2.
\end{align}
The EOM for this action is 
\begin{align}\label{Stuckelberg}
   \frac{1}{2} \frac{d}{d\t}\frac{\partial(\dot{X}^2)}{\partial \dot{X}}=\ddot{X}=0.
\end{align}
So the  EOM for this action is equivalent to \eqref{p.e.o.m.2} (provided $c\neq0$ in \eqref{p.e.o.m.2}) indicating that the action \eqref{schildpt} is equivalent to \eqref{relpar} for particles traversing in timelike and spacelike geodesics. However, \eqref{Stuckelberg} implies $\dot{X}^2$ is arbitrary constant in $\t$, which can take positive, zero, or negative value. Hence \eqref{schildpt} action allows all possible geodesics, namely timelike, spacelike and null geodesics. Based on this observation Schild argued that the action \eqref{schildpt} is superior to the action \eqref{relpar} to describe relativistic point particles.

\medskip
However, there are some important drawbacks of this action. The applicability of this new action \eqref{schildpt} comes at a cost.  As later pointed out in \cite{Karlhede:1986wb}, the action \eqref{schildpt}, unlike \eqref{relpar}, is not invariant under $\tau$ reparametrization. Since \eqref{schildpt} describes particles traversing all possible geodesics, there is no way of knowing beforehand whether the particle it describes is massless or massive. These issues can be taken care of by using the $\tau$ reparametrization invariant ``einbein" action discussed later in section~\ref{Hamiltonformalism}.

\subsection{String tracing a null worldsheet}
We now consider a string propagating through a $D$ dimensional flat spacetime. The worldsheet of the string is parametrized by $\tau$ and $\sigma$. The Nambu-Goto action \cite{Nambu:1969se,Goto:1971ce} of the usual string theory is the invariant area of the worldsheet:
\begin{align}\label{N.G.}
    S_{NG}=T\int d^2\s \sqrt{-\det{\gamma_{\alpha\beta}}},
\end{align}
where $\s^{\alpha}=\{\tau,\sigma\}$ are coordinates on worldsheet (i.e. $\a=0,1$) and $\gamma_{\alpha\beta}$ is the induced metric on the worldsheet given by
\begin{align}\label{inducedmetric}
   \gamma_{\alpha\beta}=\partial_{\alpha}X^{\mu}\partial_{\beta}X^{\nu}\eta_{\mu\nu}.
\end{align}
 Here $X^{\mu}=X^{\mu}(\tau,\sigma),~~\mu=0,1,\cdots,D-1$ are the coordinates of the $D$ dimensional target spacetime and $\eta_{\mu\nu}$ is the target spacetime (Minkowski) metric. Taking $T\to 0$ limit on this action naively would make it vanish. 

\medskip
In \cite{Schild:1976vq}, an important drawback of the Nambu-Goto action was noted. The EOM for Nambu-Goto action \eqref{N.G.} is given by
 \begin{align}\label{ctb6}
\Dot{\Pi}_{\mu}+K'_{\mu}=0,
\end{align}
where $\Pi_{\mu}$ (the canonical momentum) and $K_{\mu}$ are given by
\begin{subequations}\label{ctb7}
\begin{equation}
    \Pi_{\mu}=\frac{\partial{\mathcal{L}}}{\partial{\Dot{X}^{\mu}}}=T\frac{(\Dot{X}\cdot X')X'_{\mu}-X'^2\Dot{X}_{\mu}}{\sqrt{(\Dot{X}\cdot X')^2-\Dot{X}^2X'^2}},
    \end{equation}
    \begin{equation}
    K_{\mu}=\frac{\partial{\mathcal{L}}}{\partial{X'^{\mu}}}=T\frac{(\Dot{X}\cdot X')\Dot{X}_{\mu}-\Dot{X}^2X'_{\mu}}{\sqrt{(\Dot{X}\cdot X')^2-\Dot{X}^2X'^2}}.
    \end{equation}
\end{subequations}
In the above equations, $X'=\partial_{\sigma}X$ and $\Dot{X}=\partial_{\tau}X$. \eqref{ctb6} and \eqref{ctb7}, imply that \eqref{ctb6} is valid only if
\begin{align}\label{ctb77}
    \sqrt{(\Dot{X}\cdot X')^2-\Dot{X}^2X'^2}\neq 0,~~\implies~~-\det{\gamma_{\alpha\beta}}\neq 0.
\end{align}
This means that the worldsheet area traced by string must be non-zero. However, since Minkowski spacetime allows null submanifolds, there should be special case where string worldsheet is null. Nevertheless, \eqref{ctb6} and \eqref{ctb77} clearly shows that Nambu-Goto action cannot describe such strings. As we shall see in the later part, null worldsheet is directly connected to tensionless strings in the same way null geodesics are connected to massless particles.


\subsection{Schild Strings}
In \cite{Schild:1976vq}, Schild generalised St\"uckelberg's analysis to strings. The action proposed was string analogue of \eqref{schildpt}:
\begin{align}\label{schildst}
    S=\frac{1}{4}\int~d^2\s~ \Gamma^{\mu\nu}\Gamma_{\mu\nu}, ~~~\Gamma^{\mu\nu}=\epsilon^{\a\b}\partial_{\a}X^\m\partial_{\b}X^\n=\dot{X}^\mu X'^{\nu}-X'^\mu\dot{X}^\nu.
\end{align}
For future use, we also note the following
\begin{align}\label{Gamma^2}
   \Gamma^2=\Dot{X}^2X'^2-(\Dot{X}\cdot X')^2=\det{\gamma_{\alpha\beta}}.
\end{align}
EOM for this action are
\begin{align}\label{s.e.o.m.}
    \frac{1}{2}\partial_\a\Big(\frac{\partial~\Gamma^2}{\partial(\partial_\a X^\mu)}\Big)=\partial_{\a}(\epsilon^{\a\b}\Gamma_{\m\n}\partial_{\b}X^\n)=0
\end{align}
Using $\partial_\b X^\mu \frac{\partial~\Gamma^2}{\partial_\a X^\mu}=\delta^{\a}_{~\b}\Gamma^2$, the EOM reduces to
\begin{align}
   \partial_\alpha\Gamma^2=0.
\end{align}

Hence EOM for \eqref{schildst} implies $\Gamma^2$ is arbitrary constant (which can be zero as well) in the worldsheet coordinates. Since EOM allows $\Gamma^2=0$ as a possibility, this action allows null strings as a solution. 
\medskip

But again, \eqref{schildst} does not specify the nature of the string worldsheet and suffers from the lack of reparametrization invariance, similar to the St\"uckelberg action\footnote{The easiest way to see this is is to use \eqref{Gamma^2} to rewrite the reparametrization invariant Nambu Goto action \eqref{N.G.} as
\begin{align}
    S_{NG}=T\int d^2\s \sqrt{-\Gamma^2}.
\end{align}
In the light of the fact that this action is invariant area element, squaring the integrant in \eqref{schildst} would mean integrating $\Gamma^2=-\det{\gamma_{\alpha\beta}}$ throughout the worldsheet which won't allow the action to be invariant any more.}. A better way to construct a reparametrization invariant action will be to use Hamiltonian formalism \cite{Isberg:1993av} (see also \cite{Lindstrom:1990qb}) outlined in the upcoming sections. We shall see that, on the action formulated in this way for tensile strings we can take a sensible $T\to 0$ limit.
\medskip

However, Schild in his work provided some important insights about the geometric nature of the worldsheet of null strings which will continue to hold even for the new reparameterization invariant action for null strings.

\subsection{Geometry of the Null Worldsheet}\label{GPNWS}

Following Schild, we now discuss the geometric nature of the null string worldsheet. We  first look into the tangent surface to the worldsheet. 

\medskip

The null worldsheet has $\Gamma^2=0$ and obeys EOM \eqref{s.e.o.m.}. For a string with $\Gamma^{\mu\nu}$ given in \eqref{schildst} and $\Gamma^2=0$, $\Gamma^{\mu\nu}$ can be expressed as 
\begin{equation}\label{gamma munu}
\Gamma^{\mu\nu}=A^{[\m}B^{\n]}, \qquad \qquad A^2>0,~  B^2=0~~ \text{and}~~ A\cdot B=0,
\end{equation}
where $A^\m$ and $B^\m$ are tangent vectors to the null worldsheet. The tangent 2d surface to the worldsheet is spanned by $A^\m$ and $B^\m$ and hence a generic vector in the tangent space is given by 
\begin{equation}\label{tangentvector}
V=c_1A+c_2B
\end{equation} 
The above equation \eqref{tangentvector} along with $B^2=0$ in \eqref{gamma munu} implies $V^{2}=c_1^2A^2>0$, which in turn implies that for $c_1^2\neq0$ (i.e. $V$ doesn't coincide with $B$), $V$ is spacelike. Also, the fact that $A\cdot B=0$ from \eqref{gamma munu} and \eqref{tangentvector} together imply that $V$ is orthogonal to $B$. The takeaway from here is that whenever we take two independent vectors in a tangent space of a null worldsheet, at least one of them must be spacelike. Before we proceed we also note the following identities which will be useful later
\begin{align}\label{nullgeo}
    \Gamma^{\m\n}V_\n=c_1 A^2 B^\m,~~~\Gamma^{\mu\nu}B_\n=0.
\end{align}
That means as long as $c_1\neq 0$, for any vector $V$ in the tangent space, $\Gamma^{\m\n}V_\n$ will give us a non-zero null vector. However, for any null tangent vector (where $V$ coincides with $B$ ) to the null worldsheet we have $\Gamma^{\mu\nu}B_\n=0$.
\medskip

Now let us consider a null worldsheet given by $X^\m(\t,\s)$. The derivatives $\partial_\tau X^{\m}$ (or $\dot{X}^{\m}$) and $\partial_\s X^{\m}$ (or $X'^\mu$) are essentially two independent tangents to the worldsheet, and hence, according to the previous argument at least one of them has to be spacelike. Since in our notation $\s$ is the spacelike worldsheet coordinate, we choose $X'^{\m}$ to be the spacelike vector, i.e. $X'^{\m} X'_{\m}>0$. Let us also consider a one parameter family of null curves on this null worldsheet 
\begin{equation}
\Phi(\t,\s)=\textit{constant} \qquad \qquad \text{with} \qquad \Phi'\neq 0.
\end{equation}
From the perspective of the target spacetime these null curves will be expressed as $X^\m=X^\m(s)$ where $s$ is the affine parameter defined on a given null curve. Now we want to construct a null vector field tangential to the null curves $X^\m=X^\m(s)$. Since these null curves reside on the worldsheet, the vector field tangent to them is also a vector field tangent to the worldsheet. Since we know that $X'^{\m}$ is spacelike, using \eqref{nullgeo}, we can define a null vector field tangential to the worldsheet in the following way 
\begin{align}\label{Pro4}
    N_\m=\frac{1}{\Phi'}\Gamma_{\mu\nu}X'^{\n}.
\end{align}
\begin{figure}[ht]
    \centering
    \includegraphics[width=0.8\linewidth]{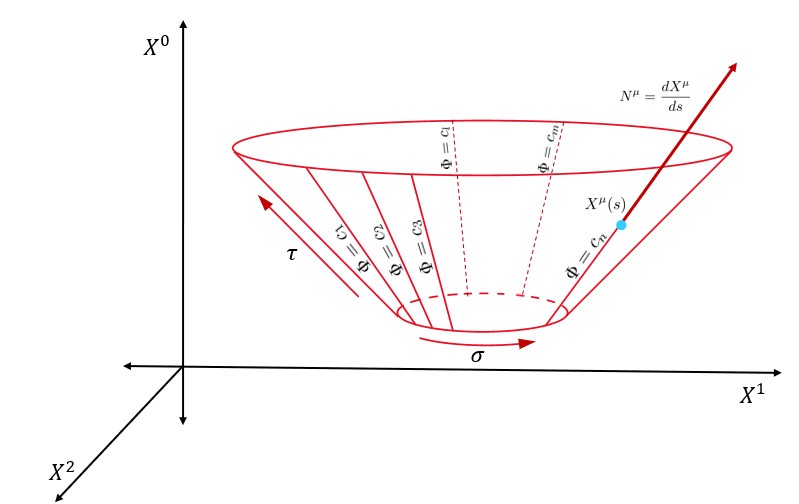}
    \caption{Heuristic depiction of a closed string null worldsheet with one parameter family of null curves $\Phi(\t,\s)=constant$ with different values of constants. Any of these curves also can be expressed in target spacetime as $X^\m=X^\m(s)$ (here the curve $\Phi=c_n$ is expressed in this way). Null vector tangential to the curve $N^\m$ at a given point $s$ on the curve $\Phi=c_n$ (depicted as sky-blue dot in the figure) is shown by the bold red arrow.}
    \label{schild worldsheet}
\end{figure}
We want this vector field to be tangential to the curve $X^\m(s)$ which means we must ensure that the vector field $N$ coincides with $dX/ds$. After that we can choose the affine parameter on the null curve $X^\m(s)$ such that $dX/ds=N$. Hence we have the following two equations which need to be simultaneously satisfied by $N$
\begin{subequations}
    \begin{equation}\label{Guddu}
        N^\m=\frac{dX^\m}{ds}=\dot{X}^\m\frac{d\t}{ds}+X'^\mu\frac{d\s}{ds},
    \end{equation}
        \begin{equation}\label{Guddu2}
      \frac{dX^\m}{ds}  =N^\m=\frac{1}{\Phi'}\Gamma^{\mu\nu}X'_{\n}=\frac{1}{\Phi'}\Big(\dot{X}^\mu X'^2 -X'^\m(\dot{X}\cdot X')\Big).
    \end{equation}
\end{subequations}
Comparing coefficients of $\dot{X}$ from both \eqref{Guddu} and \eqref{Guddu2} one can see that 
\begin{align}\label{Pro1}
    \frac{d\t}{ds}=\frac{1}{\Phi'}X'^2\neq 0.
\end{align}
Now, since $N^\nu$ is null tangent vector to the worldsheet, from \eqref{nullgeo} one can see 
\begin{align}\label{Pro2}
    \Gamma^{\mu\nu}N_\n=\Gamma^{\mu\nu}\frac{dX_\n}{ds}=0,\implies \Gamma^{\mu\nu}\dot{X}_\n\frac{d\t}{ds}+\Gamma^{\mu\nu}X'_\nu\frac{d\s}{ds}=0,
\end{align}
Now, since the null curve has been defined on the worldsheet as $\Phi(\t,\s)=\textit{constant}$, this leads to another identity
\begin{align}\label{Pro3}
    0=\frac{d\Phi}{ds}=\dot{\Phi}\frac{d\t}{ds}+\Phi'\frac{d\s}{ds}.
\end{align}
Using \eqref{Pro1}, \eqref{Pro2} and \eqref{Pro3} one arrives at $\Gamma^{\m\n}\dot{X}_\n=\dot{\phi}N^\m$, and together with the definition in \eqref{Pro4} this gives
\begin{align}\label{pro5}
    \Gamma^{\m\n}\partial_\a X_\n=N^\m \partial_\a\Phi .
\end{align}
Now, let us look back into the EOM of the Schild action \eqref{s.e.o.m.}. The l.h.s. of that equation can be rewritten using \eqref{pro5} as 
\begin{align}
    \partial_{\a}(\epsilon^{\a\b}\Gamma_{\m\n}\partial_{\b}X^\n)=\partial_{\a}(\epsilon^{\a\b}\partial_\b\Phi N_\m)=\epsilon^{\a\b}\partial_\b\Phi \partial_{\a} N_\m=\dot{\Phi}N'_\m-\Phi'\dot{N}_\m.
\end{align}
In the above we have used the fact that $\epsilon^{\a\b}(\partial^2_{\a\b}\Phi) N_\m=0$ due to antisymmetry of $\epsilon^{\a\b}$. Now, applying \eqref{Pro3} in the above we see that
\begin{align}
    \partial_{\a}(\epsilon^{\a\b}\Gamma_{\m\n}\partial_{\b}X^\n)=-\frac{\Phi'}{\frac{d\t}{ds}}N'_\m\frac{d\s}{ds}-\Phi'\dot{N}_\m=-\frac{\Phi'}{\frac{d\t}{ds}}\Bigg(N'_\m\frac{d\s}{ds}+\dot{N}_\m\frac{d\t}{ds}\Bigg)=-\frac{\Phi'}{\frac{d\t}{ds}}\frac{dN_\m}{ds}.
\end{align}
Since $\Phi'$ and $d\t/ds$ are non-zero finite numbers, the EOM \eqref{s.e.o.m.} reduces to $\frac{dN^\m}{ds}=0$, which, in turn, gives the following
\begin{align}\label{geodesic}
    \frac{d^2X^\m}{ds^2}=0.
\end{align}
The above equation \eqref{geodesic} is the EOM of null geodesic in the flat spacetime where $s$ is the affine parameter. One can also choose the affine parameter to be the worldsheet coordinate $\t$ and \eqref{geodesic} becomes $\ddot{X}=0$. Hence one can say that the worldsheet of string governed by EOM \eqref{s.e.o.m.} consists of null geodesics. We shall later see that this property appears as EOM for null strings irrespective of the action for the strings.
\medskip

Schild also discovered another property inherent to the null worldsheet which would appear as constraints for null strings irrespective of the actions proposed in future studies.  We already know from previous analysis that the tangent space to the null worldsheet at any point can always be spanned by a spacelike vector and a null vector which are normal to each other. Since on each point on the worldsheet there are spacelike tangent vectors it is always possible to consider a spacelike curve (i.e. tangent to the curve is always spacelike) on a null worldsheet. Let us consider a spacelike curve on the null worldsheet parametrized by $\s$ (i.e. $X^\m=X^\m(\s)$). Since the tangent space to the null worldsheet also essentially contain a null direction orthogonal to the spatial direction, whenever we define such spacelike curve on a null worldsheet, we can also define a null vector field $N^\m(\s)$ orthogonal to the curve at any point $\s$, i.e.  
\begin{align}
    N^2=0,~~~N_\m\frac{dX^\m}{d\s}=N_\m X'^\m=0.
\end{align}
Tangent to the curve at $\s$ (i.e. $\frac{dX^\m}{d\s}$) itself is the spacelike vector. Hence at each point $\s$ there must be a null vector $N^\m(\s)$ which belongs to the tangent space and also orthogonal to $\frac{dX^\m}{d\s}$. Since $N^\m(\s)$ belongs to the tangent space to the null worldsheet, one can define null curves along the worldsheet with $N^\m$ as tangent. From the previous analysis we know that these curves are null geodesics defined on the worldsheet. If the affine parameter on the worldsheet is chosen to be $\t$, then we have 
\begin{align}
    N^\m=\frac{\p X^\m}{\p\t}=\dot{X}^\m,~~~X^\m=X^\m(\s,\t).
\end{align}
They lead to following identities
\begin{align}\label{nullconstraints}
    \dot{X}^2=0,~~~\dot{X}\cdot X'=0.
\end{align}
As we will see in the later parts, these equations will appear as constraints in the study of any action valid for null strings.

\subsection{Subsequent Developments}
The action proposed by Schild is not reparametrization invariant and this is an important drawback. Looking at the history of string theory, we see that in 1970s and 1980s, various actions were being proposed as alternatives to the Nambu-Goto action. Schild's action \eqref{schildst} itself was not meant to be an action just for null strings, but as a better alternative to the Nambu-Goto action for relativistic strings, which, unlike Nambu-Goto action, was applicable for null strings too\footnote{In fact, for a while, Schild action \eqref{schildst} was seriously considered to be the replacement of Nambu-Goto action and was used to construct quantum theory of relativistic strings \cite{Eguchi:1979qk}. Nambu himself offered a Hamilton-Jacobi formulation of Schild action in \cite{Nambu:1980kz}. For further studies of relativistic string dynamics based on Schild action \eqref{schildst} one can see \cite{Luscher:1980fr, Kastrup:1981yu, Eguchi:1982fy}. A supersymmetric version of Schild action was proposed by Ishibashi et al in their seminal work on IKKT matrix theory \cite{Ishibashi:1996xs}. There they showed that the supersymmetric Schild action proposed by them is classical equivalent of the Green-Schwarz superstring action in its Nambu-Goto form. Then they used the Schild-type action to construct a matrix model for the Green-Schwarz superstring.}. The now well known Polyakov action, which was introduced just a few months before Schild's work \cite{Brink:1976sc, Deser:1976rb}, was yet to be accepted universally. It was only accepted in 1981 when Polyakov would demonstrate the usefulness of this action in quantisation of string theory \cite{Polyakov:1981rd}. We will see later that Polyakov action is the key to obtain an action of null string with reparametrization invariance. In this section however, we will review the important developments in the study of null strings in 1980s and 1990s which would help us understand how our understanding of null strings evolved.
        \subsection*{Attempts to construct null string actions}
      In \cite{Karlhede:1986wb}, Karlhede and Lindstrom delved deeper into strings with null worldsheet. As pointed out before, they identified the problem of St\"uckelberg action and Schild action not being reparametrization invariant. They proposed a new reparametrization invariant action for tensile bosonic string mimicking the einbein action for massive point particles (discussed later in \eqref{I(x,e)})
      \begin{align}
          S_{KL}=\int d^2\s~\phi~(\Gamma^2-T^2\phi^{-2}).
      \end{align}
Here, $S_{KL}$ denotes Karlhede and Lindstrom action. In the above $\Gamma^2$ is same as \eqref{Gamma^2} and $T$ is the string tension. They identified that null strings in flat target spacetime are identical to tensionless ($T\to 0$) strings much in the same way particles travelling in null directions are identical to massless ($m\to 0$) particles. The null string action, hence, becomes
\begin{align}
    S_{KLN}=\int d^2\s~\phi~\Gamma^2,
\end{align}
where $S_{KLN}$ represents Karlhede and Lindstrom action for null string. The EOM for this action gives $\Gamma^2=0$, ensuring that the worldsheet is null. Following the arguments presented in \eqref{GPNWS}, the equations \eqref{geodesic} and \eqref{nullconstraints} will automatically follow. However, we shall see that in later works Lindstrom et al would study null strings using an even more improved formalism.
\medskip

        In  \cite{Lizzi:1986nv}, Lizzi et al proposed a different formulation for null strings. In an earlier work, \cite{Balachandran:1986vf}, Balachandran et al proposed their own brand of action for both bosonic strings as well as superstrings. The bosonic action proposed by them is 
        \begin{align}\label{BLS}
            S_{BLS}=\int d^2\s~\Sigma_{\m\n}\Gamma^{\m\n},
        \end{align} 
        where $\Gamma^{\m\n}$ is same as \eqref{schildst}. They showed that for appropriate choice of $\Sigma$ one can obtain the EOM of both Nambu-Goto strings \eqref{ctb6} as well as of Schild's null strings (\eqref{geodesic} and \eqref{nullconstraints}).  A Hamiltonian formalism for the action similar to \eqref{BLS} for null string case was performed by Zheltukhin in \cite{Zheltukhin:1988rys}.

\subsection*{Attempts to quantise bosonic null strings}
        
    The earliest attempt to quantise null strings can be found in \cite{Lizzi:1986nv}, which built on the formalism outlined in \cite{Balachandran:1986vf} and attempted to develop a quantum theory for bosonic null strings in the light-cone gauge. The authors used Weyl ordering instead of normal ordering as ordering of operators. This choice resulted into a surprising finding; the quantum theory was consistent in any target spacetime dimension i.e. there was no critical dimension! They also found that even after quantisation, null string's spectrum remains continuous, a result which is difficult to interpret physically. Same conclusion was drawn in some subsequent studies such as \cite{Amorim:1987bk, Barcelos-Neto:1989pjq}. However, in later studies \cite{Gamboa:1989px,Gamboa:1989zc} Gamboa et al performed quantisation of bosonic null strings using normal ordering of operators and found that the critical dimension of the quantum theory still remains $D=26$. 
    The apparent contradiction between the results in \cite{Lizzi:1986nv} and \cite{Gamboa:1989px} was later reconciled in \cite{Gamboa:1989zc}, where it was shown that these choices of normal ordering correspond to two different definition of vacuum, and consequently, two different quantum theories. We will see later that the underlying quantum structure is even richer and leads to three different quantum theories starting out from a single classical theory \cite{Bagchi:2020fpr}. 
     \medskip
     
    The vacuum corresponding to the formalism in \cite{Lizzi:1986nv} was later identified as induced vacuum \cite{Bagchi:2019cay}, which arises at the tensionless limit of usual tensile string vacuum \cite{Bagchi:2019cay,Bagchi:2020fpr}. Another important finding in \cite{Lizzi:1986nv} which would later play a crucial role in the study of null strings is that, for closed null strings the Fourier modes of the constraints \eqref{nullconstraints} give rise to an algebra, which was later identified as BMS$_3$ algebra\footnote{However, the algebra found by Lizzi et al. did not have any central extension. The reason behind this was again the Weyl ordering they chose to work with. As already pointed out, this ordering correspond to Induced vacuum and it has been explicitly shown in \cite{Bagchi:2020fpr} that for this vacuum the worldsheet algebra doesn't have central extension. We shall review this in later part.}, the symmetry algebra of null infinity of 3d  asymptotically flat spacetime \cite{Barnich:2006av}.
    \medskip

    The vacuum proposed in \cite{Gamboa:1989px} is now known as the flipped vacuum \cite{Bagchi:2020fpr}. Quantum theory constructed on this vacuum has a truncated spectrum having only finite number of massless states. Later it was observed in \cite{Casali:2016atr} that this spectrum exactly coincides with the spectrum of ambitwistor string theory, which was originally proposed by Mason and Skinner in \cite{Mason:2013sva} to explain the CHY formulae for massless particles.
    \subsection*{Supersymmetric null strings}
    The study of supersymmetric null strings began with a series of works \cite{Barcelos-Neto:1989pjq, Gamboa:1989px, Gamboa:1989zc}. The work \cite{Barcelos-Neto:1989pjq} followed Green-Schwarz (GS) formalism to study null superstrings.
    The works \cite{Gamboa:1989px, Gamboa:1989zc} followed Ramond–Neveu–Schwarz (RNS) formalism to study the same. The studies in \cite{Gamboa:1989px, Gamboa:1989zc} played important role in the subsequent understanding of null bosonic strings and superstrings. We list some of their key results:
    \begin{itemize}
        \item In \cite{Gamboa:1989zc}, actions for bosonic null string were identified which would be the basis of subsequent studies of null strings. It was later found out that this action emerges after gauge fixing the reparametrization invariant action of bosonic null strings later constructed in \cite{Lindstrom:1990ar}. The gauge fixed action for null superstring was also found here which was later identified as the homogeneous tensionless limit of Ramond-Neveu-Schwarz (RNS) superstring action \cite{Bagchi:2016yyf}. 
    
        \item In \cite{Gamboa:1989px}, the constraint algebra for the null superstring action was constructed. This algebra would be later identified as homogeneous version of Supersymmetric Conformal Carrollian Algebra. This algebra has applications beyond the study of null strings, e.g. it appears as the asymptotic symmetry algebra of 3d flatspace supergravity \cite{Barnich:2014cwa}.
        
        \item As already highlighted, the studies \cite{Gamboa:1989px, Gamboa:1989zc} developed a quantum theory for null strings having critical dimension $D=26$. Here the null superstring theory was also quantised in light-cone gauge and the critical dimension was calculated be $D=10$. The spectrum obtained quantising the null superstrings contained finite number of massless states. This would later be connected to the Ambitwistor string \cite{Casali:2016atr}.
  \end{itemize}
However, the actions Gamboa et al used in their studies of bosonic strings and superstrings were gauge fixed actions, and not ones which were  reparametrization invariant. Subsequent works \cite{Lindstrom:1990ar, Lindstrom:1990qb} filled this gap for both bosonic null strings and null superstrings. \cite{Lindstrom:1990qb} followed the construction of reparametrization invariant Polyakov action for tensile string given in \cite{Brink:1976sc} and constructed the same for null string action by taking tensionless ($T\to 0$) limit on the Polyakov action. The reparametrization invariant action in \cite{Lindstrom:1990qb} at suitable gauge choice leads us to the bosonic action used in \cite{Gamboa:1989zc}. Similarly for the action for null superstrings too, the reparametrization invariant action in \cite{Lindstrom:1990qb} under appropriate gauge choice reduces to the null superstring action used in \cite{Gamboa:1989zc}.

\subsection*{ILST Formalism}
In 1993, Isberg et al wrote a seminal paper on null strings which provided a complete and concrete classical formulation of bosonic null strings \cite{Isberg:1993av}. Following their earlier work \cite{Lindstrom:1990ar}, they first constructed reparametrization invariant action for null string. As advertised earlier, this formulation would be the basis of the more recent advances of null tensionless string theory and will be central to this review. We will consider this at length in the main body of the review and in particular in Part \ref{II}. 

\subsection*{Other works}
We now briefly mention other works which provided useful insights in the studies of null strings.
In \cite{Isberg:1992ia}, Isberg et al studied spacetime conformal invariance for quantum null strings. From the tensionless limit on the Polyakov action as shown in \cite{Lindstrom:1990qb} one can clearly see how tensionless limit is directly connected to the non-Riemannian structures on the string worldsheet. In \cite{DeVega:1992tm} it was further shown that the sting tension is proportionate to the worldsheet speed of light $c$, which means null strings correspond to the $c\to 0$ limit (which is now called to be Carrollian limit) on the worldsheet. The \cite{deVega:1994hu} also studied the quantum null stings in de Sitter background. Further, \cite{Lizzi:1994rn} showed how at tensionless limit the worldsheet residual gauge symmetry of tensile strings (Virasoro algebra) reduces to the same for tensionless strings (BMS$_3$ algebra). Although Schild's formalism for null strings was later abandoned, a generalisation of Schild's study in curved background can be found in \cite{Kar:1995br}.

\bigskip

This concludes our recapitulation of the history of the development of the theory of null strings. 

\newpage

\section{String Theory: Collecting ingredients to go null}\label{String Theory}

Our review is about tensionless null strings and superstrings. Before we go into the specific details of the null string, it is important to pause and collect various things from the well-known tensile theory. This section can be easily skipped for experts, but for the ease readability of non-experts, this is an amalgamation of tensile string theoretic methods which we will adopt to the tensionless world later. In order to make this section of a manageable length and for better readability, we will focus solely on the closed bosonic theory and leave the details of open strings and superstrings to appendices. Appendix \ref{Open Strings} briefly covers bosonic open strings and Appendix \ref{Superstring} summarises the RNS superstring. 

\subsection{Hamiltonian Formalism}\label{Hamiltonformalism}
We saw in previous section that the standard relativistic point particle action \eqref{relpar} fails to describe dynamics of massless (or lightlike) point particles. We encountered similar problem in the Nambu-Goto action \eqref{N.G.} while studying strings with null worldsheet. While the St\"uckelberg action \eqref{schildpt} and Schild action \eqref{schildst} solve this problem for massless particles and null strings respectively, they fail to be reparametrization invariant. However, in 1976 Brink et al provided an altogether different formalism for relativistic point particles \cite{Brink:1976sz}. By employing Hamiltonian formulation, they managed to construct a reparametrization invariant action where massless limit could be taken. In the same year, \cite{Brink:1976sc}\footnote{However, they didn't provide the details of the steps in their work which can be found in later works such as \cite{Lindstrom:1990qb,Isberg:1993av}.)} employed the same Hamiltonian formalism to formulate a new reparametrization invariant action for strings. The string action they constructed is now famous as Polyakov action\footnote{This action was independently proposed by Deser et al in \cite{Deser:1976rb}. However, they did not follow the Hamiltonian formulation to obtain the action.}. In this section we present the Hamiltonian formulation for both relativistic point particles and bosonic strings which will provide us with the necessary tools to construct the actions for massless point particles and null strings.

\paragraph{Relativistic point particles:}
Looking at the relativistic point particle action \eqref{relpar}, one immediately identifies two things: first, the action is designed in such a way that the canonical momentum of this action automatically satisfies the relativistic mass-shell constraint
\begin{align}\label{massshell}
    P_{\m}=\frac{\delta I}{\delta \dot{X}^{\mu}}=\frac{\eta_{\mu\nu}m\dot{X}^{\nu}}{\sqrt{-\dot{X}^2}}, ~~~\implies~~~ P^2+m^2=0.
\end{align}
Since this constraint is an immediate consequence of the definition of canonical momentum, this constraint is called primary constraint. The second thing to note is: the canonical Hamiltonian of this system vanishes
\begin{align}
    H_c= P_{\mu}\dot{X}^{\mu}-L=0,\qquad L=-m\sqrt{-\dot{X}^2}.
\end{align}
In order to perform a Hamiltonian formalism for this theory, Dirac proposed a new Hamiltonian which is a Lagrange's multiplier times mass shell constraint \eqref{massshell}
\begin{align}\label{total}
    H_{T}=\frac{1}{2}e\big(P^2+m^2\big),\qquad e=e(\tau).
\end{align}
More generally in order to deal with such systems with vanishing canonical Hamiltonian, we need to define a total Hamiltonian $(H_T)$ which is the primary constraints of the system times Lagrange multipliers. The new Lagrangian corresponding to the Hamiltonian in \eqref{total} in terms of $X$ and $P$ would be
\begin{align}\label{phsplagpar}
    L_T=P_{\mu}\dot{X}^{\mu}-\frac{1}{2}e\big(P^2+m^2\big).
\end{align}
Since in a space with coordinates $X$, a Lagrangian is expressed as $f(X,\dot{X})$, \eqref{phsplagpar} cannot be a Lagrangian in the Minkowski spacetime where the particle is propagating. However, if one considers the phase space, where both position $X$ and momentum $P$ are space coordinates, \eqref{phsplagpar} is allowed to be a Lagrangian \footnote{This kind of a Lagrangian should not be confused with a Routhian. Here $P_{\mu}$'s too, are treated as position coordinates in phase space, not momentum coordinates. This form of action has been used in the study of non-Lorentzian limit of relativistic point particles \cite{Bergshoeff:2014jla,Batlle:2017cfa}.}. Hence we call \eqref{phsplagpar} the Phase Space Lagrangian of the point particle and the corresponding action
\begin{align}\label{phspacpar}
    I_T=\int d\tau~\left\{P_{\mu}\dot{X}^{\mu}-\frac{1}{2}e\big(P^2+m^2\big)\right\},
\end{align}
the phase space action. Now, let us translate this into an action that is defined in the Minkowski spacetime. In order to do that, we first find the EOM for $P$ in the phase space action. That is $P^\mu=e^{-1}\dot{X}^\mu$. Replacing $P$'s in the phase space action gives us the following action
\begin{align}\label{I(x,e)}
    I(X,e)=\frac{1}{2}\int d\tau~(e^{-1}\dot{X}^2-em^2).
\end{align}
It is a simple task to verify that eliminating $e$ from this action using EOM of $e$ gives the original action \eqref{relpar}.
\paragraph{Generalisation to strings:}
We now generalise our construction to string theory. We begin with the Nambu Goto action:
\begin{align}\label{nambugoto}
    S_{NG}=T\int d^2\s \sqrt{-\det{\gamma_{\alpha\beta}}}.
\end{align}
The EOM for the above action takes the following form
 \begin{equation}
\Dot{\Pi}^{\mu}+K'^{\mu}=0,
\end{equation}
where $\Pi_{\mu}$ (the canonical momentum) and $K_{\mu}$ are given in \eqref{ctb7}. For strings propagating in Minkowski space, we see that $\Pi_{\mu}$ satisfies two primary constraints
\begin{equation}\label{ctb8}
    \Pi_{\mu}X'^{\mu}=0,~~~~~~~~~~
    \Pi^2+T^2X'^2=0.
\end{equation}
Just like point particles, we see that the canonical Hamiltonian for this action vanishes
\begin{align}
    \mathcal{H}_c = \Pi_{\mu}\dot{X}^{\mu}-\mathcal{L}=0,
\end{align}
implying worldsheet diffeomorphism invariance of the Nambu-Goto action. Here again, we use the primary constraints in \eqref{ctb8} to construct a Hamiltonian 
\begin{align}
    \mathcal{H}_{T}=\rho~\Pi_{\mu}X'^{\mu}+\frac{\lambda}{2}\left(\Pi^2+T^2X'^2\right),
\end{align}
where $\rho(\tau,\sigma)$ and $\lambda(\tau,\sigma)$ are Lagrange multipliers. Using this Hamiltonian we can obtain the phase space Lagrangian for bosonic strings. 
\begin{align}\label{phsplagst}
    \mathcal{L}_{P}=\Pi_{\mu}\dot{X}^{\mu}-\frac{\lambda}{2}\left(\Pi^2+T^2X'^2\right)-\rho~\Pi_{\mu}X'^{\mu}.
\end{align}
The corresponding phase space action for bosonic strings is given by
\begin{align}\label{phspacst}
    S_{PS}=\int~d^2\s~\left\{\Pi_{\mu}\dot{X}^{\mu}-\frac{\lambda}{2}\left(\Pi^2+T^2X'^2\right)-\rho~\Pi_{\mu}X'^{\mu}\right\}.
\end{align}
This form of the action is important for non-Lorentzian limits of string theory \cite{Cardona:2016ytk,Batlle:2016iel,Gomis:2016zur,Casalbuoni:2024jmj,Bagchi:2024rje}. Integrating out the canonical momentum from this phase space Lagrangian \eqref{phsplagst}, we obtain
\begin{align}
    \mathcal{L}_{P}=\frac{1}{2\lambda}\left[\Dot{X}^2-2\rho(\Dot{X}\cdot X')+(\rho^2-\lambda^2T^2)X'^2\right].
\end{align}
Identifying the coefficients of $\Dot{X}^2$, $\Dot{X}\cdot X'$ and $X'^2$ as metric components
\begin{align}\label{lagmul}
    g^{\alpha\beta}=\begin{bmatrix}1 & -\rho \\ -\rho & ~~\rho^2-\lambda^2T^2\end{bmatrix},\hspace{5mm}\sqrt{-g}=\sqrt{-\frac{1}{\text{det}g^{\alpha\beta}}}=\frac{1}{\lambda T}~,
\end{align}
we obtain the Polyakov form of the string action  
\begin{align}\label{Polyakov}
   S_{P}=\frac{T}{2}\int  d^2\s \sqrt{-g}g^{\alpha\beta}\partial_{\alpha}X^{\mu}\partial_{\beta}X^{\nu}\eta_{\mu\nu}.
\end{align}
We recall that for point particle case, the action \eqref{I(x,e)} could be interpreted as scalar field theory in the worldline (parametrized by $\tau$) of the particle where the Lagrange multiplier $e$ plays the role of square root of worldline dynamical metric. Here too, one can see that the action \eqref{Polyakov} is a scalar field theory on two dimensional worldsheet (parametrized by $\tau,\sigma$) where $g^{\alpha\beta}$ (which is constructed from Lagrange multipliers $\rho$ and $\lambda$) plays the role of worldsheet dynamical inverse metric.
\subsection{Classical analysis: Closed string}

The Polyakov action \eqref{Polyakov} is known to display Poincare symmetry in the target spacetime, worldsheet diffeomorphism invariance given by
    \begin{align}
         \s^\a\to\s'^\a=\s'(\s),
    \end{align}
    and Weyl invariance given by 
    \begin{align}
        g_{\a\b}\to g'_{\a\b}=e^{2\phi}g_{\a\b},\qquad \phi=\phi(\t,\s).
    \end{align}
Since the worldsheet diffeomorphism invariance and the Weyl invariance are gauge symmetries, one can fix the gauge to simplify the action. Since there are two coordinates to fix and the woldsheet metric has three independent components, by choosing $\s,\t$ appropriately we can always fix the metric to be conformally flat, i.e.
\begin{align}
   g_{\a\b} =e^{2\phi}\eta_{\a\b}.
\end{align}
Since the action \eqref{Polyakov} is Weyl invariant, the action constructed on $g_{\a\b} =e^{2\phi}\eta_{\a\b}$ always equals the action constructed on $g_{\a\b} = \eta_{\a\b}$. Hence, we can make the following gauge choice for the worldsheet metric
\begin{align}\label{cg}
   g_{\a\b} = \eta_{\a\b}.
\end{align}
This gauge choice is called conformal gauge.
\paragraph{Residual gauge symmetry:}
Even after this gauge fixing, \eqref{Polyakov} is found to be invariant under the following gauge transformation
\begin{align}\label{ResGauge}
    \s^+\to\s^+ + f(\s^+),\qquad \s^-\to\s^- + g(\s^-),\qquad \s^\pm=\t\pm\s.
\end{align}
Under the gauge transformation, a function $G(\s^+,\s^-)$ transforms as
\begin{align}\label{fg}
    \delta G=f(\s^+)\p_{+}G+g(\s^-)\p_{-}G~=~[\mathcal{L}(f)+\bar{\mathcal{L}}(g)]G.
\end{align}
In the above $\partial_{\pm}=\frac{1}{2}(\partial_\tau\pm\partial_\sigma)$. The generators of these transformation \eqref{fg} can be Fourier expanded as
\begin{subequations}\label{FfFg}
    \begin{align}
    \mathcal{L}(f)&=f(\s^+)\p_{+}=\sum_n a_n e^{in\s^+}\p_{+}=-i\sum_n a_n\mathcal{L}_n\\
    \bar{\mathcal{L}}(g)
    &=g(\s^-)\p_{-}=\sum_n b_n e^{in\s^-}\p_{-}=-i\sum_n b_n\bar{\mathcal{L}}_n.
\end{align}
\end{subequations}
In the above, we have used the Fourier expansions $f(\s^+)=\sum_n a_n e^{in\s^+}$ and $g(\s^-)=\sum_n b_n e^{in\s^-}$. From \eqref{FfFg} one can see that the symmetry transformations \eqref{ResGauge} are generated by the infinite number of local generators $\{\mathcal{L}_n,\bar{\mathcal{L}}_n\}$ given by
\begin{align}\label{Vircyl}
    \mathcal{L}_n=ie^{in\s^+}\p_{+},\qquad \bar{\mathcal{L}}_n=ie^{in\s^-}\p_{-}.
\end{align}
 These generators satisfy the Witt algebra given by 
 \begin{equation}\label{witt}
      [\mathcal{L}_n,\mathcal{L}_m]=(n-m)\mathcal{L}_{n+m},\qquad 
      [\bar{\mathcal{L}}_n,\bar{\mathcal{L}}_m]=(n-m)\bar{\mathcal{L}}_{n+m},\qquad 
      [\mathcal{L}_n,\bar{\mathcal{L}}_m]=0. 
\end{equation}
Upon quantisation of this theory, we encounter the centrally extended version of this algebra called the Virasoro algebra given by
\begin{equation}\label{Virasoro}
  \begin{split}
      [\mathcal{L}_n,\mathcal{L}_m]=&(n-m)\mathcal{L}_{n+m}+\frac{c}{12}n(n^2-1)\delta_{n+m},\\
      [\bar{\mathcal{L}}_n,\bar{\mathcal{L}}_m]=&(n-m)\bar{\mathcal{L}}_{n+m}+\frac{\bar{c}}{12}n(n^2-1)\delta_{n+m},\\
      [\mathcal{L}_n,\bar{\mathcal{L}}_m]=&0,
  \end{split}  
\end{equation}
where $c$ and $\bar{c}$ are the central charges of the Virasoro algebra. This is famously also the symmetry algebra of a relativistic 2d CFT. 

\paragraph{Equations of motion and constraints:}
Polyakov action \eqref{Polyakov} gives us the following EOM of the fields $X^\m$
\begin{align}
     \square X^\m=0. 
\end{align}
In the conformal gauge \eqref{cg} this equation simplifies to the free wave equation
\begin{align}\label{fwe}
    \p_\a\p^\a X^\m=0.
\end{align}
The Polyakov action \eqref{Polyakov} also contains the worldsheet metric $g_{\a\b}$ and there will be EOM of these fields. Since these fields are not dynamic, these EOM act as constraints on $X^\m$. Since the variation of the action with respect to the metric gives us the stress energy tensor $T_{\a\b}$, the constraints are given by
\begin{align}\label{emt}
   T_{\a\b}= -\frac{2}{T}\frac{1}{\sqrt{-g}}\frac{\de S_P}{\de g^{\a\b}}=0.
\end{align}
Solving \eqref{emt} for Polyakov action and applying the conformal gauge $g_{\a\b}=\eta_{\a\b}$ one gets the following
\begin{subequations}\label{tensileem}
    \begin{align}
        T_{00}=&T_{11}=\frac{1}{2}(\dot{X}^2+X'^2)=0\\
        &T_{01}=\dot{X}\cdot X'=0
    \end{align}
\end{subequations}
The constraints in \eqref{tensileem} can be rewritten in terms of worldsheet lightcone coordinates as
\begin{align}\label{constlight}
    T_{++}=(\p_{+}X)^2=0,\qquad T_{--}=(\p_{-}X)^2=0.
\end{align}
\paragraph{Mode expansion of closed strings:} Now, we look into the most generic solution to the EOM \eqref{fwe}. The solution of \eqref{fwe} can be Fourier expanded as below 
\begin{equation}\label{JackReacher}
\begin{split}
     X^{\mu}(\tau,\sigma)=x^{\mu}+\sqrt{\frac{\alpha'}{2}} \alpha_0^\mu(\tau+\s)&+\sqrt{\frac{\alpha'}{2}} \tilde\alpha_0^\mu(\tau-\s)\\+i\sqrt{\frac{\alpha'}{2}}&\sum_{n\neq0}\frac{1}{n}\left[\alpha^{\mu}_{n}e^{-in(\tau+\sigma)}+\tilde\alpha^{\mu}_{n}e^{-in(\tau-\sigma)}\right].
\end{split}
\end{equation}
We are interested in closed strings. So $\s$ and $\s+2\pi$ need to be identified, which implies boundary condition $X^{\m}(\t,\s+2\pi)=X^\m(\t,\s)$ and $\alpha_0=\tilde\alpha_0$. We end up with 
\begin{align}\label{tensilemodeexp}
    X^{\mu}(\tau,\sigma)=x^{\mu}+2\sqrt{\frac{\alpha'}{2}} \alpha_0^\mu\tau+i\sqrt{\frac{\alpha'}{2}}\sum_{n\neq0}\frac{1}{n}\left[\alpha^{\mu}_{n}e^{-in(\tau+\sigma)}+\tilde\alpha^{\mu}_{n}e^{-in(\tau-\sigma)}\right].
\end{align}
In the above $\alpha^\m_0=\tilde\alpha^\m_0=\sqrt{\frac{\a'}{2}}k^\m$, $k^\m$ being the momentum of the closed string. In order to see this more plainly, we first look into the momentum conjugate of $X^\m$ is given by 
\begin{equation}\label{chamchike}
\begin{split}
   &\Pi_{\mu}=\frac{\partial\mathcal{L}}{\partial\Dot{X}^{\mu}}=\frac{1}{2\pi \a'}\Dot{X}^{\nu}\eta_{\nu\mu}, 
\end{split}
\end{equation}
The total momentum of the string can be found by integrating the canonical momentum over the entire string i.e.,
\begin{equation}\label{totalm}
\begin{split}
     k^\mu=\int_{0}^{2\pi} d\s~\Pi^{\mu}=\frac{1}{2\pi \a'}&\int_{0}^{2\pi} d\s~\Dot{X}^{\mu}=\frac{1}{2\pi \a'}~ 2\pi\sqrt{2\a'}\a^{\mu}_{0},\\
     \implies & \a^{\mu}_{0}=\sqrt{\frac{\a'}{2}}k^{\mu}.
\end{split}
\end{equation}
Now, we look into the Poisson brackets between target spacetime coordinates $X^\m$ and there momentum conjugate $\Pi^\m$. They are
\begin{subequations}\label{P.B.}
\begin{equation}
    \{X^{\mu}(\s,\t),X^{\nu}(\s',\t)\}_{P.B.}=\{\Pi^{\mu}(\s,\t),\Pi^{\nu}(\s',\t)\}_{P.B.}=0,
    \end{equation}
    \begin{equation}
        \{X^{\mu}(\s,\t),\Pi^{\nu}(\s',\t)\}_{P.B.}=\eta^{\mu\nu}\delta(\s-\s').
\end{equation}
\end{subequations}
Using the mode expansion of $X^\m$ as given in \eqref{tensilemodeexp} and the expression of canonical momentum density $\Pi^\m$ as given in \eqref{chamchike}, one can translate \eqref{P.B.} to the Poisson bracket between the modes $\a,\tilde{\a}$
\begin{equation}\label{alphaalgebra}
\begin{split}
    \{\alpha_m^\mu,\alpha_n^\nu\}=-im\delta_{m+n,0}\eta^{\mu\nu},~~~\{\tilde \alpha_m^\mu,\tilde \alpha_n^\nu\}=-im\delta_{m+n,0}\eta^{\mu\nu},~~~\{\alpha_m^\mu,\tilde \alpha_n^\nu\}=0.
\end{split}
    \end{equation}
Using the mode expansion in \eqref{tensilemodeexp}, the constraints\eqref{constlight} can be rewritten as
\begin{subequations}
    \begin{align}
    \label{ambagan}
    T_{++}=(\p_{+}X)^2=\a'\sum_{n}\mathcal{L}_n e^{-in\s^+},
   \qquad  T_{--}=(\p_{-}X)^2=\a'\sum_{n}\bar{\mathcal{L}}_n e^{-in\s^-}.
\end{align}
\end{subequations}
In the above, we have defined
\begin{align}\label{Review3}
    \mathcal{L}_n=\frac{1}{2}\sum_m \a_{m}\cdot\a_{n-m} ,~~~\bar{\mathcal{L}}_n=\frac{1}{2}\sum_m \tilde\a_{m}\cdot\tilde\a_{n-m}.
\end{align}
The Poisson brackets of $\mathcal{L}_n$ and $ \bar{\mathcal{L}}_n$ can be calculated from the Poisson brackets of $\a,\tilde\a$
\begin{align}\label{VirasoroPoisson}
    \{\mathcal{L}_n,\mathcal{L}_m\}=-i(n-m)\mathcal{L}_{n+m},~~~\{\bar{\mathcal{L}}_n,\bar{\mathcal{L}}_m\}=-i(n-m)\bar{\mathcal{L}}_{n+m},~~~\{\mathcal{L}_n,\bar{\mathcal{L}}_m\}=0.
\end{align}
This is of course the classical version of the Virasoro algebra, now arising out of the mode expansion of the solutions of the EOM and the constraints. The methods of relativistic 2d CFT, built out of the Virasoro algebra, would be central to the understanding of the quantum aspects of the closed bosonic string, which we now go on to describe. 
\subsection{Quantum Theory}
In the quantum theory, all the observables are promoted to operators and consequently the Poisson brackets between those observables are replaced with commutators. The equal time commutator between $X(\s,\t)$ and $\Pi(\s,\t)$
\begin{equation}
 \begin{split}     \left[X^\mu(\tau,\sigma),\Pi^\nu(\tau,\sigma')\right]=i\eta^{\mu\nu}\delta (\sigma-\sigma').
     \end{split}
 \end{equation}
Similarly the Poisson brackets \eqref{alphaalgebra} are replaced by
\begin{equation}\label{alphaalgebra2}
\begin{split}
    [\alpha_m^\mu,\alpha_n^\nu]=m\delta_{m+n,0}\eta^{\mu\nu},\qquad [\tilde \alpha_m^\mu,\tilde \alpha_n^\n]=m\delta_{m+n,0}\eta^{\mu\nu},\qquad [\alpha_m^\mu,\tilde \alpha_n^\nu]=0.
\end{split}
    \end{equation}
Consequently the Poisson brackets in \eqref{VirasoroPoisson} will be replaced by Virasoro algebra described in \eqref{Virasoro}. Now, let us look into the Hilbert space of the theory. The ground state is defined as
\begin{subequations}
    \begin{align}
&\a^\m_0\ket{0,k^\m}_{\a'}=\tilde{\a}^\m_0\ket{0,k^\m}_{\a'}=\sqrt{\frac{\a'}{2}}k^\m\ket{0,k^\m}_{\a'}\\&\a^\m_n\ket{0,k^\m}_{\a'}=\tilde{\a}^\m_n\ket{0,k^\m}_{\a'}=0,~~~~\forall n>0.
\end{align}
\end{subequations}
One can construct the basis of the Hilbert space by acting the creation operators $\{\alpha_{-n}^\mu,\tilde \alpha_{-n}^\mu\}$ with $n>0$ on the vacuum. A generic state in this basis will be
    \begin{align}\label{Dadhich1}
    &\ket{n_L,n_R}=\sum_{j}\rho_{j}\Bigg(\prod_{i=1}^{p}\a^{a_{i}}_{-m_{i}}\prod_{j=1}^{q}\tilde{\a}^{b_{j}}_{-n_{j}}\Bigg)_{j}\ket{0,k^{\mu}}_{\a'}, \quad n_L=\sum_{i}^{p}a_{i}m_{i}, \, n_R=\sum_{i}^{q}b_{i}n_{i}.
\end{align}
 In the above $n_L$ and $n_R$ respectively are the eigenvalues of the number operator which determines the level of the state
 \begin{align}
     N=\sum_{m>0}\a_{-m}\cdot \a_{m};\hspace{5mm} \widetilde{N}=\sum_{m>0}\tilde{\a}_{-m}\cdot \tilde{\a}_{m}.
 \end{align}
Now, we need to impose the quantum version of the constraints \eqref{constlight} on this Hilbert space in order to find the physical states. The most general constraint one can impose is a quantum version of the vanishing of the stress tensor, where the corresponding matrix elements vanish. This is given by
\begin{equation}\label{Dadhich3}
     \bra{phys'}T_{++}\ket{phys}=\bra{phys'}T_{--}\ket{phys}=0.
 \end{equation}
Rewriting \eqref{Review5} in terms of $\mathcal{L}_n$ and $\bar{\mathcal{L}}_n$ one gets the following
\begin{equation}\label{Dadhich4}
\bra{phys'}\mathcal{L}_n\ket{phys}=0,~~~\bra{phys'}\bar{\mathcal{L}}_n\ket{phys}=0, ~~\forall n\in \mathbb{Z}.
 \end{equation}
However, in usual tensile string theory one chooses to impose a more stringent constraint. We demand that the physical states belong to the highest weight representation of Virasoro algebra\footnote{Although string theory constructed on this constraint is the most studied, there are string theories defined on other kind of constraints which are consistent with the sandwich condition \eqref{Dadhich3}. In later part of the review we shall see that there is a variant of string theory which classically has the same Polyakov action but the quantum theory of which is constructed on a different representation of Virasoro algebra. A recent work \cite{Bagchi:2024tyq} has aimed towards more general studies on the Virasoro sandwich constraint.}.
\begin{align}\label{Pataliputra}
    (\mathcal{L}_n-a\delta_{n,0})\ket{phys}=(\bar{\mathcal{L}}_n-a\delta_{n,0}
    )\ket{phys}=0.
\end{align}
In the above, $a$ is the normal ordering constant of both $\mathcal{L}_0$ and $\bar{\mathcal{L}}_0$. Value of $a$ is $1$ and it can be shown from the Lorentz invariance of this theory in the target spacetime\footnote{The same calculation also simultaneously shows that the target spacetime dimension should be $D=26$ (critical dimension). We do not provide the details of the calculations here. However, in upcoming sections we shall see that tensionless null strings too, have same critical dimensions which will be discussed in detail.}. The operators $\mathcal{L}_0$ and $\bar{\mathcal{L}}_0$ can be expressed in terms of the number operators as 
\begin{align}\label{Virzeromodes}
\mathcal{L}_0=N+\frac{1}{2}\a^2_0=N+\frac{\a'}{4}k^2,~~~~~\bar{\mathcal{L}}_0=\widetilde{N}+\frac{1}{2}\tilde{\a}^2_0=\widetilde{N}+\frac{\a'}{4}k^2.
\end{align}
In the above, we have used the fact that $\alpha^\m_0=\tilde\alpha^\m_0=\sqrt{\frac{\a'}{2}}k^\m$. Applying \eqref{Virzeromodes} to the physical state condition \eqref{Pataliputra}, and recalling that $m^2=-k^2$, one can see that the mass for a string state $\ket{n_L,n_R}$ is given by 
\begin{align}
m^2_L=\frac{4}{\a'}(n_L-1),~~~~m^2_R=\frac{4}{\a'}(n_R-1).
\end{align}
In the above, $m^2_L$ is the mass-squared calculated from the $\mathcal{L}_0$ physical state condition and $m^2_R$ is the mass-squared calculated from the $\bar{\mathcal{L}}_0$ physical state condition. For sake of consistency we must have $m^2_L=m^2_R=m^2$ which essentially leads us to level matching condition
\begin{align}\label{tenlevmat}
    n_L=n_R=n.
\end{align}
Hence, we end up with the following mass spectrum for tensile closed bosonic strings
\begin{align}
    m^2=\frac{4}{\a'}(n-1).
\end{align}
From here one can deduce the nature of particles emerging from the bosonic string mass spectrum. Ground state ($n=0$) is tachyonic while first excited state $n=1$ gives massless spin two particles which lead to gravitons, antisymmetric Kalb-Ramond background fields and dilatons.
\medskip

For a quick recap of tensile open strings and RNS formalism of tensile superstrings one can see Appendices \ref{Open Strings} and \ref{Superstring}.
\bigskip

\newpage

\section{A Summary of Carrollian Symmetry}\label{A Summary of Carrollian Symmetry}

In the introduction, we mentioned that on the worldsheet of the tensionless null string, the relativistic Virasoro algebra is replaced by the Conformal Carrollian algebra (CCA). As with the Virasoro algebra, these residual gauge symmetries which form the CCA, now dictate the structure of the theory of the null string. In this section, we will review the basic ingredients of Carrollian and Conformal Carrollian symmetries and focus on two dimensional Carrollian CFTs. These methods would be crucial for our analysis of the null string. 

\subsection{Non-Lorentzian limits: Galilean symmetry}

The isometry group of flat Minkowski spacetimes in $D$ dimensions is the Poincare group $ISO(D-1,1)$: 
\begin{align}\label{chha1}
[P_{\mu},P_{\nu}]=0,  \quad [J_{\mu\nu},P_{\rho}]=-2\eta_{\rho[\mu}P_{\nu]}, \quad [J_{\mu\nu},J_{\rho\sigma}]=4\eta_{[\mu[\rho}J_{\sigma]\nu]}.
\end{align}
Here $P_{\mu}=-\partial_{\mu}, J_{\mu\nu}=x_{\mu}\partial_{\nu}-x_{\nu}\partial_{\mu}$ are generators of translations and Lorentz transformations respectively. We will be interested in understanding the non-relativistic limit of this algebra and this gives the Galilei algebra, the commutators of which are given by
\begin{align}\label{chha2}
    [J_{ij},J_{kl}]=4\delta_{[i[k}J_{l]j]},~~ \, [J_{ij},P_{k}]=-2\delta_{k[i}P_{j]},~~\, [J_{ij},G_{k}]=2\delta_{k[i}G_{j]}, ~~\, [G_{i},H]=-P_{i}.
\end{align}
The process of getting to the Galilean algebra from the Poincare algebra is called a In\"on\"u-Wigner contraction \cite{Inonu:1953}. As an illustrative example, the Lorentz boosts contract to the Galilean boosts as follows:
\begin{align}
J_{0i}=t\partial_{i}+x_{i}\partial_{t}\to \frac{1}{\epsilon}t\partial_{i}+\epsilon x_{i}\partial_{t} \quad \implies G_{i}=\lim_{\epsilon\to 0}\epsilon J_{0i}=t\partial_{i}.
\end{align}
This algebra admits a central extension where there is another non-vanishing commutator $[G_i,P_j]=\delta_{ij} M$, where $M$ is a constant which can be identified with the non-relativistic mass. The centrally extended Galilean algebra is sometimes called the Bargmann algebra. 

\medskip

Gravity in the non-relativistic world or Newtonian gravity admits a geometric formulation known as Newton-Cartan geometry \cite{Cartan:1923zea}. (Pseudo)-Riemannian structures of general relativity are replaced by Newton-Cartan manifolds in Newtonian gravity. These structures form the basis of a geometric description of Galilean symmetry. A Newton-Cartan manifold is described by a doublet $\{h^{ab}, \tau_a\}$, where $h^{ab}$ is an inverse degenerate metric and $\tau_a$ is a vector that generates its kernel:
$$h^{ab}\tau_a=0.$$
The isometry of the flat version of this structure
\begin{eqnarray}\label{CLa}
		\pounds_{\xi}{h}^{ab}=0,~~~~\quad \pounds_{\xi}\tau_{a}=0.
\end{eqnarray}
is the Galilean algebra defined above. In the above $\pounds$ denotes a Lie derivative. 

\subsection{Non-Lorentzian limits: Carrollian Symmetry}
While this $c\to\infty$ limit is the most well explored singular limit of relativistic physics, there is another singular limit which has been of great interest in recent years. This limit where $c\to 0$, initially introduced by Levy-Leblond \cite{Leblond65} and Sen Gupta \cite{SenGupta:1966qer}. This leads to the Carroll algebra, the non-vanishing commutators of which are given by
\begin{align}\label{chha3}
    [J_{ij},J_{kl}]=4\delta_{[i[k}J_{l]j]},~\, [J_{ij},P_{k}]=-2\delta_{k[i}P_{j]},~\, [J_{ij},C_{k}]=-2\delta_{k[i}C_{j]},~\,
 [P_{i},C_{j}]=\delta_{ij}H.
\end{align}
To contrast it from the Galilean contraction, it is instructive to note that the Carrollian contraction ($t\to \epsilon t$, $x^{i}\to x^{i}$, $\epsilon\to 0$) of the Lorentz boosts leads to
\begin{align}
     J_{0i} =t\partial_{i}+x_{i}\partial_{t}\to \e t\partial_{i}+ \frac{1}{\epsilon}x_{i}\partial_{t} \quad \implies C_{i}=\lim_{\epsilon\to 0}\epsilon J_{0i}=x_{i}\partial_{t}.
\end{align}
The Carroll algebra has certain peculiarities, e.g. $H=-\partial_{t}$, the Hamiltonian, is a central element. This in particular means Carroll boost commute with the Hamiltonian $[C_i, H]=0$, which means boosting a system does not change its energy. 

\medskip

We briefly discuss the geometry behind the Carroll algebra, starting with Minkowski spacetime, and taking the limit $c\to0$. For a more in-depth review of the geometric structures associated with Carroll symmetry, the reader is pointed to \cite{Ciambelli:2025unn}. The Minkowski metric in the Carroll limit gives: 
\begin{eqnarray}\label{FlC}
\eta_{\mu\nu} \rightarrow h_{\mu\nu} = \begin{bmatrix}
			0 & 0 \\
			0 & \quad \mathbb{1}_{D-1}
		\end{bmatrix}, 
		\qquad
		-c^2 \eta^{\mu\nu} \rightarrow v^\mu v^\nu = 	\begin{bmatrix}
			1 & 0 \\
			0 & \quad \mathbb{0}_{D-1}
		\end{bmatrix}   \, .
\end{eqnarray}   
So we see that in the Carroll limit, $D$-dimensional Lorentzian metric has become a degenerate rank $D-1$ Carroll metric $h_{\m\n}$ while The inverse metric $\eta^{\m\n}$ has evolved into degenerate inverse metric of rank $1$, and hence has been expressed as $v^\m v^\n$. These satisfy: 
$$h_{\mu\nu}v^{\nu}=0.$$
A generic Carrollian manifold, as opposed to the flat Carroll manifold obtained from Minkowski spacetime above, is defined as a manifold of dimension $D$ endowed with a symmetric, degenerate rank $(D-1)$ metric $h_{ab}$ and a vector $v^a$ that generates the kernel of $h$. The structure defined is a fibre bundle with the base space is formed by $D-1$ spatial dimensions and a one-dimensional fibre which is the null direction. The isometry algebra of flat Carroll manifolds 
\begin{eqnarray}\label{CLaa}
		\pounds_{\xi}h_{\mu\nu}=0, ~~~\quad \pounds_{\xi}v^{\mu}=0,
	\end{eqnarray}
is identical to the Carroll algebra we have discussed above. 

\medskip

The Carrollian manifolds and their non-relativistic cousins, the Newton-Cartan manifolds are very similar to each other and are related to each other by a base $\leftrightarrow$ fibre flip. Interestingly, for $D=2$, the base and fibre are both one dimensional and for the flat case, the symmetry algebras, the Galilean and Carrollian algebras are isomorphic. The isomorphism extends to the conformal variants, viz. the Galilean and Conformal Carrollian Algebras. We will see that this will play an important role in the discussion of tensionless strings.

In recent years, Carroll symmetry has found applications in various corners of the relativistic world. Generic null hypersurfaces exhibit Carrollian symmetry, and this fact plays a key role in null/tensionless string theory where the worldsheet of the string becomes null and we will observe Carrollian geometry on the worldsheet. Some of the classic examples of null hypersurface are light-cones, black hole event horizons, and null infinity of asymptotically flat spacetime. For a detailed review of Carrollian physics, the reader is referred to the recent review \cite{Bagchi:2025vri}.

\subsection{Conformal Field Theory and Carroll Limit}
One can generalize the above discussion of non-Lorentzian structures to involve conformal symmetries. In relativistic spacetimes of generic dimensions conformal transformations are those which transform the metric by a scale: 
\begin{align}\label{Relcondef}
    \pounds_{\xi}g_{\mu\nu}=\Lambda g_{\m\n}.
\end{align}
Conformal field theories are the field theories invariant under these transformations. The conformal isometries of flat spacetimes yields the relativistic conformal algebra. In two dimensions, this symmetry algebra enhances to form two copies of the Virasoro algebra: 
\begin{equation}
 \begin{split}
     [\mathcal{L}_n,\mathcal{L}_m]=&(n-m)\mathcal{L}_{n+m}+\frac{c}{12}n(n^2-1)\delta_{n+m},\\
     [\bar{\mathcal{L}}_n,\bar{\mathcal{L}}_m]=&(n-m)\bar{\mathcal{L}}_{n+m}+\frac{\bar{c}}{12}n(n^2-1)\delta_{n+m},\\
     [\mathcal{L}_n,\bar{\mathcal{L}}_m]=&0.
 \end{split}  
\end{equation}
Here $c$ and $\bar{c}$ are the central charges of the Virasoro algebra.
We recall that, we have already encountered the generators \eqref{Vircyl} as the residual symmetry generators of bosonic closed string theory.

\paragraph{Carrollian Conformal Algebra:}
The conformal isometries of Carrollian manifolds are given by: 
\begin{eqnarray}\label{con-car}
		\pounds_{\xi}h_{\mu\nu}=\lambda_1 h_{\mu\nu},\qquad \pounds_{\xi}v^{\mu}= {\lambda_2}v^{\mu}, \qquad \frac{\lambda_1}{\lambda_2}=-\frac{z}{2}
	\end{eqnarray}
In the above, $z$ is the anisotropy parameter, also called a dynamical critical exponent, which when set to $z=1$ gives an isotropic scaling of spatial and temporal directions. In this review, we will only be concerned with $z=1$. The conformal isometries of flat Carrollian manifolds generates the Conformal Carroll algebra. We will focus on dimensions $D=2$. Here the solution to the above equation reads: 
\begin{equation}  
\xi=[g(x^1)+x^0f^\prime(x^1)]\partial_0+f(x^1)\partial_1.
\end{equation} 
For later purposes it is useful to rewrite the above as
\begin{equation}
L(f)=x^0\,f^\prime(x^1)\,\partial_{0}+f(x^1)\,\partial_1, \quad\quad M(g)=g(x^1)\partial_{0}.
\end{equation}
In cylindrical coordinates $(x^0, x^1)=(\t,\s)$ with $\s \sim \s+2\pi$, the functions $f, g$ are periodic and can be Fourier expanded. In terms of these modes, the generators of the Conformal Carroll algebra on the cylinder become: 
\begin{align}\label{cylinder}
    L_n &=i\,e^{in\sigma}(\partial_\sigma+in\tau\,\partial_\tau),\quad M_n= i\,e^{in\sigma}\partial_\tau.
\end{align}
Here $\tau$ is the null temporal direction and $\sigma$ parametrizes the angular coordinate of $S^1$.
These generators \eqref{cylinder} satisfy the $2d$ Carrollian Conformal Algebra (CCA$_2$): 
\begin{subequations}\label{bms3}
  \begin{align}
    \left[L_n,L_m\right]&=(n-m)L_{n+m} +\frac{c_L}{12}(n^3-n)\delta_{n+m}, \\
    \left[L_n,M_m\right]&=(n-m){M}_{n+m}+\frac{c_M}{12}(n^3-n)\delta_{n+m},\\
    \left[M_n, M_m\right]&=0.
\end{align}  
\end{subequations}
Here $c_L$ and $c_M$ are the central charges for the algebra. The 2d CCA is isomorphic to $3$-dimensional Bondi-van der Burgh-Metzner-Sachs 
(BMS$_3$) algebra \cite{Duval:2014uva}, which is the symmetry algebra of asymptotically flat 3d spacetimes at the null boundary \cite{Bondi:1962px,Sachs:1962wk, Ashtekar:1996cd, Barnich:2006av}. This fact plays a pivotal role in the Carrollian framework of holography of asymptotically flat spacetime \cite{Bagchi:2010zz, Bagchi:2012cy, Duval:2014uva, Bagchi:2016bcd, Bagchi:2022emh, Donnay:2022aba}. This algebra, as advertised before, will play the central role in the study of null strings since it appears as the worldsheet residual gauge symmetry algebra of the null strings. 
\begin{figure}[t]
\centering
    \includegraphics[scale=0.38]{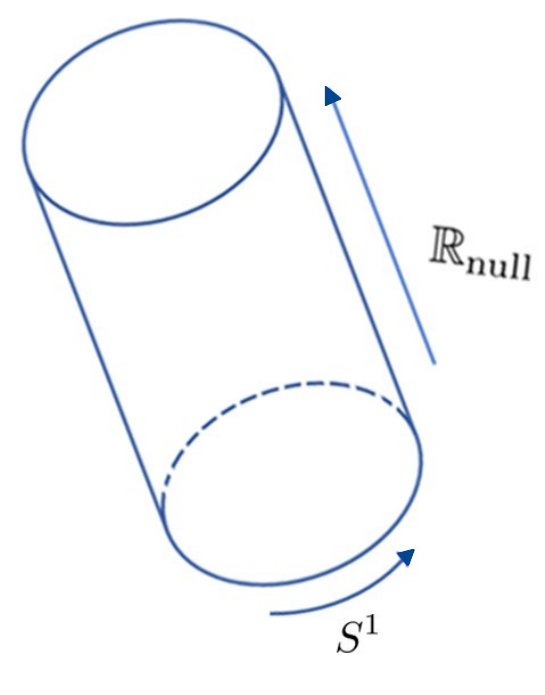}
    \caption{Null cylinder}
    \label{fig2}
\end{figure}

\medskip
 
The 2d CCA can be obtained from two copies of the Virasoro algebra by 
the following contraction
\begin{equation}\label{Review4}
    L_n=\mathcal{L}_n-\mathcal{\bar{L}}_{-n},\quad\quad M_n=\epsilon(\mathcal{L}_n+\mathcal{\bar{L}}_{-n}), \qquad \epsilon \to 0.
\end{equation}
It is also important to note that we can impose the Carroll limit as 
\begin{align}\label{car-cyl-lim}
    \t \to \epsilon \t, \quad \s \to \s, \quad \epsilon \to 0
\end{align}
on the cylinder coordinates. Applying this, together with \eqref{Review4} together with  on the expression of the vector fields representations of the Virasoro generators on the cylinder, one obtains \eqref{cylinder}. 

\paragraph{Conformal Carrollian stress tensors:}\label{CCST}
We will be interested in understanding stress tensors for the residual symmetry algebra of the null string in our attempts to quantise the theory and hence it is important to first understand the notion of stress tensors in 2d Carrollian CFTs. Stress tensors in Carroll CFTs have been obtained intrisically \cite{Dutta:2022vkg, Saha:2022gjw}. Here, we derive this in the Carroll limit following \eqref{Review4} of a 2d relativistic CFT following \cite{Bagchi:2015wna, Bagchi:2013bga}. The holomorphic and antiholomorphic stress tensors for a 2d relativistic CFT defined on the cylinder are
\begin{align}\label{Tcyl}
    T_{cyl} = \sum_n \L_n e^{in \omega} - \frac{c}{24}, \quad \bar{T}_{cyl} = \sum_n \bL_n e^{in \omega} - \frac{\bar{c}}{24}.
\end{align}
We now impose the  limit \eqref{Review4} along with \eqref{car-cyl-lim}
 and define\footnote{This kind of linear combinations of stress tensors were first considered for the non-relativistic case \cite{Bagchi:2010vw}.}
\begin{align}\label{EMccft}
    & T_1 (\sigma, \t) = \lim_{\e\to0} \left( T_{cyl} - \bar{T}_{cyl} \right), \quad  T_2 (\sigma, \t) = \lim_{\e\to0} \e \left( T_{cyl} + \bar{T}_{cyl} \right).
\end{align}
This results in
\begin{align}\label{Carr-T}
    T_1 (\sigma, \t) = \sum_n \left(L_n - in \t M_n\right) e^{-in\sigma} + \frac{c_L}{24}, \quad T_2 (\sigma) = \sum_n M_n e^{-in\sigma} + \frac{c_M}{24}. 
\end{align}
$T_1$ and $T_2$ are the stress tensors of the 2d Carroll CFT (on the cylinder). We can now inverting the above relations to rewrite the generators of the algebra in terms of the stress tensors as follows
\begin{align}
    L_n = \int d\sigma \, (T_1 + in\t T_2) e^{in\sigma}, \quad M_n = \int d\sigma \, T_2 e^{in\sigma}.
\end{align}
The central terms in \eqref{Carr-T} results in shifting of the zero modes $L_0$ and $M_0$. The expressions for the stress tensors will show up prominently in our analysis of the constraints on the null worldsheet in the following sections.

\subsection{Representation Theory of CCA}
We now look at representations of CCA$_2$ and their relations with representations of Virasoro algebra. This would be central to our discussions of the quantum null strings. 

\paragraph{Induced Representation:}
The most extensively studied representation of Virasoro algebra is the highest weight representation (HWR). Here primary states are defined as 
\begin{align}\label{h.w.r.}
\mathcal{L}_0\ket{h,\bar{h}}=h\ket{h,\bar{h}},~~~\bar{\mathcal{L}}_{0}\ket{h,\bar{h}}=\bar{h}\ket{h,\bar{h}},~~~\mathcal{L}_n\ket{h,\bar{h}}=\bar{\mathcal{L}}_{n}\ket{h,\bar{h}}=0,~~~\forall n>0.
\end{align}
Descendant states are constructed by operating $\mathcal{L}_{n}$s and $\bar{\mathcal{L}}_{n}$s (where $n<0$) on primary states. Together, the primary and its descendants form a Verma module, and different Verma modules build the entire Hilbert space.
\medskip

The Carroll contraction \eqref{Review4} maps Virasoro highest weight states to the rest-frame states of the BMS$_3$ (or CCA$_2$) induced representation \cite{Campoleoni:2016vsh}. Let us assume that in the Carroll limit, the Virasoro primary $\ket{h,\bar{h}}$ maps to $\ket{M,s}$, where
\begin{align}\label{induced state}
    L_{0}\ket{M,s}=s\ket{M,s},~~~M_{0}\ket{M,s}=M\ket{M,s}.
\end{align}
Using \eqref{h.w.r.} in \eqref{Review4} and \eqref{induced state} leads us to the following
\begin{align}
\Bigg(\frac{M_{\pm n}}{\epsilon}\pm L_{\pm n}\Bigg)\ket{M,s}=0, ~~~~s=h-\bar{h}~~~~M=\epsilon(h+\bar{h}).
\end{align}
At $\epsilon\to 0$ limit this leads to the following property of $\ket{M,s}$
\begin{align}\label{induced}
M_n\ket{M,s}=0~~~\forall n\neq 0.
\end{align}
Such states are termed as rest-frame states of the induced representation of BMS$_3$ algebra \cite{Campoleoni:2016vsh}. A BMS induced module is constructed by operating $L_n$s (for all $n\neq 0$) on this state. For more details, the reader is referred to \cite{Barnich:2014kra,Oblak:2016eij,Campoleoni:2016vsh}.

\paragraph{Highest Weight Representation:} The CCA$_2$ also admits a highest weight representation. Here, following the ideas of the Virasoro algebra, Carroll primary states are defined as
\begin{subequations}\label{BMSHWR}
    \begin{align}
L_0&\ket{h_L,h_M}=h_L\ket{h_L,h_M},~~~M_0\ket{h_L,h_M}=h_M\ket{h_L,h_M},\\&~~~~~~~L_n\ket{h_L,h_M}=M_n\ket{h_L,h_M}=0~~~\forall n>0.
\end{align}
\end{subequations}
Together with its descendants, created by acting both $L_n$s and $M_n$s with $n<0$, this primary state construct a highest weight module module. As already highlighted, Virasoro HWR at Carroll limit leads to the induced representation and not the HWR of CCA$_2$. However, Virasoro algebra also admits a different representation which at Carroll limit would lead us to the HWR of CCA$_2$.

\paragraph{Virasoro Flipped Representation:}
To understand the relativistic origin of the Carrollian HWR, we note that the Virasoro algebra allows an automorphism
\begin{align}\label{automorphism}
    \bar{\mathcal{L}}'_{n}=-\bar{\mathcal{L}}_{-n}.
\end{align}
Applying automorphism on the antiholomorphic sector, and keeping the holomorphic sector untouched, if we consider HWR of \{$\mathcal{L}_n,\bar{\mathcal{L}}'_{n}$\}, we are led to a new representation, since HWR of $\bar{\mathcal{L}}'_{n}$ are lowest weights in $\bar{\mathcal{L}}_{n}$. Primary states of the HWR in \{$\mathcal{L}_n,\bar{\mathcal{L}}'_{n}$\} is given by 
 \begin{align}\label{l.w.r.}
\mathcal{L}_0\ket{h,\bar{h}}=h\ket{h,\bar{h}},~~\bar{\mathcal{L}}'_{0}\ket{h,\bar{h}}=\bar{h}\ket{h,\bar{h}},~~&\mathcal{L}_n\ket{h,\bar{h}}=\bar{\mathcal{L}}'_{n}\ket{h,\bar{h}}=0~~\forall n>0.
\end{align}
In terms of $\{\mathcal{L}_n,\bar{\mathcal{L}}_{n}\}$ the definition \eqref{l.w.r.} becomes
\begin{align}\label{flipped}
    \mathcal{L}_0\ket{h,\bar{h}}=h\ket{h,\bar{h}},~~\bar{\mathcal{L}}_{0}\ket{h,\bar{h}}=-\bar{h}\ket{h,\bar{h}},~~&\mathcal{L}_n\ket{h,\bar{h}}=\bar{\mathcal{L}}_{-n}\ket{h,\bar{h}}=0~~\forall n>0.
\end{align}
This representation is called the flipped representation of Virasoro algebra. Tensile string theory constructed on Polyakov action has a cousin theory which classically shares the same Polyakov action but the quantum theory is constructed on this representation. The theory is called ``twisted" string theory and has been studied in \cite{Lee:2017utr,Casali:2017mss,Lee:2017crr}.
\paragraph{Carroll Limit of Flipped Representation:}
We will now see that applying the Carroll contraction \eqref{Review4} on the flipped representation of Virasoro algebra \eqref{flipped} gives HWR of CCA$_2$. Performing the contraction \eqref{Review4} on the generators \{$\mathcal{L}_n,\bar{\mathcal{L}}'_{n}$\}, we get:  
\begin{align}
     L_{n}=\mathcal{L}_{n}-\bar{\mathcal{L}}'_{-n},\hspace{5mm}M_{n}=\epsilon(\mathcal{L}_{n}+\bar{\mathcal{L}}'_{-n}).
\end{align}
Rewriting this in terms of $\{\mathcal{L}_n,\bar{\mathcal{L}}_{n}\}$ we get
\begin{align}\label{nrcon}
L_{n}=\mathcal{L}_{n}+\bar{\mathcal{L}}_{n},\hspace{5mm}M_{n}=\epsilon(\mathcal{L}_{n}-\bar{\mathcal{L}}_{n}).
\end{align}
This contraction maps HWR of Virasoro to HWR of CCA$_2$.  To see this explicitly, let us assume that the state $\ket{h,\bar{h}}$ evolves to $\ket{h_L,h_M}$, the eigenstate of $L_0$ and $M_0$ in Carroll limit $\epsilon\to 0$. Then one can see that 
\begin{subequations}\label{flimit}
  \begin{align}
    &L_n\ket{h_L,h_M}=(\mathcal{L}_{n}+\bar{\mathcal{L}}_{n})\ket{h_L,h_M}=0~~~\forall n>0,\\
    &M_n\ket{h_L,h_M}=\epsilon(\mathcal{L}_{n}-\bar{\mathcal{L}}_{n})\ket{h_L,h_M}=0~~~\forall n>0,\\
    &h_L=h+\bar{h},~~~h_M=\epsilon(h-\bar{h}).
\end{align}  
\end{subequations}
One can easily identify \eqref{flimit} as the HWR of CCA$_2$ \eqref{BMSHWR}. We shall see in later sections that this representation is important in the study of the quantum theory constructed on the flipped vacuum.  

\subsection{BCFT and Carroll limit}\label{BCFT and Carroll limit}
In a previous section we discussed CFTs on the entire 2d complex plane and on a cylinder. While CFTs defined on a cylinder are useful in the study of closed string theory, there are also open strings which have two ends and consequently, the worldsheets have two different boundaries, which also necessitates the study of CFTs on manifolds with boundaries. In the above context, one needs to study Boundary Conformal Field Theories (BCFTs). We begin by briefly looking at BCFTs defined on a cylinder with boundaries at $\s=0$ and $\s=\pi$, and then proceed to understand their Carrollian version, since they will be relevant in the study of open null strings. Our approach, based on \cite{Bagchi:2024qsb}, is a bit different from standard literature and in particular would be helpful for the Carrollian counterpart of BCFT.

\paragraph{Relativistic BCFT:}
When boundaries are introduced on the manifold where a CFT is defined, some of the conformal symmetry generators distort the boundaries and can no longer be considered a part of the symmetry of the theory. The generators preserving the boundary conditions form the symmetry of the underlying theory with boundary. With the help of conformal transformations, complicated boundaries can be mapped onto the upper half plane (UHP). For 2d relativistic CFT on the UHP, the boundary preserving symmetries form one copy of Virasoro algebra. Let us begin with the generators of 2d CFTs defined on a cylinder as given in \eqref{Vircyl}. One can rewrite these generators in a different basis, which will help us in segregating the boundary compatible generators from boundary incompatible ones.
\begin{equation}\label{boldl}
    \mathbb{L}_n=\mathcal{L}_n+\bar{\mathcal{L}}_n \quad \text{\textcolor{ForestGreen}{\ding{51}}}\qquad\qquad \widetilde{\mathbb{L}}_n=\mathcal{L}_n-\bar{\mathcal{L}}_n \quad \text{\textcolor{red}{\ding{55}}}
\end{equation}
As demonstrated in \cite{Bagchi:2025jgu}, on a cylinder $\mathbb{R}\times S^1$, the generators $\widetilde{\mathbb{L}}_n$ distort the boundaries $\s=0,\pi$, and hence incompatible with them. The surviving generators $\mathbb{L}_n$ constitute one copy of Virasoro algebra. Throughout this section, we will denote the boundary compatible generators with {\textcolor{ForestGreen}{\ding{51}}} and the incompatible generators with a 
{\textcolor{red}{\ding{55}}}. 

\begin{figure}[t]
\centering
    \includegraphics[scale=0.65]{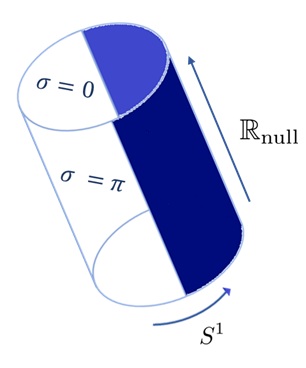}
    \caption{Null cylinder with boundaries}
    \label{nullcyl}
\end{figure}

\paragraph{Boundary Carroll Conformal Algebra:}\label{BCCA}
We now consider CCA$_2$ with boundaries at $\s=0$ and $\s=\pi$ on a null cylinder $\mathbb{R}_{null}\times S^1$ (see Fig.\ref{nullcyl}). From \eqref{cylinder}, we see that with boundaries at $\s=0,\pi$, $L_n$ are no longer allowed symmetry generators because of $\partial_\s$ terms at both the boundaries. We rewrite these generators as: 
\begin{align}
     \mathcal{O}_n=L_n-L_{-n} \quad \text{\textcolor{ForestGreen}{\ding{51}}}   \qquad \qquad \mathcal{Q}_n=L_n+L_{-n} \quad \text{\textcolor{red}{\ding{55}}}
\end{align}
The explicit forms of the generators are:
\begin{align}
\mathcal{O}_n=-2\sin(n\sigma)\,\partial_\sigma-2n\tau\cos(n\sigma)\,\partial_\tau\,,   \qquad \mathcal{Q}_{n}=2i\cos{(n\s)}\partial_{\s}-2in\t\sin{(n\s)}\partial_{\t}.
     \label{eq:whatever}
\end{align}
As obvious from the above, the generators $\mathcal{O}_n$ leave the $\s=0,\pi$ boundaries undisturbed while $\mathcal{Q}_n$ have $\partial_\s$ at $\s=0,\pi$ and hence, can no longer be allowed. In terms of the limit, 
\begin{align}
    \mathcal{O}_n=\mathbb{L}_n-\mathbb{L}_{-n}\quad \text{\textcolor{ForestGreen}{\ding{51}}}  \qquad \qquad \mathcal{Q}_n=\widetilde{\mathbb{L}}_n+\widetilde{\mathbb{L}}_{-n} \quad \text{\textcolor{red}{\ding{55}}}
\end{align}
We again see that $\mathcal{O}_n$s arise from allowed BCFT generators $\mathbb{L}_n$, while $\mathcal{Q}_n$s come from the discarded ones $\widetilde{\mathbb{L}}_n$. We also write generators $M_n$ in a new basis
\begin{align}
    \mathcal{P}_n=M_n+M_{-n}, \qquad \qquad \mathcal{R}_n=M_n-M_{-n}.
\end{align}
In explicit coordinate form, these are 
\begin{align}
\mathcal{P}_n =  2 \cos(n\sigma)\partial_\tau \qquad \mathcal{R}_{n}=-2\sin{(n\s)}\partial_{\t}.
     \label{eq:whatever2}
\end{align}
A priori both seem allowed since there are no errant $\partial_\s$ terms. But from the limit, we see $\mathcal{R}_n$s arises from the discarded generators $\widetilde{\mathbb{L}}_n$, while $\mathcal{P}_n$ comes allowed ones $\mathbb{L}_n$s: 
\begin{align}
    \mathcal{P}_n=\e(\mathbb{L}_n+\mathbb{L}_{-n}) \quad \text{\textcolor{ForestGreen}{\ding{51}}}   \qquad \qquad \mathcal{R}_n=\e(\widetilde{\mathbb{L}}_n-\widetilde{\mathbb{L}}_{-n}) \quad \text{\textcolor{red}{\ding{55}}}
\end{align}
So we keep $\mathcal{P}_n$ and discard $\mathcal{R}_n$. The boundary compatible generators constitute a hitherto unknown algebra called 2d Boundary Conformal Carrollian  Algebra (BCCA$_2$).
\begin{subequations}\label{bdyCarr}
  \begin{align}
    \left[\mathcal{O}_n,\,\mathcal{O}_m\right]&=(n-m)\mathcal{O}_{n+m}-(n+m)\mathcal{O}_{n-m},\\
    \left[\mathcal{O}_n,\,\mathcal{P}_m\right]&=(n-m)\mathcal{P}_{n+m}+(n+m)\mathcal{P}_{n-m}+\frac{c_M}{12}(n^3-n)(\delta_{n+m}+\delta_{n-m}),\\
    \left[\mathcal{P}_n,\,\mathcal{P}_m\right]&=0.
\end{align}  
\end{subequations}
In the upcoming sections we shall encounter this algebra as the worldsheet residual symmetry algebra of open null strings. 
\medskip

\subsection*{Superconformal Carrollian algebras}
We have looked at conformal Carrollian symmetries without and with boundaries so far in our recap of Carrollian theories. We will see that these structures would appear on the worldsheet of closed and open bosonic null strings respectively. When we will move to tensionless superstring theory, we will encounter supersymmetric cousins of Carroll CFTs. 

\medskip

In Appendix C, we give details of Superconformal Carrollian symmetries. Among the important points are the existence of two different types of Superconformal Carrollian algebras, which we call Homogeneous and Inhomogeneous, and these algebras arise from different contractions of the parent Super Virasoro algebra. We give some details of this. We will find that these algebras are realised on the worldsheet of two different varieties of closed null superstrings. We then go on to describe Carrollian superconformal theories with boundaries and the Boundary Superconformal Carrollian algebras that would be important for open null superstring. 

\medskip

A homogeneous supersymmetric extension of BCCA$_2$ has been discussed in appendix \ref{BasantaBiswas}. We shall see later that this algebra appears as the residual gauge symmetry algebra of homogenous open null superstrings.  

\bigskip

\bigskip

Armed with our history lesson and the brief encounters with string theory and Carrollian symmetries, we now venture into the heart of our review which will contain the details of the construction of the null tensionless string, first the classical aspects and then their quantum avatars. 

\newpage
\part{Classical Null Strings}\label{II}

\bigskip

The main body of our review starts with the description of the classical aspects of the null tensionless string. We study the theories of the closed and open null (super)string in both the intrinsic and the limiting approach. We will see that the answers match very well in both approaches. In terms of the figure below, the square closes nicely in the classical regime. 

\begin{figure}[ht]
    \centering
    \includegraphics[width=0.8\linewidth]{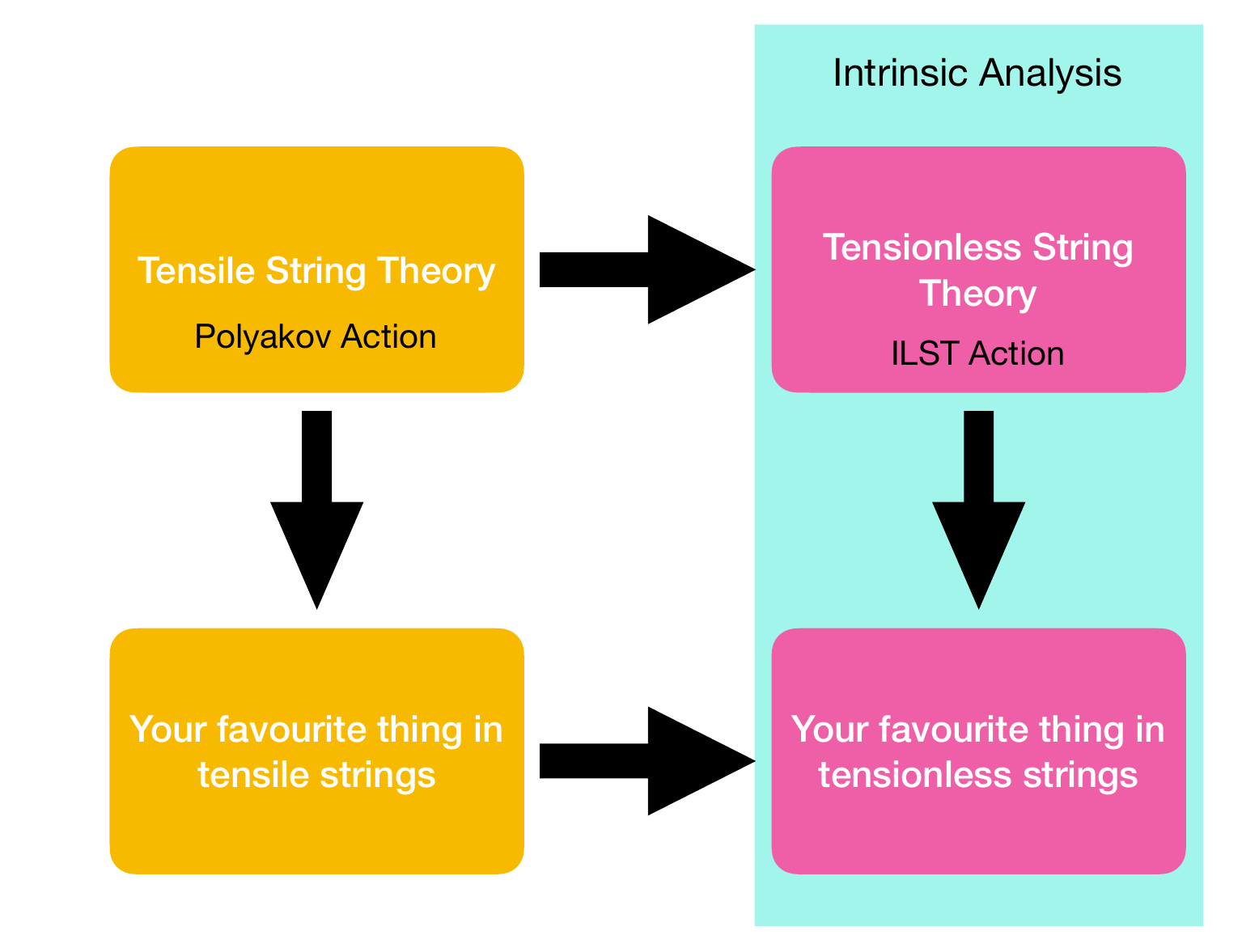}
    \caption{Classical analysis of the tensionless string}
    \label{fig:4sq}
\end{figure}

This part of our review is divided into sections as follows:

\begin{itemize}
    \item {\em{\hyperref[Bosonic closed null strings]{Bosonic Closed Null Strings:}}} Here we discuss the derivation of the ILST action which will form the basis of our intrinsic construction of the null string. The symmetries of the ILST action are discussed and the emergence of Conformal Carroll symmetries as residual gauge symmetries of the gauge fixed action is dealt with in some detail. We then discuss the tensionless limit on the worldsheet which allows us to check our intrinsic analysis. 

    \item{\em{\hyperref[Bosonic open null strings]{Bosonic Open Null Strings:}}} After bosonic closed null strings, we turn our attention to open null strings. There are three options here and apart from the usual Dirichlet or Neumann boundary conditions, there is also a null boundary condition that can be chosen. We focus on the Dirichlet open string. The symmetry analysis yields the newly discovered Boundary Conformal Carrollian algebra on the worldsheet. We also address the limit from the tensile open string. 

    \item {\em{\hyperref[Closed null superstrings]{Closed Null Superstrings:}}} In close analogy with the bosonic null strings, closed null superstrings are then constructed. There are two variants here called the Homogeneous and Inhomogeneous null superstrings. Symmetries and mode expansions are addressed both from the intrinsic and limiting perspective. 

    \item{\em{\hyperref[Open null superstrings]{Open Null Superstrings:}}} Finally we discuss open null superstrings and connect to Boundary Superconformal Carrollian symmetries. 
\end{itemize}

\newpage

\section{Bosonic closed null strings}\label{Bosonic closed null strings}
In this section we will discuss classical aspects of tensionless (null) limit of closed bosonic string theory. We first construct the action for bosonic null strings from the tensionless limit of Polyakov action following \cite{Isberg:1993av, Lindstrom:1990qb}. Then we go on to formulate the classical theory for closed null strings from first principles. We also complement this intrinsic viewpoint by taking appropriate limits on the corresponding results of the tensile closed string theory, a method introduced in \cite{Bagchi:2015nca}, which in turn implies that the contraction of the residual gauge symmetry of tensile strings leads to the conformal Carrollian symmetry. 

\subsection{Constructing Action of Null Strings}
In Sec.~\ref{Hamiltonformalism}, we saw how the Hamiltonian formalism can be used to construct the einbein action for particles and Polyakov action for strings. In the current discussion we will see how these new actions are helpful to study the massless limit for point particles and consequently, the null limit of strings.
\paragraph{Massless limit of point particle action:} A salient feature of the einbein action for point particles \eqref{I(x,e)} is that unlike \eqref{relpar}, this action does not vanish at $m=0$. This action can also be interpreted as action of scalar fields in one dimensional space (worldline parametrised by $\tau$) coupled to gravity where $m^2$ is the cosmological constant. The Lagrange multiplier $e(\tau)$ here plays the role of the square root of the metric on the worldline. Putting $m=0$ in this action, one obtains an action for massless point particles  
\begin{align}
    I_{m=0}=\frac{1}{2}\int d\tau~e^{-1}\dot{X}^2.
\end{align}
The EOM for the dynamical variable $e$ from this action turns out to be $\dot{X}^2=0$, confirming that the action indeed describes a massless relativistic point particle. A generalisation of this analysis was performed for string theory in \cite{Isberg:1993av}, which subsequently allows one to write a tensionless limit of the string action. 
\paragraph{Generalisation to strings -- tensionless limit:}\label{cup}
As outlined above, our objective now is to construct a string action that does not vanish in the limit $T=0$. To this end, we recall the phase space action for bosonic strings given in \eqref{phspacst}.  
Integrating out the canonical momentum from this action yields
\begin{align}
    S_{PS}=\int~d^2\s~\frac{1}{2\lambda}\left[\Dot{X}^2-2\rho(\Dot{X}\cdot X')+(\rho^2-\lambda^2T^2)X'^2\right].
\end{align}
Now once we fix the phase space action, looking at \eqref{lagmul} together with \eqref{Polyakov}, one can easily understand how to take the tensionless limit on the Polyakov action in a consistent manner. Equation \eqref{lagmul} makes it clear that the worldsheet inverse dynamical metric $g^{\alpha\beta}$ of the string itself also evolve along with tension of the string. Imposing a careful tensionless limit ($T\to\epsilon T$, $\epsilon\to 0$) on $g^{\a\b}$ as given in \eqref{lagmul} will leave us with the following
\begin{align}\label{null}
    g^{\alpha\beta} \xrightarrow{T\to 0} \begin{bmatrix}1 & -\rho \\ -\rho & ~~\rho^2\end{bmatrix}.
\end{align}
Consequently, the Polyakov action \eqref{Polyakov} can be equivalently written in terms of a tensor density, so that:
\begin{align}\label{ctb26}
    \frac{T}{2}\sqrt{-g}g^{\alpha\beta} \xrightarrow{T\to 0} V^{\alpha}V^{\beta}, \qquad V^{\alpha}=\frac{1}{\sqrt{\lambda}}(1,-\rho),
\end{align}
where $V^{\alpha}$ is a vector density connected to the geometry of the worldsheet. The action thus obtained is 
\begin{align}\label{ILST}
    S=\int d^2\s \hspace{1mm}V^{\alpha}V^{\beta}\partial_{\alpha}X^{\mu}\partial_{\beta}X^{\nu}\eta_{\mu\nu}.
\end{align}
This action is known as Isberg-Lindstrom-Sundborg-Theodoridis (ILST) action. Equation \eqref{ctb26} implies that at tensionless limit the inverse metric density on the worldsheet has become a rank one degenerate matrix i.e. the string worldsheet acquires a null structure at tensionless limit. Hence, for flat target spacetime, the tensionless strings are synonymous to null strings\footnote{As mentioned in the introduction, one could have tensionless strings in AdS without null worldsheets, because of the existence of another scale, viz. the AdS radius.}.
\paragraph{Symmetries of bosonic null strings:} The worldsheet symmetries as we move from tensile to tensionless strings will be more explicit when we will look into the mode expansion of tensionless strings. Note that, just like the tensile string action, the null string action too is Poincare invariant in target spacetime. It is also worldsheet diffeomorphism ($\s^\alpha\to\s^{\alpha}+\epsilon^{\alpha}$) invariant where $X$ and $V^{\alpha}$ transform in the following way
\begin{equation}\label{B3}
  \delta_{\xi} X^{\mu} =-\xi^{\b} \partial_{\b} X^{\mu}, \qquad 
\delta_{\xi} V^{\a} =-V^{\b} \partial_{\b} \xi^{\a}+\xi^{\b} \partial_{\b} V^{\a}+\frac{1}{2}\left(\partial_{\b} \xi^{\b}\right) V^{\a}.
\end{equation}
Since worldsheet diffeomorphism is a gauge symmetry, we now need to perform a gauge fixing. We choose to work with is the temporal gauge for $V^\a$, given by 
\begin{align}\label{Review1}
    V^{\alpha}=(v,0), \quad \quad \text{where} \quad  v=\sqrt{\frac{1}{4\pi c'}} = \text{const.}
\end{align}
The above choice of $v$ is made for convenience. Equation \eqref{Review1} is the tensionless counterpart of conformal gauge in usual string theory \footnote{However, it is important to note that the gauge choice in \eqref{Review1} is far from unique; ILST action \eqref{ILST} can also be studied in different gauge choices as discussed in \cite{Bagchi:2022eav}. among the other gauge choices, the Ambitwistor gauge choice \cite{Casali:2016atr} is particularly important,  about which we shall discuss later.}, where $c'$ is a constant with dimension $[c']=[L]^2$. We will see later that this $c'$ plays an important role in taking the tensionless limit.
\paragraph{Stress tensor and equations of motion:} The stress tensor of action \eqref{ILST} can be calculated as
\begin{equation}
\begin{split}
    T^{\alpha}_{\beta}&=\frac{\partial\mathcal{L}}{\partial(\partial_{\alpha}\phi)}\partial_{\beta}\phi-\delta^{\alpha}_{\beta}\mathcal{L}=V^\alpha V^\gamma\partial_\gamma X^\mu\partial_\beta X_\mu-\frac{1}{2}V^\rho V^\gamma \partial_\rho X^\mu\partial_\gamma X_\mu \delta^\alpha _\beta.
\end{split}
\end{equation}
The gauge choice \eqref{Review1} results in the following energy momentum tensor
\begin{equation}\label{emtensor}
    T^0_0=-T^1_1=\frac{v^2}{2}\dot{X}^2\equiv T_2(\sigma,\tau),\qquad T^0_1=v^2\dot{X}\cdot X'\equiv T_1(\sigma,\tau).
\end{equation}
It is straightforward to calculate the EOM for the action \eqref{ILST}, which are:
    \begin{equation}\label{tsc1} 
\partial_\alpha(V^\alpha V^\beta\partial_\beta X^\mu)=0, \qquad V^\beta\gamma_{\alpha\beta}=0,
\end{equation}
where $\gamma_{\alpha\beta}$ is the induced metric on the worldsheet defined in \eqref{inducedmetric}.
In the above, the equations on the left and right correspond to the EOM and the constraints, respectively. The gauge fixing \eqref{Review1} simplifies \eqref{tsc1} to the following EOM:
\begin{equation}\label{tsc3}
    \ddot {X}^\mu=0.
\end{equation}
In the same gauge, the constraints reduce to
\begin{align}\label{constraints}
   \dot{X}^2=0,\qquad \dot{X}\cdot X'=0.
\end{align}
We identify them as the components of the stress-energy tensor $T^{0}_{0}$ and $T^{0}_{1}$ in \eqref{emtensor}. 

\paragraph{Residual symmetry algebra:}
Just like tensile string theory, for the null worldsheet as well, we can find a remnant gauge symmetry after the gauge fixing carried out in the previous section. We see that the vielbein \eqref{Review1} remains invariant under the following transformation
\begin{align}\label{Review2}
  \s^\alpha\to\s^{\alpha}+\xi^{\alpha},\qquad \xi^\alpha=\left(f'(\sigma)\tau+g(\sigma),f(\sigma)\right).  
\end{align}
Where $f,g$ are arbitrary functions of worldsheet variables.
A generic function $G(\s^\alpha)$ thus transforms under \eqref{Review2} as:
\begin{equation}
\delta G=\left[f'(\sigma)\tau\partial_\tau+f(\sigma)\partial_\sigma+g(\sigma)\partial_\tau\right]G~=\left[L(f)+M(g)\right]G.
\end{equation}
Consequently, the generators can be defined as:
    \begin{subequations}
    \begin{align}    L(f)=f'(\sigma)\tau\partial_\tau+f(\sigma)\partial_\sigma&=\sum_n a_ne^{in\sigma}(\partial_\sigma+in\tau\partial_\tau)=-i\sum_na_nL_n, \\
    M(g)=g(\sigma)\partial_\tau&=\sum_n b_n e^{in\sigma}\partial_\tau=-i\sum_n b_nM_n.
    \end{align}
    \end{subequations}
In the above, the Fourier modes of $f$ and $g$ can be taken as $f=\sum_n a_ne^{in\sigma},~ g=\sum_n b_n e^{in\sigma}$. Combining all these we can read off the differential generators of the residual gauge symmetry
\begin{align}\label{BMS}
    L_n=ie^{in\sigma}(\partial_\sigma+in\tau\partial_\tau),\qquad M_n=ie^{in\sigma}\partial_\tau.
\end{align}
It is straightforward to check the algebra satisfied by these generators is given by
\begin{equation}\label{bmsalgebraclassical}
    [L_m,L_n]=(m-n)L_{m+n}, \qquad [L_m,M_n]=(m-n)M_{m+n}, \qquad [M_m,M_n]=0,
\end{equation}
which can be identified as the classical part of the BMS$_3$ algebra where the central charges $c_{L}=c_{M}=0$.

\subsection{Mode expansions}

The EOM for the null string may look like that of a point particle, but the pullback fields are dependent on both worldsheet coordinates.
Hence the most generic solution of the EOM is given by 
\begin{equation}\label{fcb1}
  X^{\mu}(\tau,\sigma)=x^{\mu}+\sqrt{\frac{c'}{2}}A^{\mu}_{0}\sigma+\sqrt{\frac{c'}{2}}B^{\mu}_{0}\tau+i\sqrt{\frac{c'}{2}}\sum_{n\neq 0}\frac{1}{n}(A^{\mu}_{n}-in\tau B^{\mu}_{n})e^{-in\sigma}.
\end{equation}

We further need to apply the closed string boundary condition given by 
\begin{equation}
    X^\mu(\tau,\sigma)=X^\mu(\tau, \sigma+2\pi).
\end{equation} 
This implies vanishing of the linear zero mode term in $\sigma$ in \eqref{fcb1} i.e. $A^{\mu}_{0}=0$. Also the fact that $X^\mu(\tau,\sigma)$ is real implies the hermiticty of the modes:
\begin{align}
   \big( A^{\mu}_{n}\big)^{\dagger}= A^{\mu}_{-n}, \qquad \big( B^{\mu}_{n}\big)^{\dagger}= B^{\mu}_{-n}.
\end{align}
We can now use the mode expansion \eqref{fcb1} on the two constraints \eqref{constraints}:
\begin{subequations}\label{constraints2}
   \begin{align}
       T_{1}(\tau,\sigma)= &X'\cdot\dot X=  \sum_{n}\left[L_{n}-in\tau M_{n}\right]e^{-in\sigma}=0,\\
         T_{2}(\sigma)= &\dot X^{2}= \sum_{n}M_{n}e^{-in\sigma}=0,
    \end{align}
\end{subequations}
where $L_{n}$ and $M_{n}$ can be read off using the mode expansions 
\begin{equation}\label{tsc9}
    L_n=\frac{1}{2}\sum_m A_{-m}\cdot B_{m+n}, \qquad M_n=\frac{1}{2}\sum_m B_{-m}\cdot B_{m+n}.
\end{equation}
We identify that the mode expansion of the constraints in \eqref{constraints2} as the mode expansion of stress tensor components of generic CCFTs as given earlier in \eqref{Carr-T}.  
Now, let us look into the momentum conjugate of $X^\mu$ (i.e. the canonical momentum density) from the action \eqref{ILST} with gauge choice \eqref{Review1}. 
\begin{equation}\label{chamchike2}
\begin{split}
   \Pi_{\mu}=\frac{\partial\mathcal{L}}{\partial\Dot{X}^{\mu}}=\frac{1}{2\pi c'}\Dot{X}^{\nu}\eta_{\nu\mu},  \qquad
   \implies  \Pi^{\mu}=\frac{1}{2\pi c'}\Dot{X}^{\mu}.
\end{split}
\end{equation}
The total momentum of the string can be found in terms of the zero modes by integrating the canonical momentum over the entire string i.e.,
\begin{equation}\label{totalm2}
     k^\mu=\int_{0}^{2\pi} d\s~\Pi^{\mu}=\frac{1}{2\pi c'}\int_{0}^{2\pi} d\s~\Dot{X}^{\mu}=\frac{B^{\mu}_{0}}{\sqrt{2c'}} \implies B^{\mu}_{0}=\sqrt{2c'}k^{\mu}.
\end{equation}
It is now imperative to understand the canonical structures of these modes.
Since the target spacetime for this theory remains flat, the equal time Poisson brackets satisfied by $X^\mu$ and canonical momentum $\Pi^\mu$ will be same as \eqref{P.B.}. Using these Poisson brackets, the relation \eqref{chamchike2} and the mode expansion of $X^\mu$ given in \eqref{fcb1}, one can determine Poisson brackets for the oscillator modes ($A,B$). They are
\begin{equation}\label{tsc5}
    \{A_m^\mu,A_n^\nu\}_{PB}~=~\{B_m^\mu,B_n^\nu\}_{PB}=0,\qquad \{A_m^\mu,B_n^\nu\}_{PB}=-2im\delta_{m+n}\eta^{\mu\nu}.
\end{equation}
Using \eqref{tsc5}, we can calculate the Poisson brackets for the generators $L_{n}$ and $M_{n}$ 
\begin{equation}
    i\{L_m,L_n\}_{PB}=(m-n)L_{m+n},\, i\{L_m,M_n\}_{PB}=(m-n)M_{m+n}, \, \{M_m,M_n\}_{PB}=0.
\end{equation}
This is of course the classical version of the residual symmetry algebra in \eqref{bmsalgebraclassical} for the ILST action \eqref{ILST}.
\medskip

As we can see in \eqref{tsc5}, the modes ($A,B$) do not lead to the usual harmonic oscillator algebra. In order to get harmonic oscillator algebra let us consider the following combination of modes 
\begin{equation}\label{tsc7}
    C_n^\mu=\frac{1}{2}(A_n^\mu+B_n^\nu), \qquad \tilde {C}_n^\mu=\frac{1}{2}(-A_{-n}^\mu+B_{-n}^\mu).
\end{equation}
The Poisson brackets between ($C,\tilde{C}$) now gives two decoupled sets of the oscillator algebra
\begin{equation}\label{calgebra}
\begin{split}
    \{C_m^\mu,C_n^\nu\}=-im\delta_{m+n,0}\eta^{\mu\nu}, \quad \{\tilde C_m^\mu,\tilde C_n^\nu\}=-im\delta_{m+n,0}\eta^{\mu\nu},\quad \{C_m^\mu,\tilde C_n^\nu\}=0.
\end{split}
    \end{equation}
Since these oscilators are more useful, we can go ahead and rewrite the mode expansion \eqref{fcb1} in terms of ($C,\tilde{C}$) as
\begin{equation}\label{fcb2}
\begin{split}
    X^{\mu}(\tau,\sigma)=x^{\mu}+&\sqrt{\frac{c'}{2}}(C^{\mu}_{0}-\tilde{C}^{\mu}_{0})\sigma+\sqrt{\frac{c'}{2}}(C^{\mu}_{0}+\tilde{C}^{\mu}_{0})\tau\\ &+i\sqrt{\frac{c'}{2}}\sum_{n\neq0}\frac{1}{n}\left[(C^{\mu}_{n}-\tilde{C}^{\mu}_{-n})-in\tau(C^{\mu}_{n}+\tilde{C}^{\mu}_{-n})\right]e^{-in\sigma}.
    \end{split}
\end{equation}
Here again one can see that reality condtion of $X^\mu$ needs
\begin{align}
     \big( C^{\mu}_{n}\big)^{\dagger}= C^{\mu}_{-n},~~~~~~\big( C^{\mu}_{n}\big)^{\dagger}= C^{\mu}_{-n}.
\end{align}
Applying closed string boundary condition on this mode expansion leads to $C_0^\mu=\tilde C_0^\mu$. Also redoing the steps in \eqref{totalm2} using this mode expansion will lead us to 
\begin{align}
    C_0^\mu=\tilde C_0^\nu=\sqrt{\frac{c'}{2}}k^\mu .
\end{align}
It is also illumination to re-express \{$L_{n},M_{n}$\} in terms of \{$C,\tilde{C}$\} modes, which ultimately give:
\begin{subequations}\label{TSQR5}
\begin{align}\label{TSQR5a}
    L_{n}&=\frac{1}{2}\sum_{m}\left[C_{-m}\cdot C_{m+n}-\tilde{C}_{-m}\cdot \tilde{C}_{m-n}\right],\\
    \label{TSQR5b}
    M_{n}&=\frac{1}{2}\sum_{m}\left[C_{-m}\cdot C_{m+n}+\tilde{C}_{-m}\cdot \tilde{C}_{m-n}+2C_{-m}\cdot \tilde{C}_{-m-n}\right].
    \end{align}
\end{subequations}
One should note the intriguing structure of the modes above. The generator $M_n$ has a mixture of the $C,\tilde{C}$ oscillators, that makes its action on a state non-trivial. We will come back to this when we discuss the quantum structure of this theory. 

\subsection{Taking tensionless limit on the worldsheet}
Now that we have understood the intrinsic perspective on tensionless strings and associated degenerate structures in the last couple sections, its time to reaffirm those using consistent limits on the tensile worldsheet theory, as advertised before.

\medskip

The infinite tension limit is the point-particle limit of string theory where the fundamental string shrinks to zero size. The tensionless limit, on the other hand, makes the string long and floppy. In terms of worldsheet coordinates (on the cylinder) this would be where the spatial direction becomes very large $\sigma \to \infty$ keeping $\tau$ fixed. For the closed string, this is equivalently done by rescaling 
\begin{equation}\label{eps}
    \tau \rightarrow \epsilon \tau , 
    \quad 
    \sigma \rightarrow \sigma,  \quad \epsilon \to 0.
\end{equation}
This is a Carrollian limit on the worldsheet where the worldsheet speed of light goes to zero and hence Carrollian structures appear as the worldsheet itself becomes a null surface. 

\begin{figure}[ht]
    \centering
    \includegraphics[width=0.85\linewidth]{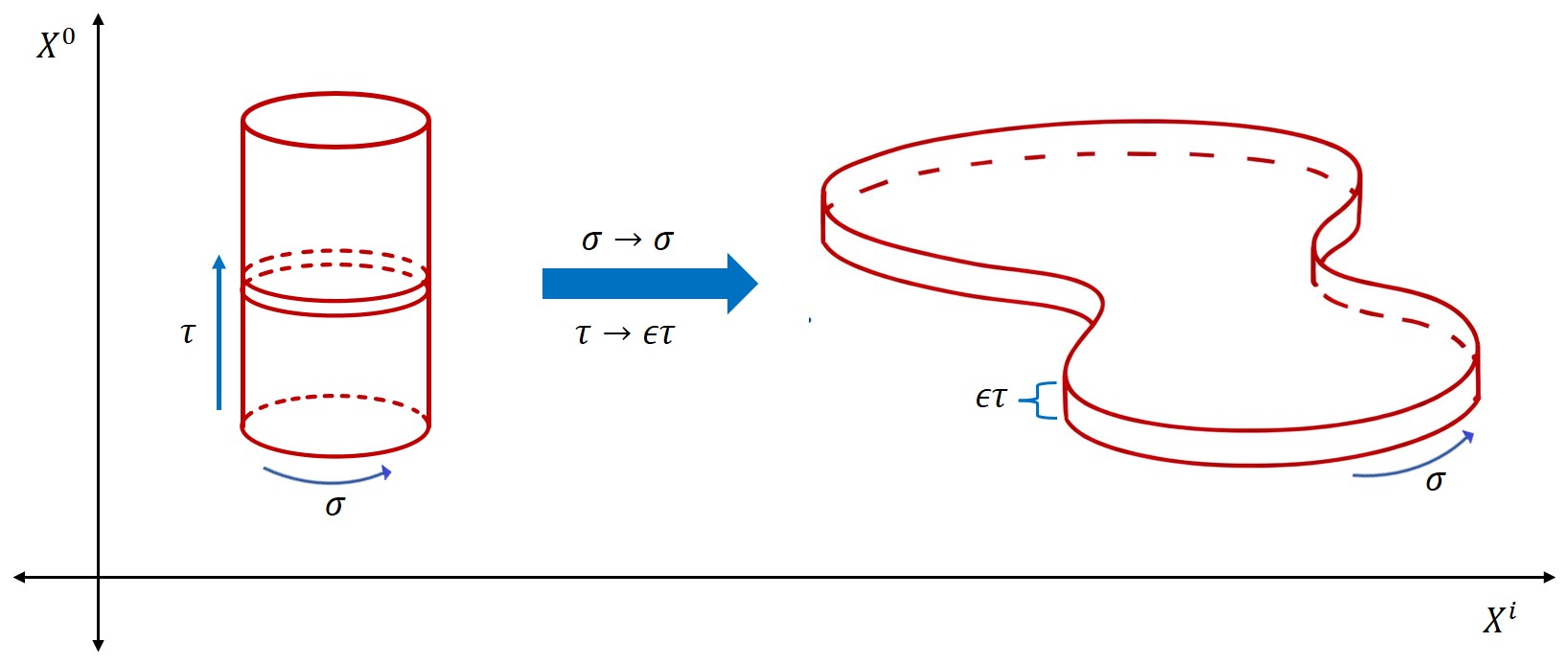}
    \caption{A heuristic description of the contraction on the worldsheet coordinates as the spatial coordinate becomes infinitely long, making the cylinder degenerate. }
    \label{fig5}
\end{figure}
\paragraph{Limit on symmetries:}
To see the effect of this scaling \eqref{eps} on the underlying symmetries, let us remind the reader the form of the tensile Virasoro generators on cylinder \eqref{Vircyl}:
\begin{align}
    \mathcal{L}_n
    &= \frac{i}{2} e^{in(\tau + \sigma)} \left( \partial_\tau + \partial_\sigma \right), 
    \quad \bar{\mathcal{L}}_n = \frac{i}{2} e^{in(\tau - \sigma)} \left( \partial_\tau - \partial_\sigma \right).
\end{align}
Under \eqref{eps}, the Virasoro generators expand as
\begin{align} 
    \mathcal{L}_n 
    = \frac{i}{2} e^{in\sigma} \left( 1 + in \epsilon \tau \right)
       \left( \frac{\partial_\tau}{\epsilon} + \partial_\sigma \right) + \mathcal{O}(\epsilon^2), \quad 
    \bar{\mathcal{L}}_{-n}
    = \frac{i}{2} e^{in\sigma} \left( 1 - in \epsilon \tau \right)
       \left( \frac{\partial_\tau}{\epsilon} - \partial_\sigma \right) + \mathcal{O}(\epsilon^2). \nonumber
\end{align}
We now extract the leading and subleading terms of the contracted generators by defining the combinations: 
\begin{align}
M_n= \epsilon \left( \mathcal{L}_n + \bar{\mathcal{L}}_{-n} \right)
         = i e^{in\sigma} \partial_\tau, \quad 
L_n= \mathcal{L}_n - \bar{\mathcal{L}}_{-n} 
         = i e^{in\sigma} \left( \partial_\sigma + in \tau \,\partial_\tau \right).  
\end{align}
These generators are identical to the residual symmetry algebra of tensionless strings we encountered earlier in the intrinsic analysis \eqref{BMS} and as we know, they satisfy the BMS$_3$ algebra \eqref{bmsalgebraclassical} without the central extension. 
\medskip

\paragraph{Mode expansion from limit:}
The tensionless limit on the worldsheet via the scalings can now be used to match up with the mode expansions from the EOM we derived earlier. It should be clear to the reader that when imposed on the closed tensile EOM in conformal gauge, i.e. $\Box X = 0$, the scaling \eqref{eps} simply gives rise to $\ddot{X}=0$, the tensionless version as has been derived intrinsically.
To investigate the mode expansions closely, let us recall the mode expansion of tensile closed bosonic string: 
\begin{equation}
     X^{\mu}(\tau,\sigma)=x^{\mu}+2\sqrt{\frac{\alpha'}{2}} \alpha_0^\mu\tau+i\sqrt{\frac{\alpha'}{2}}\sum_{n\neq0}\frac{1}{n}\left[\alpha^{\mu}_{n}e^{-in(\tau+\sigma)}+\tilde\alpha^{\mu}_{n}e^{-in(\tau-\sigma)}\right].
\end{equation}
Here $\alpha'(\sim l_s ^2)$ is related to string tension $T$ as $\alpha'=\frac{1}{2\pi T}$.
We take the tensionless limit on the worldsheet coordinates \eqref{eps} along with a scaling on the string coupling:
\begin{equation}\label{ur}
    \alpha'\to \frac{c'}{\epsilon},
\end{equation}
where $c'$ is same dimensionful constant earlier introduced in \eqref{Review1}. Clearly $\alpha'\to c'/\epsilon$ implies the small tension regime, i.e. $T\to\epsilon T$, ($\epsilon\to 0$), or equivalently the infinite limit on the string length.
In this limit the tensile mode expansion takes the following form:
\begin{equation}\label{tsc4}
     X^{\mu}(\tau,\sigma)=x^{\mu}+\sqrt{2\epsilon c'} \alpha_0^\mu\tau+i\sqrt{\frac{c'}{2}}\sum_{n\neq0}\frac{1}{n}\left[\frac{\alpha_n^\mu-\tilde\alpha_{-n}^\mu}{\sqrt{\epsilon}}-in\tau\sqrt{\epsilon}(\alpha_n^\mu+\tilde\alpha_{-n}^\mu)\right]e^{-in\sigma}.
\end{equation}
Comparing the limiting perspective in \eqref{tsc4} and the intrinsic one in \eqref{fcb1}, we find the relation between ($\alpha,\tilde\alpha$) and ($A$,$B$) as:
\begin{equation}\label{tsc8}
    A_n^\mu=\frac{1}{\sqrt{\epsilon}}(\alpha_n^\mu-\tilde\alpha_{-n}^\mu),\qquad B_n^\mu=\sqrt{\epsilon}(\alpha_n^\mu+\tilde\alpha_{-n}^\mu).
\end{equation}
Applying the canonical Poisson brackets between ($\alpha,\tilde\alpha$) to \eqref{tsc8} and dropping higher order terms, we retrieve the Poisson brackets between the modes $(A,B)$ of the null string \eqref{tsc5}. 
We recall the expression of worldsheet Virasoro generators for the tensile string Polyakov action in terms of ($\alpha,\tilde\alpha$) from \eqref{Review3}
\begin{align*}
    \mathcal{L}_{n}=\frac{1}{2}\sum_{m}\alpha_{-m}\cdot\alpha_{m+n},\qquad \bar{\mathcal{L}}_{n}=\frac{1}{2}\sum_{m}\tilde\alpha_{-m}\cdot\tilde\alpha_{m+n}.
\end{align*}
Applying \eqref{tsc8} on the expression of $L, M$ in terms of the oscillators $A,B$ given by \eqref{tsc9}, we recover the In{\"o}n{\"u}--Wigner contraction of the Virasoro algebra to the BMS algebra, now from the underlying oscillator modes:
\begin{align}\label{Review41}
    L_{n}=\mathcal{L}_{n}-\bar{\mathcal{L}}_{-n},\qquad M_{n}=\epsilon(\mathcal{L}_{n}+\bar{\mathcal{L}}_{-n}).
\end{align}
All of the above serve as checks of the proposal of the tensionless limit in terms of a Carrollian limit on the worldsheet coordinates. 
In passing, we also mention the relation between the oscillators ($C,\tilde{C}$) defined in \eqref{tsc7} to the tensile oscillators: 
\begin{subequations}\label{chh4}
\begin{align}
    &C^{\mu}_{n}=\frac{1}{2}\Big(\sqrt{\epsilon}+\frac{1}{\sqrt{\epsilon}}\Big)\alpha^{\mu}_{n}+\frac{1}{2}\Big(\sqrt{\epsilon}-\frac{1}{\sqrt{\epsilon}}\Big)\tilde{\alpha}^{\mu}_{-n},\\
    &\tilde{C}^{\mu}_{n}=\frac{1}{2}\Big(\sqrt{\epsilon}-\frac{1}{\sqrt{\epsilon}}\Big)\alpha^{\mu}_{-n}+\frac{1}{2}\Big(\sqrt{\epsilon}+\frac{1}{\sqrt{\epsilon}}\Big)\tilde{\alpha}^{\mu}_{n}.
    \end{align}
    \end{subequations}
Since the transformation from ($\alpha,\tilde\alpha$) basis to ($C,\tilde C$) basis in \eqref{chh4} preserves the Poisson brackets (or canonical commutation relations), the transformation is a \textit{Bogoliubov transformation} on the string worldsheet. This Bogoliubov transformation will play a crucial role in quantum aspects, especially the vacuum structure, of tensionless string theory, as we will see in later sections. 

\newpage

\section{Bosonic open null strings}\label{Bosonic open null strings}
We now move our attention to open null strings constructed very recently in \cite{Bagchi:2024qsb}. For the intrinsic analysis, we begin with the same ILST action constructed in \eqref{ILST}. Understandably, for closed null strings boundary terms were not an issue, but for open strings, one has to be careful about boundary contributions when considering variation of the action. In general the vanishing condition for the boundary term reads:
\begin{equation}
\int_{\partial B}d\tau\,n_\alpha V^\alpha V^\beta(\partial_\beta X^\nu)\eta_{\mu\nu}\,\delta X^\mu = 0.
\label{eq:beom}
\end{equation}
Here $\partial B$ is the boundary of the worldsheet located at $\sigma=0,\pi$, whose normal direction is $n_\alpha$, see \eqref{nullcyl} for a visual description. 
The above boundary term vanishes under the following boundary conditions:
\begin{subequations}
\label{unusual}
\begin{align}
    \delta X^\mu \big|_{\sigma=0,\pi}&=0 \qquad \text{Dirichlet,}\label{eq:Dirichlet}\\ 
    V^\beta\partial_\beta X^\nu \big|_{\sigma=0,\pi}&=0 \qquad \text{Neumann,}\\
    n_\alpha V^\alpha\big|_{\sigma=0,\pi} &=0 \qquad \text{Null.} \label{eq:DerDritteMann}
    \end{align}
\end{subequations}
The first two conditions correspond, respectively to the usual Dirichlet and Neumann boundary conditions familiar from the tensile open string. The third condition \eqref{eq:DerDritteMann}, however, is a new possibility that emerge only in the null string limit. In this review, however, we will focus on the Dirichlet boundary condition. The Neumann and, in particular, the null boundary conditions are intriguing and require further investigation.
\paragraph{Mode expansion and residual symmetry:}
To analyse the residual worldsheet symmetry algebra for open bosonic null strings, we take a sightly different approach compared to the closed null strings. We recall the most general solution to the gauge fixed EOM \eqref{tsc3} is
\begin{equation}
\begin{split}
    X^{\mu}(\tau,\sigma)=x^{\mu}+&\sqrt{\frac{c'}{2}}(C^{\mu}_{0}-\tilde{C}^{\mu}_{0})\sigma+\sqrt{\frac{c'}{2}}(C^{\mu}_{0}+\tilde{C}^{\mu}_{0})\tau\\ &+i\sqrt{\frac{c'}{2}}\sum_{n\neq0}\frac{1}{n}\left[(C^{\mu}_{n}-\tilde{C}^{\mu}_{-n})-in\tau(C^{\mu}_{n}+\tilde{C}^{\mu}_{-n})\right]e^{-in\sigma}.
    \end{split}
\end{equation}
We now apply the Dirichlet boundary condition at $\sigma=0$ and $\pi$, which requires 
\begin{equation}\label{nullbosiden}
C^\mu_n = -\tilde{C}^\mu_n.
\end{equation}
This results in the following mode expansion for the null open string, 
\begin{align}
    X^{\mu}(\tau,\sigma)=x_0^{\mu}+\sqrt{2c'}&C^{\mu}_{0}\sigma +i\sqrt{\frac{c'}{2}}\sum_{n\neq0}\frac{1}{n}\big[(C^{\mu}_{n}+C^{\mu}_{-n}) -in\tau(C^{\mu}_{n}-C^{\mu}_{-n})\big]e^{-in\sigma}.
    \label{modeexpansion1}
\end{align}
Here $(C_{n}^\mu)^{\dagger}=C_{-n}^\mu$ ensures that the modes are real. To determine the algebra satisfied by the generators of constraints, we first compute $\dot X^\mu, X'^\mu$ from \eqref{modeexpansion1} and substitute them into the constraint equations \eqref{constraints}. This results in 
\begin{subequations}\label{eq:con}
\begin{align}
\dot{X}^2&=\sum_{m} \mathcal{P}_m e^{-im\sigma} = 0, \\
 \dot{X}\cdot{X}^\prime &= \sum_{m}\big(\mathcal{O}_m-im\tau \mathcal{P}_m\big)e^{-im\sigma}=0\,, 
\end{align}
\end{subequations}
where the generators $(\mathcal{O}_n, \mathcal{P}_m$) are defined as follows
\begin{subequations}
    \label{eq:bi}
\begin{align}
\mathcal{O}_{n}&=\sum_{m}\frac{1}{2}(C_m^\mu-C_{-m}^\mu)(C_{n-m}^\nu+C_{m-n}^\nu)\eta_{\mu\nu},\\
 \mathcal{P}_m&=\sum_{n}\frac{1}{4}(C_n^\mu-C_{-n}^\mu)(C_{m-n}^\nu-C_{n-m}^\nu)\eta_{\mu\nu}.
\end{align}
\end{subequations}
Using the Poisson brackets between $(C,\tilde{C})$ oscillators, given in \eqref{calgebra}, it is straightforward to verify that the generators of the constraints satisfy 
\begin{subequations}\label{eq:osc}
  \begin{align}
    i\left\{\mathcal{O}_n,\,\mathcal{O}_m\right\}&=(n-m)\mathcal{O}_{n+m}-(n+m)\mathcal{O}_{n-m},\\
    i\left\{\mathcal{O}_n,\,\mathcal{P}_m\right\}&=(n-m)\mathcal{P}_{n+m}+(n+m)\mathcal{P}_{n-m},\\
    i\left\{\mathcal{P}_n,\,\mathcal{P}_m\right\}&=0.
\end{align}  
\end{subequations}
The above algebra (in commutator form) corresponds to the classical part of the Boundary Conformal Carrollian Algebra (BCCA) \eqref{bdyCarr} shown earlier from the field theoretic arguments.
\paragraph{Limiting analysis:}\label{Openbosonlimper}
As expected, the intrinsic formulation of the open null string can be recovered from the tensile theory by taking an ultra-relativistic limit on the tensile open string theory. We recall the standard mode expansion of the tensile open string, 
\begin{align}
\label{opentensilemode}
     X^{\mu}(\tau,\sigma)=x^{\mu}_0+\sqrt{2\alpha'}\alpha_0^\mu\sigma+i\sqrt{\frac{\alpha'}{2}}\sum_{n\neq0}\frac{1}{n}
\Big[\alpha^{\mu}_{n} e^{-in(\tau+\sigma)} + \alpha^{\mu}_{-n}e^{in(\tau-\sigma)}\Big],
\end{align}
and perform the same ultra-relativistic scaling as we did for the closed case,
\begin{equation}
\label{contraction}
    \tau\to\epsilon\tau\qquad\sigma\to\sigma \quad \text{and} \quad\alpha'\to c'/\epsilon\qquad\epsilon\to 0\,.
\end{equation}
Expanding~\eqref{opentensilemode} to first order in $\epsilon$ and comparing with \eqref{modeexpansion1}, we obtain the following Bogoliubov transformation
\begin{equation}
\label{openbv}
C^{\mu}_{n}=\frac{1}{2}\Big(\sqrt{\epsilon}+\frac{1}{\sqrt{\epsilon}}\Big)\,\alpha^{\mu}_{n}-\frac{1}{2}\Big(\sqrt{\epsilon}-\frac{1}{\sqrt{\epsilon}}\Big)\,{\alpha}^{\mu}_{-n}.
\end{equation}
We recall the expression of single copy of Virasoro generators (in a different basis)\footnote{In this review, for open tensile string theory, we use a different basis for the generators of the constraints. Dirichlet and Neumann boundary conditions identify ($\alpha$ and $\tilde\alpha$) oscillators resulting in only single copy of Virasoro generators.} given in appendix \eqref{singlecopyvir}
\begin{align}
    \mathbb{L}_n=\frac{1}{2}\sum_m \a_{m}\cdot\a_{n-m}.
\end{align}
Using \eqref{openbv} in \eqref{eq:bi} and applying the above expression of single copy of Virasoro generator we obtain the following ultra-relativistic contraction
\begin{equation}
    \mathcal{O}_n=\mathbb{L}_n-\mathbb{L}_{-n}, \qquad \mathcal{P}_n=\e (\mathbb{L}_n+\mathbb{L}_{-n}).
\end{equation}
Using the above In\"on\"u-Wigner contraction on a single Virasoro, we can obtain the BCCA for bosonic open null strings. This concludes our discussion on bosonic null strings, where we heavily relied on the ILST formalism. 

\newpage

\section{Closed null superstrings}\label{Closed null superstrings}
This section covers the supersymmetric versions of closed null strings. The discussion in this section is mainly based on works \cite{Bagchi:2016yyf, Bagchi:2017cte}. Earlier studies on the supersymmetric extensions of the BMS$_3$ algebra showed that there could be two possibilities, namely the \textit{homogeneous} and \textit{inhomogeneous} version of the algebra. The earliest mention of homogeneous super-BMS$_3$ algebra, albeit in a different guise, can be traced back to \cite{Gamboa:1989px}, the inhomogeneous super-BMS$_3$ algebra was proposed in \cite{Bagchi:2017cte} {\footnote{The Galilean (isomorphic) version of super BMS$_3$ was first constructed in \cite{Mandal:2010gx}.}}. In what follows we will review the classical aspects of both homogeneous and inhomogeneous null superstrings which can also be understood directly from judiciously taken tensionless limit of tensile RNS superstrings.

\subsection{Homogeneous null superstring}
We begin this section by discussing how to appropriately introduce fermions on the null worldsheet. We follow the formalism of supersymmetrisation of bosonic theory in tensile RNS superstring case and introduce $\psi^\mu$ as the fermionic partners\footnote{To be precise $\psi^\mu$ is fermionic density on the worldsheet with weight $-\frac{1}{4}$.} of bosonic field $X^\mu$ and $\chi$ as the fermionic partner of the vector density $V^\a$ in the ILST action.  This $\chi$ is related to a field analogous to gravitino $\chi_\a$ through the relation $\chi=V^\a\chi_\a$. In case of the tensile superstring, the fermionic field $\psi^\mu$ is a massless Majorana fermion, and the corresponding worldsheet action takes the form of a massless Dirac action written using a Majorana representation of the two-dimensional gamma matrices. In the present case, the purely fermionic sector of the action retains the Dirac-like structure, with the gamma matrices now satisfying a modified Clifford algebra appropriate for the Carrollian (degenerate) worldsheet geometry\footnote{See \cite{Bagchi:2022eui} for a detailed analysis of Carroll Clifford algebras in general dimensions. We will see that the representation of gamma matrices for homogenous and inhomogeneous null superstring are different.}. The action thus obtained for null superstring following \cite{Lindstrom:1990ar}

\begin{equation}\label{hcfer1}
S=\int d^{2} \xi ~~ \left[\left(V^{\a}\partial_{\a} X^{\mu}+i \chi \psi^{\mu}\right) \cdot\left(V^{\b} \partial_{\b} X_{\mu}+i \chi \bar{\psi}_{\mu}\right)+i\bar\psi^{\mu}\rho^{\a}  \partial_{\a}\psi_{\mu}\right].
\end{equation}
The above action displays diffeomorphism invariance as well as supersymmetry on the worldsheet. Variation of the fields under the worldsheet diffeomorphism ($\s^\alpha\to\s^{\alpha}+\xi^{\alpha}$) is given below
\begin{subequations}\label{opodartho}
\begin{align}\label{chhagol}
&\delta_{\xi} X^{\mu}  =-\xi^{\b} \partial_{\b} X^{\mu},\qquad 
\delta_{\xi} V^{\a} =-V^{\b} \partial_{\b} \xi^{\a}+\xi^{\b} \partial_{\b} V^{\a}+\frac{1}{2}\left(\partial_{\b} \xi^{\b}\right) V^{\a},\\
&\delta_{\xi}\psi^{\mu}  =-\xi^{\b} \partial_{\b} \psi^{\mu},\qquad~\hspace{.4mm}
\delta_{\xi} \chi  =-\xi^{\a} \partial_{\a} \chi+\frac{1}{4}\left(\partial_{\a} \xi^{\a}\right) \chi.
\end{align}
\end{subequations}
Under supersymmetry transformation parametrized by $\eps$, the fields vary as followed
\begin{subequations}\label{hcfer4}
\begin{align}
\delta_{\eps} X^{\mu} & =i \eps \psi^{\mu},\qquad\delta_{\eps} \psi^{\mu}  =-\eps V^{\a} \partial_{\a} X^{\mu}-\frac{1}{2} i \eps\left(\psi^{\mu} \chi\right) , \\ \label{chhagol1}
\delta_{\eps} V^{\a}& =i V^{\a}\left(\eps \chi\right),\qquad~~~~~ 
\delta_{\eps} \chi =V^{\a} \partial_{\a} \eps.
\end{align}
\end{subequations}
Just like its bosonic counterpart, the worldsheet diffeomorphism invariance necessitates gauge fixing, which is the null equivalent of the conformal gauge for the tensile superstring: 
\begin{equation}
\label{gauge}
V^{\a}=(1,0),\quad \quad \chi=0. 
\end{equation}
The gamma matrices on the worldsheet of tensile superstrings belong to the Majorana representation of the relativistic Clifford algebra with $\eta^{\a\b}$ as the gauge fixed worldsheet inverse metric
\begin{align}\label{Cliffal}
    \{\rho^{\a}, \rho^\b\} = 2 \eta^{\a\b}\mathbb{1}.
\end{align}
However, since on the null worldsheet the metric is degenerate, in order to study fermions in such worldsheet, one needs a modified version of Clifford algebra. In \eqref{Cliffal}, replacing $\eta^{ab}$ with a suitable combination of the gauged fixed vector densities $V^\a=(1,0)$, one can formulate a modified Clifford algebra on the null worldsheet
\begin{eqnarray}\label{modcliff}
\{\rho^{\a}, \rho^\b\} = 2 V^\a V^\b\mathbb{1},\qquad V^\a=(1,0).
\label{Clifftless}
\end{eqnarray}
The most obvious representation of \eqref{Clifftless} is a simple choice $\rho^\a = V^\a \mathbb{1}$, i.e., with the gauge choice $V^\a=(1,0)$ we have the following gamma matrices
\begin{eqnarray}
\label{rhonull}
\rho^0 = \mathbb{1}, \qquad \rho^1 = \mathbb{O}.
\end{eqnarray}
With these gauge choices and the simple representation of the Clifford matrices, the action \eqref{hcfer1} reduces to the following
\begin{equation}\label{hcfer5}
S=\int d^{2} \xi\left[\dot{X}^{2}+i \psi \cdot \dot{\psi}\right].
\end{equation}

Just like its bosonic counterpart, the null superstring gauge fixed action also has residual gauge symmetries. Applying the gauge choices in \eqref{opodartho} and \eqref{hcfer4} one finds that the action \eqref{hcfer5} is invariant under the following worldsheet gauge transformation parametrized by $\xi$ and $\eps$
\begin{subequations}\label{homosusy}
\begin{align}
\delta_\xi X&=\xi^\a\p_\a X, \qquad\delta_\xi \psi=\xi^\a\p_\a \psi +\frac{1}{4}(\p_\a\xi^\a)\psi, \\
\delta_\e X&=\bar{\eps}\psi, \qquad~~~~~\hspace{.7mm}\delta_\eps \psi^\mu=-i\rho^\a\p_\a X \eps.
\end{align}
\end{subequations}
In \eqref{homosusy}, the parameters $\xi$ and $\eps$ follow certain restrictions which is found by applying the gauge \eqref{gauge} on  \eqref{chhagol} and \eqref{chhagol1}
\begin{equation}\label{hcfer9}
\dot{\xi^{0}}=\left(\xi^{1}\right)^{\prime}, \qquad \dot{\xi^{1}}=0, \qquad \dot{\eps^{\alpha}}=0.
\end{equation}
Solving the above differential equations one finds how the diffeomorphism and supersymmetry transformations are restricted due to gauge fixing
\begin{equation}\label{hcfer10}
\xi^{0}=f^{\prime}(\sigma) \tau+g(\sigma), \qquad \xi^{1}=f(\sigma), \qquad \eps^{ \pm}=\eps^{ \pm}(\sigma).
\end{equation}
As demonstrated in \cite{Bagchi:2016bcd}, the generators of this transformation in superspace satisfy\footnote{Details of superspace formalism of homogenous null superstrings are presented in Appendix \eqref{homosuper}}
\begin{equation}\label{HSBMSAlgebra}
    \begin{split}
       &[L_n,L_m]=(n-m)L_{n+m},~~~~[L_n,M_m]=(n-m)M_{n+m},\\
        &[L_n,Q^{\alpha}_{r}]=\Big(\frac{n}{2}-r\Big)Q^\alpha_{n+r},~~~\{Q^{\alpha}_{r},Q^{\alpha'}_{s}\}=\delta^{\alpha\alpha'}M_{r+s},\\
       &[M_m,M_n]=[M_m,Q^{\alpha}_{r}]=0,
   \end{split}
\end{equation}
which forms the homogeneous super‑Conformal Carrollian algebra (SCCA$_H$), without the central extension. The centrally extended algebra is given in Appendix \eqref{sgcah}. 
\medskip

We now turn to deriving the above symmetry algebra using mode expansions. After gauge fixing, the EOM and the constraints obtained from \eqref{hcfer1} simplify considerably and take the following form 
\begin{subequations} \label{hcfer7}
\begin{align}\label{hcfer71}
&\ddot{X}^{\mu}=0, \qquad \dot{\psi}^{\mu}=0,\\ \label{hcfer72}
&\dot{X}^{2}=0, \qquad \dot{X} \cdot X^{\prime}+\frac{i}{2} \bar{\psi} \cdot \psi^{\prime}=0, \qquad \psi \cdot \dot{X}=0. 
\end{align}
\end{subequations}
Now, let us look into the solutions of the EOM \eqref{hcfer7} in the NS-NS sector. The Fourier expansion of the bosonic field solution remains same as \eqref{fcb1}. The mode expansion for solutions for the equation $\dot{\psi}^{\mu}=0$ is given by
\begin{align}
\label{tlessmode}
     \psi^\mu_+(\tau,\sigma)=\sqrt{c'}\sum_{r\in\mathbb{Z}+\frac{1}{2}}\beta^{~\mu+}_{r}e^{-ir\sigma},\qquad     \psi^\mu_-(\tau,\sigma)=\sqrt{c'}\sum_{r\in\mathbb{Z}+\frac{1}{2}}\beta^{~\mu-}_{r}e^{-ir\sigma}.
\end{align}
The fermionic fields satisfy the Poisson brackets,
\begin{align}\label{fpb}
     \quad \{ \psi^\mu_{\a}(\sigma),\psi^\nu_{\a'}(\sigma') \}=\eta^{\mu\nu}\delta_{\a\a'}\delta(\sigma-\sigma'). 
\end{align}
This leads to the following anticommutation relation between the fermionic modes $\b^\pm$
\begin{align}
    \{\beta_{r}^{\mu\a},\beta_{s}^{\nu\a'} \} =\delta^{\a\a'} \delta_{r+s}\eta^{\mu\nu}.
\end{align}
Using the mode expansions of bosonic fields in \eqref{fcb1} and fermionic fields in \eqref{tlessmode} we determine the mode expansion of the constraints as
\begin{subequations}\label{constantine}
\begin{align}
\dot{X}^2&
=4c'\sum_{n}M_ne^{-in\sigma}, \\
\dot{X}\cdot X'+\frac{i}{2}\bar{\psi}\cdot\psi'
&= 4c'\sum_{n} \Big[L_n-in\tau M_n \Big]e^{-in\sigma},\\
\psi_\pm\cdot\dot{X}&
=4c'\sum_{r}Q^\pm_r e^{-ir\sigma}.
\end{align}
\end{subequations}
In \eqref{constantine}, the expressions of the symmetry generators $L_n$, $M_n$ and $Q^\pm$ in terms of modes are given by
\begin{subequations}\label{flyingladdu}
\bea{}
L_n&=&\frac{1}{2}\sum_{m}  A_{-m}\cdot B_{m+n}+\frac{1}{4}\sum_{r}(2r+n)\Big(\beta^+_{-r}\cdot\beta^+_{r+n}+\beta^-_{-r}\cdot\beta^-_{r+n}\Big),  \\
M_n&=&\frac{1}{2}\sum_{m} B_{-m} \cdot B_{m+n}, \\
Q^\pm_r&=&\frac{1}{2}\sum_{m} B_{-m} \cdot \beta^\pm_{m+r}.
\eea
\end{subequations}
Using the Poisson brackets \eqref{tsc5} and \eqref{fpb} one can find the Poisson brackets satisfied by $L_n$, $M_n$ and $Q^\pm$. The Poisson brackets correspond to the residual symmetry algebra in \eqref{HSBMSAlgebra}.
\paragraph{Limiting approach:}\label{homotenlim}
We have formulated homogenous null superstring from first principles so far. We now derive the same results by taking appropriate limit on the tensile RNS superstrings. We already know how the bosonic fields behave in the tensionless limit. Looking at the tensile mode expansion \eqref{tensed1} of the worldsheet fermions one can see that for homogeneous case, at tensionless limit ($\a' \rightarrow \frac{c'}{\e},~\e\to 0$) there should be an overall scaling in the fermions in order to avoid divergence and the scaling is $\psi_{\text{tensionless}} = \sqrt{\epsilon} \, \psi_{\text{tensile}}$. Let us recall the the mode expansion for tensile RNS superstring
\begin{align}
    \psi^{\mu}_{+}(\tau,\sigma)=\sqrt{\alpha'}\sum_{r\in\mathbb{Z}+\frac{1}{2}}b^\mu_re^{-ir(\tau+\sigma)},\quad \psi^\mu_{-}(\tau,\sigma)=\sqrt{\alpha'}\sum_{r\in\mathbb{Z}+\frac{1}{2}}\tilde{b}^\mu_re^{-ir(\tau-\sigma)}.
\end{align}
Under Carroll limit this mode expansion evolves into \eqref{tlessmode} in the following way
\begin{subequations}
\bea{}
\psi_+^\mu(\sigma,\tau)&=&\sqrt{\e}\sqrt{2\a'}\sum_r b^{\mu}_r e^{-ir\sigma}(1-i\e r\tau) \approx\sqrt{2c'}\sum_r b^{\mu}_r e^{-ir\sigma} \\
\psi_-^\mu(\sigma,\tau)&=&\sqrt{\e}\sqrt{2\a'}\sum_r \tilde{b}^{\mu}_{r} e^{+ir\sigma}(1-i\e r\tau) \approx \sqrt{2c'}\sum_r \tilde{b}^{\mu}_{-r} e^{-ir\sigma}.
\eea
\end{subequations}
In the above, the tensionless fermionic modes are connected to the tensile fermionic modes in the following way
\bea{}\label{oscmap}
\beta^{\mu+}_r=b^{\mu}_r, \qquad \beta^{\mu-}_r={\tilde{b}}^{\mu}_{-r}. 
\eea
We recall that in tensile RNS superstrings, (reviewed in Appendix \ref{RNSSuperstring}), the energy-momentum tensor components and the supercurrents of the tensile superstrings lead us to the following expressions of the super-Virasoro generators in terms of oscillator modes (see \eqref{supvir})
\begin{align*}
    &\mathcal{L}_n=\frac{1}{2}\sum_m \alpha_{-m}\cdot \alpha_{m+n}+\frac{1}{4}\sum_{r\in\mathbb{Z}+\frac{1}{2}} (2r+n)b_{-r}\cdot b_{r+n},\\
    &\mathcal{Q}_{r}=\frac{1}{2}\sum_{m}\alpha_{-m}\cdot b_{r+m},
\end{align*}
and similar expression for $\bar{\mathcal{L}}_n$s and $\bar{\mathcal{Q}}_r$s in terms of the right-moving modes. Now, using \eqref{oscmap} along with the relation between the tensile and tensionless bosonic modes as given in \eqref{tsc8} into the tensile constraints in \eqref{flyingladdu}, one can check that the connection between tensionless SCCA$_H$ generators and super-Virasoro generators are connected through the In\"on\"u-Wigner contraction identical to 
\bea{}
L_n=\mathcal{L}_n-\bar{\mathcal{L}}_{-n}, \quad M_n = \e (\mathcal{L}_n+\bar{\mathcal{L}}_{-n}), \quad Q^+_r=\sqrt{\e}\mathcal{Q}_r,\quad Q^-_r = \sqrt{\e}\bar{\mathcal{Q}}_{r}.
\eea
This confirms the consistency of the intrinsic analysis of homogenous null superstrings with the tensionless limit of tensile RNS superstrings.
\subsection{Inhomogeneous null superstring}
As discussed in \cite{Bagchi:2017cte,Bagchi:2018wsn}, there is another variant of null superstrings called the inhomogeneous null superstring. In order to formulate action of this theory one needs to recall the modified 2d Clifford algebra for the Carrollian case given in \eqref{modcliff}. One can see that the representation \eqref{rhonull} is not unique, and other representations are possible keeping the nilpotency of one of the matrices. To this end, let us consider the following alternative
\be{Inhomogamma}
\rho^0=\begin{bmatrix}1 & 0 \\ 0 & -1\end{bmatrix},\qquad \rho^1=\begin{bmatrix}0 & 0 \\ 1 & 0\end{bmatrix}.
\ee
The inhomogeneous spinors corresponding to these gamma matrices are not necessarily real and as shown in \cite{Bagchi:2017cte}, the tensionless limit of RNS superstring will lead us to the null strings with complex spinors. Let us begin with the following complex spinors 
\bea{inhomofer}\psi^\mu=\begin{bmatrix}\psi^\mu_0 \\ \psi^\mu_1 \end{bmatrix}.
\eea 
One can construct a null fermionic string action by adding this fermion to the gauge fixed ILST action. The fermionic part of the action remain same as that of the massless Dirac action with \eqref{Inhomogamma} as gamma matrices . The final action we obtain is
\be{action_inhom}
S  =\int d^2 \sigma\left[\dot{X}^2+ i(\psi_0^{*}\cdot\dot{\psi_0}+\psi_1^{*}\cdot\dot{\psi_1}-\psi_1^{*}\cdot\psi_0')\right].
\ee
As one can see, unlike the action \eqref{hcfer5}, this action is asymmetric in its two spinor components. This action is invariant under the following worldsheet gauge transformations\begin{subequations}\label{inhomdiff}
\bea{}
\delta_\xi X\hspace{.5mm}&=&\xi^\a\p_\a X,\qquad
\delta_\xi \psi_0=\xi^\a\p_\a \psi_0+\frac{1}{4}\p_\a\xi^\a \psi_0, \\
\delta_\xi \psi_1&=&\xi^\a\p_\a \psi_1~+~\frac{1}{4}\p_\a\xi^\a \psi_1~+~\frac{1}{2}(\p_1\xi^0)\psi_0\\
\delta_\eps X\hspace{.5mm} &=&i(\eps^{1*} \psi_{0}+\eps^{0*}\psi_{1}),\quad 
\delta_\eps \psi_{0} = -\eps^1\dot{X},\quad
\delta_\eps \psi_{1} = -\eps^0\dot{X}-\eps^1X' .
\eea
\end{subequations}
One can see that in \eqref{inhomdiff}, the two spinor components transform in two different ways. Their transformation comes from the transformation of the two-component fermion \eqref{inhomofer} under worldsheet diffeomorphism and Weyl transformation
\be{}
\delta_\xi \psi=\xi^\a\p_\a\psi+\frac{1}{4}(\p_\a\xi^\a)\psi-\frac{1}{4}(\eps^{\a\b}\p_\a\xi_\b)\rho^0\rho^1\psi.
\ee
In the above $\eps^{\a\b}$ is the antisymmetric tensor on the worldsheet. The parameters $\xi$ and $\eps$ in the transformations \eqref{inhomdiff} follow the following restrictions in order for the action \eqref{action_inhom} to be invariant
\begin{subequations}\label{inhomocondt}
\begin{align}
 &\partial_0\xi^0 =\partial_1\xi^1, \qquad~~\hspace{.5mm}\partial_0\xi^1=0,\\ &\partial_0\eps^{1*} =\partial_0\eps^0=\partial_1\eps^1=\partial_1\eps^{0*},\\ &\partial_0\eps^{0*} =\partial_0\eps^1=0,~~\hspace{1mm} \psi_0=\psi_1^{*}.
\end{align}
\end{subequations}
As shown in \cite{Bagchi:2017cte}, generators of these residual symmetries in superspace satisfy inhomogeneous Superconformal Carrollian Algebra (SCCA$_I$)
\bea{agcai1} 
&&[L_m,L_n]=(m-n)L_{m+n},~~~[L_m,M_n]=(m-n)M_{m+n},~~~ [L_m,G_r]=\Big(\frac{m}{2}-r\Big)G_{m+r}, \nonumber \\
&&[L_m,H_r]=\Big(\frac{m}{2}-r\Big)H_{m+r},~~~[M_m,G_r]=\Big(\frac{m}{2}-r\Big)H_{m+r}, \,\nonumber \\
&&\{G_r,G_s\}=2L_{r+s}, \qquad \{G_r,H_s\}=2M_{r+s}.
\eea
The centrally extended SCCA$_I$ is given in \eqref{sgcai}. We now turn to obtain the same algebra using different approach. The EOM corresponding to the action \eqref{action_inhom} are given by
\be{NSsoln}
\ddot{X}^\mu=0, \qquad \dot{\psi}_0^\mu=0, \qquad \dot{\psi}_1^\mu={\psi'}_0^\mu. 
\ee
Solution for $X^\m$ remains same as \eqref{fcb1}. In NS-NS sector, solution of \eqref{NSsoln} can be mode expanded as
\bea{}\label{NSsolnmode}
\psi_0^\mu(\sigma,\tau)=\sqrt{c'}\sum_r \beta^{\mu}_r e^{-ir\sigma}, ~~~
\psi_1^\mu(\sigma,\tau)=\sqrt{c'}\sum_r [\gamma^\mu_r-ir\tau\beta^{\mu}_r] e^{-ir\sigma}. 
\eea
The fermionic Poisson brackets in \eqref{fpb} essentially lead us to the following non-vanishing anticommutators between the fermionic modes
\be{abb}
\{\gamma_{r}^{\mu},\beta_{s}^{\nu} \} = 2\delta_{r+s}\eta^{\mu\nu}.
\ee
The constraints of this theory, which can be obtained from the energy momentum tensor components and the supercurrents of the action \eqref{action_inhom}, are given by
\bea{}\label{inhomoconst}
\dot{X}\cdot X'+\frac{i}{4}&&\Big[\psi_0'\cdot\psi_1+\psi'_0\cdot\psi_1\Big]=0,\qquad\dot{X}^2+\frac{i}{2}\psi_0'\cdot\psi_0=0\nonumber\\
&&\psi_0\cdot X'+\psi_1\cdot\dot{X}=0,\qquad
\psi_0\cdot\dot{X}=0.
\eea
Using the mode expansion \eqref{fcb1} and \eqref{NSsoln} one finds
\begin{subequations}\label{const}
\bea{}
\dot{X}\cdot X'+\frac{i}{4}\Big[\psi_0'\cdot\psi_1+\psi'_0\cdot\psi_1\Big]&=& 4c'\sum_{n} \Big[L_n-in\tau M_n \Big]e^{-in\sigma},\\
\dot{X}^2+\frac{i}{2}\psi_0'\cdot\psi_0&=&4c'\sum_{n}M_ne^{-in\sigma}, \\
\psi_0\cdot X'+\psi_1\cdot\dot{X}&=&4c'\sum_{r} \Big[G_r-ir\tau H_r \Big]e^{-ir\sigma}, \\
\psi_0\cdot\dot{X}&=&4c'\sum_{r} H_re^{-ir\sigma},
\eea
\end{subequations}
where $L_n$, $M_n$, $G_r$ and $H_r$ are given by
\begin{subequations}\label{constantinople}
\bea{}
L_n&=&\frac{1}{2}\sum_{m}  A_{-m}\cdot B_{m+n}+\frac{1}{4}\sum_{r}(2r+n)\Big(\beta_{-r}\cdot\gamma_{r+n}+\gamma_{-r}\cdot\beta_{r+n}\Big),  \\
M_n&=&\frac{1}{2}\sum_{m} B_{-m} \cdot B_{m+n}+\frac{1}{4}\sum_{r}(2r+n)\beta_{-r}\cdot\beta_{r+n},\\
G_r&=&\frac{1}{2}\sum_{m} (A_{-m} \cdot \beta_{m+r} +B_{-m} \cdot \gamma_{m+r}), \\
H_r&=&\frac{1}{2}\sum_{m} (B_{-m} \cdot \beta_{m+r}). 
\eea
\end{subequations}
Using the algebra between the modes $A$, $B$, $\b$ and $\g$, one can find that these generators satisfy the SCCA$_I$ given in \eqref{agcai1}.
\paragraph{Limiting approach:}
Let us now see the same theory emerge from a limit of tensile superstrings. Looking into the gauge fixed action of tensile superstrings from 
\eqref{RNSSuperstring} one can see that as long as $\psi_0=\psi_1^{*}$ (see \eqref{inhomocondt}), the limit on the fermionic side which will take us from RNS superstring to the inhomogeneous tensionless superstring action \eqref{action_inhom} is $\psi_{0}\to\epsilon\psi_{0},~\psi_{1}\to\psi_{1}$, where
\bea{iscaling}
\psi_0 =\frac{\psi_++i\psi_-}{\sqrt{2}}, \quad \psi_1=\frac{\psi_+-i\psi_-}{\sqrt{2}}.
\eea
The fermionic mode expansion for tensile string (see \eqref{tensed1}) scale at tensionless limit as below
\begin{subequations}\label{istanbul}
\bea{}
\psi_0^\mu(\sigma,\tau)&=&\frac{\e}{\sqrt{2}}\Big(\psi^\mu_++i\psi^\mu_-\Big)=\sqrt{c'}\sum_r \Big[\sqrt{\e}(b^\mu_r+i\tilde{b}^\mu_{-r})+\e^{3/2}(\hdots)\Big] e^{-ir\sigma}, \\
\psi_1^\mu(\sigma,\tau)&=&\frac{1}{\sqrt{2}}\Big(\psi_+^\mu-i\psi_-^\mu\Big)=\sqrt{c'}\sum_r \Big[\frac{b^{\mu}_r-i\tilde{b}^{\mu}_{-r}}{\sqrt{\e}}-ir\tau\sqrt{\e}(b^\mu_r+i\tilde{b}^\mu_{-r})\Big]e^{-ir\sigma}.
\eea
\end{subequations}
From here one can easily deduce how the fermionic modes are being scaled at tensionless limit
\bea{} \label{fermosc}
\gamma^{\mu}_r=\frac{1}{\sqrt{\e}}\Big(b^{\mu}_r-i{\tilde{b}}^{\mu}_{-r}\Big), \quad \beta^{\mu}_r=\sqrt{\e}(b^{\mu}_r+i{\tilde{b}}^{\mu}_{-r}).
\eea
Using \eqref{fermosc} along with \eqref{tsc8} on \eqref{constantinople} one can find that the tensionless constraint modes in \eqref{constantinople} are connected to the tensile super-Virasoro generators outlined in \eqref{supvir} through the same In\"on\"u-Wigner contraction connecting homogeneous Superconformal Carrollian generators to the super-Virasoro generators given in \eqref{ingenscal}. 
\paragraph{Real fermion from exotic tensile parents:}
In continuation of the previous discussion, let us mention another intriguing result on tensionless superstrings in brief. In \cite{Bagchi:2018wsn}, it was shown that there could be two cousins of tensile string theory which at tensionless limit reduce to inhomogeneous null strings with real fermions. One of them had a gauge fixed action identical to the usual tensile superstring action (see \eqref{RNSSuperstring} for example), with a slight twist: one of the fermion components $\psi_-$ is anti-hermitian. Making this choice for $\psi_-$ does not affect the hermiticity of the action. And in fact from the mode expansion one could show this anti-hermiticity of $\psi_-$ seamlessly translates into the hermiticity properties of inhomogeneous tensionless oscillators that appear in mode expansion of  $\psi_0$ and $\psi_1$ i.e. we could have $\gamma^{\dagger}_{-r}=-\gamma_r$ and $b^\dagger_{-r}=-b_r$. 
\medskip

Apart for this twisted superstring, \cite{Bagchi:2018wsn} also discussed another version of tensile superstring theory which reduces to inhomogeneous null superstrings in tensionless limit. This theory can be written as a supersymmetric extension of tensile twisted string theory \cite{Casali:2016atr, Lee:2017utr} which is constructed on the flipped representation of Virasoro algebra (see around \eqref{automorphism} for a discussion). Note that just like Virasoro algebra, the super-Virasoro algebra too, allows the same automorphism \eqref{automorphism}, and the parent theory consequently leads to twisted tensile superstring theory. The consequences of this parent theory are very interesting, but we will not go into that for brevity.

\section{Open null superstrings}\label{Open null superstrings}
In this section, we review the open null superstring theory satisfying Dirichlet boundary conditions at both the endpoints of the string. This analysis is based on the recent work \cite{Bagchi:2025jgu}. Here we only cover the homogeneous null open superstrings. The open version of inhomogeneous null superstrings is yet to be formulated. 
\medskip

We begin with the boundary term of \eqref{hcfer1}. Its variation is given by
\begin{equation}
\label{varaction}
\begin{split}
\delta S_{bdy} = -\int d^2\sigma\, \Big[\,\partial_{\a}\,& \left(V^\a\, V^\b\,\partial_{\b} X_{\mu}\delta X^{\mu}+iV^\a V^\b\chi_{\b}\psi_\mu\delta X^{\mu}\right)  -i\partial_\a\left(\bar{\psi}^\mu \rho^\a\delta \psi_\mu\right)\Big].
\end{split}
\end{equation}
The variation of the bosonic part of \eqref{varaction} was discussed in Sec.~ \ref{Bosonic open null strings}. The fermionic part can be rewritten as
\begin{eqnarray}
\label{fbc}
\int d^2\sigma \,\partial_\a\left(\bar{\psi}^\mu \rho^\a\,\delta \psi_\mu\right) = \oint_c dl ~ (\,\bar{\psi}^\mu\, n_\a \rho^\a\,\delta\psi_\mu\,).
\end{eqnarray}
This term can be put to zero by imposing a fermionic analogue of null boundary condition in \eqref{eq:DerDritteMann}
\begin{eqnarray}
n_\a \rho^\a = 0.
\end{eqnarray}
Like the bosonic string, we will here focus on Dirichlet boundary conditions. The gauge choice remains same as \eqref{gauge}, along with gamma matrices identical to \eqref{rhonull}. Consequently, residual symmetry transformation in \eqref{homosusy}, EOM and constraints in \eqref{hcfer7} remain unchanged. \eqref{gauge} and \eqref{rhonull} imply 
\begin{align}
 \oint_c dl ~ n_a (\,\bar{\psi}\,\rho^a\,\delta\psi\,) = \int d\sigma\,( \psi_+\,\delta\psi_++\,\psi_-\,\delta\psi_-)\Big|_{\tau=-\infty}^{\tau=\infty}=0.
 \end{align}
Since this boundary term trivially vanishes, we cannot obtain any boundary condition from it. However, one can still impose boundary conditions on the fermions consistent with the gauge‑fixed supersymmetry transformations \eqref{homosusy}:
\begin{equation}\label{jalebi}
\delta_\eps X\Big|_{\s=0,\pi}=0, \qquad \implies \eps^+\,\psi_+=-\,\eps^-\,\psi_-.
\end{equation}
It is important to keep in mind that like tensile open superstring, here too, existence of boundary in the worldsheet breaks half of the worldsheet supersymmetry. Following the steps outlined in \eqref{sec2.1}, one can find the Neveu-Schwarz sector of homogeneous null open superstrings satisfying the Dirichlet boundary condition
\begin{subequations} \label{tensionlessns}
\begin{align}
    & \eps^+(0)\,= \eps^-(0)\, \implies \psi_+(0)=-\,\psi_-(0),\\
    & \eps^+(\pi)= -\eps^-(\pi)\, \implies \psi_+(\pi)=\psi_-(\pi).
    \end{align}
\end{subequations}
Applying the constraints \eqref{tensionlessns} to the fermionic mode expansion in \eqref{tlessmode} we find
\begin{align}\label{strongconstraint}
   \beta^{~\mu+}_{r}=-\beta^{~\mu-}_{-r} ~~\forall r.
    \end{align}
 The above identification in \eqref{tlessmode} results into the mode expansions of $\psi_\pm$ taking the following form
 \begin{align} \label{modeexpansion}
    \psi^\mu_+(\tau,\sigma)=&\sqrt{c'}\sum_{r\in\mathbb{Z}+\frac{1}{2}}\beta^{~\mu+}_{r}e^{-ir\sigma},\quad \psi^\mu_-(\tau,\sigma)=-\sqrt{c'}\sum_{r\in\mathbb{Z}+\frac{1}{2}}\beta^{~\mu+}_{-r}e^{-ir\sigma}.
\end{align}
Using the fermionic Poisson brackets we arrive at the following anticommutators between the fermionic modes,
\be{abb2}
\{\beta_{r}^{\mu+},\beta_{s}^{\nu+} \} = \delta_{r+s}\eta^{\mu\nu}.
\ee
For bosonic field $X^\m$, the identification between $C$ and $\tilde{C}$, open string mode expansion, and the Poisson bracket remain same as \eqref{nullbosiden}, \eqref{modeexpansion1}, and \eqref{eq:osc} respectively. Substituting the mode expansions of the null open superstring in \eqref{modeexpansion1} and \eqref{modeexpansion} to the supersymmetric constraint equations in \eqref{hcfer72} we get
\begin{subequations}
\label{m}
\begin{align}
\frac{1}{2}\,\dot X^2&   
=c'\sum_{p} \mathcal{P}_pe^{-ip\sigma}\\
\dot X \cdot X'+\frac{i}{2}\,\bar{\psi}\cdot\psi' &
=c'\sum_{p}\left(\mathcal{O}_p-i\,p\,\tau \mathcal{P}_p\right)e^{-ip\sigma} \\
       \psi_+\cdot\dot{X} =\psi_-\cdot\dot{X}&    
       =c'\sum_{r\in\mathbb{Z}+\frac{1}{2}}\mathcal{H}_{r}e^{-ir\sigma}.
    \end{align}
\end{subequations}
In the above, the generators $\mathcal{O}_{p}$, $\mathcal{P}_p$ and $\mathcal{H}_{r}$ in terms of the bosonic modes $C$ and fermionic modes $\b$ become
\begin{subequations}\label{genmode}
\begin{align}
\mathcal{O}_{p} =~&  \frac{1}{2}\sum_{n}(C_n^\mu - C_{-n}^\mu)(C_{p-n}^\nu + C_{n-p}^\nu) \notag \\
&~~~~~~ + \frac{1}{4} \sum_{r\in\mathbb{Z}+\frac{1}{2}} \left[ (2r + p)\,(\beta^{\mu+}_{-r} \beta^{\nu+}_{r+n}) - (2r - p)\,(\beta^{\mu+}_{-r} \beta^{\nu+}_{r-n}) \right] \eta_{\mu\nu} \\
\mathcal{P}_p =~&  \frac{1}{4}\sum_{n}(C_n^\mu - C_{-n}^\mu)(C_{p-n}^\nu - C_{n-p}^\nu)\, \eta_{\mu\nu} \\
\mathcal{H}_{r} =~& \frac{1}{2} \sum_{n} (C_n^\mu - C_{-n}^\mu)\, \beta^{\nu+}_{r+n} \eta_{\mu\nu}.
\end{align}
\end{subequations}
The commutation and anticommutation relations in \eqref{eq:osc} and \eqref{abb2} can be used to show that the generators in \eqref{genmode} satisfy
\begin{equation}\label{tintin}
    \begin{split}
        &[\mathcal{O}_n,\mathcal{O}_m]=(n-m)\mathcal{O}_{n+m}-(n+m)\mathcal{O}_{n-m},\\
    &[\mathcal{O}_n,\mathcal{P}_m]=(n-m)\mathcal{P}_{n+m}+(n+m)\mathcal{P}_{n-m},\\
    &[\mathcal{O}_r,\mathcal{H}_{s}]=\Big(\frac{r}{2}-s\Big)\mathcal{H}_{r+s}+\Big(\frac{r}{2}+s\Big)\mathcal{H}_{s-r},\\
    &\{\mathcal{H}_{r},\mathcal{H}_{s}\}=\mathcal{P}_{r+s},~~~
    [\mathcal{H}_{r},\mathcal{P}_s]=[\mathcal{P}_r,\mathcal{P}_s]=0,\\
    \end{split}
\end{equation}
which forms the classical part of Homogeneous Boundary Superconformal Carrollian Algebra (BSCCA$_H$). The centrally extended algebra is given in \eqref{tintin2}. 
\paragraph{Limiting approach:}
Now that we looked into the theory from intrinsic perspective, let us turn our attention to the homogenous Carroll limit on the results in the parent tensile theory and see whether they match with the corresponding results derived in the previous section. Limiting analysis of the bosonic field mode expansion remains identical to the same covered in section \ref{Openbosonlimper}. Let us begin by recalling mode expansion of the fermionic field $\psi^{\mu}(\tau,\sigma)$ on the tensile worldsheet from \eqref{tensi1}
\begin{align}
    \psi^{\mu}_{+}(\tau,\sigma)=\sqrt{\alpha'}\sum_{r\in\mathbb{Z}+\frac{1}{2}}b^\mu_re^{-ir(\tau+\sigma)},\quad \psi^\mu_{-}(\tau,\sigma)=-\sqrt{\alpha'}\sum_{r\in\mathbb{Z}+\frac{1}{2}}b^\mu_re^{-ir(\tau-\sigma)}.
     \end{align}
Applying the limit $ \tau\to\epsilon\tau,~~\alpha'\to\frac{c'}{\epsilon}~~\sigma\to\sigma,~~\epsilon\to0$ along with the homogeneous limit $\psi_{\text{tensionless}} = \sqrt{\epsilon} \, \psi_{\text{tensile}}$ on this mode expansion takes us to the following
\begin{subequations}\label{hukomukhohangla}
\begin{align}
\label{tensi2}
     \psi^\mu_{+}(\tau,\sigma)&=\sqrt{\e\alpha'}\sum_{r\in\mathbb{Z}+\frac{1}{2}}b^\mu_re^{-ir\sigma}(1-i\epsilon r\tau)\approx\sqrt{c'}\sum_{r\in\mathbb{Z}+\frac{1}{2}}b^\mu_re^{-ir\sigma}+\mathcal{O}(\epsilon)\\ \label{tensi21}
    \psi^\mu_{-}(\tau,\sigma)&=-\sqrt{\e\alpha'}\sum_{r\in\mathbb{Z}+\frac{1}{2}}b^\mu_re^{ir\sigma}(1-i\epsilon r\tau)\approx-\sqrt{c'}\sum_{r\in\mathbb{Z}+\frac{1}{2}}b^\mu_{-r}e^{-ir\sigma}+\mathcal{O}(\epsilon)
\end{align}
\end{subequations}
Comparing \eqref{hukomukhohangla} with the fermionic mode expansion in \eqref{modeexpansion}, we make the following identification   
\begin{align}
\label{beta}
    \beta^{\mu+}_{r}=b^\mu_r.
\end{align}
We recall that in tensile string theory (briefly reviewed in Appendix \ref{sec2.1}), expressions of super-Virasoro generators in terms of tensile oscillators as obtained from the tensile string energy momentum tensor components are given by
\begin{align*}
    &\mathbb{L}_n=\frac{1}{2}\sum_m \alpha_{-m}\cdot \alpha_{m+n}+\frac{1}{4}\sum_{r\in\mathbb{Z}+\frac{1}{2}} (2r+n)b_{-r}\cdot b_{r+n},\\
    &\mathbb{Q}_{r}=\frac{1}{2}\sum_{m}\alpha_{-m}\cdot b_{r+m}.
\end{align*}
Substituting \eqref{openbv} and \eqref{beta} into \eqref{genmode}, we see that the generators $\mathcal{O}_{n}$, $\mathcal{P}_n$, and $\mathcal{H}_{r}$ are connected to their tensile counterparts $\mathbb{L}_n$ and $\mathbb{Q}_{r}$ through the In\"on\"u-Wigner contraction identical to 
 contraction \eqref{HIWC}
\begin{align}
    \mathcal{O}_n=\mathbb{L}_n-\mathbb{L}_{-n},~~~P_n=\epsilon~(\mathbb{L}_n+\mathbb{L}_{-n}),~~~\mathcal{H}_
    r=\sqrt{\epsilon}\mathbb{Q}_{r},~~~\e\to 0.
\end{align}
From here one can see that the intrinsic analysis provides us with the results consistent with appropriate limit from tensile superstring theory.

\newpage
\part{Quantum Null Strings}\label{III}

We now begin our journey into the strange world of quantum null strings. Here, unlike the more straight-forward case of the classical null string, we encounter surprises at almost every step. We first see that upon careful canonical quantisation, not one, but three consistent theories emerge from the ILST action. We first account for the various novel features of the quantum null string before addressing the spectrum of each quantum theory. 

\begin{figure}[h!]
    \centering
    \includegraphics[width=0.9\linewidth]{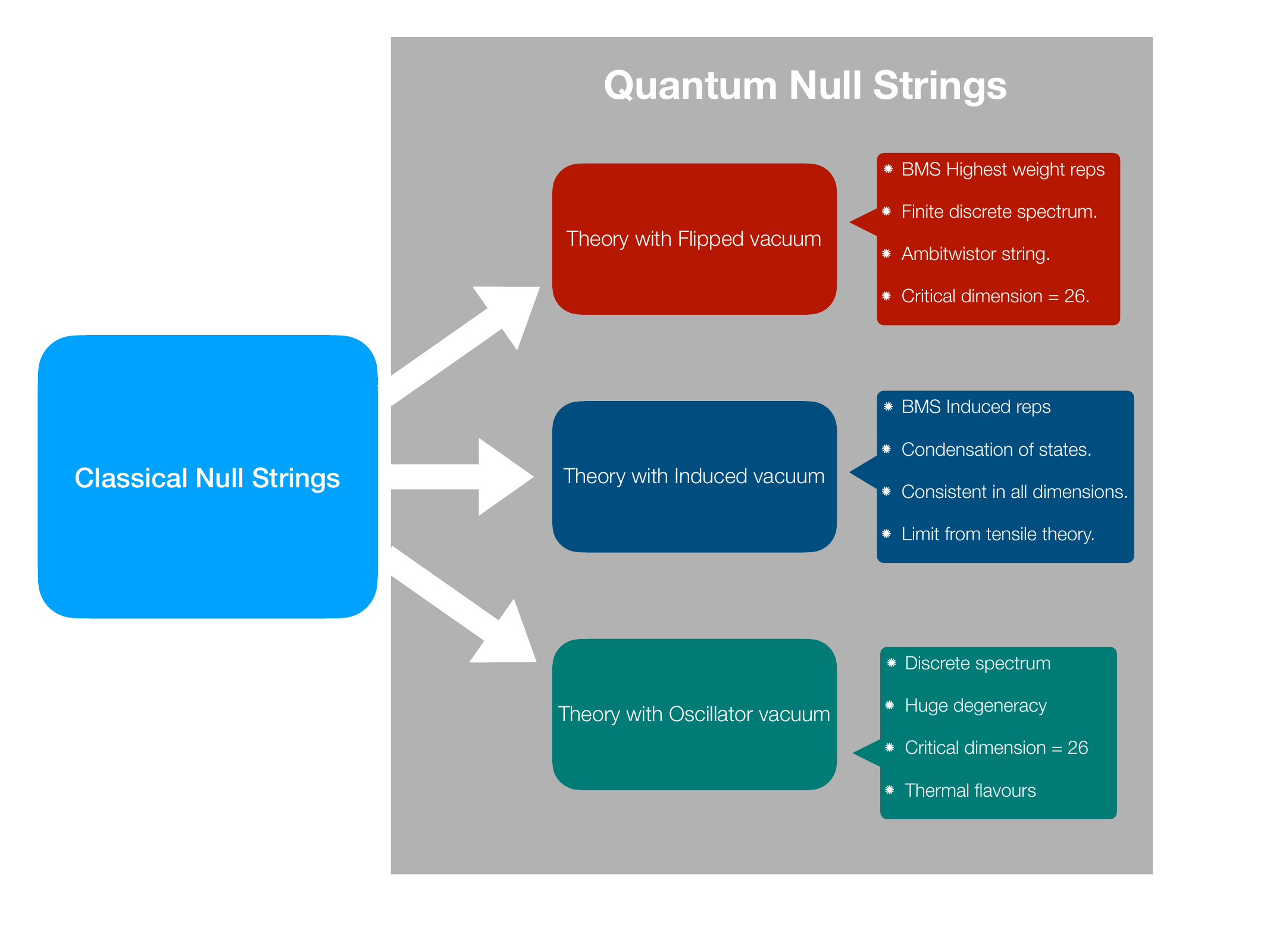}
    \caption{The menagerie of quantum null string theories}
    \label{fig:taleof3}
\end{figure}

This part of our review is divided into sections as follows:

\begin{itemize}
    \item {\em \hyperref[From one classical to three quantum theories]{From one classical to three quantum theories}}: This section examines the canonical quantisation of the classical null string given by by ILST action. The careful imposition of constraints on the worldsheet leads to three consistent quantum null strings. We describe how this comes about and then consider quantisation in the Light Cone coordinates. This helps us figure out the critical dimension of these three theories by considering the closure of the Lorentz algebra in the target spacetime. 
    
    \item  {\em \hyperref[Novelties of Quantum Null Strings]{Novelties of Quantum Null Strings}}: The three quantum null strings come with a lot of surprising properties. In this section, we collect the most intriguing of these properties. We show how an open string arises from a closed string when the tension is dialled to zero and discuss null string complementarity, a rather unexpected observer dependent phenomenon of the emergence of the open string in which one can either end up with a D-instanton or a spacefilling D-25 brane from the closed string in the null limit. We show how perturbative closed string degrees of freedom condense in the tensionless limit to form the open string. We then look at Rindler physics on accelerated string worldsheets which is a natural setting for explaining some of these aparently bizairre features. 
    
    \item {\em\hyperref[Physical Hilbert spaces for three vacua]{Physical Hilbert spaces for three vacua}}: As mentioned above and also shown in Fig~\ref{fig:taleof3}, from a single classical theory, we arrive at three different quantum null string theories. These are built of separate vacua called the Induced, Flipped and Oscillator vacua. In this section, we consider in detail the spectra of these three theories. 
    
    \item{\em \hyperref[Compactification: Circle and Torus]{Compactification: Circle and Torus}}: This section studies all three quantum theories first in target spacetimes compactified on circle and then generalises the study to compactification on a $d$ dimensional torus. While studying compactification on torus we also introduce a constant Kalb-Ramond background along the torus. We compare the results in these three theories with tensionless limit of the corresponding results in their tensile counterparts. This section concludes with a discussion on the role of T-duality (for circle compactification) and $O(d,d,\mathbb{Z})$ duality (for toroidal compactification with background field) in null strings.
\end{itemize}

\newpage

\section{From one classical to three quantum theories}\label{From one classical to three quantum theories}

The classical theory of null strings that we have encountered in the previous part of our review is based on the ILST action. We saw that various quantities in the classical null string could be derived intrinsically from the ILST action, and could be equivalently derived in a Carroll limit of the corresponding tensile quantities. 

\medskip

We now attempt to consistently quantise the theory of null strings based on the ILST action. We will take recourse to the method of Old Covariant quantisation (OCQ). We will see that careful canonical quantisation from the ILST string leads to three different quantum null strings. In the second part of this section, we will consider Light Cone quantisation (LCQ) of these three theories and compute their critical dimensions. Interestingly these three theories admit different central charges for their residual symmetry algebra which has been demonstrated in the Appendix \ref{ApA1}.
\paragraph{A quick reminder of basics:}
Our discussion of quantised tensionless bosonic string theory would be based on \cite{Bagchi:2020fpr,Bagchi:2021rfw}. In order to quantise the tensionless theory we need to promote the dynamical variables $X^{\mu}(\tau,\sigma)$ and their canonical momenta $\Pi^{\mu} (\tau,\sigma)$ to operators. We have seen that these fields were expressed in terms of tensionless oscillator modes ($A,B$) or alternatively ($C,\tilde{C}$). As a result of quantisation, these modes ($A,B$) or ($C,\tilde{C}$) would be promoted to operators. As in the tensile string, here too, equal time commutators between $X^{\mu}$ and $\Pi^{\mu}$ will be
 \begin{equation}
 \begin{split}     \left[X^\mu(\tau,\sigma),\Pi^\nu(\tau,\sigma')\right]=i\eta^{\mu\nu}\delta (\sigma-\sigma').
     \end{split}
 \end{equation}
Consequently the Poisson brackets between the ($A,B$) as given in \eqref{tsc5} will be replaced by the following commutator brackets: 
\begin{align}\label{ABcom}
    [A^{\mu}_{m},A^{\nu}_{n}]=[B^{\mu}_{m},B^{\nu}_{n}]=0,\hspace{8mm}[A^{\mu}_{m},B^{\nu}_{n}]=2m\delta_{m+n}\eta^{\mu\nu}
\end{align}
Or, alternatively 
\begin{align}\label{TSQR3}
    [C^{\mu}_{m},C^{\nu}_{n}]=[\tilde{C}^{\mu}_{m},\tilde{C}^{\nu}_{n}]=m\delta_{m+n},\hspace{8mm}[C^{\mu}_{m},\tilde{C}^{\nu}_{n}]=0.
\end{align}
We note again that $(A, B)$ are not harmonic oscillator like, while $(C,\tilde{C})$ are.
After quantisation, the worldsheet residual symmetry algebra becomes the centrally extended BMS$_3$ algebra
\begin{equation}\label{BMSalgebra}
    \begin{split}
        [L_m,L_n]&=(m-n)L_{m+n}+\frac{c_{L}}{12}m(m^2-1)\delta_{m+n,0},\\ 
        [L_m,M_n]&=(m-n)M_{m+n}+\frac{c_{M}}{12}m(m^2-1)\delta_{m+n,0}, \\
        [M_m,M_n]&=0.
    \end{split}
\end{equation}
As we will see, depending on the physical state conditions we are going to impose, there will be multiple versions of quantum tensionless string theory and the values of ($c_{L},c_{M}$) will be different for each of those theories.

\subsection{Three quantum null string theories}\label{Three quantum null string theories}
Let us recall the constraints on $X^\mu$ from our classical analysis as given in \eqref{constraints2} which state that the components of energy momentum tensor $T_{1}$ and $T_{2}$ vanish. Along with $X^\mu$, $T_{1}$ and $T_{2}$ too become operators and hence, in quantum theory, the constraints have to be imposed on these operators. However, imposing the entire operator to be zero is too strong constraint to build a quantum theory. The most general constraint that could lead to sensible quantum theories is 
 \begin{equation}\label{Review5}
     \bra{phys'}T_1\ket{phys}=\bra{phys'}T_2\ket{phys}=0.
 \end{equation}
Rewriting \eqref{Review5} in terms of $L_{n}$ and $M_{n}$ one gets
\begin{equation}\label{constraint}
     \bra{phys'}L_n\ket{phys}=0,~~~\bra{phys'}M_n\ket{phys}=0, ~~\forall n\in \mathbb{Z}.
 \end{equation}
We call the above {\em{sandwich conditions}}. There are three distinct conditions on an operator $F_{n}$, which are consistent with the sandwich condition on $F_{n}$. They are listed below:
\begin{subequations}\label{Review6}
    \begin{equation}
        F_{n}\ket{phys}=0,\hspace{5mm}\forall n>0,
    \end{equation}
    \begin{equation}
        F_{n}\ket{phys}=0,\hspace{5mm}\forall n\neq0,
    \end{equation}
    \begin{equation}
        F_{n}\ket{phys}\neq0,\hspace{2mm}\text{but}\hspace{2mm} \bra{phys'}F_n\ket{phys}=0,\hspace{5mm}\forall n.
    \end{equation}
\end{subequations}
When we try to quantise a theory with BMS$_3$ symmetry, we need to build theories where conditions \eqref{constraint} are met. Hence, following \eqref{Review6}, we can have nine possible conditions which are compatible with the sandwich conditions \eqref{constraint}. They are listed below:
\begin{subequations}\label{allconstraint}
\begin{align}
     L_n\ket{phys}=0, (n>0),~~~
     \begin{cases}
      M_m\ket{phys}=&0, (m>0)\\
       M_m\ket{phys}=&0, (m\neq0)\\
        M_m\ket{phys}=&0, (\forall~ m)
     \end{cases}
;\\
     L_n\ket{phys}=0, (n\neq0),~~~ 
     \begin{cases}
      M_m\ket{phys}=&0, (m>0)\\
       M_m\ket{phys}=&0, (m\neq0)\\
        M_m\ket{phys}=&0, (\forall~ m)
     \end{cases}\label{specialconstraint}
;\\
     L_n\ket{phys}=0, (\forall~n),~~~
     \begin{cases}
      M_m\ket{phys}=&0, (m>0)\\
       M_m\ket{phys}=&0, (m\neq0)\\
        M_m\ket{phys}=&0, (\forall~ m)
     \end{cases}.
\end{align} 
\end{subequations}
In \cite{Bagchi:2020fpr}, all this conditions has been discussed in detail. It has been shown that out of the conditions listed above, only three are consistent with BMS$_3$ algebra. These three conditions lead us to three distinct quantum theories constructed on Flipped, Induced and Oscillator vacuum respectively. They are listed below:
\begin{subequations}\label{consistent quantum theories}
    \begin{align}\label{case1}
      &\textit{Flipped:} \qquad \qquad  ~L_n\ket{phys}=~M_n\ket{phys}=0 \quad \forall~n>0,\\ \label{case2}
       &\textit{Induced:} \qquad \qquad L_n\ket{phys} \neq0,~ M_n\ket{phys}=0 ~~~\forall~n\neq 0,\\
       \label{case3}
        &\textit{Oscillator:} \qquad ~~~ L_n\ket{phys}\neq0,~~~ M_n\ket{phys}\neq 0 ~~~\forall~n \hspace{3mm} \text{(but \eqref{constraint} holds)}.
        \end{align}
\end{subequations}
The remaining possibilities listed in \eqref{allconstraint} turn out to be inconsistent. This stems from the fact that the constraints imposed by ($L_n, M_n$) cannot be chosen independently; rather, they are tied together by underlying BMS structure. Among the various choices, however, there is one notable exception. The condition specified in \eqref{specialconstraint},
\begin{equation}
    L_n\ket{phys}=0,\quad  M_n\ket{phys}=0, \qquad \qquad (\forall n,m \neq 0)
\end{equation}
corresponds to a special case, which may be interpreted as an induced vacuum with zero momentum. This choice is hence admissible. A systematic demonstration for the inconsistency of the remaining constraints can be found in \cite{Bagchi:2020fpr}. For completeness, we explicitly demonstrate the inconsistency for one representative case in appendix \ref{inconsistent constraint}. In what follows, we determine the critical dimensions of the three theories listed in \eqref{consistent quantum theories}.
\medskip

\subsection{Lightcone quantisation and critical dimensions}\label{Lightcone quantisation}
We now review light-cone gauge quantisation of these three quantum theories which will help us to determine the critical dimension of them. Light-cone gauge quantisation for induced theory was originally addressed in \cite{Lizzi:1986nv}, while the same for flipped theory was first addressed in \cite{Gamboa:1989px}. Recent work \cite{Bagchi:2021rfw} rigorously performed light-cone quantisation for all three theories including the oscillator theory. Our discussion of light-cone quantisation is mainly based on \cite{Bagchi:2021ban}. 
\medskip

We begin by introducing the light-cone coordinates $X^{\pm}$ as
\begin{align}\label{TSQR6}
    X^{\pm}=\frac{1}{\sqrt{2}}\big(X^{0}\pm X^{D-1}\big).
\end{align}
In these coordinates, the Minkowski metric becomes
\begin{align}
    \eta_{+-}=-1,~~~\eta_{ij}=\delta_{ij}.
\end{align}
We apply the light-cone gauge on $X^{+}$ in the same way it is applied for tensile string
\begin{align}\label{TSQR7}
    X^{+}=x^{+}+c'k^{+}\tau.
\end{align}
Applying the gauge \eqref{TSQR7} into the mode expansion \eqref{fcb1} implies that 
\begin{align}
    A^+_n=B^+_n=0,~~~\forall n \neq 0.
\end{align}
Applying them on the constraints \eqref{constraints} and using the mode expansion of $X$ in \eqref{fcb1} one gets the following 
\begin{equation}\label{minusdef}
A^-_m = \frac{1}{B_0^+}\sum_{n\neq 0}:A^i_n B^i_{m-n}:~,\quad B^-_m = \frac{1}{2B_0^+}\sum_n :B^i_n B^i_{m-n}:~.
\end{equation}
In the above ``:~:" denotes normal ordering. Clearly the modes $A^-_m$ and $B^-_m$ are dependent on the modes in the transverse directions. They also satisfy the following algebra called the central algebra
\begin{subequations}\label{cent}
\begin{align}\label{Dadhich5}
\left[A^-_m,A^-_n\right] &= \frac{2(m-n)}{B^+_0}A^-_{m+n}+\frac{\tilde{c}_L}{12}(m^3-m)\delta_{m+n,0},\\
\left[A^-_m,B^-_n\right] &= \frac{2(m-n)}{B^+_0}B^-_{m+n}+\frac{\tilde{c}_M}{12}(m^3-m)\delta_{m+n,0},\\
\left[B^-_m,B^-_n\right] &= 0.
\end{align}
\end{subequations}
Here $\tilde{c}_L$ and $\tilde{c}_M$ are central charges, values of which depend on the normal ordering. The zero modes $A^-_0$ and $B^-_{0}$ from earlier classical studies should be $A^-_0$ and $B^-_0=\sqrt{2c'}k^-$. However, for quantum theory one must needs to take into account the normal ordering ambiguity. Comparing \eqref{cent} with \eqref{bms3}, one can express the normal ordering of $A^-_0$ and $B^-_0$ in terms of normal orderings of $L_0$ and $M_0$ as
\be{}\label{N.O.A.}
A^-_0 = \frac{2a_L}{B^+_0},~~~~B^-_0 = \sqrt{2c'}k^- + \frac{a_M}{B^+_0}.
\ee
In the above, $a_L$ and $a_M$ respectively are the normal ordering ambiguities of $L_0$ and $M_0$ which will be fixed. We also rewrite the commutators using $\eta^{\m\n}$ in light-cone coordinates
\begin{equation}\label{modecommij}
\left[A^i_m,B^j_n\right] = 2n\delta^{ij}\delta_{m,-n}~, \quad \left[x^i,k^j\right] = i\delta^{ij}~,\quad \left[x^-,k^+\right] = -i~.
\end{equation}
Now, in order to make the null string theory to be Lorentz invariant, we need to ensure that the gauge choice \eqref{TSQR7} is invariant under the same. For infinitesimal Lorentz transformation $$\delta X^\mu = {\omega^\mu}_\nu X^\nu,$$ variation of the ILST action is given by
\begin{equation}
\begin{split}
\delta S &= \frac{1}{2\pi c'}\int d^2\sigma\left(V^aV^b{\omega^\mu}_\nu\partial_aX^\nu\partial_bX_\mu\right) = -\frac{1}2\int d\tau~\partial_\tau\left(\omega_{\mu\nu}J^{\mu\nu}\right).
\end{split}
\end{equation}
Here $J^{\mu\nu}$ is the generator of the Lorentz transformation given by
\be{}
J^{\mu\nu} \equiv  \int d\sigma \left(X^\mu \Pi^\nu-X^\nu \Pi^\mu\right).
\ee
In order to ensure the theory is Lorentz invariant, it must be ensured that $J^{\mu\nu}$ satisfy the Lorentz algebra
\begin{equation}
\left[J^{\mu\nu},J^{\rho\sigma}\right] = -i\left(J^{\mu\rho}\eta^{\nu\sigma}-J^{\nu\rho}\eta^{\mu\sigma}-J^{\mu\sigma}\eta^{\nu\rho}+J^{\nu\sigma}\eta^{\mu\rho}\right).
\end{equation}
In \cite{Bagchi:2021rfw}, all these commutators are calculated and it was found that all the commutators vanish except the $\left[J^{i-},J^{j-}\right]$. Hence, one needs to impose that $\left[J^{i-},J^{j-}\right]=0$, and that will be possible only for a certain value of normal ordering $a$ and target spacetime dimension $D$.
\medskip

The Lorentzian generator $J^{i-}$ is given by
\begin{equation}\label{Jsplit}
J^{i-} = \int_0^{2\pi}d\sigma:\left(X^i\Pi^--X^-\Pi^i\right):= L^{i-}+S^{i-},
\end{equation}
where $L^{i-}$ and $S^{i-}$ are given by
\begin{subequations}
    \begin{align}
    \label{Ldef}
        &~~~~~L^{i-} = x^ik^- - x^-k^i,\\
        \label{Sdef}
        S^{i-} = -\frac{i}2&\sum_{m\neq 0}\frac{1}m \left(:A^i_{-m}B^-_m: + :A^-_m B^i_{-m}:\right)~.
    \end{align}
\end{subequations}
Since it is straightforward to see that $\left[L^{i-},L^{j-}\right]=0$, we have
\begin{equation}\label{JJprim}
\left[J^{i-},J^{j-}\right] = \left[L^{i-},S^{j-}\right]+\left[S^{i-},L^{j-}\right]+\left[S^{i-},S^{j-}\right].
\end{equation}
\paragraph{Induced Vacuum:}
This vacuum correspond to the physical state condition \eqref{case2} where the vacuum is defined as
\be{}
B^i_m |0\>_I = 0, ~~\forall~m\neq0.
\ee
For this definition of the vacuum the normal ordering in \eqref{minusdef} becomes
\begin{equation}\label{minusdefind}
A^-_m = \frac{1}{B_0^+}\sum_{n\neq 0}A^i_n B^i_{m-n},\quad
B^-_m = \frac{1}{2B_0^+}\sum_n B^i_n B^i_{m-n},
\end{equation}
and as a result one can see that the central charges in \eqref{Dadhich5} vanishes. Normal ordering for $S^{i-}$ is given by
\begin{equation}\label{S^{i-}In}
S^{i-} = -\frac{i}2\sum_{m\neq 0}\frac{1}m \left(A^i_{-m}B^-_m + A^-_mB^i_{-m}\right).
\end{equation}
Using \eqref{S^{i-}In}, one can calculate the commutator in \eqref{JJprim} as
\begin{equation}\label{Incrit}
\begin{split}
\left[J^{i-},J^{j-}\right]
=\sum_{m\neq 0}\frac{1}{mB^+_0}\left(A^i_{-m}B^j_m+B^i_{-m}A^j_m\right)\left(B^-_0-\sqrt{2c'}k^-\right)+\sum_{m\neq 0}\frac{1}{mB^+_0}B^i_{-m}B^j_mA^-_0.
\end{split}
\end{equation}
In order to make the r.h.s. of \eqref{Incrit} vanish for all the physical states one must have the following
\bea{}
A^-_0|phys\rangle &=& 0, \\
B^-_0|phys\rangle &=& \sqrt{2c'}k^-|phys\rangle.
\eea
Comparing this with \eqref{N.O.A.} one can see that the normal ordering ambiguities $a_L$ and $a_M$ from \eqref{N.O.A.} vanish and there is no constraint on the target spacetime dimension.
\paragraph{Flipped Vacuum:} Theory constructed on this vacuum is based on the physical state condition \eqref{case1}, and the vacuum in this theory is given by
\be{ABFv}
A^i_m |0\>_F = 0, B^i_m|0\>_F = 0 ~~~\forall~m > 0.
\ee
For this definition of vacuum the normal ordering in \eqref{minusdef} becomes the following
\begin{subequations}\label{minusdeffl}
\begin{align}
A^-_m &= \frac{1}{B_0^+}\left(\sum_{n < 0}A^i_n B^i_{m-n}+\sum_{n > 0} B^i_{m-n}A^i_n\right),~~~\forall m \neq 0\\
A^-_0 &= \frac{1}{B_0^+}\left(\sum_{n < 0}A^i_n B^i_{-n}+\sum_{n > 0} B^i_{-n}A^i_n\right),\\
B^-_m &= \frac{1}{2B_0^+}\sum_n B^i_n B^i_{m-n}.
\end{align}
\end{subequations}
The central charges in \eqref{Dadhich5} for this vacuum is given by
\begin{align}
    \tilde{c}_L=\frac{8(D-2)}{(B_0^+)^2},~~~\tilde{c}_M=0.
\end{align}
Normal ordering of $S^{i-}$ in this theory is given by 
\begin{equation}\label{Sflip}
S^{i-} = -\frac{i}2\sum_{m> 0}\frac{1}m \left(A^i_{-m}B^-_m + B^i_{-m}A^-_m\right)-\frac{i}2\sum_{m< 0}\frac{1}m \left(B^-_mA^i_{-m} + A^-_mB^i_{-m}\right)~.
\end{equation}
Using \eqref{Sflip} one can calculate $\left[J^{i-},J^{j-}\right]$ 
\bea{JJflip}
&&\left[J^{i-},J^{j-}\right] = \left[L^{i-},S^{j-}\right]+\left[S^{i-},L^{j-}\right]+\left[S^{i-},S^{j-}\right] \nonumber\\
&& =\sum_{m\neq 0}\frac{1}{mB^+_0}\left(A^i_{-m}B^j_m + B^i_{-m}A^j_m\right)\left(B^-_0-\sqrt{2c'}k^-\right) +\left(\frac{D-2}{6}-4\right)\sum_{m\neq 0}\frac{m}{(B^+_0)^2}B^i_{-m}B^j_m \nonumber \\
&& \hspace{1cm} + \left(A^-_0 - \frac{D-2}{6B^+_0}\right)\sum_{m\neq 0}\frac{1}{mB^+_0}B^i_{-m}B^j_m~.
\eea
In order to fix the r.h.s. of \eqref{JJflip} to zero we need the following
\begin{subequations}\label{flippedN.O.A.}
\bea{}
D &=& 26, \\
A^-_0|phys\rangle &=& \frac{4}{B^+_0},~~{\rm \text{and}} \\
B^-_0|phys\rangle &=& \sqrt{2c'}k^-|phys\rangle.
\eea
\end{subequations}
Comparing \eqref{flippedN.O.A.} with \eqref{N.O.A.} one can easily see that $a_L=2$, $a_M=0$. Also, the critical dimension for this theory is 26.
\paragraph{Oscillator Vacuum:}
The oscillator vacuum correspond to the physical state condition described in \eqref{case3}. The vacuum in this case is defined in terms of \{$C,\tilde{C}$\} in the same line as the definition of tensile string vacuum 
\be{}\label{krishnakumarkunnath}
C^i_m |0\>_c = 0 \quad {\rm and}~~~\tilde{C}^i_m|0\>_c=0 ~~~\forall m>0.
\ee
It turns out that for this theory, it is more convenient to work in the harmonic oscillators \{$C,\tilde{C}$\}. Calculating the central algebra in \eqref{cent} for this vacuum in terms of \{$C,\tilde{C}$\} gives
\begin{subequations}\label{CC-}
\begin{align}
\left[C^-_{m},C^-_{n}\right] =& \frac{m-n}{2B^+_0}\left(3C^-_{m+n}+\tilde{C}^-_{-m-n}\right) +\frac{(D-2)}{6(B^+_0)^2}(m^3-m)\delta_{n,-m}~,\\
\left[C^-_m,\tilde{C}^-_{-n}\right] =& -\frac{(m-n)}{2B^+_0}\left(C^-_{m+n}-\tilde{C}^-_{-m-n}\right)~,\\
\left[\tilde{C}^-_{-m},\tilde{C}^-_{-n}\right] =& -\frac{(m-n)}{2B^+_0}\left(C^-_{m+n}+3\tilde{C}^-_{-m-n}\right)-\frac{(D-2)}{6(B^+_0)^2}(m^3-m)\delta_{n,-m}.
\end{align}
\end{subequations}
We also need to re-express $S^{i-}$ in terms of \{$C,\tilde{C}$\} and find the appropriate normal ordering for the oscillator vacuum. The normal ordering for $S^{i-}$ is
\begin{equation}\label{Sosc}
S^{i-} = -i\sum_{m> 0}\frac{1}m \left(C^i_{-m}C^-_m - \tilde{C}^-_{-m}\tilde{C}^i_m\right)-i\sum_{m< 0}\frac{1}m \left(C^-_mC^i_{-m} - \tilde{C}^i_m\tilde{C}^-_{-m}\right)~.
\end{equation}
Using \eqref{Sosc}, one can calculate the r.h.s. of \eqref{JJprim} and find the following
\begin{equation}\label{JJosc}
\begin{split}
\left[J^{i-},J^{j-}\right] &= \left[L^{i-},S^{j-}\right]+\left[S^{i-},L^{j-}\right]+\left[S^{i-},S^{j-}\right]\\
&=\frac{1}{\left(B^+_0\right)^2}\left(\frac{D-2}{6}-4\right)\sum_{m<0}m\left(C^i_{-m}C^j_m-\tilde{C}^i_m\tilde{C}^j_{-m}\right)\\
&+\frac{1}{B^+_0}\sum_{m\neq 0}\frac{1}m\left(C^i_{-m}C^j_m-\tilde{C}^i_m\tilde{C}^j_{-m}\right)\left(2B^-_0- 2\sqrt{2c'}k^- -\frac{D-2}{6B^+_0}\right)\\
&+\frac{1}{B^+_0}\sum_{m \neq 0}\frac{1}m B^i_{-m}B^j_mA^-_0~.\\
\end{split}
\end{equation}
The r.h.s. of \eqref{JJosc} vanishes for the following choice of $a$, $b$ and $D$
\begin{subequations}
\bea{}
D &=& 26, \\
A^-_0|phys\rangle &=& 0,~~{\text{ and}} \\
B^-_0|phys\rangle &=& \left(\sqrt{2c'}k^- + \frac{2}{B^+_0}\right)|phys\rangle.
\eea
\end{subequations}
Comparing with \eqref{N.O.A.}, one can clearly see that for this theory $a_L=0$ and $a_M=2$.
\medskip

So, in summary, in this section we have seen that there are three quantum tensionless string theories that emerge from the ILST string. We have computed their critical dimensions now by considering the string to be propagating in flat target spacetime and attempting to close the spacetime Poincare algebra. We have seen that: 
\begin{itemize}
    \item {\bf Induced theory:} This is built on the Induced vacuum and follows the induced representations of the underlying BMS$_3$ algebra on the worldsheet. Our lightcone analysis told us that there is {\em no restriction on the critical dimension}. In particular, if you start with usual tensile string theory, which is consistent only in $D=26$, the Induced null string can sit as a limit of this theory. We will see while examining the Hilbert space of the theory that this indeed is the most natural candidate when we are taking a very high energy or a tensionless limit of tensile critical string theory. 

    \item {\bf Flipped theory:} In this theory, we use the highest weight vacuum and this theory, as we will see in the coming sections, can be thought about as coming from a rather intriguing ``flipped'' tensile theory which is built on a curious vacuum which is the highest weight vacuum for the holomorphic sector and the lowest weight for the antiholomorphic sector. In our analysis above, we found that this is consistent in $D=26$. Going by this alone, we could envision the Flipped quantum null string also to appear in a corner of critical tensile quantum string theory. We should add that our analysis of the spectrum of the theory, which we will address in the coming sections, indicates that this may not be the right expectation, although we should not rule out some new duality that relates this to usual string theory and its high energy limit. 

    \item {\bf Oscillator theory:} By far the most interesting vacuum seems to be the oscillator vacuum and the theory built on it does not seem to be related to usual string theory at first glance. We will see however that we can make sense of this in terms of accelerated worldsheets in the coming sections. Again, the theory is consistent only in $D=26$ and this is indication that this could reside as a tensionless limit of a critical tensile theory. 
\end{itemize}
The above picture reflects the fact that based on the critical dimensions alone, these three quantum null string theories could arise as corners, albeit strange corners, of critical tensile string theory. 

\begin{figure}[t]
\centering
  \includegraphics[width=12cm]{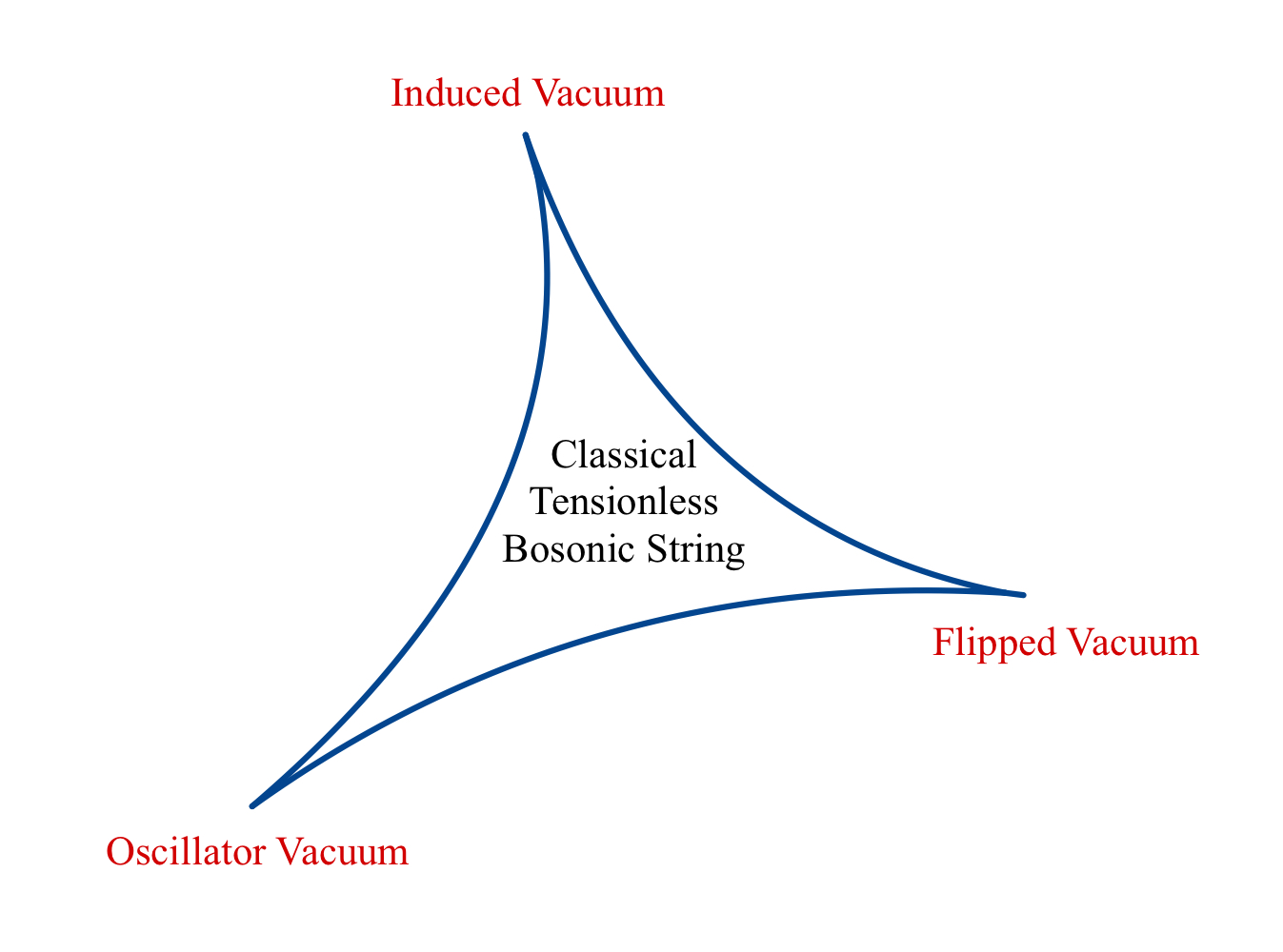}
  \caption{Three quantum avatars of a single classical theory}
  \label{amoeba}
\end{figure}

\newpage

\section{Novelties of Quantum Null Strings}\label{Novelties of Quantum Null Strings}

The quantum theory of null tensionless strings is full of surprising unexpected features at every turn. We have already seen that there are three vacua and hence three consistent quantum theories built out of a single classical theory for the bosonic null string. We now take the reader through some of the most intriguing features of the quantum null strings. 

\subsection{Open strings from closed strings in tensionless limit}

As is common knowledge in interacting (tensile) string theory, open strings can join to form closed strings and closed strings can split up into open strings. Open string amplitudes at one loop produce closed strings. In the tensionless limit, even without interactions, something remarkable happens. As tension decreases, closed strings grow long and floppy and one would expect that something akin to the figure below happens. 
\begin{figure}[ht]
\centering
  \includegraphics[width=14cm]{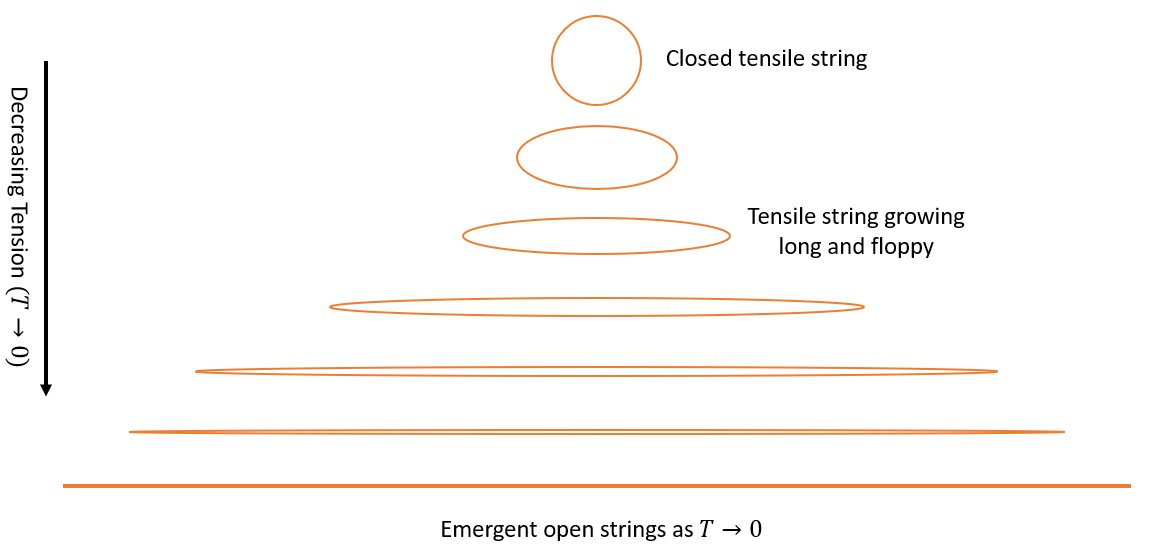}
  \caption{From closed strings to open strings - route 1}
  \label{ClosedtoOpen}
\end{figure}
So one might expect the emergence of open strings from closed string in the tensionless limit. Below, following \cite{Bagchi:2019cay}, we will show that this intuition can be made mathematically precise. 

\medskip

\paragraph{Bogoliubov transformations on the worldsheet:} The fundamental observation behind most of the material in this section is the relation between the tensile $\alpha$ oscillators and the tensionless $C$ oscillators, given by: 
\begin{equation}\label{c-oscillator}
    C^{\mu}_{n}=\cosh \theta~ \a_n^\mu+ \sinh \theta~ \tilde{\alpha}^{\mu}_{-n}, \qquad \tilde{C}^{\mu}_{n}=\sinh \theta~ \alpha^{\mu}_{-n}+\cosh \theta~ \tilde{\alpha}^{\mu}_{n},
     \end{equation}
where, the relation between $\theta$ and $\epsilon$ is as follows
\begin{equation}\label{bogocos}
    \cosh \theta =\frac{1}{2}\Big(\sqrt{\epsilon}+\frac{1}{\sqrt{\epsilon}}\Big), \qquad \sinh \theta= \frac{1}{2}\Big(\sqrt{\epsilon}-\frac{1}{\sqrt{\epsilon}}\Big).
\end{equation}
These are Bogoliubov transformations and preserve the canonical structure even as the relations become singular as $\epsilon \to 0$. It is important to stress here that the relations above only hold for very small values of $\epsilon$. We will return to the full evolution of the tensile to tensionless oscillators in a later section below. 
It is evident that the tensionless oscillators $(C,\tilde{C})$ are linear combinations of both creation and annihilation operators of the tensile theory. Consequently, the tensionless vacuum
\begin{equation}
\ket{0}_C: \qquad C^{\mu}_{\ n} \ket{0}_C = 0 = \tilde{C}^{\mu}_{\ n} \ket{0}_C \qquad \forall n>0.
\end{equation}
is not same as the tensile vacuum defined by 
\begin{equation}
\ket{0}_\a: \qquad \a^{\mu}_{\ n} \ket{0}_\a = 0 = \ta^{\mu}_{\ n} \ket{0}_\a \qquad \forall n>0.
\end{equation}
This distinction in the vacua of oscillators would play a crucial role in what follows. 

\paragraph{Boundary states and Induced vacuum:} Now we briefly recall the definition of boundary states in conventional tensile string theory. A boundary state on the worldsheet is described by a coherent state in the closed string Hilbert space and encode the CFT description of a D-brane in string theory. For a boundary at $\tau=0$, the allowed conditions are
\begin{subequations}\label{bst}
\begin{align}
\p_\t X |B_N\> &= 0 \equiv (\a_n + \ta_{-n}) |B_N\> = 0: \qquad  \text{Neumann},\\
\p_\s X |B_D\> &= 0 \equiv (\a_n - \ta_{-n}) |B_D\> = 0: \qquad  \text{Dirichlet}.
\end{align}
\end{subequations}
Solving these conditions yields Neumann and Dirichlet boundary states: 
\begin{subequations}
\bea{}
 |B_N\> &=& {\mathcal{N}_N} \prod_{n=1}^\infty \exp\left[-\frac{\a_{-n}\cdot\tilde{\a}_{-n}}{n} \right]|0\rangle_\alpha, \\
 |B_D\> &=& {\mathcal{N}_D} \prod_{n=1}^\infty \exp\left[\frac{\a_{-n}\cdot\tilde{\a}_{-n}}{n} \right]|0\rangle_\alpha.
\eea
\end{subequations}
In the above, $\mathcal{N}_N$ and $\mathcal{N}_D$ are normalization constants. 

\medskip

We remind the reader that the quantum null string had three different choices of vacuum and hence three different theories were built out of the ILST action. One of these was the so-called induced vacuum corresponding to the induced representation of the underlying BMS algebra on the worldsheet. The vacuum in the induced representation $\ket{0}_I$ is given by: 
\begin{equation}
    B_n^\mu \ket{0}_I=0, \qquad \forall n.
\end{equation}
In terms of $\{C,\tilde{C}\}$ \eqref{tsc7}, this can be rewritten as
\begin{align}\label{Neumann}
    \big(C^{\mu}_{n}+\tilde{C}^{\mu}_{-n}\big)\ket{0}_{I}=0\qquad \forall n,
\end{align}
which corresponds to the Neumann boundary state condition \eqref{bst} in the ($C,\tilde{C}$) basis:
\begin{align}\label{induced vacuum}
    \ket{0}_{I}=\mathcal{N}_c\prod_{n=1}^{\infty}\exp\Big[-\frac{1}{n}C_{-n}\cdot\tilde{C}_{-n}\Big]\ket{0}_{c}.
\end{align}
So we see that in terms of the $C$ oscillators, the induced vacuum is a Neumann boundary state. 

\paragraph{Connecting vacua:} The Bogoliubov transformations between the tensile $\alpha$ and tensionless oscillators $C$ oscillators also allows us to connect the vacua of the two theories. To see this, we express the $\a$-oscillators in terms of $C$ oscillators as  
\begin{align}
\a^{\mu}_n = e^{i F} C_{n} e^{-iF},\qquad 
\tilde{\a}^{\mu}_n= e^{i F} \tilde{C}_{n} e^{-iF},
\end{align}
where 
\begin{equation}
F = i \sum_{n=1}^{\infty} \theta \left[ C_{-n}\cdot \tilde{C}_{-n} -C_n \cdot \tilde{C}_n\right],\qquad \ \tanh \theta = \frac{\e-1}{\e+1}. 
\end{equation}
We can now express one vacuum in terms of the other:  
\begin{align}\label{vac-al-c}
|0\>_\a =  \exp~(i F)~ |0\>_c  =  \left(\frac{1}{\cosh\theta}\right)^{1+1+\cdots} \prod_{n=1}^{\infty}  \exp\left[\frac{\tanh\theta}{n} C_{-n}\tilde{C}_{-n}\right]  |0\>_c 
\end{align}
Using zeta function regularization ($1+1+1+\hdots \infty=\zeta(0)=-\frac{1}{2}$) we obtain
 \begin{equation}
 |0\>_\a= \sqrt{\cosh\theta} \prod_{n=1}^{\infty}  \exp\left[\frac{\tanh\theta}{n} \, C_{-n} \cdot \tilde{C}_{-n}\right]  |0\>_c.
 \end{equation}
 In the tensionless limit, $\e \to 0$, we have $\tanh \theta = -1$, and hence:  
\begin{equation}
\lim_{\epsilon \to 0} |0\>_\a= {\mathcal{N}} \prod_{n=1}^{\infty}  \exp\left[- \, \frac{1}{n} C_{-n} \cdot \tilde{C}_{-n}\right]  |0\>_c
\end{equation}
This is precisely the induced vacuum that we encountered earlier \eqref{induced vacuum}. So we have 
\begin{align}
  \ket{0}_{I}=  \lim_{\epsilon \to 0} |0\>_\a= {\mathcal{N}} \prod_{n=1}^{\infty}  \exp\left[- \, \frac{1}{n} C_{-n} \cdot \tilde{C}_{-n}\right]  |0\>_c
\end{align}
The induced vacuum thus corresponds to the Neumann boundary state, representing an open string free to move in all spacetime directions or a space-filling D-25 brane. Thus as the tension is dialled to zero, the closed string vacuum evolves into an \textit{open string} description  \cite{Bagchi:2019cay}\footnote{Transition from closed string to open string at tensionless limit has also been studied in presence of constant Kalb-Ramond background in a very recent work \cite{Duary:2025hdb}.}. 

\begin{figure}[ht]
\centering
  \includegraphics[width=14cm]{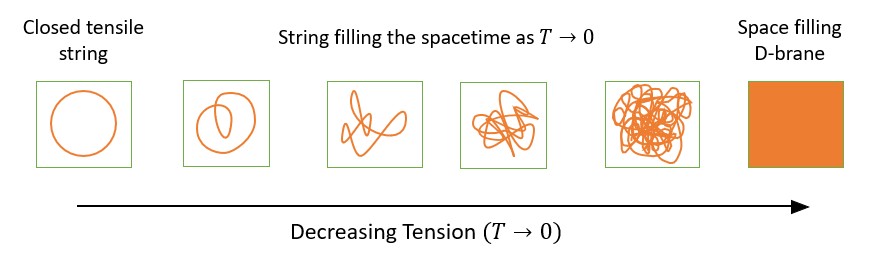}
  \caption{Emergence of spacefilling D-branes}
  \label{fig9}
\end{figure}

\paragraph{Null string complementarity:} 

The equations relating the tensionless oscillators $C$ to the tensile operators $\alpha$ can be also be used to write the tensionless vacuum in terms of the tensile vacuum, effectively inverting \eqref{vac-al-c}. This gives
\begin{align}
    |0\>_c= \mathcal{N}' \prod_{n=1}^{\infty}  \exp\left[-\frac{\tanh\theta}{n} \, \alpha_{-n} \cdot \tilde{\alpha}_{-n}\right]  |0\>_\alpha.
\end{align}
Notice that this has a crucial minus sign inside the exponential. In the limit of $\epsilon\to0$, the vacua are related as: 
\begin{align}
    |0\>_c= \mathcal{N}_D \prod_{n=1}^{\infty}  \exp\left[\frac{1}{n} \, \alpha_{-n} \cdot \tilde{\alpha}_{-n}\right]  |0\>_\alpha.
\end{align}
So we see that the tensionless vacuum in terms of the tensile theory is now a {\em{Dirichlet}} boundary state in all directions. This is thus a $D$-instanton in the tensile theory. 

\medskip

We have a rather remarkable observation at this point, which we call \textit{null string complementarity}. The picture is as follows. A tensionless observer, whose vacuum is $\ket{0}_c$, sees the tensile vacuum evolve into a spacefilling D-25 brane, i.e, an open string with Neumann boundary conditions in all directions. A tensile observer, whose vacuum is $\ket{0}_\a$, instead sees the tensionless string collapse to a  spacetime point, corresponding to D-instantons. Thus, depending on the observer, we have two very different views of the transition from closed to open strings as the tension is dialled to zero: either as D-instanton emerges or a space-filling D-25 brane. We refer to this striking observer-dependent reinterpretation as null string complementarity. 

\paragraph{Worldsheet condensation:}
 We now follow the fate of the tensile closed string perturbative states in tensionless limit \cite{Bagchi:2019cay}. Consider a level matched perturbative physical state in the tensile closed string theory:
\begin{align}\label{chh14}
    \ket{\Psi_n}=\xi_{\mu\nu}\alpha^{\mu}_{-n}\tilde{\alpha}^{\nu}_{-n}\ket{0}_{\alpha}.
\end{align}
In the above $\xi_{\mu\nu}$ is polarisation tensor. We have seen that in the tensionless limit, the tensile vacuum evolves into the induced vacuum of the tensionless string, i.e. in strict $\epsilon\to 0$ limit, $\ket{0}_\alpha\to\ket{0}_I$. We now assume that close to the tensionless point, $\ket{0}_{\alpha}$ near $\epsilon\to 0$ can be expanded as
\begin{align}\label{chh17}
\ket{0}_{\alpha}=\ket{0}_{I}+\epsilon\ket{I_{1}}+\epsilon^2\ket{I_{2}}+\cdots
\end{align}
 We now observe that in terms of the $(A, B)$ oscillators, using the above expansion and the definition of the induced vacuum $\ket{0}_{I}$, we have the following expressions: 
\begin{subequations}\label{gablu}
    \begin{align}
         &B_{n}\ket{0}_{I}=0,\hspace{5mm}\forall n\neq 0 \\
    & A_{n}\ket{0}_{I}=-B_{n}\ket{I_{1}}\hspace{8mm}A_{-n}\ket{0}_{I}=B_{-n}\ket{I_{1}}\hspace{9mm}\forall n>0 \\
    & A_{n}\ket{I_{1}}=-B_{n}\ket{I_{2}}\hspace{8mm}A_{-n}\ket{I_{1}}=B_{-n}\ket{I_{2}}\hspace{9.5mm}\forall n>0.
    \end{align}
\end{subequations}
Now, we let's look at the evolution of the perturbative physical states $\ket{\Psi_n}$. Rewriting \{$\alpha,\tilde{\alpha}$\} in terms of \{$A,B$\} and using \eqref{chh17} and \eqref{chh14}, we get
\begin{align}\label{chh16}
    \ket{\Psi_n}=\frac{1}{\epsilon}\Big(B_{-n}+\epsilon A_{-n}\Big)\Big(B_{n}-\epsilon A_{n}\Big)\Big(\ket{0}_{I}+\epsilon\ket{I_{1}}+\epsilon^2\ket{I_{2}}+\cdots\Big).
\end{align}
Using \eqref{tsc5} and \eqref{gablu}, we see that remarkably the level matched perturbative state $\ket{\Psi_n}$ will condense to the induced vacuum:  
\begin{align}\label{condensed}
    \ket{\Psi_n}\to \Xi\ket{0}_{I},\hspace{5mm} \Xi=2n\eta^{\mu\nu}\xi_{\mu\nu}. 
\end{align}
Notice that we have made no assumption on the level $n$ of the state. This establishes that for any $n$ the state $\ket{\Psi_n}$ at tensionless limit will evolve into the induced vacuum. We thus see that in the tensionless limit, all perturbative states over the tensile vacuum reduce to the tensionless vacuum resulting in a Bose-Einstein like condensation on the string worldsheet. 
\begin{figure}[ht]
  \includegraphics[width=16cm]{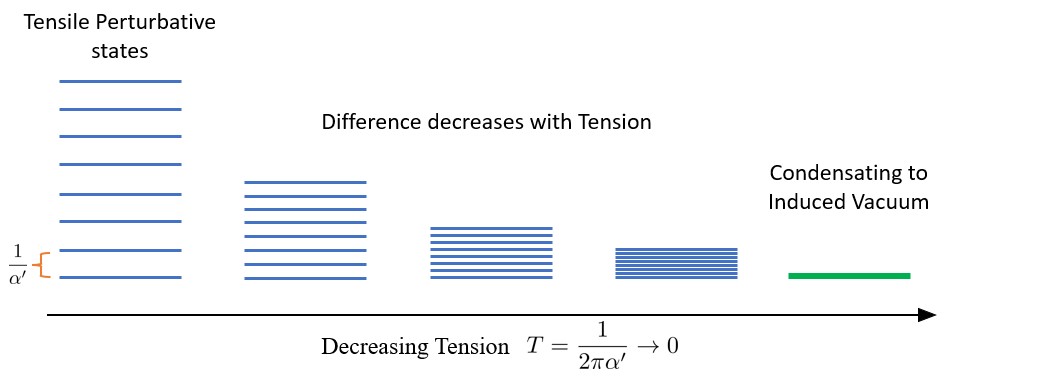}
  \caption{Bose-Einstein like condensation on the string worldsheet at $T\to 0$.}
  \label{fig10}
\end{figure}

\medskip

\paragraph{Connections to Hagedorn Physics:} It is very well known that a gas of strings is heated to higher and higher temperatures, something very interesting happens. The string partition function diverges at the Hagedorn temperature and it is believed that string theory evolves into a new phase where new degrees of freedom appear. In the theory of free strings, one can show that it becomes thermodynamically favourable to form a long open string at very high temperatures near the Hagedorn temperature. Interestingly, this very high temperature limit, the strings become effectively tensionless. If $\mathcal{T}$ is the temperature of the gas of strings and $\mathcal{T}_H$ is the Hagedorn temperature, the tension of the heated string goes as \cite{PhysRevD.26.3735,Olesen:1985ej}
\begin{align}
    T(\mathcal{T}) = T_0 \sqrt{1- \left(\frac{\mathcal{T}}{\mathcal{T}_H}\right)^2}.
\end{align}
In the above, $T_0=\frac{1}{2\pi \alpha'}$ is the original tension of unheated string. In this case, we see that in our parametrization
\begin{align}
    \epsilon = \sqrt{1- \left(\frac{\mathcal{T}}{\mathcal{T}_H}\right)^2}.
\end{align}
The previous observations of the emergence of the open string from the closed string and the condensation of the perturbative closed string states to the induced vacuum all are in keeping with the thermodynamic picture of the emergence of the long open string as the temperature of the gas of closed strings is turned up and close to the Hagedorn temperature. The picture we are advocating here is that if one keeps heating a gas of closed strings, the tension decreases and the strings become longer and longer and ultimately all closed string degrees of freedom condense on the worldsheet to form a long open string, which is the same as long anticipated from thermodynamic considerations. The tensionless limit clarifies the worldsheet phenomenon that leads to the formation of this open string from closed strings. 

\subsection{Null strings and worldsheet Rindler structures}

As we have seen in the earlier section, a lot of the most interesting new features of the quantum null strings arise from the fact that the tensile oscillators and the tensionless oscillators are connected by Bogoliubov transformations on the worldsheet. We stressed earlier that these Bogoliubov transformations were valid only very near the tensionless point where $\epsilon$ was close to zero. We now wish to understand a complete flow between a tensile and tensionless theory. As we know that the oscillators are Bogoliubov transforms of one another at the very end, it is tempting to formulate the entire flow in terms of Bogoliubov transformations which would reduce to the map \eqref{c-oscillator} as $\epsilon\to0$. 

\medskip

The most natural place where Bogoliubov transformations occur is when one thinks of accelerated observers in flat spacetimes and hence the map between Rindler and Minkowski spacetimes. In the limit of infinite acceleration, the Rindler observer hits the horizon of Rindler spacetime. For the context of the string, we could envision these structures arising on the worldsheet and hence consider accelerating worldsheets. We are thus led to an understanding of tensionless limit of string theory from the perspective of increasing acceleration of a family of worldsheets \cite{Bagchi:2020ats}. Below, following \cite{Bagchi:2020ats}, we show explicitly that the null string emerges when the Rindler horizon of the worldsheet is reached. The Rindler horizon is approached in two distinct ways: evolution in the acceleration at fixed time and evolution in time at fixed acceleration. 

\paragraph{Accelerated worldsheets and the horizon limit:} We begin with the metric described by a uniformly accelerated observer in Minkowski spacetime, which takes the Rindler form,
\begin{equation}
    dS_{Rind}^2=e^{2a \rho}(-d\eta^2+d\rho^2).
\end{equation}
Minkowski and Rindler spacetimes are related by the following coordinate transformation,
\begin{equation}
    t=\frac{1}{a}e^{a\rho} \sinh(a\eta), \qquad x=\frac{1}{a}e^{a\rho} \cosh(a\eta),
\end{equation}
where $a$ is the acceleration. We now consider a massless scalar field theory in the Rindler spacetime \cite{Birrell:1982ix}. Since the Rindler spacetime is conformally flat, the EOM for the scalar field is identical to that in flat spacetime.
\begin{equation}
    (-\partial_t^2+\partial_x^2) \phi =0=(-\partial_\eta^2+\partial_\rho^2)\phi.
\end{equation}
The corresponding Minkowski mode expansion is
\bea{}
\phi (\s, \t) = \phi_0+\sqrt{2 \alpha'} \alpha_0\tau + \sqrt{2 \pi \alpha'}\sum_{n>0}\left[\a_n u_n+\a_{-n} u_n^*+\ta_n \tilde{u}_n+\ta_{-n} \tilde{u}_n^*\right] 
\eea
where the oscillators $\alpha_n,\tilde{\a}_n$ annihilate the Minkowski vacuum and satisfy the following relations
\begin{equation}
    [ \alpha_n, \alpha_m ]=n\delta_{n+m}, \qquad  [ \tilde{\alpha}_n, \tilde{\alpha}_m ]=n\delta_{n+m}, \qquad  [ \alpha_n, \tilde{\alpha}_m ]=0.
\end{equation}
In Rindler coordinates, the mode expansion takes the form
\bea{}
\phi(\rho, \eta) = \phi_0+\sqrt{ 2 \alpha'}\beta_0\rho + \sqrt{2 \pi\alpha'}\sum_{n>0}[\b_n U_n+\b_{-n} U_n^*+\tilde{\b}_n \tilde{U}_n+\tilde{\b}_{-n} \tilde{U}_n^*].
\eea
Here, the oscillators $(\beta_n,\tilde{\beta_n})$ annihilate the Rindler vacuum $\ket{0}_R$. Unlike Minkowski modes, the functions $U_n$ and $\tilde{U}_n$ are defined only in the left and right Rindler wedges, respectively. So, we label them as $U_n^{(L)}$ and $\tilde{U}_n^{(R)}$. To define operators valid across the horizon, we introduce standard analytically continued combinations \cite{Unruh:1976db}:
\be{}
U_n^{(R)} - e^{-\frac{\pi n}{a}}U_{-n}^{(L)*}, \quad U_{-n}^{(R)*} - e^{\frac{\pi n}{a}}U_{n}^{(L)}.
\ee
These combinations allow us to relate $\alpha$ oscillators in Minkowski space to $\beta$ oscillators in Rindler spacetime through Bogoliubov transformations \cite{Unruh:1976db}: 
\begin{subequations}\label{R2M}
\bea{}
\b_{n} &=& \cfrac{e^{\pi n/2a}}{\sqrt{2\sinh\frac{\pi n}{a}}}~\a_{n}-\frac{e^{-\pi n /2a}}{\sqrt{2\sinh\frac{\pi n}{a}}}~\tilde{\a}_{-n},\\
\tilde{\b}_{n}&=& -\cfrac{e^{-\pi n/2a}}{\sqrt{2\sinh\frac{\pi n}{a}}}~\a_{-n}+\cfrac{e^{\pi n/2a}}{\sqrt{2\sinh\frac{\pi n}{a}}}~\tilde{\a}_{n}.
\eea
\end{subequations}
In the limit of large acceleration, \eqref{R2M} reduces to
\begin{subequations}\label{binfty}
\begin{align}
\b_{n}^{\infty} = \frac{1}{2}\left(\sqrt{\frac{\pi n}{2a}}+\sqrt{\frac{2a}{\pi n}}  \right) \a_{n}+\frac{1}{2}\left(\sqrt{\frac{\pi n}{2a}}-\sqrt{\frac{2a}{\pi n}}\right) \tilde{\a}_{-n},\\
\tilde{\b}_{n}^{\infty}=\frac{1}{2}\left(\sqrt{\frac{\pi n}{2a}}-\sqrt{\frac{2a}{\pi n}}\right)\a_{-n}+ \frac{1}{2}\left(\sqrt{\frac{2a}{\pi n}}+\sqrt{\frac{\pi n}{2a}}\right)\tilde{\a}_{n}.
\end{align}
\end{subequations}
Comparing the above transformations in the large acceleration limit \refb{binfty} with the Bogoliubov transformations between $\a$ oscillators and $C$ oscillators in the tensionless limit \eqref{chh4}, we can identify 
\be{atoe}
C_n = \b_{n}^{\infty}, \qquad  \C_n= \tilde{\b}_{n}^{\infty},  \qquad \text{where}~~~\e = \frac{\pi n}{2a}.
\ee
Thus, the large-acceleration limit of the Rindler worldsheet reproduces precisely the tensionless oscillators. The map between increase in the acceleration and decrease in tension is summarised below
\begin{subequations}
\begin{align}
a=0 &: \qquad \{ \b_n, \tilde{\b}_n \} \to \{ \a_n, \tilde{\a}_n \} \\
0<a<\infty &: \qquad \{ \b_n (a) , \tilde{\b}_n(a) \} \\
a\to \infty &: \qquad \{ \b_n, \tilde{\b}_n \} \to \{C_n, \C_n\}
\end{align}
\end{subequations}
\paragraph{The vacuum -- emergent boundary state:}
 Now that we have obtained the $\beta$ oscillators for accelerated worldsheet, we define the vacuum ($\ket{0}_R$) as: 
\bea{bcond}
&& \b^\mu_n|0\rangle_R = (\alpha^\mu_n +\tanh \theta_n\ \tilde{\alpha}^\mu_{-n})|0\rangle_R=0,\ \qquad n>0; \nonumber \\ 
&& \tilde{\b}^\mu_n|0\rangle_R = (\tilde{\alpha}^\mu_n+\tanh\theta_n\ \alpha^\mu_{-n})|0\rangle_R=0.
\eea 
In the limit ($a\to \infty$), $\tanh{\theta}_n=-1$. This maps to the tensionless limit ($\e \to 0$), resulting in a squeezed state
\bea{zc2za}
\zc = \lim_{a\to\infty} |0\rangle_R = \frac{1}{\mathcal{N}_\a} \prod_{n=1}^\infty \exp\left[\frac{1}{n} \a_{-n}\cdot\tilde{\a}_{-n}\right]|0\rangle_\alpha. 
\eea
This is exactly the Dirichlet boundary state in all directions defined in \eqref{bst}. Hence, as stated earlier, from the point of view of the usual string vacuum ($\ket{0}_\a$), tensionless vacuum is a \textit{D-instantons}. This picture fits very well with the Rindler worldsheets as we will explain below.

\begin{figure}[ht]
\centering
  \includegraphics[width=11cm]{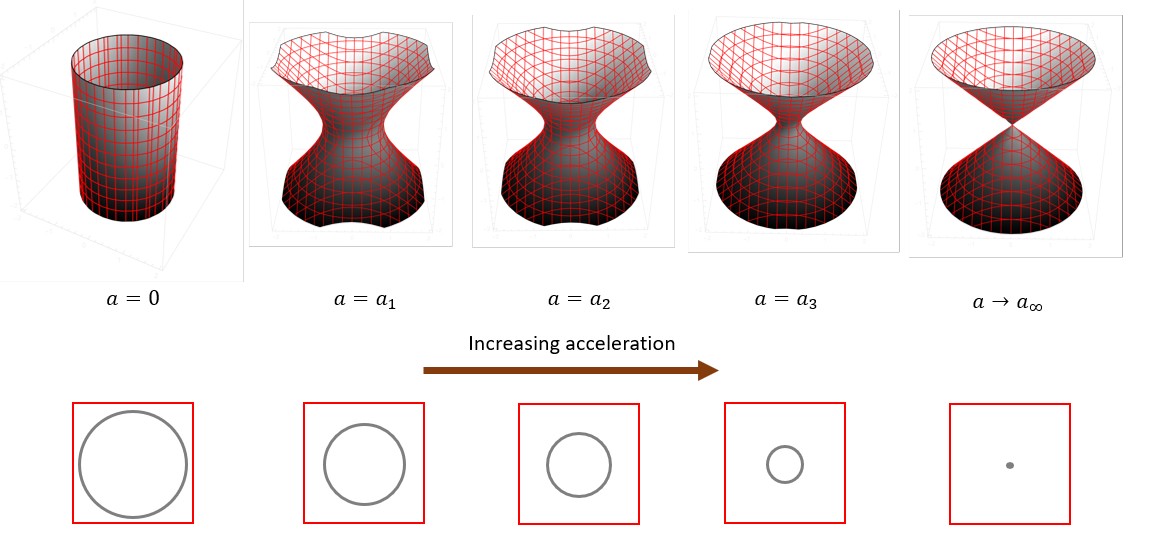}
  \caption{ Increasing accelerated worldsheets.}
  \label{fig11}
\end{figure}

\medskip

An inertial closed string worldsheet corresponds to a cylinder, which is recovered in the limit of zero acceleration, $a\to 0$. As the acceleration is increased, the worldsheet deviates from this cylindrical shape and is successively mapped to a family of curved (hyperbolic) surfaces, each more distorted than the previous one. In the limit $a\to \infty$, the worldsheet degenerates to lightcone (a null surface). Since the boundary states in \eqref{bst} are defined at $\tau=0$, it is useful to examine how $\t=0$ slices behave along this trajectory. For small acceleration, $\tau=0$ section is a circle of finite radius. As the acceleration increases, the radius steadily decreases. In the extreme limit of $a\to \infty$, the circle shrinks to a point. This limiting pointlike configuration is the spacetime interpretation of the D-instanton discussed earlier. The complementary phenomenon, the emergence of the open string arises when the Rindler horizon is reached at fixed acceleration, a view point we now turn to.

\medskip

\paragraph{Rindler time evolution and the horizon limit:} As discussed above, the Rindler horizon can be approached in two independent ways. The first involved increasing the acceleration at fixed Rindler time. We now focus on the second route: keeping the acceleration constant while allowing the Rindler time $\eta$ to evolve. In this picture, the horizon is reached in the $\eta\to \infty$ limit. This limit may be implemented equivalently through the scaling
\begin{equation}
    \eta \to \eta, \qquad \rho\to \e \rho, \qquad \e\to 0.
\end{equation}
Under this scaling, the 2d conformal generators of the Rindler theory  reduce to 
\bea{}
 L_k &=& \L_k - \bL_{-k} = i^k e^{-k \eta}(\p_\eta - k \rho \p_\rho), \cr
 M_k &=& \e(\L_k + \bL_{-k}) = - i^k e^{-k \eta} \p_\rho
\eea
which closes to form the classical part of BMS$_3$ algebra \eqref{BMSalgebra} (with $c_L=c_M=0$), the symmetry algebra characterizing the null strings \cite{Bagchi:2021ban}. 
We now describe a complementary viewpoint of the emergence of the open string from the closed string associated with this time-evolution picture.
\medskip

\begin{figure}[ht]
\centering
  \includegraphics[width=8cm]{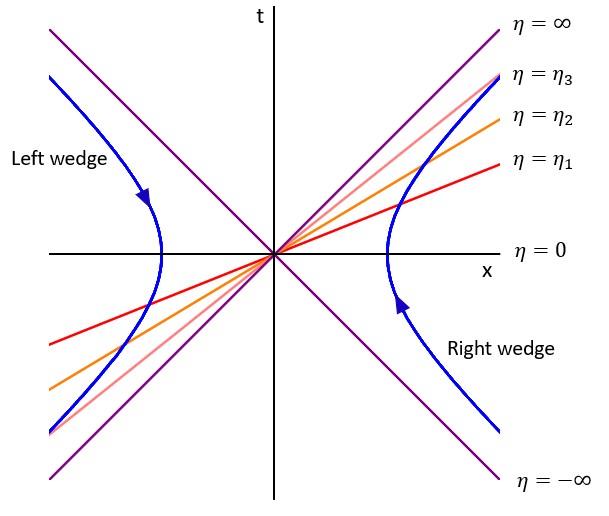}
  \caption{Equal time slices in Rindler spacetimes}
  \label{fig12}
\end{figure}

In Rindler spacetime, surfaces of constant Rindler time are straight lines passing through the origin, distinguished by their slope. As shown in Fig.~\eqref{fig12}, the right Rindler wedge contains a family of such lines with slopes increasing for $\eta_1<\eta_2<\eta_3$. In the limit $\eta_\infty$, these lines approach the Rindler horizon. In the left wedge, the Rinlder time runs in the opposite direction, and the continuation of these constant-$\eta$ lines extends into the third quadrant. We now relate this with the string worldsheet at constant acceleration. Increasing $\eta$ planes intersects the constant hyperboloid at increasing angles. The $\eta$ evolution of the closed string progressively distorts the spatial section of the closed string worldsheet. At early Rindler time, the ($\eta=0$) slice describes an ordinary closed string of finite radius. As time evolves, the effective tension decreases and the spatial profile of the string becomes elongated. The shape continuously stretches and in the limit $\eta \to \infty$, the spatial cross-section degenerates into a straight line. At this point the worldsheet becomes null and the closed string effectively transitions into an open, tensionless configuration. This picture is identical to the original one we presented as Fig~\ref{ClosedtoOpen}.
\begin{figure}[t]
\centering
  \includegraphics[width=11cm]{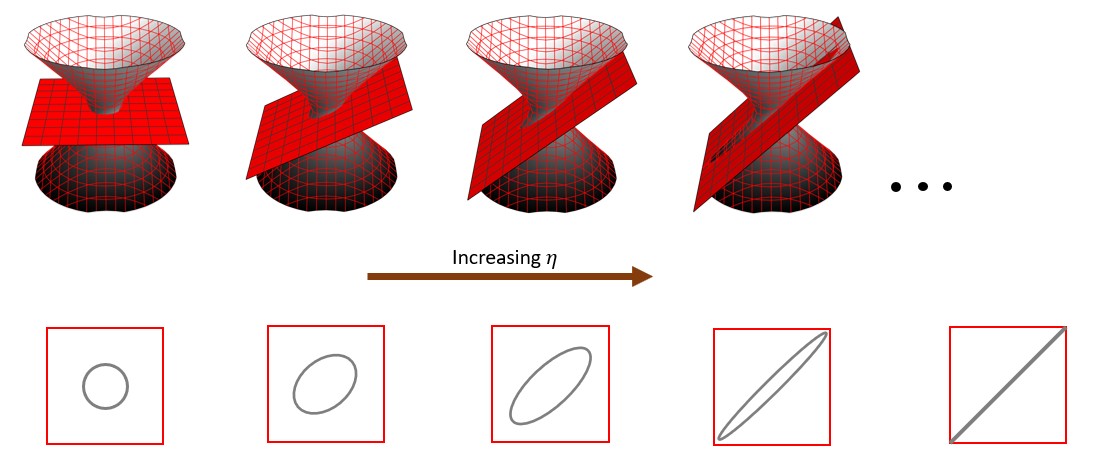}
  \caption{Equal time-slices of a Rindler worldsheet}
  \label{fig13}
\end{figure}
\medskip

Thus approaching the worldsheet horizon by evolving in Rindler time provides gives a very clear but different picture of how closed strings can appear as open strings.\footnote{An analysis with time evolving strings and the Milne patch on the worldsheet was constructed with in direct analogy with this construction of accelerated worldsheets in \cite{Karan:2024jgn}.} 

\subsection*{Rindler spacetime and Carroll strings}
In the previous subsections, we saw that introducing a Rindler structure on the closed string worldsheet leads to a number of remarkable phenomena, including the emergence of null strings and the observer-dependent picture captured by null string complementarity. A natural question then arises: \textit{under what physical circumstances does such a Rindler structure arise on the worldsheet of a string propagating in spacetime?}
\medskip

The answer comes from a simple observation: the near-horizon region of any non-extremal black hole is described by a Rindler spacetime. As a string moves closer to the black hole horizon, it propagates in a region whose geometry is approximately Rindler. This Rindler structure of the target spacetime in turn induces a corresponding Rindler structure on the worldsheet \cite{Bagchi:2021ban}. From the worldsheet perspective, approaching the spacetime horizon translates into an increase in the effective worldsheet acceleration, causing the worldsheet to deviate more and more from its cylindrical shape. In the limiting process where the string reaches the horizon, the worldsheet becomes null, precisely the situation encountered in our Rindler-worldsheet analysis, and the string becomes effectively tensionless.
\medskip

In string theory, Carrollian geometry appears in two complementary contexts. Our review as so far been about Carrollian structures appearing on the worldsheet. Strings can also propagate in target Carrollian spacetimes. This is what we will focus on in the part~\ref{IV} of our review. Bringing these insights together, we see the following unified picture.
\begin{itemize}
\item A string falling toward a black hole horizon encounters a Rindler region.

\item This induces a Rindler worldsheet, whose approach to the worldsheet horizon mirrors the tensionless limit.

\item In this null limit, the worldsheet naturally acquires a Carrollian structure, placing the near-horizon dynamics of the string within the framework of Carroll strings.
\end{itemize}

In the part \ref{IV}, we make this connection explicit by analysing closed strings near black hole horizons and demonstrating that their residual worldsheet symmetry (electric) becomes the BMS$_3$ algebra, precisely the residual worldsheet symmetry of tensionless strings. We also study the classical string solutions in these near-horizon regimes.

\medskip

Now we go back to the three quantum null strings and their formulation in flat target spacetimes. We will focus on understanding the spectra of these three theories in the coming sections. 

\bigskip

\newpage

\section{Physical Hilbert spaces for three vacua}\label{Physical Hilbert spaces for three vacua}
In this section we discuss the physical Hilbert spaces corresponding to the oscillator, induced and flipped vacuum, primarily based on \cite{Bagchi:2020fpr}. We have already established how general physical state conditions need to be imposed on the states built out on these vacua. In what follows we use these conditions accordingly.
\subsection{Oscillator Vacuum}
\label{subsecosc}
We start from the physical state condition for the oscillator vacuum, where the physical state condition that works \eqref{consistent quantum theories} is the full sandwich condition, which is the weakest among all three conditions elucidated before. For the algebra generators, the sandwich condition can be written as:
\begin{subequations}\label{TSQR1}
    \begin{align}
    \label{TSQR1a}
    \bra{phys'}L_{n}-a_{L}\delta_{n,0}\ket{phys}&=0\\
    \label{TSQR1b}
    \bra{phys'}M_{n}-a_{M}\delta_{n,0}\ket{phys}&=0.
\end{align}
\end{subequations}
As already shown via light-cone quantisation in the earlier section, the normal ordering constants for oscillator vacuum are $a_L=0$ and $a_M=2$. This provides us with the recipe to construct physical states in this case.
\paragraph{Construction of Hilbert space:}
Before discussing the Hilbert space for this theory, let us first recall the expressions of $L_n$s and $M_n$s in terms of \{$C,\tilde{C}$\} from \eqref{TSQR5}, which we will use in the sandwich condition \eqref{TSQR1}. Of course the subtlety in this case is in the zero modes, the expression of which in terms of \{$C,\tilde{C}$\} becomes,
\begin{subequations}
    \begin{align}\label{TSQR9}
    &L_{0}=\mathcal{N}-\widetilde{\mathcal{N}},~~~    M_{0}=c'k^2+\mathcal{N}+\widetilde{\mathcal{N}}+X+X^{\dagger},\\
\label{TSQR42}
     \mathcal{N}=\sum_{m>0}&C_{-m}\cdot C_{m};\hspace{5mm} \widetilde{\mathcal{N}}=\sum_{m>0}\tilde{C}_{-m}\cdot \tilde{C}_{m};\hspace{5mm}X=\sum_{m>0}C_{m}\cdot \tilde{C}_{m}.
\end{align}
\end{subequations}
Note the structure of $X$ which contains a mixture of the \{$C,\tilde{C}$\} modes. 
As discussed in \eqref{krishnakumarkunnath}, the oscillator vacuum is, by definition, annihilated by all the $C_{n}$s and $\tilde{C}_{n}$s for $n>0$. It also is eigenstate of $C_{0}$ and $\tilde{C}_{0}$:
\begin{align}\label{TSQR2}
C^{\mu}_{0}\ket{0;k}_{c}=\tilde{C}^{\mu}_{0}\ket{0;k}_{c}=\sqrt{\frac{c'}{2}}k^\mu\ket{0;k}_{c},~~~C^{\mu}_{n}\ket{0;k}_{c}=\tilde{C}^{\mu}_{n}\ket{0;k}_{c}=0,\hspace{5mm}\forall n>0.
\end{align}
Like tensile string vacuum, this vacuum too, is normalised as 
\begin{align}\label{normalised}
    \braket{0;k'|0;k}=(2\pi)^D\delta^{D}(k'-k).
\end{align}
It is important to note that the algebra satisfied by \{$C,\tilde{C}$\} is same as the algebra satisfied by the tensile oscillators \{$\alpha,\tilde{\alpha}$\}. Hence, looking at \eqref{TSQR2} one can clearly see that, just like the tensile case, $\mathcal{N}$ and $\widetilde{\mathcal{N}}$ are number operators, while \{$C_n,\tilde{C}_n$\} for $n>0$ ($n<0$) play the role of annihilation (creation) operators on the eigenstates of $\mathcal{N}$ and $\widetilde{\mathcal{N}}$.   The Hilbert space of the theory can be spanned by the eigenstates of $\mathcal{N}$ and $\widetilde{\mathcal{N}}$ as basis. A generic eigenstate of $\mathcal{N}$ and $\widetilde{\mathcal{N}}$ can be constructed from the oscillator vacuum defined in \eqref{TSQR2} as
\begin{subequations}
    \begin{equation}\label{TSQR11}
    \ket{r,s}=\sum_{j}\rho_{j}\Bigg(\prod_{i=1}^{p}C^{a_{i}}_{-m_{i}}\prod_{j=1}^{q}\tilde{C}^{b_{j}}_{-n_{j}}\Bigg)_{j}\ket{0,k^{\mu}}_{c}.
\end{equation}
\begin{equation}\label{TSQR12}    r=\sum_{i}^{p}a_{i}m_{i},\hspace{8mm}s=\sum_{i}^{q}b_{i}n_{i}.
\end{equation}
\end{subequations}
In the above, $a_{i}$ and $b_{j}$ respectively are powers of the $C_{-m_{i}}$ and $\tilde{C}_{-n_{j}}$. The eigenvalues of state $\ket{r,s}$ with $\mathcal{N}$ and $\widetilde{\mathcal{N}}$ are $r$ and $s$ respectively. $\rho_{j}$ is polarisation tensor appropriate for the state. We define level $l$ of such state as $l=r+s$. These states are orthonormal by construction i.e. $\braket{r',s'|r,s}=\delta_{r'r}\delta_{s's}$.
\paragraph{Physical Hilbert space:}
After constructing the Hilbert space, we would like to impose the physical state conditions \eqref{TSQR1} on them. Here we make the usual assumption that the oscillator vacuum is a physical state \footnote{However, this assumption is not always made. We will see in our analysis of flipped vacuum theory that the vacuum itself is no longer a physical state there.} to start with. Now, we apply the $L_n$ physical state condition from \eqref{TSQR1a} on $\ket{phys}=\ket{phys'}=\ket{0,k_{0}^{\mu}}$. For $n\neq0$ the sandwich conditions \eqref{TSQR1a} trivially satisfy due to the definition of vacuum \eqref{TSQR2}. For $n=0$ we have
\begin{align}
    \bra{0,k^{\mu}_{0}}L_{0}\ket{0,k_{0}^{\mu}}=a_{L}=0.
\end{align}
Hence, the oscillator vacuum satisfies the entire set of $L_n$ physical state conditions given in \eqref{TSQR1a}. Applying the $L_{0}$ physical state condition on a generic state $\ket{r,s}$, and further putting $a_{L}=0$, we get 
\begin{equation}\label{TSQR13}
    \begin{split}
        \bra{r,s}L_{0}\ket{r,s}=\bra{r,s}\big(&\mathcal{N}-\widetilde{\mathcal{N}}\big)\ket{r,s}=r-s=0\\
        \implies & r=s.
    \end{split}
    \end{equation}
This is nothing but the level matching condition for $\ket{r,s}$, which says only level matched states are physical here. We are still left with the sandwich conditions \eqref{TSQR1a} with $n\neq0$ to be satisfied. Using the expression of $L_n$ in \eqref{TSQR5a}, it is straightforward to show that $L_n$ acts on $\ket{r,s}$ as below (see \cite{Bagchi:2020fpr})
\begin{align}\label{Yu-fan Zheng}
    L_n\ket{r,s}=\ket{r-n,s}-\ket{r,s+n}.
\end{align}
That means if we operate $L_n$ ($n\neq0$) on a level matched state $r=s$, then we get linear combination of states $\ket{r-n,s}$ and $\ket{r,s+n}$. None of these states are level matched, which means both states are orthogonal to any level-matched state and as a result, their inner product with level matched states vanish. Hence the condition \eqref{TSQR1a}, sandwiched between two separate level matched physical states, having $n\neq0$ is automatically satisfied without imposing any additional constraint. 
\medskip

Similarly, using the expression of $M_n$ from \eqref{TSQR5b} one can see that action of this operator on $\ket{r,s}$ is given by
\begin{align}
    M_n\ket{r,s}=\ket{r-n,s}+\ket{r,s+n}+\sum_{m\neq0}\ket{r+m,s+m+n}
\end{align}
Clearly, $M_n$ with $n\neq0$ after acting on the level matched states will give an infinite combination of non-level matched states only and hence, once again, the corresponding physical state conditions in \eqref{TSQR1b} are trivially satisfied.
\medskip

At this juncture, all we are left with is the $M_0$ physical state condition from \eqref{TSQR1b}, which will give us the mass of a level matched state $\ket{n,n}$. However, this part turns out to be trickier since the state $\ket{n,n}$ is not an eigenstate of $M_0$, so the action is not straightforward. Following the analysis in \cite{Bagchi:2020fpr} we split the operator $M_0$ as expressed in \eqref{TSQR9} into two parts; one part that is diagonalised in the basis $\ket{n,n}$ via the number operators (denoted by $\mathcal{D}_0$), and the other part that cannot be diagonalised in that basis (denoted by $\mathcal{K}_0$). In terms of expressions: 
\medskip
\begin{equation}\label{split}
    M_0=\mathcal{D}_0+\mathcal{K}_0,~~~\mathcal{D}_0=c'k^2+\mathcal{N}+\widetilde{\mathcal{N}},~~~\mathcal{K}_0=X+X^{\dagger}.
\end{equation}
Now, we apply the $M_0$ physical state condition as given via the sandwich in \eqref{TSQR1b} in this split form \eqref{split} on two physical level matched states on two sides 
    \begin{align}
        \bra{n',n'}\mathcal{D}_0\ket{n,n}+ \bra{n',n'}\mathcal{K}_0\ket{n,n}=a_M \braket{n',n'|n,n}.
    \end{align}
Since the diagonal part is readily evaluated, let us look at the $\mathcal{K}_0$ sandwiched part above. We can rewrite it as
\begin{align}\label{TSQR43}
    \bra{n',n'}\mathcal{K}_0\ket{n,n}=\bra{n',n'}X+X^{\dagger}\ket{n,n}=\bra{n',n'}\sum_{m\neq 0}C_{m}\cdot \tilde{C}_{m}\ket{n,n}.
\end{align}
In the last step above, we have used the expression of $X$ as given in \eqref{TSQR42}, also the fact that 
\begin{align*}
    X^{\dagger}=\sum_{m> 0}C_{-m}\cdot \tilde{C}_{-m}=\sum_{m< 0}C_{m}\cdot \tilde{C}_{m}.
\end{align*}
For $n'=n$, things are easier to understand and we see that 
\begin{align}
    \bra{n,n}\mathcal{K}_0\ket{n,n}=0.
\end{align} 
In the above we used the fact that the states $\ket{n,n}$ are orthonormal to each other, and  \begin{align}\label{TSQR44}
    \mathcal{K}_0\ket{n,n}=\sum_{m\neq 0}C_{m}\cdot \tilde{C}_{m}\ket{n,n}=\sum_{m\neq0}\ket{n-m,n-m}.
\end{align}
Now, let us consider the final bit, i.e. the matrix elements of $\mathcal{K}_0$ given in \eqref{TSQR43} with $n'\neq n$. Equation \eqref{TSQR44} implies that for $n'\neq n$, there must be a state in the summation at the r.h.s. for which $n-m=n'$. Consequently, the matrix element will be 
\begin{align}
    \bra{n',n'}\mathcal{K}_0\ket{n,n}\sim \delta^{D}(k'-k).
\end{align}
The implication is that this matrix element will vanish only if the momentum of the states $\ket{n',n'}$ and $\ket{n,n}$ are different. For that case we will have 
\begin{align}
     \bra{n',n'}\mathcal{D}_0\ket{n,n}=a_M \braket{n',n'|n,n}.
\end{align}
We know that the operator $\mathcal{D}_0$ is diagonalized in the basis $\ket{n,n}$. Taking that into account we are allowed to impose the following condition on the state $\ket{n,n}$
\begin{align}
    \mathcal{D}_0\ket{n,n}=a_M\ket{n,n}
\end{align}
Using the expression of $\mathcal{D}_0$ as given in \eqref{split} and the facts that $k^2=-m^2$ and $a_M=2$, we find the following mass of $\ket{n,n}$.
\begin{align}\label{TSQR17}
    m^2\ket{n,n}=\frac{1}{c'}(2n-2)\ket{n,n}.
\end{align}
Once we impose this mass shell condition on $\ket{n,n}$, it will automatically ensure that the states $\ket{n,n}$ and $\ket{n',n'}$ for $n\neq n'$ will essentially have different momentum (since their mass $m^2=-k^2$ are different), which consequently ensures the condition $\bra{n',n'}\mathcal{K}_0\ket{n,n}=0$. Hence, the mass shell condition \eqref{TSQR17} is equivalently consistent with the $M_0$ physical state condition.

\subsection{Induced Vacuum}

This quantum theory for tensionless strings constructed on induced vacuum is based on physical states outlined in \eqref{case2}. The physical state condition to be satisfied here can be written in terms of the genrators:
\begin{align}\label{TSQR25} 
  \bra{phys'}L_{n}\ket{phys}=0\hspace{5mm}\forall n,~~~~M_{n}\ket{phys}=0,\hspace{5mm}\forall n\neq 0.
\end{align}
One can see that the physical state condition outlined in \eqref{TSQR25} exactly aligns with the condition \eqref{induced}. This clearly shows us that the physical states in this theory, in fact, come from the direct tensionless/worldsheet Carrollian limit of the physical states of the tensile Bosonic string theory. We will clearly show this below. 

The induced vacuum comes from the direct Carrollian limit of the tensile bosonic string vacuum. To see this, recall the vacuum in tensile bosonic closed string theory defined via highest weight conditon is written as:
\begin{align}\label{TSQR26}    \alpha^{\mu}_{n}\ket{0,k^{\mu}}_{\alpha}=\tilde{\alpha}^{\mu}_{n}\ket{0,k^{\mu}}_\alpha=0\hspace{5mm}\forall n>0.
\end{align}
Now, let us further recall how oscillators $\{A,B\}$ are connected to $\{\alpha,\tilde{\alpha}\}$ through the Carroll limit from \eqref{tsc8}. Inverting \eqref{tsc8} and using the same in \eqref{TSQR26} we get the condition:
\begin{subequations}
\begin{align}\label{TSQR27}
    \Big(\sqrt{\epsilon}A_{n}^\mu+\frac{1}{\sqrt{\epsilon}}B_{n}^\mu\Big)\ket{0,k^{\mu}}_{\alpha}&=0\\
    \Big(-\sqrt{\epsilon}A_{-n}^\mu+\frac{1}{\sqrt{\epsilon}}B_{-n}^\mu\Big)\ket{0,k^{\mu}}_{\alpha}&=0 \qquad \forall n>0.
\end{align}
\end{subequations}
 Taking tensionless or worldsheet Carrollian limit ($\epsilon \to 0$) on \eqref{TSQR27} gives us the following conditions
\begin{align}\label{TSQR28}
    B^{\mu}_{n}\ket{0,k^{\mu}}_{I}=0\hspace{5mm} \forall n\neq 0,~~~
    B^{\mu}_{0}\ket{0,k^{\mu}}_{I}=\sqrt{2c'}k^{\mu}\ket{0,k^{\mu}}_{I}.
\end{align}
Using the expression of $M_n$s in terms of $B_n$s as given in \eqref{tsc9}, it is straightforward to show that this state satisfies \eqref{TSQR25}, and hence, is a physical state. The action of $M_{0}$ on this state will give us the mass of the state. In the light of the fact that $B$'s commute with each other, it is obvious that the normal ordering constant $a_{M}$ vanishes for this theory. Keeping this is mind we rewrite the $M_0$ physical state condition as:
\begin{align}\label{TSQR29}
    M_{0}\ket{0,k^{\mu}}_{I}=\sum_{n}B_{-n}\cdot B_{n}\ket{0,k^{\mu}}_{I}=\left(\sum_{n\neq 0}B_{-n}\cdot B_{n}+B^{2}_{0}\right)\ket{0,k^{\mu}}_{I}=0.\end{align}
Using \eqref{TSQR28} in \eqref{TSQR29}, we get
\begin{align}\label{inducedmass}
    B^{2}_{0}\ket{0,k^{\mu}}_{I}=2c'k^2\ket{0,k^{\mu}}_{I}=0.
\end{align}
Keeping in mind the mass shell $k^2=-m^2$, one can see that the induced vacuum described above is massless. 
We are yet to apply the sandwich physical state conditions involving $L_{n}$s on this state. Applying those conditions on the induced vacuum $\ket{0,k^{\mu}}_{I}$ one gets
\begin{align}\label{Vishnu Ganesh Pingle}
    \bra{0,k^{\mu}}L_{n}\ket{0,k^{\mu}}=\bra{0,k^{\mu}}A_{n}\cdot B_{0}\ket{0,k^{\mu}}=c'k\cdot\bra{0,k^{\mu}}A_{n}\ket{0,k^{\mu}}=0.
\end{align}
The periodicity condition, $A^{\mu}_{0}=0$, implies that the the induced vacuum trivially satisfies the $L_{0}$ sandwich physical state condition.
The fact that the induced vacuum is massless can be understood from limiting perspective as well. We recall that the tensile string vacuum is tachyonic with the following mass
\begin{align}
    m^2\ket{0,k}_{\alpha'}=-\frac{4}{\alpha'}\ket{0,k}_{\alpha'}.
\end{align}
At tensionless limit ($\alpha\to c'/\epsilon$, $\epsilon\to 0$) this mass eigenvalue vanishes, confirming the mass derived earlier intrinsically.

\subsection{Flipped Vacuum}
The quantum theory for tensionless strings constructed on the flipped vacuum has physical states outlined in \eqref{case1}. Let us rewrite that below including the $L_{0}$ and $M_{0}$ conditions
\begin{align}\label{fphysical}
     (L_n-a_L\delta_{n,0})\ket{phys}=(M_n-a_M\delta_{n,0})\ket{phys}=0 ~~~\forall~n\geq0.
\end{align}
We have already seen earlier that for this theory to be Lorentz invariant in target spacetime, we must impose $a_L=2$ and $a_M=0$. In the upcoming discussion, the same will be shown from the limiting analysis from its parent tensile counterpart.
\medskip

One can easily identify that \eqref{fphysical} is identical to the highest weight representation of BMS$_3$ as given in \eqref{BMSHWR}. This is a clear indication of the fact that the flipped vacuum theory which we are going to study here (which is constructed on the highest weight representation of BMS$_3$) comes from the tensionless ($\epsilon\to 0$) limit of the twisted string theory; a fact that will be more explicit in the upcoming discussions.
\paragraph{Hilbert Space Structure:}
The flipped vacuum defined in terms of new set of  oscillators \{$C,\tilde{\mathcal{C}}$\} is
\begin{align}\label{TSQR31}
C^{\mu}_{n}\ket{0,k}_{A}=\tilde{\mathcal{C}}^{\mu}_{n}\ket{0,k}_{A}=0\hspace{5mm}\forall n>0.
\end{align}
In the above, the oscillator $C_n$ is same as that used in the study of oscillator vacuum while $\tilde{\mathcal{C}_n}$ is defined as 
\begin{equation}\label{flipping}
    \tilde{\mathcal{C}}_{n}=\tilde{C}_{-n}.
\end{equation}
The implication of \eqref{flipping} is that, in the `tilde' sector, the creation and annihilation operators flip their roles.
The mode expansion of $X^{\mu}(\tau,\sigma)$ as given in \eqref{fcb2} can be rewritten in terms of \{$C,\tilde{\mathcal{C}}$\} as
\begin{align}\label{TSQR35}
     X^{\mu}(\tau,\sigma)=x^{\mu}+2\sqrt{\frac{c'}{2}} C^{\mu}_{0}\tau+i\sqrt{\frac{c'}{2}}\sum_{n\neq0}\frac{1}{n}\left[(C^{\mu}_{n}-\tilde{\mathcal{C}}^{\mu}_{-n})-in\tau(C^{\mu}_{n}+\tilde{\mathcal{C}}^{\mu}_{-n})\right]e^{-in\sigma}.
\end{align}
The zero modes are given by $C^{\mu}_{0}=\tilde{\mathcal{C}}^{\mu}_{0}=\sqrt{\frac{c'}{2}}k^{\mu}$. The commutation relations of these new oscillators are given by
\begin{align}\label{TSQR33}
     [C^{\mu}_{m},C^{\nu}_{n}]=m\eta^{\mu\nu}\delta_{m+n}\hspace{5mm}[\tilde{\mathcal{C}}^{\mu}_{m},\tilde{\mathcal{C}}^{\nu}_{n}]=-m\eta^{\mu\nu}\delta_{m+n},\hspace{5mm}[C^{\mu}_{m},\tilde{\mathcal{C}}^{\nu}_{n}]=0.
\end{align}

Now, we shall write the generators \{$L_{n},M_{n}$\} in terms of the oscillators \{$C,\tilde{\mathcal{C}}$\} 
\begin{equation}\label{TSQR34}
\begin{split}
    L_{n}&=\frac{1}{2}\sum_{m}\left[C_{-m}\cdot C_{m+n}-\tilde{\mathcal{C}}_{m}\cdot\tilde{\mathcal{C}}_{-m+n}\right],\\
    M_{n}&=\frac{1}{2}\sum_{m}\left[C_{-m}\cdot C_{m+n}+\tilde{\mathcal{C}}_{m}\cdot \tilde{\mathcal{C}}_{-m+n}+2C_{-m}\cdot\tilde{\mathcal{C}}_{m+n}\right].
    \end{split}
\end{equation}
Hence, $L_{0}$ and $M_{0}$ take the following form, which is closely related to the ones in \eqref{TSQR9}
\begin{equation}\label{TSQR37}
     L_{0}=\mathcal{N}+\bar{\mathcal{N}}, ~~~~~~~
    M_{0}=c'k^2+\mathcal{N}-\bar{\mathcal{N}}+X+Y.
   \end{equation}
where $\mathcal{N}, \bar{\mathcal{N}}, X$ and $Y$ are defined as
\begin{subequations}\label{TSQR38}
\begin{align}
     &\mathcal{N}=\sum_{m>0}C_{-m}\cdot C_{m},\qquad
     \bar{\mathcal{N}}=-\sum_{m>0}\tilde{\mathcal{C}}_{-m}\cdot\tilde{\mathcal{C}}_{m}\\
     &~X=\sum_{m>0}C_{-m}\cdot\tilde{\mathcal{C}}_{m}, \qquad Y=\sum_{m>0}\tilde{\mathcal{C}}_{-m}\cdot C_{m}.
     \end{align}
\end{subequations}
Here again, the operators \{$\mathcal{N},\bar{\mathcal{N}}$\} are number operators and the Hilbert space is spanned by the eigenstates of them. The eigenstates are constructed by acting creation operators \{$C_{-n},\tilde{\mathcal{C}}_{-n}$\} on the flipped vacuum defined in \eqref{TSQR31} much in the same way as in \eqref{TSQR11}.
\medskip

\paragraph{Physical Hilbert space:}\label{bhuto}
Let us first look into the $L_0$ constraint from \eqref{fphysical}. Using the expression of $L_0$ as given in \eqref{TSQR37}, we get the following
\begin{align}
    (\mathcal{N}+\bar{\mathcal{N}}-a_L)\ket{phys}=0.
\end{align}
In the light of the fact that $a_L=2$ for this theory (this comes from both intrinsic as well as limiting perspective) only states with $r+s=2$ ($r,s$ are eigenvalues of $\mathcal{N},\bar{\mathcal{N}}$ respectively) can be physical. The implication is that, for this theory, the flipped vacuum \eqref{TSQR31} itself is no longer a physical state! It is important to note, however, that, this is true only if the target spacetime is non-compact. In upcoming sections we shall see that compactification allows states of other levels to be physical state too.
\medskip

The most general state with $r+s=2$ is given by
\begin{equation}\label{TSQR40}
    \begin{split}
    \ket{2}&=a_{\mu}C^{\mu}_{-2}\ket{0}_{A}+e_{\mu\nu}C^{\mu}_{-1}C^{\nu}_{-1}\ket{0}_{A}+h_{\mu\nu}C^{\mu}_{-1}\tilde{\mathcal{C}}^{\nu}_{-1}\ket{0}_{A}\\
    &+b_{\mu}\tilde{\mathcal{C}}^{\mu}_{-2}\ket{0}_{A}+f_{\mu\nu}\tilde{\mathcal{C}}^{\mu}_{-1}\tilde{\mathcal{C}}^{\nu}_{-1}\ket{0}_{A}+j_{\mu\nu}C^{\mu}_{-1}\tilde{\mathcal{C}}^{\nu}_{-1}\ket{0}_{A}.
    \end{split}
\end{equation}
In the above we have assumed $h_{\mu\nu}$ ($j_{\mu\nu}$) to be symmetric (antisymmetric). The physical state conditions \eqref{fphysical} impose constraints on the coefficients in \eqref{TSQR40}. For any state with $r+s=2$, the conditions in \eqref{fphysical} with $n> 2$ will be trivially satisfied and hence, would not add any constraint on the coefficient. The $L_{0}$ condition  has already been used. Hence we have five more non-trivial physical state conditions left, i.e. the $M_{0}$, $L_{1}$, $M_{1}$, $L_{2}$ and $M_{2}$ conditions. 
Using these conditions, one gets the following constraints:
\begin{equation}\label{flipped phys}
    a_\mu=b_\mu=0, ~~e_{\mu\nu}=f_{\mu\nu}=\frac{1}{2}h_{\mu\nu},~~e_{\mu\nu}k^{\mu}=j_{\mu\nu}k^{\mu}=0, ~~k^2=0.
\end{equation}
Hence the final level 2 physical state becomes 
\begin{equation}\label{TSQR45}    \ket{2}=e_{\mu\nu}\Big[C^{\mu}_{-1}C^{\nu}_{-1}\ket{0}_{A}+2C^{\mu}_{-1}\tilde{\mathcal{C}}^{\nu}_{-1}\ket{0}_{A}+\tilde{\mathcal{C}}^{\mu}_{-1}\tilde{\mathcal{C}}^{\nu}_{-1}\ket{0}_{A}\Big]+j_{\mu\nu}C^{\mu}_{-1}\tilde{\mathcal{C}}^{\nu}_{-1}\ket{0}_{A}.
\end{equation}
The constraint $k^2=0$ in \eqref{flipped phys} implies that the physical state \eqref{TSQR45} is massless. This state has vanishing norm and can be identified as CCA null state having weight $h_L=2$, $h_M=0$ discussed in \cite{Bagchi:2025vri}.

\paragraph{Limiting analysis of tensile states:} We already saw evidence of the fact that the tensionless flipped theory comes from tensionless limit of twisted string theory. In this section that will be demonstrated more explicitly. As shown in \cite{Lee:2017utr}, twisted string theory too, allows only level 2 states as physical states. However, those states have non-zero norm. Hence, the idea of non-null physical states evolving into null states at tensionless limit might sound strange. However, in \cite{Bagchi:2020fpr}, this evolution has been explained. The discussion in this section follows the analysis given in \cite{Bagchi:2020fpr}.\par
It is important to highlight here first that for tensile twisted string theory too, in the right moving sector the role of creation and annihilation operators flip. The vacuum for this theory is defined as
\begin{align}    \alpha^{\mu}_{n}\ket{0,k^{\mu}}_A=\tilde{\alpha}^{\mu}_{-n}\ket{0,k^{\mu}}_A=0\hspace{5mm}\forall n>0.
\end{align}
We define the creation operator in the left moving sector of this theory as 
\begin{align}\label{flipp}
    \bar{\alpha}^{\mu}_{-n}=-\tilde{\alpha}^{\mu}_{n}.
\end{align}
Let us consider the following state from the physical state spectrum of twisted string theory. 
\begin{align}\label{TSQR46}
\ket{\Psi}=\s_{\mu\nu}\alpha^{\mu}_{-1}\bar{\alpha}^{\nu}_{-1}\ket{0}_{A}.
\end{align}
Now, we recall from \eqref{chh4} how the tensile operators \{$\alpha,\tilde{\alpha}$\} are connected to tensionless operators \{$C,\tilde{C}$\}. Using that relation and taking into account \eqref{TSQR31} and \eqref{flipp}, one can see the evolution of the state \eqref{TSQR46} at tensionless limit.
\begin{equation}\label{TSQR47}
        \lim_{\epsilon\to 0} \ket{\Psi}=\s_{\mu\nu}\Big[\cosh{\theta}\hspace{1mm}C^{\mu}_{-1}-\sinh{\theta}\hspace{1mm}\tilde{\mathcal{C}}^{\mu}_{-1}\Big]\Big[\sinh{\theta}\hspace{1mm}C^{\nu}_{-1}-\cosh{\theta}\hspace{1mm}\tilde{\mathcal{C}}^{\nu}_{-1}\Big]\ket{0}_{A}=\ket{\phi_{1}}+\ket{\phi_{2}}+\ket{\phi_{3}}.
\end{equation}
In the above, we have defined 
\begin{equation}
   \cosh\theta=\frac{1}{2}\left(\sqrt{\epsilon}+\frac{1}{\sqrt{\epsilon}}\right), \qquad  \sinh\theta=\frac{1}{2}\left(\sqrt{\epsilon}-\frac{1}{\sqrt{\epsilon}}\right).
   \end{equation}
   $\ket{\phi_{1}},\ket{\phi_{2}}$ and $\ket{\phi_{3}}$ respectively are 
\begin{subequations}\label{TSQR48}
   \begin{align}
       \ket{\phi_{1}}&=\frac{\epsilon}{2\sqrt{2}}\s_{\mu\nu}\left[C^{\mu}_{-1}C^{\nu}_{-1}-2C^{\mu}_{-1}\tilde{\mathcal{C}}^{\nu}_{-1}+\tilde{\mathcal{C}}^{\mu}_{-1}\tilde{\mathcal{C}}^{\nu}_{-1}\right]\ket{0}_{A},\\
      \ket{\phi_{2}} &=\frac{1}{2\epsilon\sqrt{2}}\s_{\mu\nu}\left[C^{\mu}_{-1}C^{\nu}_{-1}+2C^{\mu}_{-1}\tilde{\mathcal{C}}^{\nu}_{-1}+\tilde{\mathcal{C}}^{\mu}_{-1}\tilde{\mathcal{C}}^{\nu}_{-1}\right]\ket{0}_{A},\\
       \ket{\phi_{3}}&=\s_{\mu\nu}\left[C^{\mu}_{-1}\tilde{\mathcal{C}}^{\nu}_{-1}-C^{\nu}_{-1}\tilde{\mathcal{C}}^{\mu}_{-1}\right]\ket{0}_{A}. 
   \end{align}
\end{subequations}
One can check that, all these three states have vanishing norm. However, the norm of $\ket{\Psi}$ is preserved because $\braket{\phi_{1}|\phi_{2}}\neq 0$ and that is where the non-vanishing part of $\braket{\Psi|\Psi}$ comes. Taking that into account one can calculate the norm of $\ket{\Psi}$ as below
\begin{align}\label{TSQR49}
    \braket{\Psi|\Psi}=\s^{\mu\nu}\s_{\mu\nu}.
\end{align}
It is interesting to note from \eqref{TSQR48} and \eqref{TSQR45} that the state $\ket{\phi_{1}}$ doesn't correspond to allowed form of physical states \eqref{TSQR45} derived from intrinsic perspective. When we look into $\epsilon\to 0$ limit, this state $\ket{\phi_1}$ contains a $\mathcal{O}(\epsilon)$ term and hence, gets suppressed. The remaining states correspond to \eqref{TSQR45} and they form the null state.

\subsection*{Connection to ambitwistors}
The limited mass spectrum with only the level two massless states and origin in an unfamiliar twisted string theory may seem like the Flipped null string is a rather exotic and uninteresting object. It turns out to be quite the opposite. This theory is intimately connected to the bosonic version of the Ambitwistor string \cite{Mason:2013sva} and this connection was first established in \cite{Casali:2016atr}. 

\medskip

The Ambitwisor string is famously the worldsheet origin of the Cachazo–He–Yuan (CHY) formulae \cite{Cachazo:2013hca}. The CHY formulae capture the tree level amplitudes for massless particles (gluons, gravitons). The Ambitwistor partition function localizes completely and ambitwistor string amplitudes at tree level precisely give the CHY formulae. As just mentioned, the Ambitwistor string in just the null string in the flipped vacuum. With the restriction of $h_M=0$ and $c_M=0$, the 2d Conformal Carroll algebra is known to exhibit a truncation to a chiral CFT \cite{Bagchi:2009pe, Bagchi:2012yk}. The ambitwisor string seems to use this truncation to move from a null string to a chiral string \cite{Casali:2017zkz}. Since this is a discussion which will take us a bit away from our focus in this review, we will not address this in the main body of our article. However, we provide a detailed summary of this connection between the Ambitwistor string and the null string in Appendix~{\ref{Connection to Ambitwistor strings}}.  
\medskip

\subsection{Summary}

We have explored various features of the three quantum null strings in this and the previous sections. We now provide a quick summary of these theories built on the three different vacua. The table below summarises our findings and also connects to some aspects we will cover in the remainder of our review. 

\bigskip

\begin{table}[ht]\label{summary table for three vacua}
\small
\centering
\begin{tabular}{ |c||c|c|c| }
\hline
\quad & Induced & Flipped & Oscillator \\ 
\hline\hline

\multirow{2}{6.5em}{Tensile Origin} & \multirow{2}{6.5em}{Usual strings} & \multirow{2}{7em}{Twisted strings} & \multirow{2}{8.5em}{Accelerated strings}\\ &  &  &  \\
\hline

BMS  &  \multirow{2}{5em}{Induced} & \multirow{2}{7em}{Highest Weight}& \multirow{2}{3em}{?}\\ Representation& &  &   \\
\hline 

\multirow{3}{4em}{Vacuum condition}  & & & \\& $ B_n\ket{0}_I=0, ~~\forall n\neq 0 $  & $C_n\ket{0}_A=0=\tilde{C}_{-n}\ket{0}_A$& $C_n\ket{0}_c=0= \tilde{C}_n\ket{0}_c$ \\ &  & $\forall n>0$ &$~\forall n>0$ \\ 
\hline

Critical & \multirow{2}{7em}{No constraints} & \multirow{2}{4em}{$D=26$} & \multirow{2}{4em}{$D=26$}\\Dimension & & &\\
\hline

Mass & \multirow{2}{9em}{Only massless states} & \multirow{2}{9em}{Only massless states}  &  \multirow{2}{7em}{Rich spectrum}\\  Spectrum & \multirow{2}{7em}{} &  & \multirow{2}{7em}{} \\
\hline

Relation with & Bogoliubov trans & \multirow{2}{2em}{?} & Bogoliubov trans \\ other vacua &to Oscillator &  & to Induced \\
\hline

& & &\\ Features & Worldsheet BEC & Only level 2 states & Thermal string \\ and   & & &  \\ Connections  & Gross-Mende? & Ambitwistor string  & black holes \\
(Details) & (Sec.\ref{grossmende})& (Ap.\ref{Connection to Ambitwistor strings}) & (Sec.\ref{blackhole microstates})\\
\hline
\hline

\end{tabular}
\captionof{table}{Comparison of null quantum string theories}
\end{table}

We will now move beyond flat target spaces to target spaces with compactified directions where we will see even more interesting features.

\bigskip

\newpage

\section{Compactification: Circle and Torus}\label{Compactification: Circle and Torus}
In earlier sections we have discussed how the three quantum theories of closed null strings need target spacetimes of particular dimensions to be consistent. For these theories to make sense in lower dimensions, we must now consider proper compactification of such theories. In this section we look into all three quantum theories, first on a compact circle (based on \cite{Banerjee:2023ekd}), and then move on to compactification on a $d$ dimensional torus (based on \cite{Banerjee:2024fbi}) with a $B$-field. Our discussion closely adheres to the classic description of compactified tensile strings. Naturally, we also delve into another related concept: that of T-duality. We ask, whether such a duality indeed makes sense for the null string where the string length scale effectively diverges, and whether it is still a duality that connects such string theories in different compactified backgrounds? In what follows, we will answer both of these questions in the affirmative, with some added subtleties. 

\subsection{Null string compactification on $S^1$}\label{Compactification zero modes}
Let us recall the solution of the EOM of tensionless closed string from \eqref{fcb1}. Now, let us compactify the coordinate $X^{D-1}$ ($D=26$) on a circle of radius $R$. This implies that the following points are identified
\begin{equation}
    X^{25}\sim X^{25}+2\pi RW, \hspace{10mm}W\in\mathbb{Z}.
\end{equation}
Such identification in the target spacetime means that $X^{25}$ parametrises a circle of radius $R$. The integer $W$ is the winding number of the string i.e. the number of times the string winds around the compactified direction. The closed string periodicity condition along $X^{25}$ needs to be modified in this case, 
\begin{equation}\label{mod periodicity}
    X^{25}(\sigma+2\pi)=X^{25}(\sigma,\tau)+2\pi RW.
 \end{equation}
 When we quantise this theory, this gives rise to winding states, characterised by the winding number $W$.
Let's start from the mode expansion of $X^{25}(\sigma,\tau)$ for the null string:
\begin{equation}
    X^{25}(\tau,\sigma)=x^{25}+\sqrt{\frac{c'}{2}}A^{25}_{0}\sigma+\sqrt{\frac{c'}{2}}B^{25}_{0}\tau+i\sqrt{\frac{c'}{2}}\sum_{n\neq 0}\frac{1}{n}(A^{25}_{n}-in\tau B^{25}_{n})e^{-in\sigma}.
\label{compt mode expnsn}
\end{equation}
We recall from earlier discussion (see \eqref{totalm2}) that $k^{\mu}=\sqrt{\frac{1}{2c'}}B^{\mu}_{0}$. In order to keep the wave function (which contains $e^{ik^{25}X^{25}}$) single-valued, $k^{25}$ and subsequently $B^{25}_{0}$, essentially take discrete values. Interestingly, $A^{25}_{0}$ no longer vanishes here because of the modified periodicity condition (\ref{mod periodicity}). One could find: 
\begin{equation}\label{chh7}
    A^{25}_{0}=\sqrt{\frac{2}{c'}}RW, \qquad B^{25}_{0}=\sqrt{2c'}\Big(\frac{K}{R}\Big),\qquad \quad W,K\in\mathbb{Z}.
\end{equation}
We now rewrite the mode expansion (\ref{compt mode expnsn}) using \eqref{chh7}
\begin{align}
    X^{25}=x^{25}+RW\sigma+\Big(\frac{c'K}{R}\Big)\tau+i\sqrt{\frac{c'}{2}}\sum_{n\neq 0}\frac{1}{n}(A^{25}_{n}-in\tau B^{25}_{n})e^{-in\sigma}.
\end{align}
For other non-compact dimensions  $A^{\mu}_{0}=0$ continues to hold. The zero modes of redefined oscillators $(C,\tilde C)$ (see \eqref{tsc7}) for the compactified case becomes
\begin{equation}\label{chh9}
   C^{25}_{0}=\frac{1}{2}\left[\sqrt{2c'}\left(\frac{K}{R}\right)+\sqrt{\frac{2}{c'}}RW\right],\qquad \tilde{C}^{25}_{0}=\frac{1}{2}\left[\sqrt{2c'}\left(\frac{K}{R}\right)-\sqrt{\frac{2}{c'}}RW\right].
\end{equation}
For non-compact dimensions, $C^{\mu}_{0}=\tilde{C}^{\mu}_{0}=\sqrt{\frac{c'}{2}}k^{\mu}$. 
We conclude this discussion with an observation. In tensile bosonic string theory, the zero modes in the mode expansion of $X^\m$ has the following form
\begin{align}\label{C17}
   \alpha^{25}_{0}=\frac{1}{2}\left[\sqrt{2\alpha'}\left(\frac{K}{R}\right)+\sqrt{\frac{2}{\alpha'}}RW\right],\qquad \tilde{\alpha}^{25}_{0}=\frac{1}{2}\left[\sqrt{2\alpha'}\left(\frac{K}{R}\right)-\sqrt{\frac{2}{\alpha'}}RW\right].
\end{align}
It is straightforward to see that at tensionless limit ($\alpha'\to\frac{c'}{\epsilon}$) the modes in \eqref{chh9} are connected to the tensile modes \eqref{C17} in the following way
\begin{equation}\label{C18}
\begin{split}
    C^{25}_{0}&=\frac{1}{2}\Big(\sqrt{\epsilon}+\frac{1}{\sqrt{\epsilon}}\Big)\alpha^{25}_{0}+\frac{1}{2}\Big(\sqrt{\epsilon}-\frac{1}{\sqrt{\epsilon}}\Big)\tilde{\alpha}^{25}_{0},\\
    \tilde{C}^{25}_{0}&=\frac{1}{2}\Big(\sqrt{\epsilon}-\frac{1}{\sqrt{\epsilon}}\Big)\alpha^{25}_{0}+\frac{1}{2}\Big(\sqrt{\epsilon}+\frac{1}{\sqrt{\epsilon}}\Big)\tilde{\alpha}^{25}_{0}. 
\end{split}
\end{equation}
This relation turns out to be exactly identical to the Bogoluibov transformation between tensile and tensionless oscillators. This provides us with a consistency check in our study of null string compactification.
\subsection{Null strings in constant Kalb-Ramond background} \label{sec2}
 Before proceeding to toroidal compactification of null strings let us quickly review the impact of constant Kalb-Ramond background field on null strings. Let us begin with the tensile counterpart where the action is Polyakov action with a $B$-field\footnote{This action can also be formulated from a Nambu-Goto version of strings in constant $B$-field. For details about this see \cite{Gangopadhyay:2006gmx,Banerjee:2024fbi}}
\begin{align}\label{Polyakov2}
   S_{I}=\frac{T}{2}\int d\tau d\sigma \left(\sqrt{-g}g^{\alpha\beta}\partial_{\alpha}X^{\mu}\partial_{\beta}X^{\nu}\eta_{\mu\nu}+\epsilon^{\alpha\beta}B_{\mu\nu}\partial_{\alpha}X^{\mu}\partial_{\beta}X^{\nu}\right).
\end{align}
We have already seen in section \ref{cup} how the worldsheet metric evolves at tensionless limit. The $B$-field part in the action survives at the tensionless limit only when it is scaled in the following way
\begin{align}\label{ramtaru}
    B_{\mu\nu}\to \frac{B_{\mu\nu}}{\epsilon}.
\end{align}
 For our convenience, we define scaled antisymmetric $B$-field $\mathcal{B}_{\mu\nu}$ as
\begin{align}\label{ctb27}
   \mathcal{B}_{\mu\nu}=\frac{1}{\alpha'}B_{\mu\nu},\qquad T=\frac{1}{2\pi\alpha'}
\end{align}
The action of null string with a $B$ field obtained through this limit is 
\begin{align}\label{ILSTB}
    S=\int d^2\xi \left(V^{\alpha}V^{\beta}\partial_{\alpha}X^{\mu}\partial_{\beta}X^{\nu}\eta_{\mu\nu}+\frac{1}{2\pi}\epsilon^{\alpha\beta}\mathcal{B}_{\mu\nu}\partial_{\alpha}X^{\mu}\partial_{\beta}X^{\nu}\right).
\end{align}


Here $\epsilon^{01}=1$. Now let us look into the symmetries this action \eqref{ILSTB} possess. Since we already know from section \ref{cup} that ILST action (which is \eqref{ILSTB} with $\mathcal{B}_{\m\n}=0$) \eqref{ILST} is worldsheet diffeomorphism invariant and hence we are now concerned just about the variation of the $B$ field part of the action. It gives us just a total derivative term
\begin{equation}
\delta(\epsilon^{\alpha\beta}\mathcal{B}_{\mu\nu}\partial_{\alpha}X^{\mu}\partial_{\beta}X^{\nu})=\partial_{\alpha}(\epsilon^{\alpha\beta}\mathcal{B}_{\mu\nu}\epsilon^{\gamma}\partial_{\gamma}X^{\mu}\partial_{\beta}X^{\nu}).
\end{equation}
Consequently, for closed string the action \eqref{ILSTB} too, is diffeomorphism invariant. This again necessitates us to choose a particular gauge to work in and we choose the same gauge choice as in \eqref{Review1}. The gauge fixed action, once again displays the same symmetry algebra as residual gauge symmetry \eqref{bmsalgebraclassical}. 
It is straightforward to check that the EOM and constraints for this action is identical to the same of the ILST action i.e. \eqref{tsc1} and after gauge fixing they reduce to \eqref{tsc3} and \eqref{constraint}. Hence the mode expansion for the closed bosonic string is same as well. Just like the case of tensile string theory, here too, the only quantity affected by constant $B$-field is the canonical momentum as demonstrated below
\begin{align}\label{mom}
    \Pi_{\mu}=\frac{\partial\mathcal{L}}{\partial \dot{X}^{\mu}}=\frac{\dot{X}^{\nu}}{2\pi c'}\eta_{\nu\mu}+\frac{X'^{\nu}}{2\pi}\mathcal{B}_{\mu\nu}. 
\end{align}
 It is important to note here that the new expression of canonical momenta in \eqref{mom} does not change the equal time Poisson bracket between $X(\sigma,\tau)$ and $\dot{X}(\sigma',\tau)$. Consequently, the Poisson brackets between the oscillators $\{A,B\}$ (or $\{C, \tilde{C}\}$) remains same as \eqref{tsc5} (or \eqref{calgebra}). 
Integrating \eqref{mom} over the entire closed string gives the following   canonical momentum
\begin{align}\label{ctb2}
    \pi_{\mu}=\int_{0}^{2\pi}d\sigma \Pi_{\mu}=k^{\nu}\eta_{\nu\mu}+\frac{1}{2\pi}\Big(X^{\nu}(\sigma+2\pi,\tau)-X^{\nu}(\sigma,\tau)\Big)\mathcal{B}_{\mu\nu},
\end{align}
where $k^{\mu}$ is centre of mass momentum of the null string with $\mathcal{B}_{\mu\nu}=0$. The effect of the changed periodicity is obvious here.

\subsection{Compactification on torus $T^d$} 

Now, let us generalise the previous analysis to target spaces compactified on a torus $T^d$.  The metric components on the compact torus $T^d$ are chosen to be Euclidean, i.e. $G_{IJ}=\delta_{IJ}$ ($I,J\in\{1,2,....,d\}$ are indexes for compact dimensions).
 In order to compactify $d$ dimensions, we make the following identification\footnote{For compactified indices $I$, repeated indices are not summed over unless explicitly mentioned.} 
\begin{align}\label{chh42}
      X^{I}\sim X^{I}+2\pi \omega^{I}R_{I}, \qquad I\in\{1,2,....,d\}.
\end{align}
This implies that the radius of circle parametrized by $X^I$ is $R_{I}$.
Following the same argument given in previous section one concludes that
\begin{align}\label{CTBC1}
     X^{I}(\sigma+2\pi,\tau)=X^{I}(\sigma,\tau)+2\pi \omega^{I}R_{I}, \qquad I\in\{1,2,....,d\},
\end{align}
where $\omega^I$s are winding numbers along $I$th direction. For convenience we rescale the compactified coordinates as below
\begin{align}\label{nakgol}
    \widetilde{X}^{I}=\frac{X^{I}}{R_{I}}.
\end{align}
$G_{IJ}$ and $\mathcal{B}_{IJ}$, being covariant rank 2 tensors transform in the following way under the rescaling \eqref{nakgol}
\begin{subequations}\label{BBS1}
\begin{align}
     &G_{IJ}=\frac{\partial X^{K}}{\partial \widetilde{X}^{I}}\frac{\partial X^{L}}{\partial \widetilde{X}^{J}}\delta_{KL}=R_{I}R_{J}\sum_{K=1}^{d}\sum_{L=1}^{d}\delta^{K}_{I}\delta^{L}_{J}\delta_{KL}=R^{2}_{I}\delta_{IJ}\label{BBS11}\\
     &\mathbb{B}_{IJ}=\frac{\partial X^{K}}{\partial \widetilde{X}^{I}}\frac{\partial X^{L}}{\partial \widetilde{X}^{J}}\mathcal{B}_{KL}=R_{I}R_{J}\sum_{K=1}^{d}\sum_{L=1}^{d}\delta^{K}_{I}\delta^{L}_{J}\mathcal{B}_{KL}=R_{I}R_{J}\mathcal{B}_{IJ}\label{BBS12}
\end{align}
\end{subequations}
For future use, let us also write the inverse metric $G^{IJ}$, which simply reads $G^{IJ}=\frac{1}{R^{2}_{I}}\delta^{IJ}$.
After this rescaling we rewrite the identification $ \widetilde{X}^{I}\sim \widetilde{X}^{I}+2\pi \omega^{I}$, and
 consequently,
\begin{align}\label{CTBC2}
    \widetilde{X}^{I}(\sigma+2\pi,\tau)-\widetilde{X}^{I}(\sigma,\tau)=2\pi \omega^{I}.
\end{align}
Let us now rewrite the gauge fixed action in \eqref{ILST} in terms of redefined coordinates
\begin{align}\label{agol}
    \int d^2\xi \left(v^2\dot{\widetilde{X}}^{I}\dot{\widetilde{X}}^{J}G_{IJ}+\epsilon^{\alpha\beta}\mathbb{B}_{IJ}\partial_{\alpha}\widetilde{X}^{I}\partial_{\beta}\widetilde{X}^{J}\right).
\end{align}
$v$ is constant given in \eqref{Review1}. Let us look into the canonical momentum components (see \eqref{mom}) along the compactified directions
\begin{align}
    \widetilde{\Pi}_{I}=\frac{1}{c'}\sum_{J=1}^{d}\dot{\widetilde{X}}^{J}G_{IJ}+\sum_{J=1}^{d}\mathbb{B}_{IJ}\widetilde{X}'^{J}.
    \end{align}
Integrating over string length as done in \eqref{ctb2}, we find the following
\begin{align}\label{ctb24}
    \widetilde{\pi}_{I}=\frac{1}{2\pi}\int_{0}^{2\pi}d\sigma~ \widetilde{\Pi}_{I}=\sum_{J=1}^{d}K^{J}G_{IJ}+  \sum_{J=1}^{d}\mathbb{B}_{IJ}\omega^{J}.
\end{align}
In the above $K^{I}$ is the actual center of mass momentum of the string given by
\begin{align}
    K^{I}=\frac{1}{2\pi c'}\int_{0}^{2\pi}d\sigma~\dot{\widetilde{X}}^{I}.
\end{align}
At this point it is important to recall that canonical momentum is the translation generator, which means the quantum mechanical wave function should contain the factor $\exp\left[{i\sum_{I=1}^{d}\widetilde{\pi}_{I}\widetilde{X}^{I}}\right]$. In order to keep this function periodic in $\widetilde{X}^I\sim \widetilde{X}^I+ 2\pi \omega^I$, 
we must have $\omega^{I}\widetilde{\pi}_{I}\in\mathbb{Z}$. Since $\omega^I\in\mathbb{Z}$, we end up with $\widetilde{\pi}_{I}=k_{I},\hspace{2mm}k_{I}\in\mathbb{Z}$. Now, for further convenience, we define dimensionless versions of the pullback field :
\begin{align}\label{chh49}
    \widetilde{X}^{I}=\sqrt{\frac{c'}{2}}Y^{I}.
\end{align} 
The mode expansion \eqref{fcb1} then transforms into the following 
\begin{align}\label{modeexp}
    Y^{I}=y^{I}+A^{I}_{0}\sigma+B^{I}_{0}\tau+i\sum_{n\neq 0}\frac{1}{n}(A^{I}_{n}-in\tau B^{I}_{n})e^{-in\sigma}.
\end{align}
Using \eqref{ctb24}, \eqref{CTBC2} and recalling that $\widetilde{\pi}_{I}=k_{I}$ one can easily determine the zero modes in the above mode expansion
\begin{align}\label{B2}
  (B_{0})_I=\sqrt{2c'}K_{I}=\sqrt{2c'}\Big(k_{I}-\sum_{J=1}^{d}\mathbb{B}_{IJ}\omega^{J}\Big),\hspace{10mm}A^{I}_{0}=\sqrt{\frac{2}{c'}}\omega^{I}.
\end{align}
Mode expansion of $Y^{I}$ in terms of harmonic oscillators $C$ and $\tilde{C}$ becomes:
\begin{equation}\label{ctb10}
\begin{split}
   Y^{I}_{L}&=y^{I}_{L}+K^{I}_{L}(\tau+\sigma)+i\sum_{n\neq0}\frac{1}{n}(C^{\mu}_{n}-in\tau C^{\mu}_{n})e^{-in\sigma},\\
    Y^{I}_{R}&=y^{I}_{R}+K^{I}_{R}(\tau-\sigma)+i\sum_{n\neq0}\frac{1}{n}(\tilde{C}^{\mu}_{n}-in\tau \tilde{C}^{\mu}_{n})e^{in\sigma}. 
\end{split}
\end{equation}
Here $(K^{I})_{L,R}$ are the left and right moving momenta of the string along compactified directions, given by:
\begin{align}\label{ctb15}
    (K_I)_{L,R}=\frac{1}{\sqrt{2}}\bigg\{\sqrt{c'}\Big(k_{I}-\sum_{J=1}^{d}\mathbb{B}_{IJ}\omega^{J}\Big)\pm \frac{1}{\sqrt{c'}}\sum_{J=1}^{d}G_{IJ}\omega^{J}\bigg\}.
\end{align}
Having ourselves equipped with the necessary tools, we now proceed to discuss all three quantum theories of null strings and the impact of compactification on the spectra. 
\subsection{Effect of compactification}
For this section, we only briefly cover the results discussed in \cite{Banerjee:2023ekd,Banerjee:2024fbi} with emphasis on the common aspects of impact of compactification on all the three theories. For more detailed coverage the reader is directed towards the references. Let us begin with the effect of compactification on the physical Hilbert spaces of the theory. We will follow the notations of section \ref{Physical Hilbert spaces for three vacua} in what follows.
\subsection*{Physical Hilbert space}
We recall from section \ref{Physical Hilbert spaces for three vacua} that $L_0$ physical state conditions supplies us with important restriction on the physical states which is the level matching condition. From the earlier discussion of quantum null strings we see that this condition plays crucial role in oscillator vacuum and flipped vacuum. We now discuss how compactification modifies the $L_0$ physical state condition for the three quantum theories.
\paragraph{Oscillator vacuum:} For circle compactification of oscillator vacuum rewriting $L_0$ using \eqref{chh9} gives us the following
\begin{align}
     L_{0}=\mathcal{N}-\widetilde{\mathcal{N}}+KW,
\end{align}
which lead us to the modified level matching condition: $s=r+KW$. Similarly for toroidal compactification of oscillator vacuum with constant $B$-field, using \eqref{ctb15} we arrive at the following
\begin{equation}\label{ctb18}
\begin{split}
    L_{0}=\mathcal{N}-\widetilde{\mathcal{N}}+\frac{1}{2}\sum_{I=1}^{d}\big(K^{I}_{L}&K_{I\hspace{.25mm}L}-K^{I}_{R}K_{I\hspace{.25mm}R}\big)
    =\mathcal{N}-\widetilde{\mathcal{N}}+\sum_{I,J=1}^{d}\Big(k_{I}-\mathbb{B}_{IJ}\omega^{J}\Big)\omega^{I},\\
    &=\mathcal{N}-\widetilde{\mathcal{N}}+\sum_{I=1}^{d}k_{I}\omega^{I},
\end{split}
\end{equation}
leading us to the following generalised level matching condition
\begin{align}
    s-r=\sum_{I=1}^{d}k_{I}\omega^{I}.
\end{align}
Clearly the constant $B$-field doesn't have any impact on the level matching condition. 
\paragraph{Induced vacuum:} In case of induced vacuum, the states constructed other than the induced vacuum are not well understood and the role of $L_0$ physical state condition for those states is not clear. So we will not go into much details about this. 
\paragraph{Flipped vacuum:} For flipped vacuum, using \eqref{C18} and the fact that $C^{\mu}_{0}=\tilde{\mathcal{C}}^{\mu}_{0}$ one can rewrite $L_0$ as
\begin{align}
    L_{0}=\mathcal{N}+\bar{\mathcal{N}}+KW,
\end{align}
which results into the following level matching condition for $a_L=2$
\begin{align}\label{flippedlevel}
    r+s+KW=2.
\end{align}
For non-compact case, the level matching condition allowed only states of level 2 to be physical states. However, here, the $KW$ term allows states of any level to be physical state. Even more interestingly, for $r+s=2$, \eqref{flippedlevel} implies that $KW=0$, which means one can have infinite tower of physical states at level 2 (for $K=0$, any value of $W$ gives physical states and vice-versa). Similarly, for toroidal compactification with $B$-field, \eqref{ctb15} implies that 
\begin{align}
    \label{flippedB1}
 L_0=
\mathcal{N}+\bar{\mathcal{N}}+\sum_{I=1}^{d}k_{I}\omega^{I},
\end{align}
which results into following level condition
\begin{align}\label{B10}
     r+s+\sum_{i=1}^{d}k_{I}\omega^{I}=2
\end{align}
Here one finds an important distinction between flipped vacuum theory in circle compactification and the same theory in toroidal compactification. When more dimensions are compactified the \eqref{B10} allows much more freedom. For each level, there can be infinite possible combinations of \{$k_I,\omega^{I}$\} satisfying \eqref{B10}, allowing infinite number of states at each level. 
\subsection*{Mass spectrum}
We have already seen from the discussion of the three quantum theories that the mass spectrum of the theory emerges from the $M_0$ physical state condition. In section \ref{Compactification zero modes} we also saw how the zero modes get modified due to circle and toroidal compactification. Using \eqref{TSQR9} and recalling that only momentum components along the non-compact dimensions contribute to the mass-squared of the string, one can rewrite $M_0$ for oscillator vacuum in \eqref{TSQR9} as 
\begin{align}
    M_{0}=c'\frac{K^2}{R^2}-c'm^2+\mathcal{N}+\widetilde{\mathcal{N}}+X+X^{\dagger}
\end{align}
In same way, applying \eqref{chh9} on expression of $M_0$ for flipped vacuum as given in \eqref{TSQR37}, we get the following
\footnote{For flipped vacuum the flipping of left moving oscillators in \eqref{flipping} i.e. $\tilde{\mathcal{C}}_{n}=\tilde{C}_{-n}$ implies that for $n=0$, $\tilde{\mathcal{C}}_{0}$ is same as $\tilde{C}_{0}$ given in \eqref{C18}.}
\begin{align}
     M_{0}=c'\frac{K^2}{R^2}-c'm^2+\mathcal{N}-\bar{\mathcal{N}}+X+Y.
\end{align}
For induced vacuum, using \eqref{chh7} on $M_0$ from \eqref{tsc9} leads us to the following
\begin{align}
    M_0=c'\frac{K^2}{R^2}-c'm^2+\sum_{n\neq 0}B_{-n}\cdot B_{n}.
\end{align}
One can see that compactification in all the above cases has resulted into a shift in $M_0$ for all the three theories, subsequently changing the mass:
\begin{align}
   m^2\to m^2-\frac{K^2}{R^2}.
\end{align}
Consequently for all the theories the mass spectrum acquires the  extra term $\frac{K^2}{R^2}$ 
\begin{align}\label{nullcomp}
    \frac{K^2}{R^2}.
\end{align}
To appreciate the origin of this shift, recall that for tensile strings, compactification results into the following additional term in the mass-spectrum
\begin{align}\label{tensilecomp}
    \frac{K^2}{R^2}+\frac{1}{\a'}W^2R^2.
\end{align}
Clearly at tensionless limit $\a'\to \infty$ \eqref{tensilecomp} reduces to the shift $\frac{K^2}{R^2}$.
\medskip

For compactification on a torus $T^d$ having metric $G_{IJ}$ and background field $\mathbb{B}_{IJ}$, using the zero modes in \eqref{B2} (or alternatively \eqref{ctb15}) and repeating the aforementioned steps for all the three theories we obtain the following extra term in the mass spectrum of all three theories
   \begin{align}\label{ctb25}
m^2=k^{T}G^{-1}k-2k^{T}G^{-1}\mathbb{B}\omega-\omega^{T}\mathbb{B}G^{-1}\mathbb{B}\omega.
    \end{align}
    In the above, we have used matrix notation, where $\mathbb{B}_{IJ}$ and $G^{IJ}$ are represented as $d\times d$ matrices and $\omega$ and $k$ are represented as column vectors. This can also be equivalently rewritten in the following $2d\times 2d$ matrix form
\begin{align}\label{Buscher5}
m^2=  \Big[\omega^{T} \hspace{2mm} k^{T}\Big]\mathcal{G} \begin{bmatrix}
    \omega \\  k
\end{bmatrix},~~~~\mathcal{G}=\begin{bmatrix}
    -\mathbb{B}G^{-1}\mathbb{B} & \mathbb{B}G^{-1}\\-G^{-1}\mathbb{B} & G^{-1}
\end{bmatrix}.
\end{align}
To obtain the complete mass spectrum, one needs to add the non-compact mass spectrum to the above. Since for non-compact case, both induced and flipped vacuum have massless spectrum (shown before in \eqref{inducedmass} and \eqref{flipped phys}), for circle compactified case, \eqref{nullcomp} and for toroidal compactified case, \eqref{ctb25} (or \eqref{Buscher5}) give the complete mass spectrum\footnote{Although for flipped vacuum, compactification allows levels other than $n=2$ as physical states, it has been shown in \cite{Banerjee:2023ekd} that for those states too, contribution to mass spectrum always comes from compactification only.}. For oscillator vacuum, however, we also need to take into account the contribution from the level of the states. For non-compact case this contribution was shown in \eqref{TSQR17}. Circle compactification of oscillator vacuum leads us to the the following spectrum 
\begin{subequations}\label{FD2}
    \begin{align}
    s=r+KW,
    \implies  m^2\ket{r,s}=\left[\frac{K^2}{R^2}+\frac{1}{c'}(r+s-2)\right]\ket{r,s}.
    \end{align}
\end{subequations}
Toroidal compactification of oscillator vacuum on $d$-dimensions leads us to the the following spectrum
\begin{subequations}\label{tsc11}
    \begin{align}
 &s-r=\sum_{I=1}^{d}k_{I}\omega^{I},\\
    m^2=k^{T}G^{-1}k-2k^{T}&G^{-1}\mathbb{B}\omega-\omega^{T}\mathbb{B}G^{-1}\mathbb{B}\omega+\frac{1}{c'}(r+s-2).
\end{align}
\end{subequations}
Let us recall that the mass spectrum for the tensile physical perturbative states in presence of constant $B$-field
\begin{align}\label{tensile}
 \Big[\omega^{T} \hspace{4mm} k^{T}\Big]\begin{bmatrix}
    \cfrac{1}{\alpha'^2}\left(G-b G^{-1}b\right) & ~~\cfrac{1}{\alpha'}~bG^{-1}\\-\cfrac{1}{\alpha'}~G^{-1}b & G^{-1}
\end{bmatrix}
\hspace{2mm}
\begin{bmatrix}
    \omega \\ \\ k
\end{bmatrix},
\end{align}

Here $G^{IJ}$ is same as \eqref{BBS11} while $b_{IJ}$ is related to the constant $B$-field $B_{IJ}$ as
\begin{equation}\label{ramgorur}
    b_{IJ}=\frac{\partial X^{K}}{\partial \widetilde{X}^{I}}\frac{\partial X^{L}}{\partial \widetilde{X}^{J}}B_{KL}=R_{I}R_{J}\sum_{K=1}^{d}\sum_{L=1}^{d}\delta^{K}_{I}\delta^{L}_{J}B_{KL}=R_{I}R_{J}B_{IJ}.
\end{equation}
Now, \eqref{ctb27} and \eqref{BBS1} implies the following
\begin{align}
    \frac{1}{\alpha'}b_{IJ}=\mathbb{B}_{IJ}.
\end{align}
Taking $\alpha'\to\frac{c'}{\epsilon}$ with $\epsilon\to 0$ on the above, the mass spectrum \eqref{tensile} reduces to \eqref{Buscher5}.

\subsection{A tensionless T-duality}\label{T-duality on null strings}
T-duality is one of the fundamental properties of string theory that relates string theories defined on different compactified target spacetime. Such duality gives rise to amazing new insights; at one hand it played a crucial role in the realisation that all string theories are different limits of one theory called M-theory, while on the other hand it provided important insights on the quantum aspect of spacetime geometry as well as new results in algebraic geometry like Mirror symmetry, etc. In this subsection, we investigate the role played by this duality in null strings.

\subsection*{Circle compactification}
 For circle compactification on null strings there was no contribution from the winding number $W$ in the mass spectrum. This is important, because the contribution of winding number in the mass-spectrum leads to a fundamental property of string theory; namely T-duality, which dictates equivalence of string theories compactified on two different target spacetime having radius $R$ and $\frac{\a'}{R}$. Complete absence of winding number in the null string mass spectrum \eqref{nullcomp} shows that one cannot find equivalence between null strings compactified on two different radii. In order to understand this better, one can have a closer look into tensile string mass spectrum
\begin{align}
    m^2=\frac{K^2}{R^2}+\frac{1}{\a'}W^2R^2.
\end{align}
The string length scale is determined by $\a'$ which is also inverse to the tension. Furthermore, when the compactification scale ($R$) is much larger than the string length scale ($\a'$), contribution of the winding (momentum) part becomes large (small). In the T-dual description, this is where the compactification radius becomes much smaller than the string length scale, i.e. the momentum part dominates. Now, in tensionless string we take the limit $\a'\to\infty$, but the compactification radius in the target spacetime $R$ is kept finite. Clearly this scenario corresponds to the case where the compactification scale is much smaller than string length scale and consequently momentum ($K$) domination in the mass spectrum is obvious. As a result, in order to find a viable sense of T-duality here, one needs to look for a different limit of string theory where the $\a'<<R$.

\subsection*{Toroidal compactification, $B$-field and $O(d,d,\mathbb{Z})$ symmetry}
Compactification of $d$ dimensions, along with $B$-field gives a much richer picture of the duality structure in null strings. Mass spectrum of tensile strings with $B$-field famously display symmetry under $O(d,d,\mathbb{Z})$ transformations of \{$\omega^I,k_I$\}\footnote{These transformation matrices are $d\times d$ matrices action on $2d$ dimensional column vector \{$\omega^I,k_I$\}.}, a group of transformations satisfying
the following
\begin{align}
    O^{T}\begin{bmatrix}
       0 & \mathbb{1}_{d}\\
        \mathbb{1}_{d} &0
   \end{bmatrix}O=\begin{bmatrix}
       0 & \mathbb{1}_{d}\\
        \mathbb{1}_{d} &0
   \end{bmatrix}.
\end{align}
As discussed in \cite{Giveon:1994fu}, $O(d,d,\mathbb{Z})$ group is generated by three class of transformations; namely the basis transformation, discrete shift symmetry and sectorised duality transformations. Behaviour of null strings under each of these transformations has been studied in detail in \cite{Banerjee:2024fbi}. In this review we briefly touch upon these dualities and see which of them survive for null strings.
\paragraph{Basis transformation:}
The basis change transformation acts on \{$k_I,\omega^I$\} in the following way
\begin{align}
    \omega\to\omega'=C^T\omega,~~~k\to k'=C^{-1}k,~~~C\in GL(d,\mathbb{Z}).
\end{align}
Now, let us recall the mass spectrum for null string theory compactified on a torus with metric $G_{IJ}$ and background field $\mathbb{B}_{IJ}$
 \begin{align*}
         m^2=k^{T}G^{-1}k-2k^{T}G^{-1}\mathbb{B}\omega-\omega^{T}\mathbb{B}G^{-1}\mathbb{B}\omega.
    \end{align*}
One can clearly see that the transformation connects this theory to a null string theory compactified on another torus with metric $G'_{IJ}$ and background field $\mathbb{B}'_{IJ}$ where
\begin{align}
    \mathbb{B}'=(C^T)^{-1}\mathbb{B}C^{-1},~~~G'=(C^T)^{-1}GC^{-1}.
\end{align}
Hence this duality maps null strings on two different target spacetimes.
\paragraph{Discrete shift symmetry:}
In order to understand this symmetry let us first recall the following relation concerning the mass spectrum
\begin{align}\label{K_L,R}
     m^2=\frac{1}{2c'}&\sum_{I,J=1}^{d}(K_{I\hspace{.25mm}L}+K_{I\hspace{.25mm}R})G^{IJ}(K_{J\hspace{.25mm}L}+K_{J\hspace{.25mm}R}),
\end{align}
where $(K_{I})_{\hspace{.25mm}L,R}$ are given below
\begin{align}
    (K_{I})_{\hspace{.25mm}L,R}=\frac{1}{\sqrt{2}}\bigg\{\sqrt{c'}\Big(k_{I}-\sum_{J=1}^{d}\mathbb{B}_{IJ}\omega^{J}\Big)\pm \frac{1}{\sqrt{c'}}\sum_{J=1}^{d}G_{IJ}\omega^{J}\bigg\}.
\end{align}
It is straightforward to see that $(K_{I})_{\hspace{.25mm}L,R}$ are invariant under the following transformations 
\begin{align}\label{ctb21}
    \mathbb{B}_{IJ}\to \mathbb{B}_{IJ}+\mathcal{N}_{IJ},\hspace{8mm} k_{I}\to k_{I}+\mathcal{N}_{IJ}\omega^{J},\hspace{8mm} \omega^I\to \omega^I, \hspace{8mm}\mathcal{N}_{IJ}=-\mathcal{N}_{JI}, 
\end{align}
These are called discrete shift transformation. Since both $(K_{I})_{\hspace{.25mm}L,R}$ are invariant under this transformation, and  $m^2$ is constructed from them as given in \eqref{K_L,R}, $m^2$ must be invariant too. This means discrete shift symmetry maps between null strings n two different target spacetimes, where both target spacetimes are compactified on metric $G_{IJ}$, but the $B$-fields in compact dimensions are different as shown in \eqref{ctb21}.

\paragraph{Sectorised T-duality:}
The sectorised T-duality acts on \{$\omega^I,k_I$\} as below
\begin{align}\label{secdual}
    &\omega'^I=\omega^I,~~~k'_I=k_I,~~~\forall I\neq i\nonumber\\
    &\omega'^i=k_i,~~~k'_i=\omega^i.
\end{align}
This means at $i$-th position there will be a flip between $\omega_i$ and $k^i$ and the rest will be unchanged. Unlike for the other transformations discussed before, we do not have any generic formula on how the sectorised duality would work on null strings. However, working on specific examples of compactification will provide us with useful insights in this regard. Here we mainly discuss the results for $d=2$ case, and for details of the higher dimensional case, the reader is referred to \cite{Banerjee:2024fbi}.
\medskip

In $d=2$ compactification we begin with the following metric and $B$-field on the torus.
\begin{align}\label{GB2}
   G_{IJ}=R_I^2\delta_{IJ},~~~ \mathbb{B}_{IJ}=b\hspace{1mm}\epsilon_{IJ},\hspace{5mm}I,J\in \{1,2\}.
\end{align}
Here $\epsilon_{IJ}$ is two dimensional antisymmetric tensor. It has been shown in \cite{Banerjee:2024fbi} that under the sectorised duality transformation $\omega^2\leftrightarrow k_2$, the mass spectrum cannot be expressed in the form \eqref{ctb25} (or, alternatively, in the form \eqref{Buscher5}) for any metric $G'_{IJ}$ and $\mathbb{B}'_{IJ}$ and hence, the other limit cannot be tensionless/null string in a Lorentzian target spacetime. This happens since null string in Lorentzian target spacetime essentially gives \eqref{Buscher5} as mass spectrum.  
Interestingly, applying sectorised T-duality for null strings leads to the generalised metric form \eqref{tensile} with the following metric and $B$-field
\begin{align}
    (G_{c'})_{IJ}=\begin{bmatrix}
         b^2R^{2}_{1} & \frac{bR_{1}}{R_{2}}\\ 
         \frac{bR_{1}}{R_{2}} & \frac{1}{R^{2}_{2}}
    \end{bmatrix}~~~~\mathbb{B}_{IJ}=0.
\end{align}
One can see that the metric $(G_{c'})_{IJ}$ is degenerate, indicating the emergence of a non-Riemannian torus in the target spacetime. One can also see that $(G_{c'})_{IJ}$ can be obtained from its tensile counterpart in a consistently taken limit where one cycle becomes much smaller than the other.
From here one can deduce that the tensionless strings in a Lorentzian target spacetime via sectorised duality connects to some degenerate limit on the target space torus where one of the compact directions is contracted. This idea of a duality between string theories with non-Lorentzian worldsheets and non-Lorentzian target spaces has been an active direction for research about which we shall briefly discuss in section \ref{Duality web}. 
\medskip

\paragraph{A special case: Inversion transformation:} This transformation is the transformation when the entire set of $\omega_I$s and $k^{I}$s are interchanged. That means it is combination of sectorised dualities along all compact directions. In order to understand whether this transformation leads to dualities between null strings, let us look into the spectrum \eqref{Buscher5}. This symmetry can work for this spectrum only if
\begin{align}\label{Buscher2}
    \mathcal{G}\to\mathcal{G'}\hspace{5mm}\mathcal{G'}=\begin{bmatrix}
    -\mathbb{B}'G'^{-1}\mathbb{B}' & \mathbb{B}'G'^{-1}\\-G'^{-1}\mathbb{B}' & G'^{-1}
\end{bmatrix}=\begin{bmatrix}
   G^{-1}  & -G^{-1}\mathbb{B}\\\mathbb{B}G^{-1} & -\mathbb{B}G^{-1}\mathbb{B}
\end{bmatrix}.
\end{align}
$G'$ and $\mathbb{B}'$ respectively are metric and $B$-field after the transformation. \eqref{Buscher2} leads us to the following 
\begin{subequations}
    \begin{equation}\label{Buscher3}
    G'^{-1}=-\mathbb{B}G^{-1}\mathbb{B}
\end{equation}
\begin{equation}\label{Buscher4}
    -G'^{-1}\mathbb{B}'=\mathbb{B}G^{-1}
\end{equation}
\end{subequations}
Together, \eqref{Buscher3} and \eqref{Buscher4} imply the following
\begin{align*}
\mathbb{B}G^{-1}\mathbb{B}\mathbb{B}'=\mathbb{B}G^{-1} 
   \implies \mathbb{B}\mathbb{B}' =\mathbb{1}_{d}
\end{align*}
Hence, the inversion transformation gives duality connection between tensionless strings only if the $B$-field applied is invertible. Since $B$-fields are invertible only if they are in even dimensions, inversion symmetry for null strings works only for even dimension compactification.

\subsection{Summary}
In this section, we have discussed all the three quantum theories for null strings in compactified target spacetimes, both circle and toroidal compactification along with constant $B$-field. We summarise the important results below.

\begin{table}[ht]\label{summarytable}
\small
\centering
\begin{tabular}{ |c|c||c|c|c| }
\hline
\multirow{2}{3em}{\quad} & \multirow{2}{4em}{Theory} & \multirow{2}{7em}{Uncompactified} & \multirow{2}{11em}{Circle Compactification} &  Toroidal Compactification \\ & & & & with $\mathbb{B}$-field \\
\hline\hline
\multirow{10}{3.5em}{Physical States Level} &  &   &  &   \\ & Oscillator & $r=s=n$ & $s-r=KW$ & $s-r = \sum_{I}k_I\omega^I$ \\ & &(level matched) & (modified level matched) & (modified level matched)\\ \cline{2-5} & & & & \\ & \multirow{3}{4em}{Flipped} & $r+s=2$ & $r+s-KW=2$ & $r+s - \sum_{I}k_I\omega^I=2$\\ & & (Only level 2) & All levels (level 2 has & All levels (all levels have\\  & & & infinite physical states) & infinite physical states)\\ \cline{2-5} & & & & \\ & Induced & level not defined & level not defined & level not defined\\
\hline
\multirow{11}{4em}{Mass Spectrum} &  &  &  & \\ & & & & $m^2=k^{T}G^{-1}k$\\ & Oscillator & $m^2=\frac{1}{c'}(2n-2)$ & $m^2=\frac{K^2}{R^2}+\frac{1}{c'}(r+s-2)$ & $-2k^{T} G^{-1}\mathbb{B}\omega -\omega^{T}\mathbb{B}G^{-1}\mathbb{B}\omega$\\ & & & & $+\frac{1}{c'}(r+s-2)$\\ \cline{2-5} &  &  &  &\\ & Flipped & $m^2=0$ & $m^2=K^2/R^2$ & $m^2=k^{T}G^{-1}k$ \\ & & & & $-2k^{T} G^{-1}\mathbb{B}\omega -\omega^{T}\mathbb{B}G^{-1}\mathbb{B}\omega$\\ \cline{2-5} & & &  \quad &\\  & Induced & $m^2=0$& $m^2=K^2/R^2$ & $m^2=k^{T}G^{-1}k$ \\ & & & & $-2k^{T} G^{-1}\mathbb{B}\omega -\omega^{T}\mathbb{B}G^{-1}\mathbb{B}\omega$\\
\hline
\end{tabular}
\captionof{table}{}
\end{table}
\begin{itemize}
    \item We saw that oscillator vacuum and flipped vacuum, compactification leads to modification in the level of the physical states. For the oscillator vacuum, compactification leads to a modified level matching condition similar to tensile string theory.
    \item For flipped vacuum, the modification is much more dramatic. For non-compact case, this theory only had level 2 states as physical states while for circle compactified case, this theory has physical states at all levels with infinite number of states at level 2. At toroidal compactification, this theory has infinite number of physical states at all levels.
    \item The modification in the mass spectrum is same for all three theories.
    \item The complete absence of winding number $W$ in the mass spectrum of compactified null strings implies absence of T-duality in this corner. However, in presence of a constant Kalb-Ramond background, part of $O(d,d,\mathbb{Z})$ duality can be observed among null string theories.
\end{itemize}
  
\medskip

\newpage
\part{Carroll string and string Carroll}\label{IV}
This last part of the review discusses the string dynamics in target spacetime with Carrollian structure. After briefly introducing Carroll strings and their formulation, we present the systematic Carroll expansion of Lorentzian spacetimes, which provides a useful framework for studying ultra-relativistic limits of particle and string theories. We then apply these ideas to black hole backgrounds by analysing the behaviour of strings in near-horizon geometries, where Carrollian structures arise naturally. We conclude this part with a brief review of some important developments related to null tensionless strings and Carroll strings.
\begin{figure}[ht]
    \centering
    \includegraphics[width=0.75\linewidth]{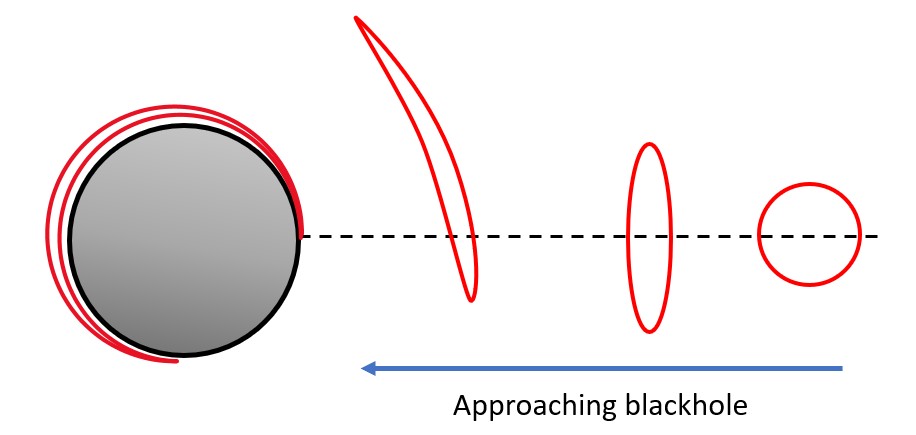}
    \caption{Heuristic depiction: A particular class of Carroll string wrapping around a black hole event horizon.}
    \label{wrapping black hole}
\end{figure}

\medskip

This part of our review is divided into sections as follows:

\begin{itemize}
    \item {\em \hyperref[Carroll Strings]{Carroll Strings}}: Here, we discuss aspects of strings propagating in Carrollian spacetimes, in contrast to Carrollian worldsheets. We show that naively taking Carroll limits on string actions lead to static solutions.
    \item {\em \hyperref[String Near Black holes]{String Near black holes}}: We use the machinery of \textit{Carroll expansions} to study $c\to 0$ regime of string theories. This, alongwith what we've learnt before about null strings, fits well into discussing string solutions near black hole horizons. 
    \item {\em\hyperref[A quick tour of related developments]{A quick tour of related developments}}: In this part we talk about other recent developments in the discussion of null strings and related topics that has not been covered in the preceding sections. 
\end{itemize}
\newpage

 \section{Carroll Strings} \label{Carroll Strings}

In the earlier sections, we reviewed null tensionless strings, where Carrollian structures emerge as residual gauge symmetries on the worldsheet. In this section, we flip things around and focus instead on Carrollian geometry in the \textit{target} spacetime. Subsequently, our main interest is understanding the dynamics of strings propagating in such Carrollian backgrounds. We begin by considering Carrollian point particles and then expand the discussion to include extended objects such as strings. 

\subsection{Carroll particles}

In order to obtain the action for the Carroll particle, we begin with the relativistic phase-space action of a massive point particle,\footnote{We can also obtain the action for free Carroll particle using the method of coadjoint orbits \cite{Duval:2014uoa} or the method of non-linear realisations \cite{Coleman:1969sm, Callan:1969sn} on the Carroll algebra.} and subsequently take the appropriate Carroll limit \cite{Bergshoeff:2014jla}. The relativistic phase-space action for a massive point particle given in \eqref{phspacpar} is as follows,
\begin{equation}\label{relaction}
    S=\int d\tau \left(P_\mu \dot{X}^\mu-\frac{e}{2}(P_\mu P^\mu+c^2m^2)\right),
\end{equation}
where $X^\mu=X^\mu(\tau)$ ($\mu=0,\cdots, d$) denotes the embedding of the particle worldline in target (Minkowski) spacetime and $\tau$ is an affine parameter along the worldline\footnote{Here we use mostly plus signature $(-,+,+,+)$, and derivatives are written as $\dot X^\mu=\partial_\tau X^\mu$.}. Here $P_\mu$ is the canonical momentum conjugate to $X^\mu$ and $e$ is the einbein.
\medskip

To impose Carroll limit, we separate the spatial and temporal components of the target-space coordinates and momenta as follows
\begin{equation}
    X^0=ct,~~P^0=\frac{E}{c},
\end{equation}
where $E$ denotes the energy conjugate to target space time $t$. Substituting in the above action \eqref{relaction}, we obtain 
\begin{equation}
    S=\int d\tau \left[-E\dot{t}+P_i \dot{X}^i-\frac{e}{2c^2}\left(-E^2+c^2 P_i P^i+c^4m^2\right)\right].
\end{equation}
    In order to obtain finite action in the Carroll limit ($c\to 0$), we rescale einbein ($e\to c^2\tilde{e}$) and mass ($m\to \frac{M}{c^2}$) followed by $c\to 0$ limit. The resulting action for Carroll particle takes the following form
    \begin{equation}\label{carrollparticle}
        S_{C_p}=\int d\tau \left(-E\dot{t}+P_i\dot X^i-\frac{\tilde{e}}{2}\left(E^2-M^2\right)\right).
    \end{equation}
    One can explicitly check, the above action \eqref{carrollparticle} is invariant under the following Carroll transformations
    \begin{equation}
        \begin{split}
            E &\to E, \qquad \qquad t \to t+ c - \vec{\beta}\cdot \vec{X}\\
            \vec{X} \to &R\vec{X} + \vec{b}, \qquad \quad 
            \vec{P} \to R\vec{P}+\vec{\beta}E.
            \end{split}
    \end{equation} 
    Here the parameter $\beta$ describes the Carroll boost, $R$ corresponds to $SO(d)$ rotation,  $b$ and $c$ describes space and time translation respectively.
    Upon varying the action \eqref{carrollparticle} w.r.t $t$, $\vec{X}$, $P$, $E$ and einbein $\tilde{e}$, we obtain the following EOM and mass-shell constraint,
    \begin{equation}
        \dot{E}=0, \qquad \dot{P}=0, \qquad \dot{X}=0, \qquad \dot{t}=-\tilde{e}E, \qquad E^2-M^2=0.
    \end{equation}
    As can be clearly seen, the above EOM describe \textit{trivial} dynamics of a free Carroll particle: the particle remains at rest and possesses constant energy, reflecting the fact that the Hamiltonian is a central element of the Carroll algebra (see \eqref{chha3}). 
    \medskip

    We now consider an alternative scaling of the phase-space action of relativistic massive particle \eqref{relaction},
    \begin{equation}
       t\to t/c, \qquad E \to cE, \qquad X^i\to cX^i,\qquad P^i\to P^i/c, 
    \end{equation}
    while keeping the scaling of the einbein ($e \to c^2\tilde{e}$) and the mass ($m\to M/c^2$) unchanged. It is straight forward to check that the above scaling preserves the canonical commutation relation. This corresponds to the following scaling of the target spacetime coordinates
    \begin{equation}
        X^0\to X^0, \qquad P^0\to P^0, \qquad X^i\to cX^i, \qquad P^i\to P^i/c. 
    \end{equation}
    Taking the Carroll limit ($c\to 0$) leads to the following action:
    \begin{equation}\label{nonrelaction}
        S=\int d\tau \left(-E\dot{t}+P_i\dot{X}^i-\frac{\tilde{e}}{2}(P_iP^i+M^2)\right).
    \end{equation}
    The above action is invariant under Carroll boosts only when $E=0$. This constraint can equivalently be imposed by introducing a Lagrange multiplier ($\rho$) directly in the action through an additional term $\rho E$ (see \cite{Bergshoeff:2022qkx} for detailed explanation). Variation of the above action \eqref{nonrelaction} yields the following EOM
    \begin{equation}
        \dot{X}^i=\tilde{e}P^i, \qquad \dot{P}_i=0
    \end{equation}
    From these equations, we conclude that in the Carroll regime, particles that move, must have zero energy, while, as seen earlier, particles that remain stationary correspond to non-zero energy. In the next subsection, we generalise this discussion on Carroll particles to Carroll strings. 
    \subsection{Carroll strings}

Following the formalism developed in the previous subsection, we now turn to the dynamics of strings propagating in Carrollian target spaces \cite{Cardona:2016ytk}. We refer to such objects as \emph{Carroll strings}. Depending on the choice of scaling limit, one can construct two distinct types of Carroll strings, one with finite tension and another without tension. Both cases can be obtained by taking appropriate Carrollian limits of the canonical phase-space action of a relativistic string. In either case, the resulting dynamics is trivial, mirroring the behaviour of Carroll particles.

\medskip

We start from the canonical phase-space action of a relativistic string in a Minkowski target space,
\begin{equation}\label{relstringaction}
    S = \int d\tau d\sigma \left\{ \Pi_\mu \dot{X}^\mu - \frac{\lambda}{2}\big(\Pi_\mu \Pi^\mu + T^2 X'^2\big) - \rho\, \Pi_\mu X'^\mu \right\},
\end{equation}
where $\lambda(\tau,\sigma)$ and $\rho(\tau,\sigma)$ are Lagrange multipliers,
$T$ is the string tension, and $\Pi_\mu$ are canonical momenta subject to the constraints
\begin{equation}
    \Pi_\mu X'^\mu = 0, \qquad \Pi_\mu \Pi^\mu + T^2 X'^2 = 0.
\end{equation}

\medskip

To obtain the Carrollian theory, we perform what is known as a \emph{stringy Carroll limit}. This procedure works as a generalisation of the Carroll limit for particles, by rescaling \textit{two} of the longitudinal target-space coordinates $X^a$ ($a=0,1$) and their conjugate momenta as
\begin{equation}
    X^a \to c\,X^a, \qquad \Pi^a \to \frac{\Pi^a}{c}.
\end{equation}
The transverse coordinates $X^i$ ($i=2,\dots,d-1$) remain unchanged. Substituting this into \eqref{relstringaction} yields
\begin{equation}
    S = \int d\tau d\sigma \left\{ \Pi_\mu \dot{X}^\mu - \frac{\lambda}{2}\left(\frac{\Pi_a \Pi^a}{c^2} + \Pi_i \Pi^i + c^2 T^2 X'_a X'^{a} + T^2 X'^2_i \right) - \rho\, \Pi_\mu X'^\mu \right\}.
\end{equation}
For the action to remain finite in the Carroll limit $c \to 0$, we rescale the Lagrange multiplier and string tension as
\begin{equation}
    \lambda = c^2 \tilde{\lambda}, \qquad T = \frac{T}{c},
\end{equation}
keeping $\rho$ fixed. The resulting Carroll string action with ($E \equiv \Pi^0$) reads
\begin{equation}\label{carrollstringaction}
    S_{C_s} = \int d\tau d\sigma \left\{ \Pi_\mu \dot{X}^\mu - \frac{\tilde{\lambda}}{2}\big(-E^2 + (\Pi^1)^2 + T^2 X'^2_i \big) - \rho\, \Pi_\mu X'^\mu \right\},
\end{equation}

\medskip

Varying \eqref{carrollstringaction} with respect to $X^a$, $X^i$, $\Pi_a$, and $\Pi_i$, we obtain the following EOM and constraints

\begin{align}
    &\dot{\Pi}^a = (\rho\, \Pi^a)', \qquad \qquad~~ \quad
    \dot{\Pi}^i = \big(\rho\, \Pi^i + \tilde{\lambda} T^2 X'^i \big)', \nonumber\\
    &\dot{X}^a = \rho\, X'^a + \tilde{\lambda}\, \Pi^a, \qquad \quad
    \dot{X}^i = \rho\, X'^i, \\
    &\Pi_a X'^a + \Pi_i X'^i = 0, \qquad \quad \Pi_a \Pi^a + T^2 X'^2_i = 0. \nonumber
    \end{align}
In a gauge analogous to the conformal gauge, $\tilde{\lambda}=1$ and $\rho=0$, the EOM simplify to
\begin{equation}
    \dot{\Pi}^a = 0, \qquad \dot{\Pi}^i = T^2 (X'^i)', \qquad \dot{X}^a = \Pi^a, \qquad \dot{X}^i = 0.
\end{equation}
We observe that the transverse coordinates remain static, indicating that the Carroll string, like its particle counterpart, does not propagate.

\medskip

Finally, we consider the tensionless limit $T \to 0$ of the Carroll string action \eqref{carrollstringaction}, which leads to
\begin{equation}
    S = \int d\tau d\sigma \left\{ \Pi_\mu \dot{X}^\mu - \frac{\tilde{\lambda}}{2} \Pi_a \Pi^a - \rho\, \Pi_\mu X'^\mu \right\}.
\end{equation}
The corresponding EOM are
\begin{equation}
\begin{split}
    \dot{\Pi}^a &= (\rho\, \Pi^a)', \qquad \dot{\Pi}^i = \rho\, \Pi'^i, \\
    \dot{X}^a &= \rho\, X'^a + \tilde{\lambda}\, \Pi^a, \qquad \dot{X}^i = \rho\, X'^i,
\end{split}
\end{equation}
which, in the conformal gauge, again show that the string remains static. With these peculiar dynamics of Carrollian particles and strings, we conclude our discussion on Carroll strings and move on to discuss the mechanism of Carrollian expansions to study strings in the Carroll regime.

        \section{String Near black holes}\label{String Near Black holes}
         
    In this section, as the finish line draws nearer, we reach a very important juncture, where we delve into a powerful application of Carrollian physics applied to strings.
    In particular, this section deals with the formalism of the string Carroll expansions, which matches with the near-horizon expansion of non-extremal black holes and thereby provides a robust framework for studying the dynamics of bosonic strings in such backgrounds. We begin with the brief introduction to particle Carroll expansion and then generalise it to the string case. Finally, we investigate classical string solutions in the background of a static BTZ black hole and find a variety of interesting string configurations.
         
    \subsection{Introducing the machinery: Carroll expansions}
    We have talked about explicit Carroll limits in the preceding scenarios. Intriguingly, a direct $c=0$ version of known particle and string phase space leads to static models only. To understand the Carroll dynamics of such objects better, we have to zoom in on the \textit{Carroll regime}, i.e. in the $c\to 0$ realm of such phase spaces. The recipe is to look for perturbative expansions around the the ultra-relativistic ($c\to 0$) theory. We will consider such expansions are analytically well defined, such that the leading order results give us the bang on $c=0$ results, but subleading corrections are also tractable.
\medskip
    
    This small-$c$ expansion, referred to as the \textit{Carroll expansion} of a Lorentzian manifold is our main focus in this section. This expansion is performed in powers of the speed of light $c$, rather than taking the strict $c \to 0$ limit\footnote{This is analogous to the non-relativistic expansion \cite{Hansen:2019pkl,Hansen:2020pqs, Hartong:2024ydv} in the inverse power of the speed of light $c$ for large $c$.}.  For this review, we will restrict ourselves to an expansion of related quantities in even powers of $c$\footnote{The non-relativistic expansion includes odd powers \cite{Ergen:2020yop, Hartong:2023ckn}.}. In what follows, we begin with the \textit{particle Carroll} expansion, where only the temporal component of the spacetime geometry the small-$c$ expansionis singled out \cite{Henneaux:2021yzg, deBoer:2021jej, Hansen:2021fxi, deBoer:2023fnj}. We then generalise this to the \textit{string Carroll} expansion, in which two special directions are singled out and scaled by factors of $c^2$.

    \paragraph{Particle Carroll expansion}
    To understand the particle Carroll expansion of a ($d+1$) dimensional flat Lorentzian target spacetime, the metric can be written as
    \begin{equation}
        \eta(c)= -c^2 dx^0 dx^0+ dx^idx^i,
    \end{equation}
    where $(i=1,\cdots,d)$ represents spatial (transverse) coordinates, and $x^0$ labels the temporal (or longitudinal) coordinate. We now generalise this to a curved manifold, where the metric and its inverse is written in terms of Vielbeins as follows 
    \begin{equation} \label{metric and inverse particle}
        g_{\mu\nu}=-c^2\mathcal{H}_{\mu}\mathcal{H}_{\nu}+ \Pi_{\mu\nu},\qquad
        g^{\mu\nu}=-\frac{1}{c^2}\mathcal{E}^{\mu}\mathcal{E}^{\nu}+ \Pi^{\mu\nu}.
    \end{equation}
    These Vielbeins are also called pre ultra-local (PUL) variables. Here $\mu,\nu=\{0,1,\cdots,d\}$ and $\Pi_{\mu\nu}$ denotes transverse symmetric tensors
    \begin{equation}
        \Pi_{\mu\nu}=\delta_{ij}\mathcal{T}^i_{\mu}\mathcal{T}^j_{\nu} \qquad \text{and} \qquad \Pi^{\mu\nu}=\delta^{ij}\mathcal{T}_i^{\mu}\mathcal{T}_j^{\nu}.
    \end{equation}
    Here $\delta_{ij}=$ diag$(1,\cdots,d)$ are the tangent space indices. These pre-Carrollian variables satisfy the following relations
    \begin{equation}
        \mathcal{H}_\mu\mathcal{E}^\mu=-1, \qquad \mathcal{H}_\mu \Pi^{\mu\nu}=0,\qquad \Pi_{\mu\nu}\mathcal{E}^\mu=0, \qquad \delta^\mu_\nu=-\mathcal{E}^\mu\mathcal{H}_\nu+\Pi^{\mu\rho}\Pi_{\nu\rho}.    \end{equation}
    These longitudinal and transverse Vielbeins expand in $c^2$ as,
    \begin{equation}
    \begin{split}
        \mathcal{H}_\mu=h_\mu +\mathcal{O}(c^2),\qquad \mathcal{E}^\mu=e^\mu+ c^2 E^\mu+\mathcal{O}(c^4),\\  \mathcal{T}^\mu_i=\tau_i^\mu+\mathcal{O}(c^2), \qquad \mathcal{T}^i_\mu= \tau^i_\mu+c^2\pi^i_\mu+\mathcal{O}(c^4). 
        \end{split}
    \end{equation}
    Substituting them into the metric and its inverse \eqref{metric and inverse particle}, we obtain 
        \begin{equation}
       g_{\mu\nu}=\Omega_{\mu\nu}-c^2 h_\mu h_\nu+c^2\Phi_{\mu\nu} + \mathcal{O}(c^4),\qquad 
    g^{\mu\nu}= -\frac{1}{c^2}T^{\mu\nu} + \Psi^{\mu\nu}+ \mathcal{O}(c^2),
    \end{equation}
    where, 
    \begin{equation}
        \Omega_{\mu\nu}=\delta_{ij}\tau^i_\mu\tau^j_\nu, \quad  T^{\mu\nu}=e^\mu e^\nu, \quad \Phi_{\mu\nu}=2\delta_{ij}\tau_{(\mu}^i\pi_{\nu)}^j, \quad \Psi^{\mu\nu}=\delta^{ij}\tau^\mu_i\tau^\nu_j-2e^{(\mu}E^{\nu)}.
    \end{equation} 
    This construction is referred to as the \textit{particle Carroll} expansion of the target spacetime geometry. 
    \medskip

    \paragraph{String Carroll expansion:}
    Now that we have established the particle Carroll expansion, we move further to discuss its string generalisation. Here, the small $c^2$ expansion is performed along two directions of the target spacetime\footnote{This is in analogy with the string generalisation of the Newton-Cartan geometry \cite{Andringa:2012uz, Bergshoeff:2019pij}. Here we closely follow the approach of \cite{Hartong:2021ekg,Hartong:2022dsx} in the non-relativistic case.}. As will be discussed later, this expansion plays a crucial role in understanding the blackhole near horizon regimes.
    \medskip
    
    A $(d+2)$ dimensional decomposition of the flat metric with string Carroll expansion in mind can be written as
    \begin{equation}
        \eta(c)=-c^2 dx^0 dx^0 +c^2 dx^1 dx^1 +dx^i dx^i.
    \end{equation}
    Here, note that $(x^0,x^1)$ labels two longitudinal coordinates, both with a $c^2$ factor out front. The Vielbeine formalism for a generic $(d+2)$ dimensional Lorentzian manifold takes the following form
    \begin{equation} \label{metric and inverse string}
        g_{\mu\nu}=c^2 \eta_{AB}\mathcal{H}^A_{\mu}\mathcal{H}^B_{\nu}+ \delta_{ij}\mathcal{T}^i_{\mu}\mathcal{T}^j_{\nu}, \qquad
        g^{\mu\nu}=\frac{1}{c^2}\eta^{AB}\mathcal{E}_A^{\mu}\mathcal{E}_B^{\nu}+ \delta^{ij}\mathcal{T}_i^{\mu}\mathcal{T}_j^{\nu}.
    \end{equation}
    Here $A,B=\{0,1\}$ and $i,j= \{2,\cdots,d\}$. The indices $\mu,\nu$ run over the entire spacetime. Further, $\eta_{AB}=$ diag$(-1,1)$ and $\delta_{ij}=$ diag$(1,\cdots,1)$ represent the tangent space metric. The Vielbeins admit the following expansion
    \begin{equation}\label{stringexpansion}
    \begin{split}
        \mathcal{H}_\mu^A&=h_\mu^A +\mathcal{O}(c^2),\qquad \mathcal{E}^\mu_A=e^\mu_A+ c^2 E^\mu_A+\mathcal{O}(c^4),\\  \mathcal{T}^\mu_i&=\tau_i^\mu+\mathcal{O}(c^2), \qquad \mathcal{T}^i_\mu= \tau^i_\mu+c^2\pi^i_\mu+\mathcal{O}(c^4), 
        \end{split}
    \end{equation}
     As shown in the particle Carroll case, following the invertibility condition of the metric, these PUL variables satisfy
     \begin{equation}
        \mathcal{H}_\mu^A\mathcal{E}^\mu_B=\delta^A_B, \quad \mathcal{T}_\mu^i\mathcal{T}_j^\mu=\delta^i_j, \quad \mathcal{H}_\mu^A \mathcal{T}^{\mu}_i=0=\mathcal{E}^\mu_A \mathcal{T}_{\mu}^i, \quad \delta^\mu_\nu=\mathcal{E}^\mu_A\mathcal{H}_\nu^A+\mathcal{T}_i^{\mu}\mathcal{T}^i_{\nu}.  
        \end{equation}
        Substituting \eqref{stringexpansion} into the metric and its inverse  \eqref{metric and inverse particle} yields, 
        \begin{equation}\label{stringcarrollmetric}
       g_{\mu\nu}=\Omega_{\mu\nu}+c^2 H_{\mu\nu}+c^2\Phi_{\mu\nu} + \mathcal{O}(c^4),\qquad 
    g^{\mu\nu}= \frac{1}{c^2}T^{\mu\nu} + \Psi^{\mu\nu}+ \mathcal{O}(c^2),
    \end{equation}
    where, 
    \begin{equation}
    \begin{split}
        \Omega_{\mu\nu}&=\delta_{ij}\tau^i_\mu\tau^j_\nu, \quad  H_{\mu\nu}=\eta_{AB}h_\mu^Ah_\nu^B,\quad  \Phi_{\mu\nu}=2\delta_{ij}\tau_{(\mu}^i\pi_{\nu)}^j,\\
        T^{\mu\nu}&=\eta^{AB}e^\mu_Ae^\nu_B,\quad \Psi^{\mu\nu}=\delta^{ij}\tau^\mu_i\tau^\nu_j-2\eta^{AB}e^{(\mu}_AE^{\nu)}_B.
        \end{split}
    \end{equation} 
    The transformation properties of these fields and the causal structure of the resulting geometry is discussed in detail in \cite{Bagchi:2024rje}. Now we turn to an important physical realisation of this framework, where we establish a correspondence between string Carroll geometry and the near-horizon geometry of non-extremal black holes. 

\subsection{Near-horizon expansion = string Carroll expansion }\label{Near-horizon expansion admits string Carroll expansion}

It is well known that the near-horizon limit of a generic $d$ dimensional non-extremal black hole reduces to a $2d$ Rindler spacetime times $S^{d-2}$. In what follows, we demonstrate the matching of this near-horizon expansion with the string Carroll expansion introduced in the earlier section. To show this correspondence, we consider the static BTZ black hole\footnote{However note that this can be proved to hold in any dimensions.},
\begin{equation}\label{static btz}
	ds^2 = -\left( -M + \frac{r^2}{\ell^2} \right) dt^2 + \left( -M + \frac{r^2}{\ell^2} \right)^{-1} dr^2 + r^2 d\phi^2.
\end{equation}
Here $r_h = \ell \sqrt{M}$ represents the event horizon of the black hole. We now perform the following transformation in \eqref{static btz}
\begin{equation}
    r = r_h + \epsilon \rho, ~~\epsilon \to 0
\end{equation} 
and expand the metric in powers of $\epsilon$ to zoom into the near horizon expansion of BTZ black hole as given below
\begin{equation}\label{static expansion 1}
	ds^2 = r_h^2 \, d\phi^2 + \epsilon \left( -\frac{\rho}{a} dt^2 + \frac{a}{\rho} d\rho^2 + 2r_h \rho \, d\phi^2 \right) + \O(\epsilon^2)\,.
\end{equation}
Here, $ a= \ell^2/2r_h$. Longitudinal directions are labelled by $\{t,\rho$\}, whereas $\phi$ denotes the transverse direction. It can be clearly seen that the longitudinal part of the metric \eqref{static expansion 1}, at $\O(\epsilon)$, under coordinate transformation $\rho=\eta^2/4 a$, takes the form of the standard $2d$ Rindler spacetime
\begin{equation}
    ds^2 = -\frac{\eta^2}{4 a^2} dt^2 + d\eta^2 \,.
\end{equation}

We now compare the above near horizon expansion \eqref{static expansion 1} of static BTZ black hole with the string Carroll expansion \eqref{stringcarrollmetric} of the metric, and identify $\epsilon$ with $c^2$ to obtain
    \begin{subequations}\label{static expansion components}
        \begin{align}
		\Omega_{\mu\nu} dx^\mu dx^\nu =&\; r_h^2 d\phi^2,\\
		\eta_{AB} h_\mu^A h_\nu^B dx^\mu dx^\nu =&\; -\frac{\rho}{a} dt^2 + \frac{a}{\rho} d\rho^2, \\
		\Phi_{\mu\nu} dx^\mu dx^\nu =&\; 2 r_h \rho d\phi^2 \,.
	\end{align}
    \end{subequations}   
    Now that the correspondence between the near-horizon expansion and the string Carroll expansion is established, we move further to the study of strings probing the near horizon regime of black holes.

\subsection{Strings near BTZ black holes}\label{Strings near BTZ black holes}
     Using string Carroll expansions, the dynamics of bosonic closed strings in the near-horizon region of $3+1$ $d$ Schwarzschild black hole was first discussed in \cite{Bagchi:2023cfp,Bagchi:2024rje}. We follow the same discussion pertaining to $2+1$ $d$ static BTZ black holes following the recent work \cite{Banerjee:2025bkg} in the current section. The analysis here aligns with Carroll expansion of relativistic field theory \cite{deBoer:2021jej,deBoer:2023fnj,Henneaux:2021yzg}, where the leading-order (LO) term gives rise to \textit{electric} Carroll theory, and the next-to-leading-order (NLO) term corresponds to the \textit{magnetic} theory. We explore similar electric and magnetic sectors in what follows, although the exact details are a bit different for string expansions. Our natural starting point is the Polyakov action \cite{Polchinski:1998rq} given by
	 
	\begin{equation}\label{Polyakov3}
		S = - \frac{T}{2}\int  \sqrt{ -\gamma }\, \gamma^{\a\b} \partial_\a X^\mu \partial_\b X^\nu  g_{\mu\nu}\,\, d\tau d\s\,.
	\end{equation} 
	Here, $T$ is the string tension,  $X^0,\dots,X^d$ label the target space coordinates and $\a,\b=\{\tau,\sigma\}$ denote the worldsheet coordinates.
	\medskip
	
	We now expand the above action \eqref{Polyakov3} in powers of the $\epsilon$ as identified in the preceding section, where the background metric admits the expansion given in \eqref{static expansion components} and $X^\mu$, $\gamma_{ab}$ and the its inverse expand as,
	\begin{subequations}
	    \begin{align}
		X^\mu(\t,\s) &= x^\mu(\t,\s) + \epsilon y^\mu(\t,\s) + \mathcal{O}(\epsilon^2)\,,\label{X expansion}\\
		\label{gamma}
		\gamma_{ab}(\t,\s) &= \gamma_{(0)ab}(\t,\s) + \epsilon\gamma_{(2)ab}(\t,\s) + \mathcal{O}(\epsilon^2)\,, \\
		\label{gamma inv}
		\gamma^{ab}(\t,\s) &= \gamma_{(0)}^{ab}(\t,\s) + \epsilon \gamma_{(2)}^{ab}(\t,\s) + \mathcal{O}(\epsilon^2) \,.
	\end{align}
	\end{subequations}	
	Here the $\gamma_{(0)}$ has Lorentzian signature. Substituting these expansions in \eqref{Polyakov3}, we obtain,
	\begin{equation}
		\mathcal{L} = \mathcal{L}_{LO} + \epsilon \mathcal{L}_{NLO} + \mathcal{O}(\epsilon^2).
	\end{equation} 
    where, the Lagrangian at different orders takes the following form 
    \begin{subequations}
    \begin{align}\label{lo polyakov}
		\mathcal{L}_{LO} = & -\frac{T}{2}\, \sqrt{ - \gamma_{(0)} }\,\gamma_{(0)}^{\a\b}(\t,\s)\,\Omega_{\a\b}\left( x\right) \,, \\
		\label{nlo polyakov}
		\mathcal{L}_{NLO} = & -\frac{T}{2} \sqrt{ - \gamma_{(0)} } \,\left[ \gamma_{(0)}^{\a\b}\hat{\Phi}_{\a\b}(x,y)- \frac{1}{2}G_{(0)}^{\a\b\rho\xi}T_{ab}(x)\gamma_{(2)\rho\xi} \right] \,.
	\end{align}
    \end{subequations}    
    Here,
    \begin{equation}
    \begin{split}
    \Omega_{\a\b}(x)=~&\Omega_{\mu\nu}(x)\partial_\a x^\mu\partial_\b x^\nu\,,~~H_{\a\b}(x)= H_{\mu\nu}(x)\partial_\a x^\mu\partial_\b x^\nu\,,~~
    \Phi_{\a\b}(x)=\Phi_{\mu\nu}(x)\partial_\a x^\mu\partial_\b x^\nu\,,\\
    \hat{\Phi}_{\a\b}(x,y)&=~H_{\a\b}(x)+ \Phi_{\a\b}(x) + 2 \Omega_{\mu\nu}(x)\partial_{(\a}x^\mu\partial_{\b)}y^\nu + \partial_\a x^\mu\partial_\b x^\nu y^\xi \partial_\xi \Omega_{\mu\nu}(x)\,,
    \end{split}
    \end{equation}
    and $G_{(0)}^{\a\b\rho\xi}$ is the Wheeler-DeWitt metric, defined as
	\begin{align} 
		G_{(0)}^{\a\b\rho\xi} =   \gamma_{(0)}^{\a\rho}\gamma_{(0)}^{\b\xi} + \gamma_{(0)}^{\a\xi}\gamma_{(0)}^{\b\rho} - \gamma_{(0)}^{\a\b}\gamma_{(0)}^{\rho\xi} \,.
	\end{align}    
  Now that we have obtained the Lagrangian at the LO \eqref{lo polyakov} and NLO \eqref{nlo polyakov}, constraints and the EOM can be derived. One should note here a subtlety: The string Carroll expansion of the Polyakov action gives rise only to the magnetic sector \footnote{In fact, the LO Polyakov action in \eqref{lo polyakov} gives rise to only trivial dynamics, with the NLO Polyakov giving the first non-trivial correction.}. 
  \medskip

  However, the relativistic phase-space action under string Carroll expansion results in two distinct sectors emerging from two different scaling of string tension and Lagrange multipliers. These two sectors, in our parlance, correspond to electric strings and magnetic strings. The equivalence for the magnetic sector both from Polyakov and phase-space formalism can be easily established. One can actually see the magnetic NLO phase-space Lagrangian corresponds to LO Polyakov Lagrangian, the NNLO phase-space Lagrangian corresponds to NLO Polyakov Lagrangian, and so on.
  \medskip
  
  Choosing specific ansatz for embedding fields, one can then find different classes of solutions for magnetic strings. For derivation of the solutions in detail see \cite{Banerjee:2025bkg} for BTZ case, and for Schwarzschild black hole see \cite{Bagchi:2023cfp, Bagchi:2024rje}.
  \medskip

\subsection*{Electric strings: Phase space perspective}\label{electricbtz}
        As already highlighted, the analysis of both electric and magnetic Carroll strings under one umbrella requires the relativistic phase space formulation. We therefore start with the phase space action for a Lorentzian target space \cite{Bagchi:2024rje},
        \begin{equation}\label{eq:relativistic phase space action}
            S=\int d\tau\oint d\sigma \left[\dot{X}^\mu \Pi_\mu-\frac{1}{2} e \left(c^2 g^{\mu\nu}(X)\Pi_\mu \Pi_\nu+(c^2T)^2g_{\mu\nu}(X)X'^\mu X'^\nu\right)-\lambda X'^\mu \Pi_\mu\right],
        \end{equation}
        where $T$ denotes the string tension, $e$ and $\lambda$ are the Lagrange multipliers and $X'$ and $\dot X$ represents differentiation with respect to $\s$ and $\t$, respectively. In addition to the embedding field $X^\mu$ expansion given in \eqref{X expansion}, $\Pi_\mu$ and  $\lambda$ also expand in powers of $c^2$:
        \begin{eqnarray}\label{Pi expansion}
    \Pi_\mu=\Pi_{(0)\mu}+c^2\Pi_{(2)\mu}+\O(c^4),\qquad \lambda=\lambda_{(0)}+c^2\lambda_{(2)}+\O(c^4).
\end{eqnarray}
    The expansion of $e$ and $T$ differ depending on the sector under consideration. For electric Carroll theory, these scale as\footnote{For magnetic theory, these parameters scale as $\hat{T}=cT,~~ \hat{e}=c^2e$ (see \cite{Bagchi:2024rje}), which reproduces the Polyakov magnetic form.} 
    \begin{equation}
        \tilde{T}=c^2T, \qquad \tilde{e}=e.
    \end{equation}
    Here $\tilde{e}$ can be expanded as $ \tilde e=\tilde e_{(0)}+c^2\tilde e_{(2)}+\O(c^4).$ Substituting these scalings and the expansions into the phase-space action \eqref{eq:relativistic phase space action} yields the following phase-space Lagrangian at leading order
    \begin{equation}\label{L:LO_electriclimit2}
        \tilde{L}_{\text{LO}} = \oint d\sigma \left[ \left(\dot x^\mu - \lambda x'^\mu\right)e^{\bar{A}}_\mu \Pi_{\bar{A}} + \frac{1}{2\tilde e} H_{\mu\nu}\left(\dot x^\mu - \lambda x'^\mu\right)\left(\dot x^\nu - \lambda x'^\nu\right) - \frac{\tilde T^2}{2}\tilde e\, \Omega_{\mu\nu} x'^\mu x'^\nu \right],
    \end{equation}
    where, we dropped ``$(0)$" from the subscripts and decomposed the momentum $\Pi_\mu$ into its longitudinal and transverse components 
        \begin{equation}
        \Pi_A = v^\mu_A \Pi_\mu\,,\qquad \Pi_{\bar{A}} = e^\mu_{\bar{A}} \Pi_\mu\,.
    \end{equation}
    This constitutes the action of the electric Carroll string. Note that in this case, the LO action itself has large sets of non-trivial solution spaces, some of which have been discussed in \cite{Bagchi:2023cfp, Bagchi:2024rje, Banerjee:2025bkg}. One of these string configurations, where the string expands and wraps the black hole horizon is depicted in Fig~\eqref{wrapping black hole}.
    \subsection{Residual symmetries on the Electric worldsheet}\label{sec:residual symmetries}
    Having obtained the Lagrangian for electric Carroll theory, we turn to investigate the residual symmetry algebra that still exists after gauge fixing\footnote{For completeness we must mention the residual gauge symmetry on magnetic string worldsheets remain two copies of Virasoro, just like the relativistic counterpart.}. To this end, we rewrite the Lagrangian in a form where worldsheet diffeomorphism invariance is manifest. This is achieved by introducing the zweibeine $(\mathbb{m}_a,\mathbb{s}_a)$ and its inverse $(\mathbb{n}^a,\mathbb{s}^a)$, defined as
    \begin{eqnarray}\label{zweibeine}
        \lambda = - \frac{\mathbb{n}^1}{\mathbb{n}^0},\qquad\tilde e = \frac{1}{(\mathbb{n}^0)^2\mathbb{s}}\,,\qquad \mathbb{s}=\det(\mathbb m_a,\mathbb s_a)\,.
    \end{eqnarray}
    The zweibeine $(\mathbb{m}_a,\mathbb{s}_a)$ and its inverse $(\mathbb{n}^a,\mathbb{s}^a)$ are related as
    \begin{equation}
        \mathbb{n}^0= -\frac{\mathbb{s}_1}{\mathbb{s}},\qquad \mathbb{n}^1= \frac{\mathbb{m}_0}{\mathbb{s}},\qquad \mathbb{s}^0 = -\frac{\mathbb{m}_1}{\mathbb{s}}, \qquad \mathbb{s}^1 = \frac{\mathbb{m}_0}{\mathbb{s}},\qquad \mathbb{s} = \mathbb{m}_0 \mathbb{s}_1 - \mathbb{m}_1 \mathbb{s}_0\,.
    \end{equation}
    Substituting \eqref{zweibeine} in \eqref{L:LO_electriclimit2} and defining $\tilde \Pi_{\bar{A}}$ as $\tilde \Pi_{\bar{A}} = -\cfrac{\Pi_{\bar{A}}}{\tilde T^2 \mathbb s_1}$, we obtain the Lagrangian for electric Carroll strings:
    \begin{equation}\label{L:general}
        \tilde L_{LO}
    = \oint  d\sigma~ \mathbb{s}
    \left[
    \tilde T^2\,\mathbb{n}^a \partial_a x^\mu\,e_\mu^{\bar{A}}\,\tilde \Pi_{\bar{A}}
    +\frac{1}{2}\,H_{\mu\nu}\,\mathbb{n}^a\partial_a x^\mu\,\mathbb{n}^b\partial_b x^\nu
    -\frac{\tilde T^2}{2}\,\Omega_{\mu\nu}\,\mathbb{s}^a\partial_a x^\mu\,\mathbb{s}^b\partial_b x^\nu
    \right]\,,
    \end{equation}
    It is interesting to note here that $\tilde{T}=0$ corresponds to  \textit{tensionless (or null) strings} \cite{Bagchi:2015nca,Isberg:1993av,Bagchi:2013bga} in $2d$ Rindler spacetime. Consequently for the non-vanishing tension, the action \eqref{L:general} may be viewed as the \textit{tensile deformations} of the tensionless theory\footnote{The same  Lagrangian \eqref{L:general} appears in \cite{Gomis:2023eav} in a decoupling limit following a sequence of T-dualities with vanishing $B$-field (see also \cite{Blair:2023noj}). T-duality of null strings has been discussed in section~\eqref{T-duality on null strings}.}. 
    \medskip
    
    We now turn to analyse the residual symmetry algebra on the worldsheet for electric Carroll strings\footnote{For detailed derivation readers are pointed to \cite{Banerjee:2025bkg} (see also \cite{Figueroa-OFarrill:2025njv}).}. 
     The Carroll boost acts on the zweibeine and on and $\tilde \Pi_{\bar{A}}$ as 
    \begin{equation}\label{carroll boost}
         \delta\mathbb m_a=\lambda\mathbb s_a\,,\quad\delta\mathbb s_a=0\,,\quad\delta \mathbb n^a=0\,,\quad\delta \mathbb s^a=\gamma\mathbb n^a\,,\quad \delta\tilde \Pi_{\bar{A}}=\lambda\mathbb{s}^a\partial_a x^\mu e_{\mu\bar A}\,,
    \end{equation}
    where $\gamma(\s,\t)$ is a function of the worldsheet coordinates. The worldsheet theory is also invariant under diffeomorphisms generated by the vector field $\xi^a$, acting as
    \begin{equation}\label{diffeomorphism}
        \delta\mathbb m_a=\mathcal{L}_\xi\mathbb m_a\,,\quad\delta\mathbb s_a=\mathcal{L}_\xi\mathbb s_a\,,\quad \delta x^\mu=\xi^a\partial_a x^\mu\,,\quad\delta \tilde \Pi_{\bar{A}}=\xi^b\partial_b\tilde \Pi_{\bar{A}}\,.
    \end{equation}
    Similarly for the inverse zweibeine. In addition, the worldsheet exhibits Weyl invariance. Using the transformation properties of $\mathbb s$ under Carroll boost \eqref{carroll boost} and diffeomorphism \eqref{diffeomorphism}, together with the gauge fixing condition $\mathbb{s}^a = (0,\mathbb{s}^1)$, we obtain
     \begin{equation}
         \gamma = \frac{\mathbb s^1}{\mathbb n^0} \partial_1 \xi^0.
     \end{equation}
    Moreover, using \eqref{zweibeine}, the worldsheet gauge transformation  of $\tilde e$ and $\lambda$ is found to be
    \begin{subequations}
        \begin{align}
        \delta \tilde e & = \xi^a \partial_a \tilde e - 2\tilde e \lambda\partial_1\xi^0 + \tilde e \left( \partial_0\xi^0 - \partial_1 \xi^1 \right)\\\delta \lambda & = \xi^a \partial_a \lambda - \lambda^2 \partial_1 \xi^0 + \lambda \left( \partial_0\xi^0 - \partial_1 \xi^1 \right) + \partial_0\xi^1.
        \end{align}
    \end{subequations}
    Now we perform the local gauge fixing by setting $\tilde e=1$ and $\lambda=0$. Solving for the residual transformations preserving this gauge we obtain,
    \begin{subequations}\label{residual diffeomorphism}
    \begin{align}
        \xi^a  = \left( f'(\s)\t + g(\s), f(\s) \right)\,,\qquad
        \gamma  = \left( \frac{\mathbb s^1}{\mathbb n^0} \right) \left( f''(\s)\t + g'(\s) \right).
    \end{align}
     \end{subequations}
    Here $f(\s),g(\s)$ are arbitrary functions. The action of $\xi^a$ on an arbitrary function $F(\tau,\sigma)$ is given as,
    \begin{eqnarray}
    \delta F=\Big[f'(\s)\t\partial_\t+f(\s)\partial_\s\Big]F +g(\s)\partial_\t F=\Big[L(f)+M(g)\Big] F \,,
    \end{eqnarray}
    where, the generators $L(f)$ and $M(g)$, expressed in terms of Fourier modes, satisfy 
    \begin{align}\label{BMS3 Algebra}
        \left[L_m, L_n\right] = (m-n)L_{m+n}\,,~~~
        \left[L_m, M_n\right] = (m-n)M_{m+n}\,,~~~
        \left[M_m, M_n\right] = 0,
    \end{align}
    which reproduces the classical part of $\text{BMS}_3$, or equivalently, the 2d CCA with $z=1$ \cite{Duval:2014uva,Hao:2021urq,Bagchi:2015nca,Figueroa-OFarrill:2025njv}. 
    \medskip
    
    The BMS$_3$ symmetry group obtained above coincides with the worldsheet residual symmetry algebra of tensionless strings presented earlier in section~\ref{Bosonic closed null strings}. The Lagrangian \eqref{L:general} describing the electric Carroll string, however, admits both tensile as well as tensionless contributions, reducing to the latter when  $\tilde{T}=0$. Consequently, the additional tension-dependent terms can be interpreted as tensile deformations of the tensionless strings, thereby generalising the class of string theories with worldsheet Carroll symmetries beyond the tensionless regime. 
    \medskip


    To conclude, in this section, we focussed on the machinery of Carrollian expansion. String Carroll expansion has been shown to naturally capture the near-horizon geometry of non-extremal black holes. This correspondence provides an important framework for exploring these near horizon regimes. We then mentioned how magnetic and electric Carroll strings in these backgrounds arise. Although we did not discuss the details of these solutions for brevity, they could be easily found in literature \cite{Bagchi:2023cfp, Bagchi:2024rje, Banerjee:2025bkg}.

\newpage

\section{A quick tour of related developments}\label{A quick tour of related developments}
In the penultimate section of our review, we take the reader through a number of new developments which are closely related to our previous discussions. In order not to make this review too long, we will provide very brief summaries of each topic and will be far from exhaustive. This section acts as a pointer to the exisiting literature and the reader is encouraged to read the origininal references for further details on the specific topic. Also, there is no underlying thread to this section except for the individual topics are related to tensionless null strings in one way or another and the section should be treated as an amalgamation of recent studies we find interesting. 
    
\subsection{Null strings as 3D black-hole microstates}\label{blackhole microstates}
We begin our tour with the recapitulation of a rather bold proposal of constructing blackhole microstates and the entropy of a black hole from a model of null strings \cite{Bagchi:2022iqb}. The Bekenstein-Hawking entropy law for black holes 
\begin{equation}
        S_{BH}=\frac{\text{Area}}{4G}
\end{equation}
is one of the cornerstones of holography and a crowning achievement for string theory was reproducing this formula from counting of microstates \cite{Strominger:1996sh}. The fuzzball programme \cite{Mathur:2005zp, Skenderis:2008qn} is a radically different approach where we think of microstates as horizon-sized stringy configurations and the classical black hole geometry as something being emergent only in the coarse grained limit. 

\medskip

In \cite{Bagchi:2022iqb}, it was proposed that black holes can be understood as a theory of fluctuations of horizon membranes, offering another way to understand black hole entropy that has similarities with the membrane paradigm \cite{Thorne:1986iy}. Since horizons are null, Carrollian structures arise naturally on them. In three dimensional spacetimes, horizons would have 2d Carrollian structures. Hence horizons in 3d can be modelled in terms of a 2d null string theory. The fluctuations of this horizon would form an effective description of a blackhole and in this 3d case, this would be a quantum null string. 

\medskip

Starting with the ILST action and considering the Oscillator theory and considering winding around the angular direction of the would-be black hole, it was shown that one can reproduce the Bekenstein-Hawking area law for the BTZ black holes, and rather amazingly, also the logarithmic corrections to it. The central assumption in \cite{Bagchi:2022iqb} is that a particular BTZ black hole is built out of a band of states in the null string spectrum the mass of which is related to the radius of the horizon in a way dictated by the first law of black hole thermodynamics in the near-horizon limit. This construction is conceptually very different from the fuzzball proposal and also differs from the so-called fluff-ball proposal \cite{Afshar:2016uax} which relies on both horizon and asymptotic data. In the approach of \cite{Bagchi:2022iqb}, in contrast, the construction is purely at the horizon and is oblivious to the embedding spacetime,-- the work can be considered to be a microscopic realization of the soft hair proposal of Hawking, Perry, and Strominger \cite{Hawking:2016msc}.

\subsection{Null branes}
The concept of null strings discussed so far can be generalised to null $p$-brane where a $p$-dimensional object traces a $p+1$ dimensional Carrollian worldvolume in the target spacetime. The action for these null brane was first obtained in \cite{Isberg:1993av, Hassani:1994rf} by taking $T \to 0$ limit on the action of tensile $p$ brane\footnote{The method to do this involves obtaining the Polyakov-like action for tensile $p$ branes using Hamiltonian formalism and then taking the tensionless limit on the resulting action.}. The null brane action is invariant under Carrollian diffeomorphism and null Weyl scaling denoted as Weyl$_\mathcal{N}$ in \cite{Dutta:2024gkc}. The invariance under Weyl$_\mathcal{N}$ is particularly striking since the analogous invariance cannot be found for tensile $p$ brane action for $p\geq 2$. For null brane action too, a convenient gauge choice to work with is temporal gauge choice which is analogous to the standard gauge choice used for ILST action \eqref{Review1}. The residual gauge symmetry algebra for this action is isomorphic to BMS$_{p+2}$ \cite{Grumiller:2019fmp}. This symmetry algebra is a generalisation of Campiglia-Laddha extended BMS$_4$ \cite{Campiglia:2015yka}.

\medskip

Quantisation of tensile $p$ brane has several difficulties which are yet to be resolved. However, \cite{Dutta:2024gkc} has initiated a possible direction towards quantising null $p$ brane. The set up used here is a toroidal null p-brane propagating in a toroidally compactified flat target spacetime. In this work, residual BMS$_{p+2}$ symmetry is partially fixed by using light-cone gauge. The sandwich quantisation scheme analogous to the one discussed for the null string quantisation has been applied on the remaining constraints to construct the physical Hilbert space. The vacuum in this physical Hilbert space is annihilated, as usual, by annihilation operators. The tower of physical states created on this vacuum using creation operators are categorised into $p+1$ classes. The sandwich quantisation scheme followed in \cite{Dutta:2024gkc} might pave way towards the still unresolved question of quantisation of the tensile $p$ brane.

\subsection{Duality web}\label{Duality web}
         
Let us recall from Sec~\ref{T-duality on null strings}, that T-duality is generally absent in tensionless strings and in presence of constant Kalb-Ramond background, only a part of $O(d,d,\mathbb{Z})$ duality can be observed. Also, while looking at the effect of $T^2$ compactification of null strings on the flat Lorentzian target spacetime we found clear indication that sectorised duality connects this theory to a different limit of string theory compactified on a degenerate torus. This result is indicative of null strings being part of a much larger picture where various limits of string theory are connected to each other through duality web.

\medskip
        
T-duality connection between non-relativistic closed string theory (NRCS) and discrete light cone quantised (DLCQ) string theory was discovered in \cite{Gomis:2000bd}. There has been further studies about these duality connections between different limits of string theories. In recent works like \cite{Gomis:2023eav,Blair:2023noj,Blair:2024aqz,Chen:2025gaz,Blair:2025prd} a vast picture of duality web connecting many known limits of string theory has been developed. Interestingly, tensionless null strings also appeared at a corner of this duality web. In this section we briefly mention only the relevant parts of their results.
\medskip

One of the important class of decoupling limits of string theory studied in \cite{Gomis:2023eav,Blair:2023noj,Blair:2024aqz,Blair:2025prd} are Matrix $p$-brane theories (denoted by M$p$T) which involve $p$-brane Newton-Cartan geometry as target spacetime for $p\geq-1$\footnote{Flat version of this geometry for $p\geq-1$ appears at the following generalised Galilean limit of Minkowski metric
\begin{equation}
    ds^3=\w~dx^\a dx^\b\eta_{\a\b}+\omega^{-1}dx^{\tilde{\a}}dx^{\tilde{\b}}\delta_{\tilde{\a}\tilde{\b}},~~~\omega\to\infty.
\end{equation}
Here $\a,\b\in\{0,...,p\}$ and $\tilde{\a},\tilde{\b}\in \{p+1,...,D-1\}$ respectively give $p+1$ longitudinal directions and $D-p-1$ transverse directions.} and Carroll like geometry for $p<-1$. In this terminology, tensionless null strings with Lorentzian target spacetime belongs to M(-1)T category, while electric Carroll strings discussed in \ref{electricbtz} belong to M(-3)T category \cite{Blair:2023noj}. These works also discussed another class of decoupling limits called Multicritical Matrix p-brane Theories (MM$p$T), which involves more complex limit in the target spacetime.
\medskip

In Matrix p-brane Theories (M$p$T) with $p\geq-1$, the light excited states give rise to Dp branes governed by corresponding Matrix gauge theories, which justifies the above nomenclature. While quantising M0T, one encounters D0-branes which are connected to the Banks-Fischler-Shenker-Susskind (BFSS) Matrix theory. Similarly light excited states of M(-1)T give D(-1) branes (i.e. D-instantons) governed by Ishibashi-Kawai-Kitazawa-Tsuchiya (IKKT) Matrix theory. It is interesting to recall in this context that while quantising tensionless null string theory (which is an example of M(-1)T) we encountered oscillator vacuum states which are indeed D-instantons from the viewpoint of tensile observer.
\medskip

The works \cite{Gomis:2023eav,Blair:2023noj,Blair:2025prd} showed that T-duality over a timelike circle compactification connects M(-1)T theories to M0T theories. This duality connection is reflected to the timelike T-duality connection between the BFSS Matrix theory and IKKT Matrix theory. T-dualising M(-1)T over spatial $p$-torus gives M(-$p$-1)T theories which are basically Carroll strings. Compactifying M(-1)T theory on a lightlike circle and applying T-duality transformation leads to MM0T theory.
The fact that tensionless null strings in presence of constant Kalb-Ramond field display part of $O(d,d,\mathbb{Z})$ duality has interesting parallels to the other M$p$T theories. It has been found that in presence of appropriate Kalb-Ramond field, T-duality maps M$p$T theories to themselves \cite{Blair:2025prd}. It is important to note that the theories M(-1)T, M0T and M(-p-1)T ($p>0$) share same worldsheet topology \cite{Gomis:2023eav,Chen:2025gaz}.

        \subsection{Path integral formulation}\label{Path integral formulation}
        While performing canonical quantisation of tensionless null strings we encountered three distinct ways to quantise the theory leading into three distinct theories, namely the oscillator, induced and flipped vacuum theories. 
        A recent work \cite{Chen:2023esw} provided a path integral formulation for the quantum theories on induced and flipped vacuum. Furthermore, this work also studied the quantum theories for both homogeneous and inhomogeneous superstrings from path integral approach. We briefly mention the results below.
        \medskip

        We recall that ILST action for bosonic null strings has CCA$_2$ as residual gauge symmetry. Hence, in order to restrict the path integral to the physically inequivalent worldsheet geometries, the work \cite{Chen:2023esw} introduced a CCA$_2$ invariant $bc$ ghost system  which gives the Faddeev-Popov determinant. Studying the energy momentum tensor of the resultant system, CCA$_2$ algebra with anomaly terms (central extension) was obtained. Performing anomaly cancellation for induced vacuum theory did not yield any constraint on the target space dimension and vanishing normal ordering constants, while doing the same for flipped vacuum led to critical dimension $D=26$ and $a_L=2$. These results are consistent with the results obtained from light-cone quantisation discussed in section \ref{Lightcone quantisation}. The authors extended their study to the null superstrings and studied various versions of homogeneous and inhomogeneous null superstrings with varying level of supersymmetry. Their study of quantum null superstring is focussed only on the flipped vacuum. By introducing a CCA$_2$ invariant version of $\b\g$-ghost system and applying anomaly cancellation, it was found that most of those versions of null superstrings have non-integer critical dimensions. However, for $\mathcal{N}=2$ homogeneous superstring and $\mathcal{N}=1$ inhomogeneous doublet superstrings, the critical dimension remains 10 (for both NS and R sector), consistent with their parent tensile theories .
        
        \subsection{Carroll string quantisation}
        The study of Carroll strings was first introduced in \cite{Cardona:2016ytk} where bosonic strings were considered in string Carroll background. As already discussed, string theory in more generalised p-brane Carrollian target spacetime has also been recently studied in \cite{Gomis:2023eav,Blair:2023noj,Blair:2024aqz,Blair:2025prd}. In section~\ref{String Near Black holes}, we have seen how near horizon region of a black hole display string Carroll geometry and we also encountered both electric and magnetic limits of string theory in that region. Carroll strings studied in the aforementioned works belong to the electric Carroll strings discussed in section \ref{String Near Black holes}.
        \medskip
        
        In a recent work~\cite{Figueroa-OFarrill:2025njv}, quantisation of electric Carroll strings in target spacetime displaying particle Carroll geometry was studied using the BRST method.  The physical states were determined by solving the corresponding consistency conditions. The authors considered only the analogue of the highest weight vacuum in the null string and in direct analogy of the tensionless string in this highest weight or flipped vacuum, the Carrollian string also has a finite number of physical states. It was also shown that the  quantum Carroll string is consistent only in 26 target-space dimensions, mirroring the critical dimension of the bosonic null flipped string or ambitwistor string.

        \subsection{High energy string scattering, Celestial CFT and null strings}\label{grossmende}

        As we have stressed throughout the review, the high energy limit of string theory corresponds to the tensionless limit. This has long been known to be dominated by classical saddle points of the string worldsheet path integral, first demonstrated in the seminal works \cite{Gross:1987kza, GROSS198973}. The fixed angle scattering amplitudes at high energies are dominated by special classical worldsheet configurations and are thus exponentially suppressed. This is in sharp contrast to the power law behaviour of quantum field theory. 

\medskip
        
        A recent important work \cite{Kervyn:2025wsb} revisits this picture and provide a detailed analysis of the saddle-point expansion, including subleading quantum corrections. It also relates high energy string tree-level string scattering amplitudes to the framework of flatspace holography in the context of the dual 2d Celestial CFTs \cite{Strominger:2017zoo}. This string-theoretic approach to celestial amplitudes was first proposed in \cite{Stieberger:2018edy} (see also \cite{Donnay:2023kvm,Stieberger:2024shv, Castiblanco:2024hnq,Bockisch:2024bia} for related works).
        \medskip
        
         Celestial holography (see \cite{Raclariu:2021zjz, Pasterski:2021rjz} for reviews) provides the following complementary way to describe scattering amplitudes. Here one performs a Mellin transform over the energy of the particle and expresses the amplitude as correlation functions of operators living on a $2d$ celestial sphere at null infinity. In this representation, the directions of external particles label points on the sphere, while their energies are replaced by conformal dimensions. Consequently, scattering processes in $4d$ can be written in a form reminiscent of a $2d$ conformal field theory. 
         \medskip
         
         In \cite{Kervyn:2025wsb}, it has been shown that open and closed tree level amplitudes admit an expansion in $1/\a'$, where the leading term reproduces the known exponential behaviour and the sub-leading term arise from the quantum fluctuations around the classical world-sheet configuration. These corrections are related to Bernoulli numbers (or zeta values at negative odd integers). This structure suggests that these corrections encode the effects of an infinite tower of light higher-spin states, which become massless in the tensionless limit.
        \medskip

        An interesting result of \cite{Kervyn:2025wsb} is the identification of a correspondence between the worldsheet saddle point expansion and the large energy limit of celestial amplitudes, obtained by Mellin transform of the scattering amplitudes to the celestial sphere. In particular, one-to-one correspondence between the stationary phase expansion of the celestial amplitudes and the saddle point expansion of the string amplitude to each order has been demonstrated. In this framework, the saddle point on the string worldsheet is mapped to a point on the celestial sphere, and each subleading correction in $1/\a'$ maps to a corresponding correction in the large energy expansion of the celestial amplitude. This establishes a link between the $2d$ worldsheet description of strings and $2d$ conformal structure underlying celestial holography.
        \medskip

        From the tensionless string theory point of view, this framework provides a clear physical interpretation of the high energy expansion. In the high energy or tensionless limit ($\a' \to \infty$), the string spectrum contains an infinite tower of states that become light, including higher spin fields, so that the usual effective field theory description breaks down. The leading  saddle-point contribution captures the classical dynamics of the tensionless string, while subleading corrections encode the contributions of these light higher-spin modes. The fact that this structure is mirrored in the celestial description suggests that 2D conformal symmetry of the string worldsheet, which contracts to BMS$_3$ (or Conformal Carrollian) symmetry in the tensionless regime may be realized as the symmetry of the celestial CFT governing flat-space scattering, thereby linking worldsheet dynamics, higher-spin physics, and celestial holography in a unified framework.

\newpage

\section{Conclusions}

We have taken the reader on a journey in this review, starting out in the 1970s with the seminal work of Schild and the early developments of the null strings, through to the fundamentally important work of ILST, which formed the basis of a lot of the more recent developments, and finally in the main body of our review, we saw that the emergent symmetries on the worldsheet where Conformal Carroll symmetries and their supersymmetric cousins and these underlying symmetry structures helped build the theories of closed and open null tensionless (super)strings and the myriad of novel quantum phenomena that emerged from them. 

\medskip

It should be very clear to anyone reading our review and familiar with usual tensile string theory that there is a vast ocean of things that one needs to address in order to get the theory of null strings anywhere close to its tensile parent. We will thus end our review with a list of the most pressing questions that we believe need to answered right away. 

\subsection*{Open questions}

\begin{itemize}
    \item {\em Quantisation of bosonic closed null strings:} The careful canonical quantisation of closed bosonic null strings yielded the three vacua of the null strings on which three different quantum theory were built. We have reviewed how all of them seem to be consistent in the correct number of dimensions by the closing of the Poincare algebra in lightcone quantisation. However, a more careful analysis in terms of path integrals and BRST methods is warranted to make our claims completely airtight. As we have also reviewed briefly in Sec.~\ref{Path integral formulation}, there has been an attempt at path integral quantisation, but this led to some odd properties like problems with defining the path integral for the oscillator vacuum and a set of curious dimensions for the null superstring, 
    which possibly indicates some subtleties about the null string in the path integrals that need reexamination. A BRST analysis would of course be even more robust. It may be useful to take a closer look at this in the light of the modern language of Carrollian worldsheet structures. 
    
    
    \item {\em Quantising bosonic open null strings:} The theory of open null strings is very new and demands a careful quantisation. It would be interesting to see if all the vacua of the closed strings remain consistent for the open strings or not. Here it is imperative to also explore the different types of open boundary conditions we encountered. We only dealt with the Dirichlet open strings and it is natural to ask what differences one encounters for the Neumann and particularly the null boundary condition. Even in the context of the Dirichlet conditions, we encountered a new algebra, the one we called the Boundary Conformal Carroll algebra. One needs to study the representation of this carefully and build the quantum aspects of a theory with this as the underlying algebra. A recent analysis of aspects of the representation theory of the BCCA can be found in \cite{Buzaglo:2025nti}. Clearly there is a lot to be understood here. 

    \item {\em D-branes in tensionless string theory:} A closely related question is what are D-branes in null string theory. Given that the different vacua arise as D-branes of the original theory, the interpretations and indeed the construction of D-branes should be a rich and layered question. Also, we have seen a closed to open transition in the tensionless limit. It would be of interest to understand what the open-closed duality means in this context. 
    
    \item {\em Quantising closed and open null superstrings:} All our comments from above about the quantisation of both closed and open strings hold when we move to superstring. The canonical quantisation itself is a difficult task as finding the various different vacua would mean hunting through many more possibilities. The previous questions about open strings and D-branes are also very important in this context. 
    
    \item {\em Dualities in tensionless strings:} The five different tensile superstrings are connected by a web of dualities. It is but natural to ponder what happens to the tensionless version of this web. From the bosonic quantum structure of the null theory, it seems that the web will get a lot more intricate.

    \item {\em Connections to high energy scattering:} One of the principle motivations for attempting to understand the tensionless limit of string theory is to get a handle on the very high energy behaviour of the theory. From the seminal work of Gross and Mende, we know of the simplifications in this limit and the structures of the string amplitudes at very high energies. What has been missing is the construction of worldsheet methods that would lead to this very high energy sector directly. The worldsheet theory of the null string holds great promise in this regard. The key question is the construction of vertex operators in the null string. Vertex operators for a Carrollian scalar in the highest weight vacuum were constructed in \cite{Hao:2021urq} (see also \cite{Chen:2025gaz}). But for the Gross-Mende analysis of the null string, one needs to consider vertex operators in the induced vacuum, which is the natural limit from the tensile theory. This should lead to interesting differences  \cite{Bagchi}. 
    
    \item {\em Connections to flat holography:} Consider a theory of tensionless strings in three dimensional asymptotically flat spacetimes (AFS). The symmetries on worldsheet and the symmetries at the null boundary of spacetime are the same, - BMS$_3$ (or 2d conformal Carroll). In a manner similar to \cite{Giveon:1998ns,Kutasov:1999xu}, one could envision connecting worldsheet symmetries to the spacetime symmetries and thereby connect null strings with holography in 3d AFS. A different path has been proposed in \cite{Kervyn:2025wsb} by connecting high string amplitudes to the 2d Celestial CFT dual for 4d AFS. See also an interesting proposal of connecting strings in AdS$_3$ to 2d Celestial CFTs \cite{Banerjee:2025tec}. 

    \item {\em{Hagedorn physics:}} Another very important place where tensionless strings would become crucial is in the context of the Hagedorn transition in string theory. We have previously speculated that the appearance of the open null string in the tensionless limit is the signature of the formation of the very long string which becomes thermodynamically favourable when heating up a gas of closed strings to near the Hagedorn temperature. Although there have been tantalizing hints of this, it would be important to pin down a concrete calculation to show the natural emergence of Hagedorn physics in the tensionless string. 
    
\end{itemize}

These are of course only the very immediate goals of the tensionless strings programme. Since there have been surprises at every turn, with each new calculation, we expect new physics to appear, especially in the magical world of the quantum null string. It seems this is an exciting journey that has only just begun. 

\bigskip

\subsection*{Acknowledgements}
We have a lot of people to thank in our tensionless adventures. AB particularly thanks Shankhadeep Chakrabortty and Pulastya Parekh who were the first two collaborators on this adventure. We are also thankful to the following people for collaborating (directly or indirectly) on this project that has now turned well over a decade old: Joan Simon, Daniel Grumiller, Mangesh Mandlik, Punit Sharma, Kedar Kolekar, Sudipta Dutta, Shahin Sheikh-Jabbari, Jelle Hartong, Emil Have, Stefan Fredenhagen, Pronoy Chakrabortty, Sharang Iyer, Sachin Grover, Amartya Saha, Arkachur Bhattacharya, Ansh Mishra. In addition to those acknowledged above, PP would like to thank Pushkar Soni for assistance with the figures and especially Nilay Kundu for his valuable guidance in presenting work related to tensionless strings in various seminars.

\medskip

AB thanks Stephan Stieberger for the invitation to write this article, which has been long overdue.  

\medskip
AB is partially supported by a Swarnajayanti Fellowship from the Science and Engineering Research Board (SERB) under grant SB/SJF/2019-20/08 and also by an ANRF grant CRG/2022/006165.
\medskip

ABan is supported in part by an OPERA grant and a seed grant NFSG/PIL/2023/P3816 from BITS-Pilani, and an early career research grant ANRF/ECRG/2024/002604/PMS from ANRF India. He also acknowledges financial support from the Asia Pacific Center for Theoretical Physics (APCTP) via an Associate Fellowship.
\medskip

RC is supported by the Mathematical Physics group of Beijing Institute of Mathematical Sciences and applications.
\medskip

PP is supported by the Infosys Endowment for the study of the Quantum Structure of Spacetime at Tata Institute of Fundamental Research.
\medskip
\newpage

\appendix
\section*{APPENDICES}
\section{Review of tensile open strings}\label{Open Strings}
In this Appendix, we provide a concise discussion on the tensile bosonic open strings, whose worldsheet has boundaries at $\s=0$ and $\s=\pi$. To study the variation of the Polyakov action \eqref{Polyakov} for open strings, one also need to consider the contributions arising from these boundaries. Working in the conformal gauge \eqref{cg}, variation of the action with respect to $X^\mu$ yields
\begin{equation}
\begin{split}
    \delta_{X}S=T\int d\t d\s~\big(-\square X\cdot \delta X
    +\nabla\cdot(\nabla X~\delta X)\big).
\end{split}
\end{equation}
The second term in the integral gives us the boundary term which takes the following form at $\s=0,\pi$ boundary
\begin{align}
    T\int_{\mathcal{B}} dl_{\mathcal{B}}~(\Vec{n}\cdot \nabla X)~\delta X=T\int d\t~X'\cdot \delta X\Big|^{\s=\pi}_{\s=0}.
\end{align}
This term vanishes in two alternative ways,
\begin{enumerate}
    \item \textbf{Dirichlet Condition:} In this case, we set $\delta X^\m=0$ at $\s=0,\pi$. Since $\s$ has fixed values at both the boundaries this essentially means
    \begin{align}\label{tensiledirichlet}
        \dot{X}^\m\big|_{\s=0,\pi}=0.
    \end{align}
    Physically this means the string endpoints are fixed to a particular value of $X^\m$.
    \item \textbf{Neumann Condition:} In this case, we set
    \begin{align}\label{tensileneumann}
        X'^\m\big|_{\s=0,\pi}=0.
    \end{align}
    Physically this means the endpoints of the string are free to move along $X^\m$ directions. 
\end{enumerate}
One can also have mixed boundary condition where in one endpoint we have Dirichlet boundary condition and other end Neumann condition. However, in this review we only cover both side Dirichlet (DD) and both side Neumann (NN) boundary conditions which will be useful later.
\medskip
 

Imposing Dirichlet condition ($\dot{X}^\m=0$) on both boundaries $\s=0,\pi$ on the solution of the wave equation as given in \eqref{JackReacher} gives the following 
\begin{align}\label{TensileDirich}
    x^\m=c^\m, \quad\a^\m_0=0,\quad\a^\m_n=-\tilde{\a}^\m_{n}\qquad\forall n\neq 0.
\end{align}

Imposing Neumann condition $X'^\m=0$ on both boundaries $\s=0,\pi$ on the solution of the wave equation as given in \eqref{JackReacher} gives the following 
\begin{align}\label{TensileNeum}
    \a^\m_n=\tilde{\a}^\m_{n}\qquad \forall n.
\end{align}
Due to the identifications in \eqref{TensileDirich} and \eqref{TensileNeum} open string worldsheet theory only has one copy of Virasoro algebra given by
\begin{align}\label{singlecopyvir}
    \mathbb{L}_n=\frac{1}{2}\sum_m \a_{m}\cdot\a_{n-m}.
\end{align}
\section{Superstring: RNS formalism}\label{Superstring}
This appendix focuses on the study of RNS formalism of the tensile superstring. We begin with the action of tensile RNS superstring in superconformal gauge 
\begin{align}\label{RNSSuperstring}
    S=\frac{T}{2}\int d^2\s~(\eta^{\a\b}\p_\a X^\m\p_\b X_\m+\bar{\psi}^\m\rho^\a\p_\a\psi_\m).
\end{align}
In the above, $\psi^\m$ are Majorana-Weyl spinors
\begin{align}\label{psi}
    \psi^\m=\begin{bmatrix}
        \psi^\m_-\\
        \psi^\m_+
    \end{bmatrix}.
\end{align}
$\rho^\a$ ($\a=0,1$) are the Dirac matrices in Majorana representation
\begin{align}\label{rhorel}
    \rho^0=\begin{bmatrix}
        0 && -1\\ 1 && 0
    \end{bmatrix},~~~\rho^1=\begin{bmatrix}
        0 && 1\\1 && 0
    \end{bmatrix}.
\end{align}
These matrices follow the Clifford algebra in two dimensions
\begin{align}\label{Cliff}
    \{\rho^\a,\rho^\b\}=2\eta^{\a\b}\mathbb{1}.
\end{align}
In the above, $\eta^{\a\b}$ is the gauge fixed worldsheet metric. For the sake of convenience, we rewrite the action \eqref{RNSSuperstring} in terms of worldsheet lightcone coordinates $\s^\pm=\t\pm\s$.
\begin{align}\label{lcsusyaction}
   S= \frac{T}{2}\int d^2\s~ \left[2\partial_{+}X^\m\partial_{-}X_\m+i(\psi^\m_+\partial_{-}\psi_{+\m}+\psi^\m_-\partial_{+}\psi_{-\m})\right].
\end{align}
In the above $\partial_{\pm}=\frac{1}{2}(\partial_\tau\pm\partial_\sigma)$. This action is invariant under the worldsheet superconformal transformation parametrized by $\xi$ and $\e$, under which the fields $X$ and $\psi$ transform as below
\begin{subequations}\label{susy}
\begin{align}
\delta_\xi X&=\xi^a\p_a X, \qquad  \delta_\xi\psi_\pm = \xi^a\p_a \psi_\pm+\frac{1}{2}\p_\pm\xi^\pm\psi_\pm, \\
\delta_\e X&=\bar{\e}\psi, \qquad \delta_\e\psi = -i\rho^a\p_aX\e. 
\label{sitabhog}
\end{align}
\end{subequations}
In the above, $\xi^a$ and $\e^a$ satisfy the following conditions
\begin{align}\label{guptadirac}
     \partial_{\pm}\xi^{\mp}=0,\qquad  \partial_{\pm}\e^{\mp}=0.
\end{align}

Using \eqref{psi} and \eqref{rhorel}, one can rewrite the supersymmetry transformation as below
\begin{equation}
\delta_\e X=\epsilon^+\,\psi_-+\epsilon^-\,\psi_+,\quad \delta\psi_{-}=2\e^-\p_-X,\quad \delta\psi_{+}=-2\e^+\p_+X.
\end{equation}
From \eqref{lcsusyaction} it is straightforward to derive the EOM for bosons as well as fermions
\begin{align}\label{susyeom}
  \p_-\p_+X^\m=0,\qquad \partial_{+}\psi^\m_{-}=0, \qquad \partial_{-}\psi^\m_{+}=0.
\end{align}
The energy momentum tensor components and the supercurrents of the RNS action also vanish, which lead us to the following constraints
\begin{subequations}\label{superconstraints}
    \begin{align}
        T_{\pm\pm}=&~\p_
        \pm X^\m\p_\pm X_\m+\frac{i}{2}\psi^\m_\pm\partial_{\pm}\psi_{\pm\m}=0,\\
 &~~J_{\pm}=\psi_{\pm}\cdot\partial_{\pm}X=0.
    \end{align}
\end{subequations}

\subsection{Closed superstrings}
The EOM for bosonic field $X^\m$ as given in \eqref{susyeom} yields the closed string mode expansion identical to \eqref{tensilemodeexp}. \eqref{susyeom} it is apparent that $\psi^{+}$ ($\psi^{-}$) are essentially functions of $\s^+$ ($\s^-$). For closed string case, there can be two possible periodic boundary conditions for fermions given by 
\begin{align}
    \psi^{\mu}_{+}(\tau,\sigma)=\pm\psi^{\mu}_{+}(\tau,\sigma+2\pi).
\end{align}
The boundary conditions with +ve sign lead to the Ramond sector while these with -ve sign lead to the Neveu-Schwarz (NS) sector. If both $\psi^{\mu}_{+}$ and $\psi^{\mu}_{-}$ belong to the NS sector then the mode expansion of them becomes
\begin{align}\label{tensed1}
    \psi^{\mu}_{+}(\tau,\sigma)=\sqrt{\alpha'}\sum_{r\in\mathbb{Z}+\frac{1}{2}}b^\mu_re^{-ir(\tau+\sigma)},\quad \psi^\mu_{-}(\tau,\sigma)=\sqrt{\alpha'}\sum_{r\in\mathbb{Z}+\frac{1}{2}}\tilde{b}^\mu_re^{-ir(\tau-\sigma)}.
\end{align}
The oscillators $(b,\tilde{b})$ satisfy the following algebra
\begin{align}
    \{b_{r}^{\mu},b_{s}^{\nu}\}=\{\tilde{b}^\m_r,\tilde{b}^\n_s\}=\delta_{r+s}\eta^{\mu\nu},\qquad \{b_{r}^{\mu},\tilde{b}^\n_s\}=0.
\end{align}
Using this mode expansion \eqref{tensed1} along with the mode expansion of $X$ given in \eqref{tensilemodeexp} in the constraints in \eqref{superconstraints} one can find the super-Virasoro constraints in terms of the modes 
\begin{subequations}\label{supvir}
\begin{align}
    &\mathcal{L}_n=\frac{1}{2}\sum_m \alpha_{-m}\cdot \alpha_{m+n}+\frac{1}{4}\sum_{r\in\mathbb{Z}+\frac{1}{2}} (2r+n)b_{-r}\cdot b_{r+n},\\
    &\mathcal{Q}_{r}=\frac{1}{2}\sum_{m}\alpha_{-m}\cdot b_{r+m}.
\end{align}
\end{subequations}
The generators $\bar{\mathcal{L}}_n$ and $\bar{\mathcal{Q}}_{r}$ depend on $(\tilde{\a},\tilde{b})$ in the same way as above. Using the oscillator algebras of $(\alpha, \tilde{\a})$ and $(b,\tilde{b})$, one obtains the (left moving) super-Virasoro algebra as, 
\begin{subequations}\label{sv2}
\begin{align}
    [\mathcal{L}_{n},\mathcal{L}_{m}]&=(n-m)\mathcal{L}_{n+m}+\frac{c}{12}(n^3-n)\delta_{n+m,0}, \\
    [\mathcal{L}_{n},\mathcal{Q}_{r}]&=\Big(\frac{n}{2}-r\Big)\mathcal{Q}_{n+r},\\
    \{\mathcal{Q}_{r},\mathcal{Q}_{s}\}&=2\mathcal{L}_{r+s}+\frac{c}{3}\Big(r^2-\frac{1}{4}\Big)\delta_{r+s,0}.
\end{align}
\end{subequations}
Similar relations are satisfied for (right moving) super-Virasoro generators $\bar{\mathcal{L}},\bar{
\mathcal{Q}}$. Together the two sectors give two copies of super-Virasoro algebra.
\subsection{Open superstrings}\label{sec2.1}
Having discussed the vanishing of the bosonic part for Dirichlet and Neumann boundary conditions earlier in Appendix~\ref{Open Strings}, here we shall see what happens in the case for fermions. We begin with the fermionic part of the action \eqref{lcsusyaction},
\begin{equation}
     S_{F} \sim \int d\tau d\s \left(\psi_+ \partial_-\psi_+ +\psi_-\partial_+\psi_-\right),
\end{equation}
where we have suppressed the Lorentz indices. The variation of this action results in 
\begin{equation}\label{bdyterm}
    \delta S_F \sim \int d\tau \left(\psi_-\delta \psi_- - \psi_+ \delta \psi_+\right)\Big|_{\s=0}^{\s=\pi}.
\end{equation}
For open strings, the boundary terms in \eqref{bdyterm} must vanish separately at boundaries $\s=0,\pi$. This is achieved for
\begin{equation}
    \psi_+(0)=\pm \psi_-(0),\qquad \psi_+(\pi)=\pm \psi_-(\pi).
\end{equation}
In order to preserve spacetime Poincare invariance, we impose Neumann or Dirichlet boundary condition in analogy with the bosonic string. Without loss of generality, we choose $NN$ and $DD$ boundary conditions for the worldsheet fermions as
\begin{subequations}
\begin{align}\label{rasagolla}
   &\text{NN:}\qquad  \psi_+(0)=\,\psi_-(0)\,, \qquad \psi_+(\pi)=\eta\,\psi_-(\pi)\,,\\ \label{rajbhog}
   & \text{DD:}\qquad \psi_+(0)=-\,\psi_-(0)\,, \qquad \psi_+(\pi)=-\eta\,\psi_-(\pi)\,,
    \end{align}
\end{subequations}
where $\eta=+1$ labels the Ramond sector and $\eta=-1$ corresponds to the NS sector of the open string. It is straightforward to show that the boundary conditions \eqref{rasagolla} break half of the worldsheet supersymmetry. From \eqref{sitabhog}, the Dirichlet condition implies
\begin{equation} \label{ns}
    \delta_\e X\Big|_{\s=0,\pi}=0, \qquad \epsilon^+(0)\,= \epsilon^-(0)\,, \qquad 
    \epsilon^+(\pi)= -\eta \epsilon^-(\pi)\,,
\end{equation}
where, $\eta=\pm 1$ label the two sectors as defined earlier. In this review, we focus on the Dirichlet boundary condition in the NS sector given by
\begin{equation}\label{khirkadam1}
    \psi_+(0)=-\,\psi_-(0),
   \qquad \psi_+(\pi)=\psi_-(\pi).
\end{equation}
 Applying this condition to the mode expansions \eqref{tensed1} yields the relation between left- and right- moving modes fermionic modes
\begin{align}\label{spoken}
    \tilde{b}_r=-b_r.
\end{align}
Consequently, the mode expansions for the bosonic field $X^\m$ and fermions $\psi^\m$ (after imposing \eqref{TensileDirich} and \eqref{spoken}) take the form
\begin{subequations}
\label{tensi0}
\begin{align}
     X^{\mu}(\tau,\sigma)=x^{\mu}+\sqrt{2\alpha'}\alpha_0^\mu\s+i\sqrt{\frac{\alpha'}{2}}\sum_{n\neq0}\frac{1}{n}\left[\alpha^{\mu}_{n} e^{-in(\tau+\sigma)} + \alpha^{\mu}_{-n}e^{in(\tau-\sigma)}\right],\\ \label{tensi1}\psi^{\mu}_{+}(\tau,\sigma)=\sqrt{\alpha'}\sum_{r\in\mathbb{Z}+\frac{1}{2}}b^\mu_re^{-ir(\tau+\sigma)},\quad \psi^\mu_{-}(\tau,\sigma)=-\sqrt{\alpha'}\sum_{r\in\mathbb{Z}+\frac{1}{2}}b^\mu_re^{-ir(\tau-\sigma)}.
     \end{align}
\end{subequations}
Using \eqref{tensi0} in the constraints \eqref{superconstraints}, one obtains the super-Virasoro generators for open superstrings. We present these generators in a different basis (also defined for relativistic BCFT \eqref{boldl}), appropriate for separating the boundary compatible generators from the incompatible ones. 
\begin{subequations}\label{supviropen}
\begin{align}
    &\mathbb{L}_n=\frac{1}{2}\sum_m \alpha_{-m}\cdot \alpha_{m+n}+\frac{1}{4}\sum_{r\in\mathbb{Z}+\frac{1}{2}} (2r+n)b_{-r}\cdot b_{r+n},\\
    &\mathbb{Q}_{r}=\frac{1}{2}\sum_{m}\alpha_{-m}\cdot b_{r+m}.
\end{align}
\end{subequations}
These generators close to form a single copy of the super-Virasoro algebra
\begin{subequations}\label{svbold}
\begin{align}
        &[\mathbb{L}_{n},\mathbb{L}_{m}]=(n-m)\mathbb{L}_{(m+n)}+ \frac{\mathbb{c}}{12}n(n^2-1)\delta_{n+m,0},\\ &[\mathbb{L}_{n},\mathbb{Q}_{r}]=\Big(\frac{n}{2}-r\Big)\mathbb{Q}_{n+r},\\&\{\mathbb{Q}_{r},\mathbb{Q}_{s}\}=2\mathbb{L}_{r+s}+\frac{\mathbb{c}}{3}\left(r^2-\frac{1}{4}\right)\delta_{r+s,0}.
        \end{align}
\end{subequations}
The generators in this new basis ($\mathbb{L},\mathbb{Q}$), are related to the standard left-right generators ($\mathcal{L},\mathcal{Q}$) as follows
\begin{equation}
         \mathbb{L}_{n}= \mathcal{L}_{n}+\bar{\mathcal{L}}_{n}, \qquad \mathbb{Q}_{r}=\mathcal{Q}_{r}+\bar{\mathcal{Q}}_{r}.
\end{equation}
With this we conclude our discussion of tensile superstring for both the closed and open sector.

\newpage

\section{Supersymmetric extension of Carroll CFT}\label{CSCFT} 
In this appendix, we briefly recapitulate supersymmetric extensions of Carroll CFTs. We first look into the Carrollian superconformal field theories in manifolds without boundaries which would be applicable to the study of closed null superstrings and then we consider Carrollian superconformal field theories with boundaries. These would find application in the study of null open superstrings. 

\subsection{Carrollian superconformal field theories}
In earlier studies, it was found that 2d CCA can be supersymmetrically extended in two different ways, namely to the Homogeneous and Inhomogeneous Superconformal Carroll Algebras (SCCA)\footnote{The homogeneous SCCA was first found by Gamboa et al in \cite{Gamboa:1989px}. Inhomogeneous SCCA was first constructed in \cite{Bagchi:2017cte}.}. These algebras arise from two different Carrollian In\"on\"u-Wigner contractions of the super-Virasoro algebra. Since the bosonic subalgebra is still the CCA in both cases, the difference in the contraction lies in the scaling of the supersymmetry and superconformal generators\footnote{An algebra isomorphic to the inhomogeneous SCCA was originally constructed in \cite{Mandal:2010gx} as Galilean limit of super-Virasoro algebra. We have seen that in two dimensions, Galilean and Carrollian algebras are isomorphic. This isomorphism extends to the superconformal algebras as well.}. To understand this, we begin with the relativistic superconformal field theories (SCFTs) whose generators $\{\mathcal{L}_{n},\mathcal{Q}_{r},\bar{\mathcal{L}}_{n},\bar{\mathcal{Q}}_{r}\}$ satisfy two copies super-Virasoro algebra as given in \eqref{sv2}. For both R-sector ($r\in\mathbb{Z}$) and NS-sector ($r\in\mathbb{Z}+\frac{1}{2}$), the algebra has the form given in \eqref{sv2}. We will now perform both contractions and construct the resulting superconformal algebras in the following. 

\paragraph{Homogeneous SCCA:}
Let us consider the following In\"on\"u-Wigner contraction on the  supe-Virasoro generators in \eqref{sv2}
\bea{homocon}
L_n=\mathcal{L}_n-\bar{\mathcal{L}}_{-n}, \quad M_n = \eps (\mathcal{L}_n+\bar{\mathcal{L}}_{-n}), \quad Q^+_r=\sqrt{\eps}\mathcal{Q}_r,\quad Q^-_r = \sqrt{\eps}\bar{\mathcal{Q}}_{r}.
\eea
This scaling is called homogeneous scaling since both $\mathcal{Q}^+_r$'s and $\mathcal{Q}^-_{r}$'s have been scaled in an identical manner. The resulting algebra is called the Homogeneous Superconformal Carrollian algebra (SCCA$_H$). Its non-zero commutation relations are given by
\bea{sgcah}
&& [L_n, L_m] = (n-m) L_{n+m} + \frac{c_L}{12} \, (n^3 -n) \delta_{n+m,0}, \nonumber\\
&& [L_n, M_m] = (n-m) M_{n+m} + \frac{c_M}{12} \, (n^3 -n) \delta_{n+m,0}, \\
&& [L_n, Q^\a_r] = \Big(\frac{n}{2} - r\Big) Q^\a_{n+r}, \quad \{Q^\a_r, Q^\b_s \} = \delta^{\a\b} \left[M_{r+s} + \frac{c_M}{6} \Big(r^2 - \frac{1}{4}\Big)  \delta_{r+s,0} \right]. \nonumber
\eea
Here $\alpha$ and $\beta$ are used to designate $\pm$. As highlighted before, this algebra also appears as the asymptotic symmetry algebra at the null infinity of conventional $\mathcal{N}=2$ supergravity in 3d asymptotically flat spacetimes \cite{Barnich:2014cwa}. In the main body of the review, we will encounter this algebra as the residual symmetry algebra of the homogeneous null superstring. 

\paragraph{Inhomogeneous SCCA:}
The super-Virasoro algebra also admits another kind of Carrollian In\"on\"u-Wigner contraction where the supersymmetry generators are scaled in a manner similar to the bosonic Virasoro generators: 
\bea{ingenscal}
L_n&=&\mathcal{L}_n-\bar{\mathcal{L}}_{-n}, \quad M_n = \eps (\mathcal{L}_n+\bar{\mathcal{L}}_{-n}), \nonumber \\
G_r&=&\mathcal{Q}_r-i \bar{\mathcal{Q}}_{-r}, \quad K_r = \eps (\mathcal{Q}_r+i\bar{\mathcal{Q}}_{-r}).
\eea
This scaling is called as inhomogeneous scaling since here the supersymmetry generators $G$ and $K$ are contracted in different manners. Notice the interesting factors of $i$ in the combination of the supersymmetry generators. The resulting algebra is the inhomogeneous super Conformal Carrollian algebra (SCCA$_I$). The non-zero commutation relations are given by
\bea{sgcai}
&& [L_n, L_m] = (n-m) L_{n+m} + \frac{c_L}{12} (n^3 -n) \delta_{n+m,0}, \nonumber\\
&& [L_n, M_m] = (n-m) M_{n+m} + \frac{c_M}{12} (n^3 -n) \delta_{n+m,0}, \\
&& [L_n, G_r] = \Big(\frac{n}{2} -r\Big) G_{n+r}, \ [L_n, K_r] = \Big(\frac{n}{2} -r\Big) K_{n+r}, \ [M_n, G_r] = \Big(\frac{n}{2} -r\Big) K_{n+r}, \nonumber\\
&& \{ G_r, G_s \} = 2 L_{r+s} + \frac{c_L}{3} \Big(r^2 - \frac{1}{4}\Big)   \delta_{r+s,0}, \ \{ G_r, K_s \} = 2 M_{r+s} + \frac{c_M}{3} \Big(r^2 - \frac{1}{4}\Big)   \delta_{r+s,0}.\nonumber
\eea
Comparing SCCA$_I$ in \eqref{sgcai} to the SCCA$_H$ in \eqref{sgcah} one sees that SCCA$_I$ has a richer structure (more non-zero commutations) in the SUSY sector. Like SCCA$_I$, SCCA$_I$ too has been found as the symmetry algebra at the null infinity of an exotic theory of supergravity in 3d asymptotically flat spacetimes \cite{Lodato:2016alv}. In the main text, this algebra arises as the residual gauge symmetries of the inhomogeneous tensionless superstring. The factors of $i$ in the linear combinations mentioned above lead to some intriguing features of the parent theory, as we will mention in passing in the body of the article.

\subsection{Superconformal Carroll CFT with boundaries}\label{BasantaBiswas}
Following the recent work \cite{Bagchi:2025jgu}, we now have a quick look into the Superconformal Carroll CFT with boundaries. 

\paragraph{Homogeneous SCCA with boundaries:}
We have seen in section \ref{BCCA} that after introducing boundaries in a system with conformal Carrollian symmetry, only a subalgebra of CCA$_2$ consisting of $\mathcal{O}_n$s and $\mathcal{P}_n$ survive. In \cite{Bagchi:2025jgu}, it has been shown that in presence of boundaries in a system with homogenous superconformal Carrollian symmetry, the following subalgebra of SCCA$_H$ survives
\begin{align}\label{combination1}
    \mathcal{O}_{n}&=L_n-L_{-n},~~~~\mathcal{P}_n=M_n+M_{-n},\quad 
    \mathcal{H}_{r}=Q^{+}_{r}+Q^{-}_{-r}.
\end{align}
They satisfy the following algebra
\begin{equation}\label{tintin2}
    \begin{split}
        &[\mathcal{O}_n,\mathcal{O}_m]=(n-m)\mathcal{O}_{n+m}-(n+m)\mathcal{O}_{n-m}\\
    &[\mathcal{O}_n,\mathcal{P}_m]=(n-m)\mathcal{P}_{n+m}+(n+m)\mathcal{P}_{n-m}+\frac{c_M}{12}(n^3-n)(\delta_{n+m}+\delta_{n-m})\\
    &[\mathcal{O}_r,\mathcal{H}_{s}]=\Big(\frac{r}{2}-s\Big)\mathcal{H}_{r+s}+\Big(\frac{r}{2}+s\Big)\mathcal{H}_{s-r}\\
    &\{\mathcal{H}_{r},\mathcal{H}_{s}\}=\mathcal{P}_{r+s}+\frac{c_M}{3}(r^2-\frac{1}{4})\delta_{r+s,0},~~~
    [\mathcal{H}_{r},\mathcal{P}_s]=[\mathcal{P}_r,\mathcal{P}_s]=0.\\
    \end{split}
\end{equation}
This is the fully centrally extended version of the symmetry algebra we encounter while studying the homogeneous null open superstring in \eqref{tintin}. This algebra can also be obtained from the the following In\"on\"u-Wigner contraction of one copy of super-Virasoro algebra given in \eqref{svbold}
\begin{align}\label{HIWC}
    \mathcal{O}_n=\mathbb{L}_n-\mathbb{L}_{-n},~~~P_n=\epsilon~(\mathbb{L}_n+\mathbb{L}_{-n}),~~~\mathcal{H}_
    r=\sqrt{\epsilon}\mathbb{Q}_{r},~~~\e\to 0.
\end{align}
  For realisation of this algebra in superspace the reader is referred to \cite{Bagchi:2025jgu}.

\paragraph{Inhomogeneous SCCA with boundaries:}
The work \cite{Bagchi:2025jgu} also looked into the inhomogeneous SCCA in presence of boundaries, where the surviving generators are given by
\begin{align}\label{inhom-comb}
\mathcal{O}_n&=L_n-L_{-n},~~~\mathcal{P}_n=M_n+M_{-n},
\quad \mathcal{K}_r=K_r+K_{-r},~~~\mathcal{Y}_r=K_r-K_{-r}.
\end{align} 
These generators satisfy the following algebra
\begin{align}\label{proredin}
    &\left[\mathcal{O}_n,\mathcal{O}_m\right]=(n-m)\mathcal{O}_{n+m}-(n+m)\mathcal{O}_{n-m} \nonumber\\
    &\left[\mathcal{O}_n,\mathcal{P}_m\right]=(n-m)\mathcal{P}_{n+m}+(n+m)\mathcal{P}_{n-m}+\frac{c_M}{12}(n^3-n)(\delta_{n+m}+\delta_{n-m}) \nonumber\\
    &\left[\mathcal{O}_n,\mathcal{K}_r\right]=\left(\frac{n}{2}-r\right)\mathcal{K}_{n+r}+\left(\frac{n}{2}+r\right)\mathcal{K}_{n-r}\\&\left[\mathcal{O}_n,\mathcal{Y}_r\right]=\left(\frac{n}{2}-r\right)\mathcal{Y}_{n+r}-\left(\frac{n}{2}+r\right)\mathcal{Y}_{n-r}\nonumber\\&
\{\mathcal{K}_r,\mathcal{K}_s\} = \{\mathcal{Y}_r,\mathcal{Y}_s\}
\;= \{\mathcal{Y}_r,\mathcal{K}_s\}
\;=\;
[\mathcal{P}_r,\mathcal{K}_s] = [\mathcal{P}_r,\mathcal{Y}_s]
\;=\;
[\mathcal{P}_n,\mathcal{P}_m] = 0.\nonumber
\end{align}
This algebra comes from the following In\"on\"u-Wigner contraction of one copy of super-Virasoro algebra
\begin{align}\label{Prolimin}
    &\mathcal{O}_n=\mathbb{L}_n-\mathbb{L}_{-n},~~~P_n=\epsilon~(\mathbb{L}_n+\mathbb{L}_{-n}), \nonumber \\&\mathcal{K}_
    r=\epsilon(\mathbb{Q}_{r}+\mathbb{Q}_{-r}),~~~~\mathcal{Y}_
    r=\epsilon(\mathbb{Q}_{r}-\mathbb{Q}_{-r}),~~~\e\to 0.
\end{align}
Dropping $\mathcal{Y}_r$ from the above algebra \eqref{proredin}, the remaining version comes from the contraction of the Homogeneous algebra \eqref{tintin2}. This makes the Inhomogeneous BSCCA less interesting compared to the its Homogeneous version.
\medskip

As of now, we do not have an open null superstring which realises this algebra, the BSCCA$_I$ as its worldsheet residual gauge symmetries. 

\newpage

\section{Superspace formalism for tensile and tensionless strings}
In this appendix we briefly discuss about the superspace realisation of tensile superstring followed by the same of homogeneous and inhomogeneous null superstrings. We also discuss how the superspace for homogeneous and inhomogeneous null superstrings can be obtained from appropriate limit of the superspace for tensile strings.
\subsection{Tensile supertrings}
 In an $\mathcal{N}=(1,1)$ superspace, which consists of worldsheet coordinates $\s^\a$ ($\a=0,1$) and Grassmanian coordinates $\theta^\pm$\footnote{In this section, for convenience, we denote $\pm$ to express holomorphic and antiholomorphic coordinates (both $\s$ and $\theta$) as well as holomorphic and antiholomorphic super-Virasoro generators.}, one can define a general superfield $Y$ as  
\begin{align}
    Y^\m(\sigma^{\pm},\theta^{\pm})=X^\m(\sigma^{\pm})+i\theta^+\psi^\m_{+}(\sigma^{\pm})+i\theta^-\psi^\m_{-}(\sigma^{\pm})+\frac{1}{2}\theta^+\theta^-B^\m(\sigma^{\pm}).
\end{align}
  In the study of on-shell supersymmetry the auxiliary field $B^\m$ can be set to zero. In this superspace, the RNS superstring action in \eqref{RNSSuperstring} can be rewritten in terms of superspace integral of a Lagrangian constructed out of the superfield $Y$
\begin{align}\label{superaction}
    S=-\frac{T}{2}\int~d^2\theta d^2\sigma \overline{D}Y^\m DY_\m.
\end{align}
In the above, the $D,\overline{D}$ are supercovariant derivative.
The gauge symmetry transformations of the superstring action as described in \eqref{susy} translate to the following transformations in terms of the superspace coordinates
\bea{}\label{sspace}
\delta\sigma^\pm &=&\xi^\pm+i\e^\pm\theta^\pm, \ \ \ \delta\theta^\pm=\e^\pm+\frac{1}{2}\theta^\pm\p_\pm\xi^\pm.
\eea
Now we consider the variation of the superfield $Y$ under the superconformal transformation which will help us finding its generators
\begin{align}\label{sfield}
\delta Y&=(\delta\sigma^+\p_++\delta\sigma^-\p_-+\delta\theta^+\p_{\theta^+}+\delta\theta^+\p_{\theta^-})Y \nonumber\\
&=\Big[\Big(\xi^+(\s^+)\p_++\frac{1}{2}\p_+\xi^+(\s^+)\theta^+\p_{\theta^+}\Big)+\e^+(\s^+)\Big(\p_{\theta^+}+i\theta^+\p_+\Big)\nonumber\\
& +\Big(\xi^-(\s^-)\p_-+\frac{1}{2}\p_-\xi^-(\s^-)\theta^-\p_{\theta^-}\Big)+\e^-(\s^-)\Big(\p_{\theta^-}+i\theta^-\p_-\Big)\Big]Y \nonumber \\
&=[\mathcal{L}^+(\xi^+)+\mathcal{Q}(\e^+)+\mathcal{L}^-(\xi^-)+\bar{\mathcal{Q}}(\e^-)]Y.
\end{align}
Here $\mathcal{L}$'s and $\mathcal{Q}$'s respectively are bosonic and fermionic generators. From \eqref{guptadirac} it is manifest that $\xi^{+}$ ($\xi^{-}$) and $\e^{+}$ ($\e^{-}$) are essentially functions of $\s^+$ ($\s^-$). They have the following Fourier expansion
\be{}\label{Fourier1}
\xi^{\pm}(\s^\pm) = \sum_{n} a_n^{\pm} e^{in\s^\pm},~~~\e^{\pm}(\s^\pm) = \sum_{r\in\mathbb{Z}+\frac{1}{2}} b_r^{\pm} e^{ir\s^\pm}.
\ee
Using \eqref{Fourier1}, one can Fourier expand the generators $\mathcal{L}^\pm$, and $\mathcal{Q}^\pm$ too
\bea{2}
\label{virvectorf}
\mathcal{L}^\pm(\xi^\pm)&=&-i\sum_n a^\pm_n \mathcal{L}^\pm_n,~~~\mathcal{Q}^\pm(\e^\pm)=\sum_{r\in\mathbb{Z}+\frac{1}{2}} b^\pm_r \mathcal{Q}^\pm_r.
\eea
Using \eqref{virvectorf} along with \eqref{sfield}, one can obtain the following vector field representation of the superconformal generators in the superspace:
\begin{subequations}\label{shankhadeepda}
    \begin{align}
        \mathcal{L}^\pm_{n}=ie^{in\sigma^\pm}\Big(\partial_\pm+\frac{in}{2}\theta^\pm\partial_{\theta^\pm}\Big)~~~~\mathcal{Q}^{\pm}_r=e^{ir\sigma^\pm}(\partial_{\theta^\pm}+i\theta^\pm\partial_\pm).
    \end{align}
\end{subequations}
These generators follow two copies of the Super-Virasoro algebra wthout central extension, non-vanishing commutators (and anti-commutators) of which are given bellow
\begin{align}\label{svalgebra}
[\mathcal{L}^\pm_{n},\mathcal{L}^\pm_{m}]=(n-m)\mathcal{L}^\pm_{n+m},~~~[\mathcal{L}^\pm_{n},\mathcal{Q}^\pm_{r}]=\Big(\frac{n}{2}-r\Big)\mathcal{Q}^\pm_{n+r},~~~\{\mathcal{Q}^\pm_{r},\mathcal{Q}^\pm_{s}\}=2\mathcal{L}^\pm_{r+s}.
\end{align}
\subsection{Homogeneous null superstrings}\label{homosuper}
In order to understand the superspace for homogeneous null superstrings, let us first recall the residual symmetry of the gauge fixed homogenous null superstring action from \eqref{hcfer10}
\begin{equation}
\xi^{0}=f^{\prime}(\sigma) \tau+g(\sigma), \qquad \xi^{1}=f(\sigma), \qquad \epsilon^{ \pm}=\epsilon^{ \pm}(\sigma).
\end{equation}
Just like for the case of tensile RNS superstrings, here too, we can clearly visualise the generators generating these transformation. In order to do so, here again, one has to realise the transformations \eqref{hcfer10} in a $\mathcal{N}=(1,1)$ Carrollian superspace, consisting of two Grassmann coordinates $\theta$ and $\bar{\theta}$ along with the two world‑sheet coordinates $\{\sigma, \t$\}. This superspace is called homogeneous Carrollian superspace and in such superspace, the transformations \eqref{hcfer10} translate to the following \cite{Bagchi:2016yyf},
\begin{subequations}\label{carrollsupertrans}
  \begin{align}
\sigma^{\a} & \rightarrow \sigma^{\prime a}=\sigma^{\a}+\xi^{\a}+\frac{i}{2} \epsilon^{+} V^{\a} \theta+\frac{i}{2} \epsilon^{-} V^{\a} \bar{\theta}  \\
\theta & \rightarrow \theta^{\prime}=\theta+\epsilon^{+}+\frac{1}{4} \theta \partial_{\a} \xi^{\a}  \\
\bar{\theta} & \rightarrow \bar{\theta}^{\prime}=\bar{\theta}+\epsilon^{-}+\frac{1}{4} \bar{\theta} \partial_{\a} \xi^{\a} 
\end{align}  
\end{subequations}
Let us define a superfield $Y\left(\sigma^{\a}, \theta, \bar{\theta}\right)$ in this superspace
\begin{align}
    Y^\m\left(\sigma^{\a}, \theta, \bar{\theta}\right)=X^\m\left(\sigma^{\a}\right)+i\theta\psi^{\m}_+\left(\sigma^{\a}, \theta, \bar{\theta}\right)+i\bar{\theta}\psi^{\m}_-\left(\sigma^{\a}, \theta, \bar{\theta}\right).
\end{align}
This superfield transforms under \eqref{carrollsupertrans} as given below, 
\begin{equation}\label{kumropotash}
\delta Y=\left(\delta \tau \partial_{\tau}+\delta \sigma \partial_{\sigma}+\delta \theta \partial_{\theta}+\delta \bar{\theta} \partial_{\bar{\theta}}\right) Y. 
\end{equation}
Using \eqref{hcfer10} and \eqref{carrollsupertrans} along with $V^\a=(1,0)$ from \eqref {gauge} in \eqref{kumropotash} we end up with the following form of superfield transformation
\begin{align}
\delta Y= & {\left[\left(f^{\prime} \tau+g+\frac{i}{2} \epsilon^{+} \theta+\frac{i}{2} \epsilon^{-} \bar{\theta}\right) \partial_{\tau}+f \partial_{\sigma}\right.} \nonumber\\
&~~~~~~~~~~~~~~~~~~ \left.+\left(\epsilon^{+}+\frac{1}{2} f^{\prime} \theta\right) \partial_{\theta}+\left(\epsilon^{-}+\frac{1}{2} f^{\prime} \bar{\theta}\right) \partial_{\bar{\theta}}\right] Y  \nonumber\\
= & {\left[L(f)+M(g)+Q^{+}\left(\epsilon^{+}\right)+Q^{-}\left(\epsilon^{-}\right)\right] Y} .
\end{align}
Looking at this explicit form, one can easily identify superfield transformation generators,
\begin{align}\label{cgen}
\begin{split}
&L(f)  =f \partial_{\sigma}+f^{\prime}\left[\tau \partial_{\tau}+\frac{1}{2}\left(\theta \partial_{\theta}+\bar{\theta} \partial_{\bar{\theta}}\right)\right] , \quad M(g)=g \partial_{\tau},  \\
Q^{+}&\left(\epsilon^{+}\right) =\epsilon^{+}\left(\partial_{\theta}+\frac{i}{2} \theta \partial_{\tau}\right), \quad Q^{-}\left(\epsilon^{-}\right)=\epsilon^{-}\left(\partial_{\bar{\theta}}+\frac{i}{2} \bar{\theta} \partial_{\tau}\right). 
\end{split}
\end{align}
In the above, $L$ and $M$ are bosonic generators and $Q^{\pm}$ are fermionic generators. Since the string is closed, the periodicty of the worldsheet in $\s$ implies that the functions $f, g$ and $\epsilon^{\pm}$ are essentially periodic functions of $\s$ and hence, can be Fourier expanded as
\begin{equation}
\label{genmode2}
f=\sum_{n} a_{n} e^{i n \sigma}, \quad g=\sum_{n} b_{n} e^{i n \sigma}, \quad \epsilon^{ \pm}=\sum_{n} \zeta_{n}^{ \pm} e^{i n \sigma}.
\end{equation}
Substituting \eqref{genmode2} back into \eqref{cgen}, one can Fourier expand the generators $L$, $M$, and $Q^{\pm}$ too
\begin{equation}
L(f)=-i \sum_{n} a_{n} L_{n}, \quad M(g)=-i \sum_{n} b_{n} M_{n}, \quad Q^{ \pm}\left(\epsilon^{ \pm}\right)=\sum_{n} \zeta_{n}^{ \pm} Q_{n}^{ \pm}.
\end{equation}
One can identify the mode operator $L_n$, $M_n$ and $Q^{\pm}$ given as , 
    \begin{align}\label{HSBMS}
    \begin{split}
&L_n=ie^{in\sigma}\Big[\partial_\sigma+in\tau\partial_\tau+\frac{in}{2}\big(\theta\partial_\theta+\bar{\theta}\partial_{\bar{\theta}}\big)\Big],~~~M_n=ie^{in\sigma}\partial_\tau\\
    &Q^{+}_{r}=e^{ir\sigma}\Big(\partial_{\theta}+\frac{i}{2}\theta\partial_{\tau}\Big),~~~~Q^{-}_{r}=e^{ir\sigma}\Big(\partial_{\bar{\theta}}+\frac{i}{2}\bar{\theta}\partial_{\tau}\Big).
    \end{split}
    \end{align}
These generators satisfy the homogeneous super‑Conformal Carrollian algebra (SCCA$_H$), without the central extension.
\begin{equation}\label{HSBMSAlgebra2}
    \begin{split}
       &[L_n,L_m]=(n-m)L_{n+m}~~~~[L_n,M_m]=(n-m)M_{n+m}\\
        &[L_n,Q^{\alpha}_{r}]=\Big(\frac{n}{2}-r\Big)Q^\alpha_{n+r}~~~\{Q^{\alpha}_{r},Q^{\alpha'}_{s}\}=\delta^{\alpha\alpha'}M_{r+s}\\
       &[M_m,M_n]=[M_m,Q^{\alpha}_{r}]=0.
   \end{split}
\end{equation}
These generators in the superspace can be derived from the limiting approaches as well. We recall from earlier discussion of homogeneous null superstrings in section \ref{homotenlim} that for this theory both components of the worldsheet fermion $\psi_\pm$ are scaled similarly i.e. $\psi_\pm\to\sqrt{\e}\psi_\pm$. The same is reflected in the superspace as well; the scaling which takes us from relativistic superspace to the homogeneous Carrollian superspace involves scaling both the Grassmannian coordinates in same manner, i.e.
\begin{align}
    \t\to\e\t,\qquad\s\to\s,\qquad\theta^+\to\sqrt{\e}\theta,\qquad\theta^-\to\sqrt{\e}\bar{\theta}.
\end{align}
Applying this scaling along with the In\"on\"u-Wigner contraction \eqref{homocon} ($\mathcal{Q},\bar{\mathcal{Q}}$ replaced by $\mathcal{Q}^{\pm}$) on the super-Virasoro generators given in \eqref{shankhadeepda} leads us to the generators in \eqref{HSBMS}. 
\subsection{Inhomogeneous null superstrings}
In order to understand the inhomogeneous tensionless superstrings from superspace perspective we first recall the residual gauge symmetry transformations of the action \eqref{action_inhom} from \eqref{inhomdiff} and \eqref{inhomcond}
\begin{align*}
    &\delta_\xi X=\xi^\a\p_\a X,\qquad
\delta_\xi \psi_0=\xi^\a\p_\a \psi_0+\frac{1}{4}\p_\a\xi^\a \psi_0, \\
&\delta_\xi \psi_1=\xi^\a\p_\a \psi_1~+~\frac{1}{4}\p_\a\xi^\a \psi_1~+~\frac{1}{2}(\p_1\xi^0)\psi_0,\\
&\delta X =i(\e^{1*} \psi_{0}+\e^{0*}\psi_{1}),\quad 
\delta \psi_{0} = -\e^1\dot{X},\quad
\delta \psi_{1} = -\e^0\dot{X}-\e^1X',\\
&\partial_0\xi^0=\partial_1\xi^1, \qquad \partial_0\xi^1=0,\\ &\partial_0\e^{1*}=\partial_0\e^0=\partial_1\e^1=\partial_1\e^{0*},\\
&\partial_0\e^{0*}=\partial_0\e^1=0, \qquad \psi_0=\psi_1^{*}.
\end{align*}

Now, since the fermions in this theory are complex, it is natural to choose complex Grassmannian coordinates in the inhomogeneous superspace.
Let us consider them to be \{$\phi,\chi$\}. The superfield for this case is given by
\begin{align}
    Y^\m(\s^\a,\phi,\chi)=X^\m(\s^\a)+i\phi\psi_0(\s^\a)+i\chi\psi_{1}(\s^\a).
\end{align}
In this superspace the  residual gauge symmetry transformations given in \eqref{inhomdiff} translate to the following transformations
\begin{subequations}\label{ihsuspacetr}
\bea{}
\delta\tau&=&\xi^0+i(\e^0\chi+\e^1\phi), \qquad 
\delta\sigma=\xi^1+i\e^1\chi, \\
\delta\phi&=&\e^0+\frac{1}{4}(\p_0\xi^0+\p_1\xi^1)\phi+\frac{1}{2}\p_1\xi^0\chi, \\
\delta\chi&=&\e^1+\frac{1}{4}(\p_0\xi^0+\p_1\xi^1)\chi.
\eea
\end{subequations}
In the above $\xi^\a(\s,\t)$ and $\e^\a(\s,\t)$ comes from the solution of the differential equations satisfied by them given in \eqref{inhomocondt}
\begin{subequations}\label{inhomcond}
\bea{}
\xi^0&=&f'(\sigma)\tau+g(\sigma),\ \ \ \ \xi^1=f(\sigma), \\
\e^0&=&e'(\sigma)\tau+h(\sigma). \ \ \ \ \e^1=e(\sigma). 
\eea
\end{subequations}
In the above, $e$ and $h$ are Grassmann-odd functions of $\s$ having periodicity $2\pi$ while $f$ and $g$ are Grassmann-even functions of $\s$ with the same periodicity. Hence, the transformations in \eqref{ihsuspacetr} ultimately become the following
\begin{subequations}{}
\bea{}
\delta\tau&=&f'\tau+g+ie'\tau\chi+ih\chi+ie\phi, \qquad
\delta\sigma=f+ie\chi, \\
\delta\phi&=&e'\tau+h+\frac{f'\phi}{2}+\frac{f''\tau\chi}{2}+\frac{g'\chi}{2}, ~\quad
\delta\chi=e+\frac{f'\chi}{2}. 
\eea
\end{subequations}
Variation of the superfield $Y$ under this residual gauge transformation gives
\be{inhomsfield}
\delta Y=(\delta\tau\p_\tau+\delta\sigma\p_\sigma+\delta\phi\p_\phi+\delta\chi\p_\chi)Y =[L(f)+M(g)+G(e)+H(h)] Y,
\ee
where $L(f)$, $M(g)$, $G(e)$ and $H(h)$ take the following form
\begin{subequations}\label{badurgopal}
\bea{}
L(f)&=&\Big[f\p_\sigma+f'\Big(\frac{\phi\p_\phi+\chi\p_\chi}{2}+\tau\p_\tau\Big)+\frac{f''}{2}\tau\chi\p_\phi\Big], \\
M(g)&=&\Big[g\p_\tau+\frac{g'}{2}\chi\p_\phi\Big], \\
G(e)&=&\Big[e\p_\chi+ie(\phi\p_\tau+\chi\p_\sigma)+e'\tau\p_\phi+ie'\chi\tau\p_\tau\Big], \\
H(h)&=&\Big[h\p_\phi+ih\chi\p_\tau\Big]. 
\eea
\end{subequations}
In the light of the fact that $f$, $g$, $e$ and $h$ are all periodic functions of $\s$ having $2\pi$ periodicity, they all can be Fourier expanded in terms of $e^{in\s}$. Applying their Fourier expansions in \eqref{badurgopal} one can find the Fourier modes of $L(f)$, $M(g)$, $G(e)$ and $H(h)$
\begin{subequations}\label{inhomgen}
\bea{}
L_n&=&ie^{in\sigma}\Big[\p_\sigma+in\Big(\frac{\phi\p_\phi+\chi\p_\chi}{2}+\tau\p_\tau\Big)-\frac{n^2}{2}\tau\chi\p_\phi\Big], \\
M_n&=&ie^{in\sigma}\Big[\p_\tau+\frac{in}{2}\chi\p_\phi\Big], \\
G_r&=&e^{ir\sigma}\Big[\p_\chi+i(\phi\p_\tau+\chi\p_\sigma)+ir\tau\p_\phi-r\chi\tau\p_\tau\Big], \\
H_r&=&e^{ir\sigma}\Big[\p_\phi+i\chi\p_\tau\Big]. 
\eea
\end{subequations}
These generators satisfy the inhomogeneous version of the superconformal Carrollian algebra
\bea{agcai} 
&&[L_m,L_n]=(m-n)L_{m+n}, \quad [L_m,M_n]=(m-n)M_{m+n}, \nonumber \\
&&[L_m,G_r]=\Big(\frac{m}{2}-r\Big)G_{m+r}, \quad [M_m,G_r]=\Big(\frac{m}{2}-r\Big)H_{m+r},\ \{G_r,G_s\}=2L_{r+s},\nonumber \\
&&[L_m,H_r]=\Big(\frac{m}{2}-r\Big)H_{m+r}, \quad \{G_r,H_s\}=2M_{r+s}.
\eea
In order to understand this formulation from limiting perspective one needs to look into define complex superspace coordinates in the tensile superspace
\begin{align}
    \phi=\frac{1}{\sqrt{2}}(\theta^+-i\theta^-),~~~\chi=\frac{1}{\sqrt{2}}(\theta^++i\theta^-).
\end{align}
To obtain inhomogeneous Carrollian superspace one needs to the following scaling on the coordinates 
\begin{align}
    \t\to\e\t,\qquad \s\to\s, \qquad \phi\to \e\phi, \qquad \chi\to\chi.
\end{align}
This scaling along with the In\"on\"u-Wigner contraction \eqref{ingenscal} of super-Virasoro generators \eqref{shankhadeepda} lead us to the generators \eqref{inhomgen}.
\newpage

\section{Connection to Ambitwistor strings}\label{Connection to Ambitwistor strings}
In the previous decade, a series of seminal works by Cachazo-He-Yuan (CHY) \cite{Cachazo:2013hca,Cachazo:2013iea,Cachazo:2013gna,Cachazo:2014xea,Cachazo:2014nsa} examined the tree-level scattering amplitudes of massless particles. Ambitwistor string theory proposed by Mason and Skinner \cite{Mason:2013sva} provided a string theoretic basis for these scattering equations. This version of string theory was originally formulated as holomorphic strings with constraint $\Pi^2=0$ ($\Pi^\m$ is the momentum of the string). Target spacetime for this theory is ambitwistor space, the space of complex null geodesics in complex Minkowski space \cite{Mason:2013sva, Geyer:2014fka}. Ambitwistor string theory is often considered to be field theory limit ($\a'\to 0$) of string theory\footnote{Ambitwistor string action has been obtained from chiral $\a'\to 0$ limit of tensile string action in \cite{Mason:2013sva}.}. Interestingly, however, the CHY formulas for Yang-Mills and gravity amplitudes bore striking resemblance to the Gross-Mende scattering amplitudes which correspond to the tensionless ($\a'\to\infty$) limit of strings \cite{Gross:1987kza,Gross:1987ar}. In \cite{Casali:2016atr}, Casali and Tourkine proposed an alternative formalism for ambitwistor strings as tensionless null strings which is consistent with the Gross-Mende limit. In subsequent work \cite{Casali:2017zkz}, they extended their formalism by complexifying both worldsheet and the target spacetime, quantised their theory and obtained the CHY formula for tree level amplitudes. Here we briefly review their formulation of ambitwistor strings as null strings.
\subsection{Ambitwistor string action from ILST action}
Let us begin with rewriting the ILST action \eqref{ILST}, using the vielbein density $V^\a$ as given in \eqref{ctb26} 
\begin{align}\label{ramdholai}
    S=\int d^2\s~\frac{1}{\lambda}(\dot{X}-\rho X')^2.
\end{align}
The conjugate momentum for this action is given by
\begin{align}
    \Pi^\m=\frac{1}{\lambda}(\dot{X}^\m-\rho X'^\m).
\end{align}
Recalling from earlier analysis that $\lambda$ and $\rho$ act as Lagrange multipliers one can see that they impose the following constraints on the action 
\begin{align}\label{bharatdholai}
    \Pi^2=0,~~~\Pi_\m X'^\m=0.
\end{align}
Hence one can define the following phase space action which is equivalent to the null string action \eqref{ILST}
\begin{align}\label{laxmandholai}
   S= \int d^2\s~(\Pi_\mu\dot{X}^\m-\rho \Pi_\mu X'^\m-\lambda\Pi^2).
\end{align}
In timelike gauge $\rho=0$ and $\lambda=1$, we retrieve the gauge fixed action we have worked in so far. For $\rho=1$ and $\lambda=0$ gauge (called ambitwistor gauge) we arrive at the ambitwistor string action\footnote{For detailed discussion of the action of the form \eqref{laxmandholai} in different gauges one can look into \cite{Bagchi:2022eav}.}
\begin{align}
    S=\int d^2\s(\Pi_\m\partial_{-}X^\m).
\end{align}
However, the constraints accompanying this action \eqref{bharatdholai} are different than that of ambitwistor strings, which are
\begin{align}\label{shatrughnadholai}
    \Pi^2=0,~~~\Pi_\m\partial_+X^\m=0.
\end{align}
Fortunately, as discussed in \cite{Casali:2016atr}, the EOM corresponding to the action \eqref{laxmandholai} is $\partial_-X=0$, using which on \eqref{bharatdholai} one can find \eqref{shatrughnadholai} in following way
\begin{align}
    \Pi_\mu\partial_+X^\m=\Pi_\mu(\partial_-X^\m+X'^\m)=0.
\end{align}
Hence, one can clearly see that ambitwistor string is on shell equivalent to the null strings.
\subsection{ILST equivalent formalism for ambitwistor strings}
Using partial gauge fixing, where $\rho=1$ with unspecified $\lambda$\footnote{It is important to note that in the phase space action fixing $\lambda=0$ works perfectly, however, in the form \eqref{ramdholai}, singularity arises.}, we can rewrite the action \eqref{ramdholai} as
\begin{align}\label{hanumandholai}
    \int d^2\s \frac{1}{\lambda}(\partial_-X)^2.
\end{align}
This action is basically ILST action \eqref{ILST} with $V^\a$ given by
\begin{align}
    V^\a=\frac{1}{\sqrt{\lambda}}(1,-1).
\end{align}
Comparing with the gauge fixed action of the null strings (i.e. \eqref{ILST} along with \eqref{Review1}) one can clearly see that the $\s^-$ coordinate for \eqref{hanumandholai} is equivalent to the coordinate $\t$ in \eqref{ILST}, while the $\s^+$ coordinate is equivalent to $\s$.  Consequently the classical analysis for the action \eqref{hanumandholai} is much the same as section \ref{Bosonic closed null strings}. This action \eqref{hanumandholai} has a the following residual gauge symmetry on the worldsheet
    \begin{align}
        \s^+\to \s^++\s^-f'(\s^+)+g(\s^+),~~~\s^-\to\s^-+f(\s^+).
    \end{align}
    Consequently the gauge symmetry algebra will be
    \begin{align}
        L_n=ie^{in\s^+}(\partial_++in\s^-\partial_{-}),~~~M_n=ie^{in\s^+}\partial_{-}.
    \end{align}
    This $L_n$s and $M_n$s satisfy the classical part of the BMS$_3$ algebra in \eqref{bmsalgebraclassical}
   \begin{equation}
    [L_m,L_n]=(m-n)L_{m+n},~~ [L_m,M_n]=(m-n)M_{m+n}, ~~[M_m,M_n]=0.
\end{equation}
 The EOM are given by
\begin{align}\label{rabondholai}
    \partial^2_{-}X^\m=0.
\end{align}
Using the fact that $\Pi=\frac{1}{\lambda}\partial_-X$ for \eqref{hanumandholai} he constraints the theory is subjected to as given in \eqref{bharatdholai} can be rewritten as
\begin{align}\label{Bibheeshondholai}
    (\partial_-X)^2=0,~~~\partial_-X\cdot\partial_\s X=0.
\end{align}
Solution to the EOM \eqref{rabondholai} with closed string periodicity $\s^\pm\sim\s^\pm+2\pi$ gives the following\footnote{This mode expansion has the same form as \eqref{fcb1}, where the summation includes the $n=0$ modes and \{$\s,\t$\} are replaces by \{$\s^+,\s^-$\}.}
\begin{equation}\label{ambitwistmodexpand}
  X^{\mu}(\s^\pm)=\sum_{n}\frac{1}{n}(A^{\mu}_{n}-in\s^- B^{\mu}_{n})e^{-in\s^+},
\end{equation}
where we need to make the following identification for the modes in order to ensure periodicity in $\s^\pm\sim\s^\pm+2\pi$
\begin{align}
    A_n\sim A_n+2\pi B_n
\end{align}
\subsection*{Classical constraints from phase space formalism}
Let us consider the phase space action in \eqref{laxmandholai}. If we partially gauge fix the action with $\rho=1$ without fixing $\lambda$. For such action the EOM for $X$ and $\Pi$ are
\begin{align}
    \partial_-X^\m=2\lambda\Pi^\m,~~~\partial_-\Pi^\m=0.
\end{align}
The constraints for this action are given in \eqref{bharatdholai}. The EOMs are solved by following mode expansions
\begin{subequations}\label{kumbhakarnadholai}
    \begin{align}
    \label{kumbhakarnadholai1}
        &~~\Pi^\m=\sum_n B^\m_n e^{-in\s^+},\\
       \label{kumbhakarnadholai2}  X^{\mu}(\s^\pm)=&\sum_{n}\frac{1}{n}(A^{\mu}_{n}-i2n\lambda\s^- B^{\mu}_{n})e^{-in\s^+}.
    \end{align}
\end{subequations}
From \eqref{kumbhakarnadholai2} one can see that in order to confirm the periodicity of $X$ $\s^{\pm}\sim \s^{\pm}+2\pi$, setting $\lambda$ to zero is essential. The constraints for ambitwistor strings as given in \eqref{shatrughnadholai} take the following form
\begin{subequations}\label{indrajitdholai}
    \begin{align}\label{indrajitdholai1}
        \Pi^2=\sum_n M_n e^{-in\s+},\\
        \label{indrajitdholai2}
        \Pi_\m \partial_+ X^\m=\sum_n (L_n-in\lambda\s^-M_n)e^{-in\s^+}, 
    \end{align}
\end{subequations}
where $L_n$s and $M_n$s are constructed on the $A,B$ oscillators as usual
\begin{equation*}
    L_n=\sum_{m}A_{-m}\cdot B_{m+n},~~~M_n=\sum_{n}B_{-m}\cdot B_{n+m}.
\end{equation*}
 From \eqref{indrajitdholai2}, we see that $\Pi_\m \partial_+ X^\m$ for $\lambda=0$ acts like a holomorphic Virasoro constraint as shown below
 \begin{align}
      \Pi_\m \partial_+ X^\m|_{\lambda=0}=\sum_n L_n e^{-in\s^+}.
 \end{align}
 It is important here to recall that ambitwistor string theory was originally formulated as holomorphic CFT on the string worldsheet with an additional constraint $\Pi^2=0$. Here, from the classical analysis, we saw that exactly same scenario is arising from null string formulation with $\lambda=0$ gauge. 
 \subsection*{Complex null strings}
 As highlighted before, the ambitwistor string theory is formulated on a space of complex null geodesics in complexified Minkowski spacetime, where the worldsheet too, is complexified. While the foregoing discussion (following \cite{Casali:2016atr}) did successfully connect the classical action of ambitwistor string to ILST action for null strings, both target spacetime and worldsheet has been considered to be real. In \cite{Casali:2017zkz}, a complex version of ILST strings was formulated in a complexified Minkowski target spacetime which filled this gap. Here the worldsheet is 2d complex manifold with two holomorphic coordinates $z$ and $\bar{z}$ and vielbein density $V^\a$ too is complexified. Here too, the gauge fixed action display the BMS$_3$ algebra with the following form of generators
\begin{align}
L_n=-z^{n+1}\partial_z-(n+1)z^n\bar{z}\partial_{\bar{z}},~~~M_n=-z^{n+1}\partial_{\bar{z}}.
\end{align}
This formalism of complex null string makes the connection between null stings and ambitwistor strings much more robust. 
\subsection{Quantisation}
It was known from earlier studies \cite{Mason:2013sva} that bosonic ambitwistor string has critical dimension 26, also that the spectrum of this theory is finite with only massless particles. Casali et al in \cite{Casali:2016atr,Casali:2017zkz} identified that quantum theory of null strings constructed on flipped vacuum has the identical features and hence, is the correct candidate for ambitwistor strings. Let us recall the definition of flipped vacuum
\begin{align*}
    A_n\ket{0}=B_n\ket{0}=0,~~~\forall n>0.
\end{align*}
As discussed in section \ref{bhuto}, only level 2 states constructed on this vacuum are physical states and they give massless spin 2 particles like graviton, $B$-field and dilaton. As shown in \cite{Casali:2017zkz}, the physical states for this theory can be identified as null states in the highest weight representation of BMS$_3$ algebra with $h_L=2$ and $h_M=0$. The central charges for this particula quantum theory has been found to be $c_L=2D$ ($D=26$ is dimension of target spacetime) and $c_M=0$ \cite{Bagchi:2020fpr}. Highest weight representation for BMS$_3$ algebra has been earlier studied in \cite{Bagchi:2009pe}, where it has been shown that for $c_M=0$, the descendant states for BMS$_3$ highest weight representation reduces to descendants of the Virasoro subalgebra of the BMS$_3$ algebra. Consequently, we are left with chiral Virasoro module. This confirms that at the quantum level ambitwistor string theory is equivalent to the null string theory constructed on flipped vacuum (since it is known that ambitwistor string theory is chiral CFT).
\medskip

\newpage

\section{What is an inconsistent constraint?}\label{inconsistent constraint}
We saw in section \ref{From one classical to three quantum theories}, only four possibilities out of nine are consistent with BMS$_3$ algebra resulting in three distinct quantum theories. In this appendix, we demonstrate the origin of inconsistency by examining one of five non-viable constraint choices listed in \eqref{allconstraint}. We consider the following constraint condition as an example
\begin{equation}
L_{n}|phys\>=0,\quad\forall\ n>0,\qquad M_{n}|phys\>=0,\quad\forall\ n\neq0. 
\end{equation}
We begin by considering a physical vacuum $|0,k^\mu\>$ which satisfies  
\be{}
  M_{n}|0,k^\mu\>=  \sum_m    B_{-m}\cdot B_{m+n}|0,k^\mu\>=0 \qquad\forall\ n\neq 0. 
\ee
Using the conditions imposed by $M_n$ and $L_n$, we obtain
\begin{subequations}
\begin{align}
    B^\mu_n|0,k^\mu\>&=0 \qquad (n\neq 0)\\
    L_{n}|0,k^\mu\>&= \sum_m    A_{-m}\cdot B_{m+n}|0,k^\mu\>=0  \qquad\forall\ n>0 \cr
 &= A_n\cdot B_0|0,k^\mu\>=0  \qquad \qquad \qquad \forall\ n>0. 
\end{align}
\end{subequations}
Since the vacuum is eigenstate of $B_0$ with eigenvalue $k^\mu$, this implies 
\begin{equation}
    A^\mu_n|0,k^\mu\>=0 \qquad\qquad \forall~n> 0.
\end{equation} 
Consequently, any excited state must be constructed by acting with negative modes of the oscillators on which in turn implies the excited states can be defined by 
\be{}
A^{\mu_1}_{-n_1}A^{\mu_2}_{-n_2}\hdots A^{\mu_j}_{-n_i}|0,k^\mu\>,\qquad \forall\ n_1,n_2,\hdots, n_i>0.
\ee
Hence, the vacuum must satisfy
\begin{subequations}
\bea{}
A^\mu_n|0,k^\mu\>&=&(C^\mu_n-\C^\mu_{-n})|0,k^\mu\>=0 \quad\forall\ n>0, \\
B^\mu_n|0,k^\mu\>&=&(C^\mu_n+\C^\mu_{-n})|0,k^\mu\>=0 \quad\forall\ n\neq 0,
\eea
\end{subequations}
We can rewrite the above conditions as
\begin{subequations}
    \bea{}
C^\mu_n|0,k^\mu\>&=&\C^\mu_{-n}|0,k^\mu\>=0 \quad\forall\ n>0, \\
C^\mu_n|0,k^\mu\>&=&-\C^\mu_{-n}|0,k^\mu\> \quad\forall\ n<0.
\eea
\end{subequations}
As we shall see later the first condition is the definition of the flipped vacuum. The second relation couples the creation modes of $C$ and $\C$ oscillators. This identification effectively collapses the two independent sets of oscillators into a single one. As a consequence, the realisation of the algebra is destroyed, since the decoupling of the modes is crucial for the closure of the BMS Algebra. This completes our brief examination on the inconsistent constraints.

\section{Central charges of three quantum theories}\label{ApA1}
The central charges arise due to quantum anomalies and normal ordering ambiguities. Their values can be computed using the Jacobi identities, which relate them to the spacetime dimension $D$. In this appendix, we determine the central charges for each of the three quantum null string theories \cite{Bagchi:2020fpr}.
\medskip

We recall the classical BMS$_3$ algebra \eqref{bmsalgebraclassical} given by
\begin{equation}
     [L_m,L_n]=(m-n)L_{m+n}, \qquad [L_m,M_n]=(m-n)M_{m+n}, \qquad [M_m,M_n]=0.
\end{equation}
We now turn to compute the quantum correction to the above algebra. At the quantum level, for $m+n=0$ on the RHS, we encounter normal ordering ambiguity which involve a $c$-number given by 
\begin{equation}\label{centrallyextended}
[L_n,L_{-n}]=2nL_0+A_L(n), \qquad [L_n,M_{-n}]=2nM_0+A_M(n).
\end{equation}
This is the centrally extended BMS$_3$ algebra where $A_L(n)$ and $A_M(n)$ are called the normal ordering ambiguities. From \eqref{centrallyextended}, it can be easily seen that $A_L(-n)=-A_L(n)$ and similarly for $A_M$. From the Jacobi identities, one can obtain a recursion relation to find all $A(n)$,
\be{central1}
A_L(n)=\frac{n(n^2-1)}{6}A_L(2)-f(n)A_L(1),
\end{equation}
where $A_L(1)$ and $A_L(2)$ are unknown constants. This leads to a general form of $A_L(n)$,
\begin{equation}
    A_L(n)=\alpha n^3 + \beta n.
\end{equation}
Similarly for $A_M(n)$. In what follows, we compute the anomaly for each of the three quantum theories. For this, we need to determine $\alpha$ and $\beta$ which can be evaluated by computing expectation values of commutator $[L_n,L_{-n}]$ in vacuum $\ket{0,0^\mu}$ with zero momentum. 
\begin{equation}
\begin{split}
\langle0,0^\mu|[L_1,L_{-1}]|0,0^\mu\rangle=2\<0,0^\mu|L_0|0,0^\mu\>+A_L(1) \\
\langle0,0^\mu|[L_2,L_{-2}]|0,0^\mu\rangle=4\<0,0^\mu|L_0|0,0^\mu\>+A_L(2)
\end{split}
\end{equation}
Similarly for $A_M(n)$, 
\begin{equation}
\begin{split}
\langle0,0^\mu|[L_1,M_{-1}]|0,0^\mu\rangle=2\<0,0^\mu|M_0|0,0^\mu\>+A_M(1)\\
 \langle0,0^\mu|[L_2,M_{-2}]|0,0^\mu\rangle=4\<0,0^\mu|M_0|0,0^\mu\>+A_M(2)
 \end{split}
 \end{equation}
We now turn to find the central charge for Oscillator, Induced and Flipped vacuum.
\medskip

\paragraph{Oscillator vacuum:} 
We begin with the oscillator vacuum $|0,0^\mu\rangle_c$. From \eqref{case3} we have,
\begin{equation} 
{}_c\langle0,0^\mu|L_0|0,0^\mu\rangle_c={}_c\langle0,0^\mu|M_0|0,0^\mu\rangle_c=0. 
\end{equation}
We now compute the following expectation value of the commutators
\begin{subequations}
\begin{align}
{}_c\langle0,0^\mu|[L_n,L_{-n}]|0,0^\mu\rangle_c&={}_c\langle0,0^\mu|L_nL_{-n}|0,0^\mu\rangle_c-{}_c\langle0,0^\mu|L_{-n}L_n|0,0^\mu\rangle_c \cr
&=A_L(n)=0.
\end{align}
\begin{align}
{}_c\langle0,0^\mu|[L_1,M_{-1}]|0,0^\mu\rangle_c&={}_c\langle0,0^\mu|L_1M_{-1}|0,0^\mu\rangle_c-{}_c\langle0,0^\mu|M_{-1}L_1|0,0^\mu\rangle_c \cr 
&=A_M(1)=2{}_c\langle0,0^\mu|c'k^2|0,0^\mu\rangle_c=0.
\end{align}
\begin{align}
{}_c\langle0,0^\mu|[L_2,M_{-2}]&|0,0^\mu\rangle_c={}_c\langle0,0^\mu|L_2M_{-2}|0,0^\mu\rangle_c-{}_c\langle0,0^\mu|M_{-2}L_2|0,0^\mu\rangle_c \cr
&=\frac{1}{4}{}_c\langle0,0^\mu|C_1\cdot C_1 C_{-1}\cdot C_{-1}|0,0^\mu\rangle_c +\frac{1}{4}{}_c\langle0,0^\mu|\C_1\cdot \C_1 \C_{-1}\cdot \C_{-1}|0,0^\mu\rangle_c \cr
&=A_M(2)=D.
\end{align}
\end{subequations}
Putting the values of $A_L(n), A_M(1)$ and $A_M(2)$ in \eqref{centrallyextended}, we obtain the following centrally extended algebra
\begin{equation}
\begin{split}
 [L_p,L_q]&=(p-q)L_{p+q},\\
 [L_p,M_q]&=(p-q)M_{p+q}+ \frac{D}{6}p(p^2-1)\delta_{p+q}, \\
 [M_p,M_q]&=0.
 \end{split}
 \end{equation}
 with central charge $c_L=0$ and $c_M=2D$.
 \medskip
 
\paragraph{Induced vacuum:}\label{A3}  
 Here we consider the Induced vacuum $|0,0^\mu\rangle_I$. The action of $L_n$ and $M_n$ on the vacuum can be read from \eqref{TSQR25} and \eqref{Vishnu Ganesh Pingle} as 
\begin{equation}
{}_I\<0,0^\mu|L_n|0,0^\mu \>_I=c'k\cdot {}_I\<0,0^\mu| A_n|0,0^\mu\>_I=0 =M_n|0,0^\mu\>_I \qquad \forall\ n.
\end{equation}

The expectation value of the commutator yields
\begin{equation}
\begin{split}
{}_I\<0,0^\mu|[L_n,L_{-n}]|0,0^\mu\>_I&=A_L(n)=0,\\
{}_I\<0,0^\mu|[L_n,M_{-n}]|0,0^\mu\>_I&= A_M(n)=0.
\end{split}
\end{equation}
From the expressions of $A_L(n)$ and $A_M(n)$ derived above, we obtain $c_L= c_M=0$. 
\medskip

\paragraph{Flipped vacuum:}\label{A2}
We recall the action of $L_n$ and $M_n$ on the flipped vacuum $|0,0^\mu\>_A$ from \eqref{fphysical} as given below, 
\begin{equation}
L_n|0,0^\mu\>_A=0, \qquad 
M_n|0,0^\mu\>_A=0\qquad\forall\ n\geq 0. 
\end{equation}
We now proceed to compute the expectation value of commutators to find central charge. 
\begin{subequations}
\begin{align}
{}_A\langle0,0^\mu|[L_2,L_{-2}]|0,0^\mu\rangle_A&=\frac{1}{4}{}_A\langle0,0^\mu|C_1\cdot C_1 C_{-1}\cdot C_{-1}|0,0^\mu\rangle_A \cr
&\quad+\frac{1}{4}{}_A\langle0,0^\mu|\tilde{\mathcal{C}}_1\cdot \tilde{\mathcal{C}}_1 \tilde{\mathcal{C}}_{-1}\cdot \tilde{\mathcal{C}}_{-1}|0,0^\mu\rangle_A= A_L(2)=D, \\
{}_A\langle0,0^\mu|[L_2,M_{-2}]|0,0^\mu\rangle_A&=\frac{1}{4}{}_A\langle0,0^\mu|C_1\cdot C_1 C_{-1}\cdot C_{-1}|0,0^\mu\rangle_A \cr
&\quad-\frac{1}{4}{}_A\langle0,0^\mu|\tilde{\mathcal{C}}_1\cdot \tilde{\mathcal{C}}_1 \tilde{\mathcal{C}}_{-1}\cdot \tilde{\mathcal{C}}_{-1}|0,0^\mu\rangle_A= A_M(2)=0.
\end{align}
\end{subequations}
Putting the above values of $A_L(2)$ and $A_M(2)$ in \eqref{centrallyextended}, we obtain the following centrally extended algebra as
\begin{equation}
\begin{split}
[L_p, L_q] &= (p-q) L_{p+q} + \frac{D}{6} p(p^2-1)\delta_{p+q,0}, \\
[L_p, M_q] &= (p-q) M_{p+q}, \\
[M_p, M_q] &= 0.
\end{split}
\end{equation}
with central charge $c_L=2A_L(2)=2D$ and $c_M=2A_M(2)=0$. With this we conclude the discussion of central charges for all the three quantum theories.

\newpage

\bibliographystyle{JHEP}
\bibliography{References}

\end{document}